\documentclass{article}
\usepackage{amsmath}
\usepackage{amsfonts}
\usepackage{amssymb}

\def\dint{\mathop{\displaystyle \int}}
\def\dsum{\mathop{\displaystyle \sum}}
\def\doint{\mathop{\displaystyle \oint}}
\def\diint{\mathop{\displaystyle \iint}}
\def\dprod{\mathop{\displaystyle \prod }}

\newcommand{\apsection}[2]{%
 \renewcommand{\thesection}{\Alph{section}}%
 \refstepcounter{section}
 \section*{\large Appendix \Alph{section}. #1}\label{#2}%
 \addcontentsline{toc}{section}{Appendix \Alph{section}. #1}
 \setcounter{equation}{0}\setcounter{mtheorem}{1}\setcounter{mremark}{1}\setcounter{mdefinition}{1}\setcounter{mexample}{1}\setcounter{mexercise}{1}}

\def \R {{\mathbb R}}
\def \C {{\mathbb C}}
\def \Z {{\mathbb Z}}
\def \N {{\mathbb N}}
\def \Sr {{\mathbb S}}
\def \M {\overline{M}}
\def \E {\mathbb{E}}
\def \H {\mathbb{H}}
\def \PP {\mathbb{P}}
\def \spr {\cdot}
\def \di {\partial}

\newcommand{\tdd}[1]{\Txfrac{\partial}{\partial #1}}

\def \dlt {\mathcal{D}}
\newcommand{\OP}[1]{\mathcal{O}_{\hspace{-1pt}#1}}
\newcommand{\DOP}[1]{\mathfrak{D}_{\hspace{-1pt}#1}}
\def \dirz {z \hspace{-5pt} \text{\small $/$}}

\def \diru {u \hspace{-5pt} \text{\small $/$}}
\def \dirv {v \hspace{-5pt} \text{\small $/$}}
\def \dirdi {\di \hspace{-5.5pt} /}
\def \vbeta {\beta \hspace{-4.5pt} \raisebox{8pt}{\tiny $\vee$}}
\def \vgamma {\gamma \hspace{-4.5pt} \raisebox{6pt}{\tiny $\vee$}}
\DeclareSymbolFont{ltrs}     {OT1}{pzc}{m}{it}
\DeclareSymbolFont{ltrsa}     {OMS}{cmsy}{m}{n}
\DeclareSymbolFont{ltrsb}     {OMS}{cmsy}{b}{n}
\DeclareMathSymbol{\Ss}{\mathord}{ltrsa}{"53}
\DeclareMathSymbol{\confgr}{\mathord}{ltrsa}{"43}
\DeclareMathSymbol{\compgr}{\mathord}{ltrsa}{"4B}
\DeclareMathSymbol{\compgr}{\mathord}{ltrsa}{"4B}
\DeclareMathSymbol{\cnfalg}{\mathord}{ltrs}{"63}
\def \confalg {\text{\large $\cnfalg$}}
\def \cconfalg {\confalg_{\text{\tiny $\C$}}}
\def \VA {{\mathcal V}}
\def \Alg {{\mathcal A}}
\def \hh {\mathfrak{h}}
\def \hhc {\mathfrak{h}_{\C}}
\def \hhh {\widehat{\mathfrak{h}}}
\def \hhhc {\widehat{\mathfrak{h}}_{\C}}
\def \LAT {Q}
\def \DIM {\mathit{dim}}
\def \Spin {\mathit{Spin}}
\def \spin {\mathit{spin}}
\def \Cliff {\mathit{Cliff}}
\def \diag {\mathit{diag}}
\def \SO {\mathit{SO}}
\def \SU {\mathit{SU}}

\def \MOD {\mathit{mod}}
\def \ord {\mathit{ord}}
\def \SL {\mathit{SL}}

\def \RE {\mathit{Re}}
\def \IM {\mathit{Im}}
\def \Span {\mathit{Span}}
\def \TR {\mathit{tr}}
\def \cotg {\text{\rm cotg}}
\def \hcom {\mathfrak{H}}
\newcommand{\cnj}[1]{{#1}^{\hspace{1pt} +}}

\def \sphi {{\phi\raisebox{7.5pt}{\hspace{-4pt}\scriptsize $*$}}}

\def \spsi {{\psi\raisebox{7.5pt}{\hspace{-4pt}\scriptsize $*$}}}
\def \sPsi {{\Psi\raisebox{8pt}{\hspace{-6pt}\scriptsize $*$}\hspace{1pt}}}
\def \bsPsi {{\Psi\raisebox{8pt}{\hspace{-8pt}{\tiny (}\hspace{-0.5pt}\scriptsize $*$\hspace{-0.5pt}{\tiny )}}\hspace{0pt}}}
\def \trn {t}
\def \mtrx {{\raisebox{-3pt}{\hspace{-2.5pt}}}}
\def \MINK {M}
\def \wfun {\wp}
\def \pfun {p}

\def \zfun {\mathfrak{Z}}
\def \ID {\mathbb{I}}
\def \W {\mathcal{W}}
\def \Rw {\mathfrak{w}}
\def \czeta {\zeta}
\DeclareMathSymbol{\mmzeta}{\mathord}{ltrsa}{"5A}
\DeclareMathSymbol{\bmzeta}{\mathbf}{ltrsb}{"5A}
\def \mzeta {\text{\scriptsize $\mmzeta$}}
\def \bzeta {\text{\scriptsize $\bmzeta$}}
\newcommand{\VEC}[1]{{\mathop{#1}\limits^{\to}}}
\def \mxi {\xi}
\def \xx {\VEC{\mxi}}
\def \ee {\VEC{e}}
\def \cmu {\alpha}
\def \cnu {\beta}

\def \aa {A}
\def \bb {B}

\def \DOM {\mathcal{V}}
\def \SCT {C}
\def \mgrt {\gg}
\DeclareMathSymbol{\Vol}{\mathord}{ltrs}{"56}
\newcommand{\VOL}[1]{\Vol_{\hspace{-0.5pt}#1}}
\def \MED {\text{\small $\mathcal{E}$}}

\def \Su {\mathop{\sum}\limits}

\def \lvac {\left\langle 0 \! \left| \right. \right.  \!}
\def \rvac {\!\! \left. \left. \right| \! 0 \right\rangle}

\def \ralpha {\!\! \left. \left. \right| \! \alpha \right\rangle}

\def \rbeta {\!\! \left. \left. \right| \! \beta \right\rangle}

\newcommand{\rv}[1]{\!\! \left. \left. \right| \! {#1} \right\rangle}
\def \La {\left\langle \!\!{\,}^{\mathop{}\limits_{}}_{\mathop{}\limits^{}}\right.}
\def \Ra {\left. \!\!{\,}^{\mathop{}\limits_{}}_{\mathop{}\limits^{}}\right\rangle}
\def \Vl {\left. \!\!{\,}^{\mathop{}\limits_{}}_{\mathop{}\limits^{}} \right|}
\def \la {\left( \!\!{\,}^{\mathop{}\limits_{}}_{\mathop{}\limits^{}}\right.\hspace{-1.5pt}}
\def \ra {\hspace{-1.5pt}\left. \!\!{\,}^{\mathop{}\limits_{}}_{\mathop{}\limits^{}}\right)}

\newcommand{\BINOMIAL}[2]{{#1 \choose #2}}
\newcommand{\gvspc}[1]{\raisebox{#1}{$\,$}$\!$}
\newcommand{\mgvspc}[1]{\raisebox{#1}{$\,$}\!}

\newcounter{tmpc}

\newenvironment{plist}{%
\setcounter{tmpc}{0}
\begin{list}{{\rm (\alph{tmpc})}}{\usecounter{tmpc}
\setlength{\leftmargin}{16pt}
\setlength{\rightmargin}{0cm}
\setlength{\itemsep}{2.5pt}
\setlength{\topsep}{5pt}
\setlength{\labelsep}{2pt}
\setlength{\labelwidth}{13pt}
\setlength{\listparindent}{18pt}}
}{\end{list}}

\newcounter{mtheorem}
\newcounter{mremark}
\newcounter{mdefinition}
\newcounter{mexample}
\newcounter{mexercise}

\newenvironment{mtheorem}[1][\bf Theorem \thesection.\arabic{mtheorem}]{%
	\medskip\noindent\textbf{#1.}\it ${}$ ${}$}{\addtocounter{mtheorem}{1}\medskip}
\newenvironment{mproposition}[1][\bf Proposition \thesection.\arabic{mtheorem}]{%
	\medskip\noindent\textbf{#1.}\it ${}$ ${}$}{\addtocounter{mtheorem}{1}\medskip}

\newenvironment{mcorollary}[1][\bf Corollary \thesection.\arabic{mtheorem}]{%
	\medskip\noindent\textbf{#1.}\it ${}$ ${}$}{\addtocounter{mtheorem}{1}\medskip}
\newenvironment{mremark}[1][\it Remark \thesection.\arabic{mremark}]{%
	\medskip\noindent\textbf{#1.} }{\addtocounter{mremark}{1}\medskip}

\newenvironment{mexercise}[1][\bf Exercise \thesection.\arabic{mexercise}]{%
	\medskip\noindent\textbf{#1.} }{\addtocounter{mexercise}{1}\medskip}

\newenvironment{proof}[1][\rm\it Proof]{%
	\noindent\textbf{#1.} }{\ $\Box$

\medskip

}
\newenvironment{Skproof}[1][\rm\it Sketch of the proof]{%
	\noindent\textbf{#1.} }{\ $\Box$

\medskip

}

\newcommand{\prlabel}[1]{\label{#1}}
\newcommand{\rmlabel}[1]{\label{#1}}

\newcommand{\beq}{\begin{equation}}
\newcommand{\eeq}{\end{equation}}
\newcommand{\beqa}{\begin{eqnarray}}
\newcommand{\eeqa}{\end{eqnarray}}
\newcommand{\nn}{\nonumber \\}

\newcommand{\mbf}[1]{\ensuremath{\mathchoice
                    {\mbox{\boldmath$\displaystyle\mathbf{\mathit{#1}}$}}
                    {\mbox{\boldmath$\textstyle\mathbf{\mathit{#1}}$}}
                    {\mbox{\boldmath$\scriptstyle\mathbf{\mathit{#1}}$}}
                    {\mbox{\boldmath$\scriptscriptstyle\mathbf{\mathit{#1}}$}}}}
\newcommand{\Mbf}[1]{\ensuremath{\mathchoice
                    {\mbox{\boldmath$\displaystyle\mathbf{#1}$}}
                    {\mbox{\boldmath$\textstyle\mathbf{#1}$}}
                    {\mbox{\boldmath$\scriptstyle\mathbf{#1}$}}
                    {\mbox{\boldmath$\scriptscriptstyle\mathbf{#1}$}}}}

\DeclareSymbolFont{stmry}{U}{stmry}{m}{n}
\DeclareMathDelimiter\llbracket{\mathopen}{stmry}{"4A}
					  {stmry}{"71}
\DeclareMathDelimiter\rrbracket{\mathclose}{stmry}{"4B}
					   {stmry}{"79}
\newcommand{\Dbrackets}[1]{\left\llbracket{#1}\right\rrbracket}

\newcommand{\txfrac}[2]{\frac{\raisebox{1pt}{$#1$}}{\raisebox{-3pt}{$#2$}}}
\newcommand{\Txfrac}[2]{\frac{\raisebox{2pt}{$#1$}}{\raisebox{-5pt}{$#2$}}}

\newcommand{\nfrc}[2]{%
\text{\raisebox{1pt}{\(#1\)}\hspace{0pt}$/$\hspace{-1.5pt}\raisebox{-4pt}{\(#2\)}}}
\newcommand{\INTG}[2]{\mathop{\dint}\limits_{\hspace{-6pt}{#1}}^{\hspace{6pt}{#2}}}

\newcommand{\Brkts}[2]{\left\{\raisebox{#1}{\hspace{-2pt}}\right. {#2}
\left.\raisebox{#1}{\hspace{-2pt}}\right\}}

\newcommand{\rcoset}[2]{%
\raisebox{-5pt}{$#1$}\raisebox{-2pt}{$\backslash$}\raisebox{0pt}{$#2$}}

\def \podr {\hspace{-15pt}}
\def \ppodr {\hspace{-7pt}}

\def \spc {\hspace{1pt}}

\title{{\bf\normalsize
LECTURES \spc ON \spc ELLIPTIC \spc FUNCTIONS \spc AND \spc MODULAR \spc FORMS \spc
IN \spc CONFORMAL \spc FIELD \spc THEORY}}
\author{\normalsize Nikolay M. Nikolov$^{\hspace{1pt}1,a}$ \ \ and \ \
Ivan T. Todorov$^{\hspace{1pt}1,2,b}$}
\date{\begin{itemize}
\item[\small $^{1}$]
{\small Institute for Nuclear Research and Nuclear Energy,
Tsarigradsko Chaussee 72, BG-1784 Sofia, Bulgaria}
\item[\small $^{2}$]
{\small Institut f\"ur Theoretische Physik, Universit\"at G\"ottingen,
Friedrich-Hund-Platz~1,
\linebreak
D--37077 G\"ottingen, Germany}
\end{itemize}
{\small \today}}

\begin{document}

\maketitle

\renewcommand{\thefootnote}{\alph{footnote}\hspace{2pt}}
\footnotetext[1]{e.mail: mitov@inrne.bas.bg}
\footnotetext[2]{e.mail: todorov@inrne.bas.bg,
itodorov@theorie.physik.uni-goettingen.de}
\renewcommand{\thefootnote}{\arabic{footnote}}

\begin{abstract}
A concise review of the notions of elliptic functions,
modular forms, and $\vartheta$--functions is provided,
devoting most of the paper to applications to
Conformal Field Theory (CFT), introduced within the
axiomatic framework of quantum field theory.
Many features, believed to be peculiar to chiral 2D ($=$ two dimensional)
CFT, are shown to have a counterpart in any (even dimensional) globally conformal
invariant quantum field theory.
The treatment is based on a recently introduced higher
dimensional extension of the concept of vertex algebra.
\end{abstract}

\newpage

{\small
\tableofcontents}

\newpage

\section{Introduction}
\setcounter{equation}{0}\setcounter{mtheorem}{1}\setcounter{mremark}{1}\setcounter{mdefinition}{1}\setcounter{mexample}{1}\setcounter{mexercise}{1}

Arguably, the most attractive part of Conformal Field Theory (CFT) is that
involving elliptic functions and modular forms. Modular inversion, the
involutive $S$--transformation of the upper half--plane
\beq
S \, = \, \left(\hspace{-4pt} \begin{array}{rr} 0 \hspace{-4pt} & -1 \\
1 \hspace{-4pt} & 0 \end{array} \hspace{-3pt} \right) \in \SL (2,\Z)
; \quad
\tau \rightarrow \allowbreak -\frac{1}{\tau}
\quad
(\IM \,
\tau >0)
\, , \  \label{1.1}
\eeq
relates high and low temperature behaviour, thus providing the oldest and
best studied\footnote{%
For a physicist oriented review of modular inversion~--~see \cite{DK02}.}
example of a duality transformation \cite{KW41}.

The aim of these lectures is threefold:
\begin{plist}
\item[(1)] To offer a brief introduction to the mathematical background,
including a survey of the notions of elliptic functions, elliptic curve (and
its moduli), modular forms and $\vartheta$--functions.
(The abundant footnotes are designed to provide some historical background.)
\item[(2)] To give a concise survey of axiomatic CFT in higher (even)
dimensions with an emphasis on the vertex algebra approach developed in~\cite{N03}
\cite{NT03}.
\item[(3)] To give an argument indicating
that \textit{finite temperature correlation functions} in a
globally conformal invariant (GCI) quantum field theory in any even number
of space--time dimensions are (doubly periodic) \textit{elliptic functions}
and to study the modular properties of the corresponding temperature
mean values of the conformal Hamiltonian.
\end{plist}

Two-dimensional (2D) CFT models provide basic known examples in which
the chiral energy average in a given superselection sector is a modular from of
weight 2. In a rational CFT these energy mean values span a finite
dimensional representation of $\SL (2,\Z)$.
We demonstrate that modular transformation properties can also be
used to derive high temperature asymptotics of thermal energy densities
in a 4--dimensional CFT.

We include in the bibliography some selected texts on the mathematical
background briefly annotated in our (half page long)
``Guide to references'' at the end of the lectures.
(Concerning modular forms we follow the notation of Don Zagier in \cite{FNTP}.)
A detailed exposition of the authors' original results can be found in~\cite{NT03}.

\section{Elliptic functions and curves}
\setcounter{equation}{0}\setcounter{mtheorem}{1}\setcounter{mremark}{1}\setcounter{mdefinition}{1}\setcounter{mexample}{1}\setcounter{mexercise}{1}

The theory of elliptic functions has been a centre of attention of the 19th
and the early 20th century mathematics (since the discovery of the double
periodicity by N. H. Abel in 1826 until the work of Hecke\footnote{%
Erich Hecke (1887--1947) was awarded his doctorate under David Hilbert (1862--1943)
in 1910 in G\"ottingen for a dissertation on modular forms and their application
to number theory.}
and Hurwitz's\footnote{%
Adolf Hurwitz (1859-1919).}
book \cite{HC} in the 1920's~--~see \cite{K26} for an engaging
historical survey). This is followed by a period of relative dormancy
when E. Wigner ventured to say that it is ``falling into oblivion''\footnote{%
E.~Wigner, \textit{The limits of science}, Proc.~Amer.~Phil.~Soc.~\textbf{94}
(1950) 422;
see also his collection of Scientific Essays,
\textit{Symmetries and Reflections} p.~219
(Eugene Paul Wigner, 1902--1995, Nobel Prize in physics, 1963).}.
(Even today physics students rarely get to learn this chapter of mathematics
during their undergraduate years.) The topic experiences a renaissance in
the early 1970's, which continues to these days (see the guide to the
literature until 1989 by D. Zagier in \cite{FNTP} pp. 288--291). The
proceedings \cite{FNTP} of the 1989 Les Houches Conference on Number
Theory and Physics provide an excellent shortcut into the subject and
further references.
The subject continues to be a focus of mathematical physicists' attention
(for a recent application to noncommutative
geometry~--~see~\cite{CM03} \cite{CD03}).

\subsection{Elliptic integrals and functions}

If we did not know about trigonometric functions when first calculating
the integral
\(z=\INTG{0}{x}
\Txfrac{dt}{\sqrt{1-t^{2}}}\),
we would have come out
with a rather nasty multivalued function $z(x)$. Then an unprejudiced young
man might have discovered that one should instead work with the inverse
function $x(z)=\sin z$, which is a nice \textit{single valued entire
periodic function.} This is more or less what happened for elliptic
integrals\footnote{%
After nearly 200 years of study of elliptic integrals, starting with the
17th century work of John Wallis (1616--1703) and going through the
entire 18th century with contributions from Leonard Euler (1707--1783) and
Adrien--Marie Legendre (1752--1833), a 23--years old Norwegian, the
pastor's son, Niels Henrik Abel (1802--1829) had the bright idea to look
at the inverse function
and prove that it is
single valued,
meromorphic and doubly periodic.
As it often happens with 19th century
discoveries, Carl Friedrich Gauss (1777--1855) also had developed this idea
in his notebooks~--~back in 1798~--~on the example of the
lemniscate (see \cite{McKM} Sects.~2.3 and 2.5).},
say, for an integral of the type
\beq
z=
\INTG{x}{\infty}
\frac{1}{\sqrt{4\xi^{3}-g_{2}\xi-g_{3}}} \
d\xi
\, , \quad
g_2^3 - 27 g_3^2 \neq 0
. \,  \label{2.1}
\eeq
The inverse function
\(x=x(z)\)
can be written
in the Weierstrass' notation\footnote{%
Karl Theodor Wilhelm Weierstrass (1815--1897);
the $\wfun$-function appeared in his Berlin lectures in 1862.
Series of the type (\ref{2.2}) were, in fact, introduced by another young
deceased mathematician
(one of the precious few appreciated by Gauss~--~whom
he visited in G\"ottingen in 1844)
Ferdinand Gotthold Eisenstein (1823--1852)~--~see \cite{Weil}.}
as a
manifestly
\textit{meromorphic} (\textit{single valued}),
\textit{doubly periodic function}
(see Exercise~2.2~(\textit{a})):
\begin{equation}
x(z) \, = \,
\wfun (z;\, \omega_1,\, \omega_2) \
( \, = \, \wfun (z) \hspace{1pt} ) \, = \,
\frac{1}{z^{2}} +
\dsum\limits_{\omega \, \in \, \Lambda \backslash \{0\}}
\left( \frac{1}{(z+\omega)^{2}}-\frac{1}{\omega^{2}} \right)
\, , \
\label{2.2}
\end{equation}
where
$\Lambda $ is the 2--dimensional lattice of periods,
\begin{eqnarray}
&
\Lambda =
\Brkts{12pt}{
\omega =m\omega_{1}+n\omega_{2}
\, :\
m,n \in \Z,\
\IM \, \frac{\omega_1}{\omega_2}>0
}
\, , \
&
\nn &
g_{2}=60 \dsum\limits_{\omega \, \in \, \Lambda \backslash \{0\}}
\omega^{-4}
\, , \quad
g_{3}=140 \dsum\limits_{\omega \, \in \, \Lambda \backslash \{0\}}
\omega^{-6}.
&
\label{2.3}
\end{eqnarray}
Indeed, knowing the final answer~(\ref{2.2}) it is easy to chek
that $x(z)$ satisfies a
first order differential equation
(Exercise~2.2~(\textit{b})):
\beq
y^{2} \, = \, 4x^{3}-g_{2}x-g_{3} \quad \text{for} \quad x \, = \, \wfun (z)
\, , \quad
y \, = \, \wfun' \left( z\right)
\, . \  \label{2.4}
\eeq
This is the counterpart of the equation
\(y^{2}=1-x^{2}\) for \(x(z)=\sin (z)\).
The condition that the third degree polynomial $y^{2}$ (\ref{2.4}) has no
multiple zero can be expressed by the nonvanishing of the discriminant,
proportional to \(g_{2}^{3}-27g_{3}^{2}\)
(in the case of coinciding roots, the integral~(\ref{2.1}) reduces
to a trigonometric one).

\begin{mremark}\label{nrem2.1}
More generally, elliptic integrals are integrals over \textit{rational
functions} $R(x,y)$, when $y^{2}$ is a \textit{third or a fourth degree
polynomial} in $x$ with different roots. A \textit{fourth--degree curve},
\(
\widetilde{y}^{2}=a_{0}\widetilde{x}^{4}+a_{1}\widetilde{x}^{3}+a_{2}
\widetilde{x}^{2}+a_{3}\widetilde{x}+a_{4}
\)
can be brought to the Weierstrass canonical form (\ref{2.2}) by what may be
called a ``\textit{M\"obius}\footnote{%
Augustus Ferdinand M\"obius (1790--1868).}
\textit{phase space transformation}'':
\(
\widetilde{x} = \txfrac{ax+b}{cx+d}
\),
\(
\widetilde{y} = \txfrac{A}{(cx+d)^{2}} \, y
\),
\(
ad-bc\neq 0 \neq A
\)
i.~e., if $y$ transforms as a derivative
(with a possible dilation of the independent variable $z$
in accord with the
realization (\ref{2.2})). We have, in particular, to equate
$\txfrac{a}{c}$\textit{\
}to one of the zeroes of the polynomial $\widetilde{y}^{2}(\widetilde{x})$,
thus killing the coefficient of $x^{4}$
(see Exercise~2.3).
An example of such type of integrals is the \textit{Jacobi's}\footnote{%
Carl Gustav Jacob Jacobi (1804--1851) rediscovers in
1828 the elliptic functions (by inverting the elliptic integrals)
and is the first to apply them to
number theory.
Jacobi, himself, says that the theory of elliptic functions was
born when Euler presented to the Brelin Academy (in January 1752) the
first series of papers, eventually proving the addition and multiplication
theorems for elliptic integrals (see \cite{Weil84}).
Bourbaki (in particular, Jean Dieudonn\'e) have taken as a motto his
words (from a letter to Legendre of 1830,
deploring the worries of Fourier about applications):
``le but unique de la science, c'est l'honneur de l'esprit humain.''}
``\textit{sinus amplitudinus}'',
\beq\label{Jacobi}
x \, = \, \mathrm{sn} \left( z, k^2 \right)
\, , \quad
z \, := \,
\INTG{1}{x}
\frac{d\xi}{\sqrt{\left( \xi^2-1 \right) \left( 1 - k^2 \xi^2 \right)}}
\, , \quad k^2 \, \neq \, 0,1
\eeq
which is proven to be a doubly periodic meromorphic function\footnote{%
It is not difficult to show that the solution of the Newton equation
of a length $L$ and mass $m$ \textit{pendulum},
$m \frac{d^2\theta}{dt^2}$ $+$ $m \frac{G}{L} \sin \theta$ $=$ $0$
($G$ being the Earth gravitational acceleration),
is expressed in terms of the elliptic sinus~(\ref{Jacobi})~--~see
\cite{McKM} Sect.~2.1 Example~4 and p.~77.}
with periods $4K$ and $2iK'$, where
\(
K \! := \! \INTG{0}{1} \!
\Txfrac{dx}{\sqrt{\left( 1-x^2 \right) \left( 1-k^2 x^2 \right)}}
\, ( \, = \txfrac{\pi}{2} \, F(\txfrac{1}{2},\txfrac{1}{2};1;k^2)\),
$F$ being the hypergeometric function)
and
\(
K' \! := \! \INTG{1}{\frac{1}{k}} \!
\Txfrac{dx}{\sqrt{\left( x^2-1 \right) \left( 1-k^2 x^2 \right)}}
\)
(see, e.g.~\cite{McKM} Sects.~2.1 and~2.5;
concerning the other Jacobi functions,
$\mathrm{cn} (z, k^2)$ $= \sqrt{1-\mathrm{sn}(z,k^2)}$ and
$\mathrm{dn} (z,k^2)$ $= \sqrt{1-k^2\mathrm{sn}(z,k^2)}$ see Sect.~2.16
of~\cite{McKM}).
\end{mremark}

We proceed to displaying some simple properties of elliptic functions,
defined as \textit{doubly periodic meromorphic functions on the
complex plane}.
Basic facts of complex analysis, such as Liouville's and Cauchy's\footnote{
Augustin--Louis Cauchy (1789--1857); Joseph Liouville (1809--1882).}
theorems, allow one to establish far reaching non--obvious results
in the study of elliptic functions.
\begin{plist}
\item[(1)]
Periodicity implies that an elliptic function $f(z)$ is
determined by its values in a basic parallelogram, called a
\textit{fundamental domain:}
\begin{equation*}
F=\left\{ \alpha\,\omega_{1} + \beta\,\omega_{2}
\, ; \
0 \leqslant \alpha
, \,
\beta < 1\right\}
\, . \
\end{equation*}
\item[(2)]
\textit{If $f$ is bounded in $F$, then it is a constant}. Indeed,
periodicity would imply that $f$ is bounded on the whole complex plane. The
statement then follows from Liouville's
theorem.
Thus a non--constant elliptic function must have a pole in $F.$
\item[(3)]
\textit{The sum of the residues of the simple poles of $f$ in $F$ is zero}.
This follows from Cauchy's theorem, since the integral over the boundary
$\partial F $ of $F$ vanishes:
\begin{equation*}
\doint\limits_{\partial F} f(z) \ dz \, = \, 0
\, , \
\end{equation*}
as a consequence of the periodicity. (By shifting, if necessary, the boundary on
opposite sides we can assume that $f$ has no poles on $\partial F$.) It
follows that $f$ has at least \ $2$ poles $\ $in\ $F$ (counting
multiplicities).
\item[(4)]
\textit{Let $\left\{ a_i \right\}$ be the zeroes and poles of $f$ in $F$
and $n_i$ be the multiplicity of \(a_i\)
(\(n_i >0 \) if $a_i$ is a zero, \(n_i < 0\) if $a_i$ is a pole).
Then applying (3) to
$\Txfrac{f'\left( z \right)}{f \left( z \right)}$
gives \(\sum n_i = 0\)}.
\end{plist}
More properties of zeroes and poles of an elliptic function in a
fundamental domain are contained in Theorem (1.1.2) of Cohen in
\cite{FNTP}, p.~213 (see also \cite{McKM} Sect.~2.7). The above list
allows to write down the general
form of an elliptic function $f \! \left( z \right)$.
If the singular part of $f \! \left( z \right)$ in $F$ has the form:
\beq\label{f_bas0}
\Su_{k \, = \, 1}^{K} \Su_{s \, = \, 1}^{S_k} \, N_{k,s} \
\frac{1}{\left( z-z_s \right)^k}
\, \
\eeq
for some \(K,S_1,\dots,S_K \in \N\), \(N,N_{s,k} \in \C\), \(z_s \in F\)
(\(k = 1,\dots, K\), \(s=1,\dots,S_k\)),
then $f \! \left( z \right)$ can be represented in a finite sum:
\beq\label{f_bas}
f \! \left( z \right) = N \, + \,
\Su_{k \, = \, 1}^{K} \Su_{s \, = \, 1}^{S_k} \, N_{k,s} \
\pfun_k \! \left( z-z_s;\, \omega_1,\, \omega_2 \right)
\, \
\eeq
where $\pfun_k \! \left( z; \omega_1,\, \omega_2 \right)$
are\footnote{%
These are the ``basic elliptic functions'' of Eisenstein
according to Andr\'{e} Weil (1906--1998)
who denotes them by $E_k$~--~see \cite{Weil} Chapter III.},
roughly speaking, equal to:
\beq\label{p_rough}
\pfun_k \! \left( z;\, \omega_1,\, \omega_2 \right) \, := \,
\Su_{\omega \, \in \, \Lambda}
\frac{1}{\left( z+\omega \right)^k}
\, . \
\eeq
The series~(\ref{p_rough}) are absolutely convergent for \(k \geqslant 3\)
and \(z \notin \Lambda\), and
\beq\label{p_k}
\pfun_{k+1} \! \left( z;\, \omega_1,\, \omega_2 \right)
\, = \, - \frac{1}{k}
\left( \di_{z} \pfun_k \right) \left( z;\, \omega_1,\, \omega_2 \right)
\ \quad (\, \di_z \, := \, \frac{\di}{\di z} \, )
\, . \
\eeq
For \(k = 1,2\) one should specialize the order of summation or, alternatively,
add regularizing terms.
Such a regularization for the \(k=2\) case
has been used in fact in the definition
of the Weierstrass' $\wfun$ function~(\ref{2.2});
for \(k=1\), the function
\beq\label{zfun}
\zfun \hspace{-1pt} \left( z;\, \omega_1,\, \omega_2 \right)
\, = \,
\frac{1}{z}
+
\dsum\limits_{\omega \, \in \, \Lambda \backslash \{0\}}
\left(
\frac{1}{z + \omega}
-\frac{1}{\omega}+\frac{z}{\omega^{2}}
\right)
\, , \quad
\eeq
is known as Weierstrass' $\zfun$ function.
Note that the $\zfun$--function~(\ref{zfun}) is not elliptic
(due to the above property~(2))
but any linear combination
\(\mathop{\sum}\limits_{s \, = \, 1}^{S} N_{1,s}\) \(\zfun ( z-\) \(z_s;\)
\(\omega_1,\) \(\omega_2)\)
with
\(\Su_{s \, = \, 1}^{S} N_{1,s} = 0\) will be elliptic.
This follows from the translation property~\cite{L87}
\begin{eqnarray}\label{fn_cn2}
\zfun \hspace{-1pt} \left( z+\omega_1;\, \omega_1,\, \omega_2 \right)
\, = && \podr
\zfun \hspace{-1pt} \left( z;\, \omega_1,\, \omega_2 \right)
- 8 \hspace{1pt} \pi^2 \hspace{1pt} G_2 \! \left( \omega_1,\, \omega_2 \right)
\hspace{1pt} \omega_1
- 2 \pi i
\, , \quad
\\ \label{fn_cn3}
\zfun \hspace{-1pt} \left( z+\omega_2;\, \omega_1,\, \omega_2 \right)
\, = && \podr
\zfun \hspace{-1pt} \left( z;\, \omega_1,\, \omega_2 \right)
- 8 \hspace{1pt} \pi^2 \hspace{1pt} G_2 \! \left( \omega_1,\, \omega_2 \right)
\, , \quad
\end{eqnarray}
where
\beq\label{G_2}
- 8\pi^{2}
G_{2} \! \left( \omega_1,\, \omega_2 \right)
\, = \,
\mathop{\sum}\limits_{n \, \in \, \Z \backslash \left\{ 0 \right\}}
\frac{1}{\left( n\,\omega_2 \right)^{2}} \, +
\mathop{\sum}\limits_{m \, \in \, \Z \backslash \left\{ 0 \right\}} \!\!
\left(\hspace{1pt}
\mathop{\sum}\limits_{n \, \in \, \Z}
\frac{1}{\left( m\,\omega_1 + n\,\omega_2 \right)^{2}}
\right)
\eeq
will be considered in more details in Sect.~3.2.

\begin{mexercise}\label{nwe2.1}
Prove the absolute convergence of the series~(\ref{G_2})
using the \textit{Euler's formulae}
\beq\label{Euler}
\mathop{\lim}\limits_{N \to \infty} \Su_{n \, = \, -N}^{N}
\frac{1}{z+n} \, = \, \pi \, \cotg \, \pi z
\, , \quad
\mathop{\lim}\limits_{N \to \infty} \Su_{n \, = \, -N}^{N}
\frac{\left( -1 \right)^n}{z+n} \, = \, \frac{\pi}{\sin \pi z}
\, \
\eeq
(with a subsequent differentiation).
\end{mexercise}

It is convenient to single out a class of elliptic functions
$f(z;\omega_{1},\omega_{2})$,
which are \textit{homogeneous} in the sense that
$\rho^{k}$
$f(\rho z;\rho \omega _{1},\rho \omega_{2})=$
$f(z;\omega_{1},\omega_{2})$ for $\rho \neq 0$
and some $k=1,2,\dots$.
The Weierstrass function (\ref{2.2}) provides an example of a homogeneous
function of degree $2$.
In the applications to GCI QFT a natural system of
basic elliptic functions is
\beq\label{2.5nnn}
\pfun^{\kappa\lambda}_{k}(z;\, \omega _{1},\, \omega _{2})
\, = \,
\mathop{\lim}\limits_{M \to \infty}
\mathop{\sum}\limits_{m = -M}^{M} \,
\mathop{\lim}\limits_{N \to \infty}
\mathop{\sum}\limits_{n = -N}^{N} \,
\frac{(-1)^{\kappa m+\lambda n}}{\left( z+m\,\omega_{1}+n\,\omega_{2} \right)^k}
, \quad
\kappa,\lambda = 0, 1
\eeq
(cp. with Eq.~(\ref{Euler})),
characterized by the (anti)periodicity condition
\begin{eqnarray}\label{tans1}
\pfun^{\kappa\lambda}_{k}(z+\omega_1;\, \omega _{1},\, \omega _{2})
\, = && \hspace{-15pt}
\left( -1 \right)^{\kappa} \,
\pfun^{\kappa\lambda}_{k}(z;\, \omega _{1},\, \omega _{2})
\quad \text{for} \quad
k + \kappa + \lambda > 0
\, , \quad
\\ \label{tans2}
\pfun^{\kappa\lambda}_{k}(z+\omega_2;\, \omega _{1},\, \omega _{2})
\, = && \hspace{-15pt}
\left( -1 \right)^{\lambda} \,
\pfun ^{\kappa\lambda}_{k}(z;\omega _{1},\omega _{2})
\quad \text{for} \quad
\kappa,\, \lambda = 0,\, 1
\, . \quad
\end{eqnarray}
The functions $p_k$ are encountered in a family of examples,
described in Sect.~4.4;
$p_1^{\kappa\lambda}$ with \(\kappa+\lambda>0\) appear in the study of Gibbs
states of a chiral 2D Weyl field (Sect.~5.3);
the thermal $2$--point function of a free massless scalar field
in $4$--dimensions is presented as a difference of two $p_1^{00}$
functions~--~see~(\ref{6.11}).
The functions~(\ref{2.5nnn}) are connected for different $k$ by:
\beq\label{k_ind}
\pfun^{\kappa\lambda}_{k+1} \! \left( z;\, \omega_1,\, \omega_2 \right)
\, = \, - \frac{1}{k}
\left( \di_{z} \pfun^{\kappa\lambda}_k \right)
\left( z;\, \omega_1,\, \omega_2 \right)
\eeq
and we will set
\beq\label{pfun_1}
\pfun_k (z;\, \omega _{1},\, \omega _{2}) \, \equiv \,
\pfun_k^{00} (z;\, \omega _{1},\, \omega _{2})
\, . \quad
\eeq

\begin{mexercise}\label{exr:1}
(\textit{a})
Prove that
$\wfun \hspace{-1pt} \left( z;\, \omega_1,\, \omega_2 \right)$~(\ref{2.2})
is doubly periodic in $z$ with periods $\omega_1$ and $\omega_2$.
\textit{Hint}: prove that the derivative $-\txfrac{1}{2}\,\di_z \wfun$
is the elliptic function $\hspace{1pt}p_3$~(\ref{p_rough}) so that
$\wfun (z+\omega)$ $-$ $\wfun (z)$ is a constant for \(\omega \in \Lambda\);
show that the constant is zero by taking $z$ $=$ $-\txfrac{\omega}{2}$.

\vspace{0.05in}

\noindent
(\textit{b})
Prove that
$\wfun \hspace{-1pt} \left( z;\, \omega_1,\, \omega_2 \right)$~(\ref{2.2})
satisfy the equation~(\ref{2.4}).
\textit{Hint}: prove that the difference between the two sides of Eq.~(\ref{2.4})
is an entire elliptic function vanishing at \(z=0\).

\vspace{0.05in}

\noindent
(\textit{c})
Prove the relations
\vspace{-18pt}

\beqa
\label{eqnA.22a}
& \hspace{-52pt}
\pfun^{10}_{k} \! \left( z;\, \omega_{1},\, \omega_{2} \right) \hspace{-7pt} & = \,
2\hspace{1pt}\pfun_k \! \left( z;\, \omega_{1},\, 2\omega_{2} \right) -
\pfun_k \! \left( z;\, \omega_{1},\, \omega_{2} \right)
\\ \label{eqnA.23a}
& \hspace{-52pt}
\pfun^{01}_{k} \! \left( z;\, \omega_{1},\, \omega_{2} \right) \hspace{-7pt} & = \,
2\hspace{1pt}\pfun_k \! \left( z;\, 2\omega_{1},\, \omega_{2} \right) -
\pfun_k \! \left( z;\, \omega_{1},\, \omega_{2} \right)
\, , \
\\ \label{eqnA.24a}
& \hspace{-52pt}
\pfun^{11}_{k} \! \left( z;\, \omega_{1},\, \omega_{2} \right) \hspace{-7pt} & = \,
2\hspace{1pt}\pfun_k \! \left( z;\, \omega_{1}+\omega_2,\, 2\omega_{2} \right) -
\pfun_k \! \left( z;\, \omega_{1},\, \omega_{2} \right)
\, , \
\\ \label{eqnA.19n}
& \hspace{-52pt}
\pfun_1 \hspace{-1pt} \left( z;\, \omega_{1},\, \omega_{2} \right) \hspace{-7pt} & = \,
\zfun \hspace{-1pt} \left( z;\, \omega_{1},\, \omega_{2} \right)
+
8 \pi^2 G_2 \! \left( \omega_1,\, \omega_2 \right) z
\, , \
\\ \label{eqnA.21a}
& \hspace{-52pt}
\pfun_2 \hspace{-1pt} \left( z;\, \omega_{1},\, \omega_{2} \right) \hspace{-7pt} & = \,
\wfun \hspace{-1pt} \left( z;\, \omega_{1},\, \omega_{2} \right)
-
8 \pi^2 G_2 \! \left( \omega_1,\, \omega_2 \right)
\, , \
\eeqa
\protect\vspace{-40pt}

\beqa
\label{eqnA.19n1}
& \hspace{-30pt} \pfun_1 \hspace{-1pt} \left( z+\omega_1;\,
\omega_{1},\, \omega_{2} \right) \hspace{-7pt} & = \,
\pfun_1 \hspace{-2pt} \left( z;\, \omega_1,\, \omega_2 \right)
- 8 \hspace{1pt} \pi^2 \hspace{1pt} G_2 \! \left( \omega_1,\, \omega_2 \right)
\left( \omega_1 - \omega_2 \right)
- 2 \pi i
\, , \
\\ \label{eqnA.19n2}
& \hspace{-30pt} \pfun_1 \hspace{-1pt} \left( z+\omega_2;\,
\omega_{1},\, \omega_{2} \right) \hspace{-7pt} & = \,
\pfun_1 \hspace{-2pt} \left( z;\, \omega_1,\, \omega_2 \right)
\, . \
\eeqa
\textit{Hint}:
to prove Eqs.~(\ref{eqnA.22a})--(\ref{eqnA.24a}) take even $M$ and $N$
in Eq.~(\ref{2.5nnn}) and split appropriately the resulting sum;
proving Eqs.~(\ref{eqnA.19n})--(\ref{eqnA.19n2})
one can first show that the difference between the two sides of Eq.~(\ref{eqnA.19n})
is an entire, doubly periodic, odd function and therefore, it is zero
(see also Appendix~\ref{ap:1}).
\end{mexercise}

\begin{mcorollary}\label{mcor:1}
Every elliptic function $f \hspace{-1pt} \left( z \right)$
satisfying the periodicity conditions
\beqa\label{f_per}
f (z+\omega_1;\, \omega _{1},\, \omega _{2})
\, = && \hspace{-15pt}
\left( -1 \right)^{\kappa} \,
f (z;\, \omega _{1},\, \omega _{2})
\, , \
\nn
f (z+\omega_2;\, \omega _{1},\, \omega _{2})
\, = && \hspace{-15pt}
\left( -1 \right)^{\lambda} \,
f (z;\omega _{1},\omega _{2})
\, , \
\eeqa
for some \(\kappa, \lambda = 0,\, 1\) admits unique (nontrivial) expansion
\beq\label{f_exp}
f \! \left( z \right) = N \, + \,
\Su_{k \, = \, 1}^{K} \Su_{s \, = \, 1}^{S_k} \, N_{k,s} \
\pfun^{\kappa\lambda}_k \! \left( z-z_s;\, \omega_1,\, \omega_2 \right)
\, \
\eeq
where \(K,S_1,\dots,S_K \in \N\), \(N,N_{s,k} \in \C\), \(z_s \in F\)
(\(k = 1,\dots, K\), \(s=1,\dots,S_k\)).
In the case \(\kappa=\lambda=0\) the coefficients $N_{1,k}$ satisfy
\beq\label{N_cond}
\Su_{s \, = \, 1}^{S_1} N_{1,s} \, = \, 0
\, . \
\eeq
\end{mcorollary}

\begin{mexercise}
\label{newex2.3}
Transform the forth degree equation
$y^2$ $=$ $(x-e_0)$ $(x-e_1)$ $(x-e_2)$ $(x-e_3)$
(with different roots $e_{\nu}$) into a third degree one
$y^2$ $=$ $4 \, (x-e_1')$ $(x-e_2')$ $(x-e_3')$,
using the M\"obius transformation of
Remark~2.1.
(\textit{Answer}: the transformation is
$x$ $\mapsto$ $e_0$ $+$ $(x-a)^{-1}$ and
$y$ $\mapsto$ $\Txfrac{A}{(x-a)^2} \, y$ with
$A^2$ $=$ $\txfrac{1}{4}$ $(e_0-e_1)$ $(e_0-e_2)$ $(e_0-e_3)$;
then
$e_j'$ $=$ $a$ $-$ $(e_0-e_j)^{-1}$ (\(j=1,2,3\))
where fixing $a$ $=$ $-\txfrac{1}{3}$ $\mathop{\dsum}\limits_{j=1}^3$ $(e_0-e_j)^{-1}$
is equivalent to the condition
$\mathop{\dsum}\limits_{j=1}^3$ $e_j'$ $=$ $0$ obeying the form~(\ref{2.4}).)
\end{mexercise}

\subsection{Elliptic curves}

A nonsingular projective cubic curve with a distinguished ``point at
infinity'' is called elliptic. An elliptic curve $E$ over $\mathbb{C}$
(or, more generally, over any number field of characteristic different
from $2$ and $3$) can be reduced to the Weierstrass form
(in homogeneous coordinates \(X,Y,Z\)),
\begin{equation}
E\, :\ Y^{2}Z=4X^{3}-g_{2}XZ^{2}-g_{3}Z^{3}
\quad
(g_{2}^{3}-27g_{3}^{2}\neq 0)  \label{2.5}
\end{equation}
with the point at infinity, given by
\begin{equation}
e=(X:Y:Z)=(0:1:0)
\, . \  \label{2.6n}
\end{equation}
Let $\Lambda $ be a (2--dimensional) period lattice (as in Eq. (\ref{2.3})). The
uniformization map
\begin{equation}
z \, \mapsto \,
\left\{\!\!
\begin{array}{lll}
(\wfun (z):\wfun' (z):1)
& \text{for} &
z\notin \Lambda
\\
(0:1:0) & \text{for} & z\in \Lambda
\end{array}
\right.
\label{2.7}
\end{equation}
($\wfun (z)$ $\equiv$ $\wfun (z;\omega_1,\omega_2)$,
$\wfun' (z)$ $\equiv$ $\di_z\wfun (z;\omega_1,\omega_2)$)
from $\mathbb{C}$ to the projective complex plane provides an isomorphism between
the torus $\mathbb{C}/\Lambda$ and the projective algebraic curve (\ref{2.5}).
It follows
that $E$ is a (commutative) algebraic group (as the quotient of the additive
groups $\mathbb{C}$ and~$\Lambda$).
The addition theorem for Weierstrass functions,
\beqa
\wfun (z_{1}+z_{2}) && \podr = \,
-\wfun (z_{1})-\wfun (z_{2})+\frac{1}{4}
\left(
\frac{\wfun' (z_{1})-\wfun' (z_{2})}{\wfun (z_{1})-\wfun (z_{2})}
\right)^{2}
\label{2.8}
\nn
( && \podr \rightarrow \, -2 \hspace{1pt}
\wfun (z_{1}) + \frac{1}{4}
\left(
\frac{\wfun'' (z_1)}{\wfun' (z_1)}
\right)^{2}
\quad \text{for} \quad
z_{2} \rightarrow z_{1} \ )
\eeqa
allows to express the group law in terms of the projective coordinates as follows.

The origin (or neutral element) of the group is the point at infinity $e$
(\ref{2.6n}). If ($x=$ $\txfrac{X}{Z}$,
$y=$ $\txfrac{Y}{Z}$) is a finite point of $E$
(\ref{2.5})~--~i.~e., a solution of the equation
\begin{equation}
y^{2}=4x^{3}-g_{2}x-g_{3}
\, , \ \label{2.9}
\end{equation}
then its \textit{opposite} under the group law is the symmetric point
$(x,$ $-y)$ (which also satisfies (\ref{2.9})). If $P_{1}=$ $(x_{1},$ $y_{1})$,
$P_{2}=$ $(x_{2},$ $y_{2})$ are non--opposite finite points of $E$, then their
``sum'' $P_{3}=$ $(x_{3},$ $y_{3})$ is defined by
\beqa
&
x_{3} \, = \, -x_{1}-x_{2}+\Txfrac{m^{2}}{4}
\, , \quad
y_{3} \, = \, -y_{1}-m \left( x_{3}-x_{1} \right)
& \nn &
\text{for} \quad
m = \Txfrac{y_{1}-y_{2}}{x_{1}-x_{2}}
\quad \text{if} \quad P_{1} \, \neq \, P_{2}
\, ; \quad
m \, = \, \Txfrac{12\, x_1^2 -g_{2}}{2y_{1}}
\quad \text{if} \quad
P_{1} \, = \, P_{2}
\, . & \qquad  \label{2.10}
\eeqa
The structure of rational points on an elliptic curve~--~a hot topic of
modern mathematics~--~is reviewed in \cite{RS02}.

\begin{mexercise}\label{exr:2}
Compute the sum $P+Q$ of points
$P=$ $(-\txfrac{11}{9},$
$\txfrac{17}{27})$,
$Q=$ $(0,$ $1)$ of an elliptic curve \(y^{2}=x^{3}-x+1\).
(Answer: $(x,$ $y)=$ $\txfrac{1}{121}(159,$
$-\txfrac{1861}{11})$).
\end{mexercise}

\begin{mproposition}\label{pr:2.1}
{\rm (\cite{Sh94} Proposition 4.1).}
Two elliptic curves $E:$ \(y^{2}=4x^{3}-g_{2}x-g_{3}\) and
$\widetilde{E}:$ \(\widetilde{y}^{2}=
4\widetilde{x}^{3}-\widetilde{g}_{2}\widetilde{x}-\widetilde{g}_{3}\)
are \textrm{isomorphic}
(as complex manifolds with a distinguished point)
iff there exists $\rho \neq 0$, such that \(\widetilde{g}_{2}=\rho ^{4}g_{2}\),
\(\widetilde{g}_{3}=\rho ^{6}g_{3}\);
the isomorphism between them is then realized by the relation
\(\widetilde{x}=\rho ^{2}x\), \(\widetilde{y}=\rho ^{3}y\).
\end{mproposition}

$\star$
The following text (between asterisks) is designed to widen the scope
of a mathematically oriented reader and can be skipped in a first reading.

An elliptic curve, as well as, every algebraic (regular, projective) curve $M$
can be fully characterized by its \textit{function field}
(\cite{ZS}).
This is the space $\C (M)$ of all meromorphic functions over $M$,
i.e., functions $f$ such that in the vicinity of each point \(p \in M\),
$f$ takes the form $(w-w(p))^{d}$ $(a+(w-w(p)) \, g(w))$
for some local coordinate $w$ and an analytic function $g (w)$ around $p$,
\(d \in \Z\),
and a \textit{nonzero} constant $a$ for $f$ $\neq 0$.
The number $\ord_p f$ $:=d$ is then uniquely determined for nonzero $f$,
depending only on $f$ and $p$:
it is called the \textit{order of} $f$ \textit{at} $p$.
Thus, the order is a function $\ord_p:$ $\C (M) \backslash \{0\} \to \Z$
satisfying the following natural properties
\begin{plist}
\item[($\ord_1$)] \ \(\ord_p (fg) = \ord_p f + \ord_p g\);
\item[($\ord_2$)] \
\(\ord_p (f \hspace{-1pt}+\hspace{-1pt} g) \geqslant \min \{\ord_p f, \ord_p g\}\)
\ for \ \(f \neq -g\);
\item[($\ord_3$)] \ \(\ord_p \, c = 0\) \ for \ \(c \in \C \,\backslash \{0\}\)
\end{plist}
(it sometimes is convenient to set $\ord_p 0$ $:=\infty$).
Functions $\nu:$ $\C (M) \backslash \{0\} \to \Z$ satisfying
($\ord_1$)--($\ord_3$) are called
\textit{discrete valuations} (on the field $\C (M)$).
They are in one-to-one correspondence with
the points \(p \in M\): \(p \mapsto \ord_p\).
Moreover, the \textit{regular functions at} $p$, i.e. the functions
taking finite (complex) values at $p$, are
those for which
$\ord_p f$ $\geqslant 0$; these functions form a ring $R_p$
with a (unique) maximal ideal
$\mathfrak{m}_p$ \(:= \left\{ f : \ord_p f > 0 \right\}\).
Then the value $f (p)$ can be algebraically expressed as the coset
$[f]_p$ of $f$ in the quotient ring $R_p / \mathfrak{m}_p$ $\cong \C$
(since the quotient by a maximal ideal is a field!).

On the other hand, the field $\C (M)$ of meromorphic functions
on a (compact) projective curve
can be algebraically characterized as a degree one transcendental extension
of the field $\C$ of complex numbers:
$\C (M)$ contains a non algebraic element over $\C$ and every two
elements of $\C (M)$ are algebraically dependent
(over $\C$, i.e., satisfy a polynomial equation with complex coefficients).
Such fields are called \textit{function fields}.
The simplest example is the field $\C (z)$ of
the complex rational functions of
a single variable $z$.
This is, in fact, the function field of the \textit{Riemann}\footnote{%
Georg Friedrich Bernhard Riemann (1826--1866)
introduced the "Riemann surfaces" in his Ph.D. thesis in G\"ottingen,
supervised by Gauss (1851).}
\textit{sphere}~$\PP^1$.
Summarizing the above statements we have:

\begin{mtheorem} {\rm (\cite{ZS}) Chapt.~VI}
The nonsingular algebraic projective curves are in one--to--one correspondence
with the degree one transcendental extensions of $\C$,
naturally isomorphic to the fields of meromorphic
functions over the curves.~$\star$
\end{mtheorem}

The function field of an elliptic curve $E$ $:= \C/\Lambda$ is generated
by $\wfun$ and $\wfun'$ (\cite{McKM}, Sect.~2.13),
\beq\label{nwn2.35}
\C (E) \, = \, \C (\wfun) [\wfun'] \, = \,
\C (\wfun) \hspace{-2pt}
\left[\raisebox{9pt}{\hspace{-4pt}}\right.
\sqrt{(\wfun - e_1)(\wfun - e_2)(\wfun - e_3)}
\left.\raisebox{9pt}{\hspace{-1pt}}\right]
\, ,
\eeq
where $e_1$, $e_2$ and $e_3$ are the roots of
the third order polynomial~(\ref{2.9}),
\beq\label{nwn2.36}
4\wfun^3 - g_2 \wfun - g_3 \, = \,
4(\wfun - e_1)(\wfun - e_2)(\wfun - e_3)
\, (\, = \, (\wfun')^2)
\,
\eeq
(which should be different in order to have an elliptic curve).
Thus, $\C (E)$ is a quadratic algebraic extension of the field $\C (\wfun)$
of rational functions in $\wfun$.

\begin{mexercise}\label{to-ex1}
Let $(\omega_1,\omega_2)$ be a basis of $\Lambda$.
Prove that the roots of $\wfun'$~(\ref{nwn2.36}) in the basic cell
\(\{ \lambda\,\omega_1\) $+$ $\mu\,\omega_2$ $:$
$0$ $\leqslant$ $\lambda,$ $\mu$ $<$ \(1 \}\)
are (up to ordering) $\txfrac{\omega_1}{2}$, $\txfrac{\omega_2}{2}$,
$\txfrac{\omega_1+\omega_2}{2}$, corresponding to
$e_1$ $=$ $\wfun \left( \txfrac{\omega_1}{2} \right)$,
$e_2$ $=$ $\wfun \left( \txfrac{\omega_2}{2} \right)$,
$e_3$ $=$ $\wfun \left( \txfrac{\omega_1+\omega_2}{2} \right)$.
\textit{Hint}: use the fact that $\wfun'$ is an odd function of $z$ as in
Exercise~2.2.~(a).
\end{mexercise}

\begin{mexercise}\label{to-ex2}
Show that $\wfun_j (z)$ $=$ $\sqrt{\wfun (z) -e_j}$,
\(j = 1,2,3\) have single valued branches in the neighbourhood
of the points
$z_1$ $=$ $\txfrac{\omega_1}{2}$,
$z_2$ $=$ $\txfrac{\omega_2}{2}$,
$z_3$ $=$ $\txfrac{\omega_1+\omega_2}{2}$,
respectively.
Prove that they have simple poles on the lattice $\Lambda$ and may
be standardized by fixing the residue at the origin as~$1$.
Demonstrate that they belong to
different quadratic extensions of the field $\C (E)$ corresponding
to double covers of the torus $E$ with primitive periods
$(\omega_1,2\omega_2)$, $(2\omega_1,\omega_2)$ and
$(\omega_1+\omega_2,2\omega_2)$, respectively
(we shall also meet the corresponding index 2 sublattices in
Sect.~2.4).
Deduce that,
\beqa\label{to_ex-e1}
&
\hspace{-20pt}
\wfun_1 (z) \, ( = \! \sqrt{\wfun (z) - e_1} \hspace{1pt}) \hspace{-1pt} =
p_1^{01} (z;\omega_1,\omega_2)
, \quad
\wfun_2 (z) \, ( = \! \sqrt{\wfun (z) - e_2} \hspace{1pt}) \hspace{-1pt} =
p_1^{10} (z;\omega_1,\omega_2)
\hspace{-20pt}
& \nn &
\hspace{-20pt}
\wfun_3 (z) \, ( = \! \sqrt{\wfun (z) - e_3} \hspace{1pt}) \hspace{-1pt} =
p_1^{11} (z;\omega_1,\omega_2)
\hspace{-20pt}
&
\eeqa
where $p_1^{\kappa\lambda}$ are the functions~(\ref{2.5nnn})
(see~\cite{McKM}, Sect.~2.17).
\end{mexercise}

\begin{mexercise}\label{to-ex3}
Find a relation between the sinus amplitudinus function
$\mathrm{sn} (z,k^2)$ (\ref{Jacobi}) and the functions
$\wfun_j$ of
Exercise~2.6.
\textit{Answer}:
\beq\label{wf_3}
\wfun_3 (z)
\, ( \hspace{1pt} \equiv p_1^{11} (z; \omega_1,\omega_2) \hspace{1pt}) = \,
\frac{\sqrt{e_2-e_3}}{
\mathrm{sn} \left( z \sqrt{e_2-e_3},k^2 \right) \raisebox{9pt}{}}
\, .
\eeq
\end{mexercise}

\begin{mexercise}\label{to-ex4}
Use the change of variables \(x \mapsto e_3 + \Txfrac{e_2-e_3}{x^2}\)
to convert the indefinite integral
$\dint \left[\raisebox{9pt}{\hspace{-2pt}}\right. 4$ $(x$ $-$ $e_1)$ $(x$ $-$ $e_2)$
$(x$ $-$ $e_3) \left.\raisebox{9pt}{\hspace{-2pt}}\right]^{-\frac{1}{2}}$ $dx$
into
$(e_2$ $-$ $e_3)^{-\frac{1}{2}}$
$\dint \left[\raisebox{9pt}{\hspace{-2pt}}\right. (1$ $-$ $x^2)$ $(1$ $-$
$k^2x^2) \left.\raisebox{9pt}{\hspace{-2pt}}\right]^{-\frac{1}{2}}$ $dx$
for $k^2$ $=$ $\Txfrac{e_1-e_3}{e_2-e_3}$
(as in Exercise~2.3).
Deduce as a consequence the relations:
\beq\label{to_ex-e2}
\wfun (z) \, = \, e_3 + \frac{e_2-e_3}{
\left\{ \mathrm{sn} \left( z \sqrt{e_2-e_3},k^2 \right) \right\}^2
\raisebox{9pt}{}}
\, .
\eeq
\end{mexercise}

\begin{mexercise}\label{to-ex5}
Prove that addition of half--periods and the reflection \(z \mapsto -z\)
are the only involutions of $E$ $=$ $\C / \Lambda$.
Prove that the quotient space $E/(z \sim -z)$ is isomorphic to $\mathbb{P}^1$.
Identify the quotient map $E$ $\to$ $E / (z \sim -z)$
as the Weierstrass function $\wfun (z)$.
\end{mexercise}

\subsection{Modular invariance}

Proposition~2.2
implies, in particular, that two lattices, $\Lambda$ and
$\rho\hspace{1pt}\Lambda$, with the same ratio of the periods,
\begin{equation}
\tau :=\frac{\omega _{1}}{\omega _{2}}
\, \in \hcom =\left\{ \tau
\in \mathbb{C}
\, ; \
\IM \, \tau >0\right\}  \label{2.11}
\end{equation}
correspond to isomorphic elliptic curves.
The isomorphism is given by multiplication with
a non-zero complex number $\rho$:
\beqa\label{L_isom}
& \hspace{-8pt}
\C / \Lambda \ \cong \ \C / \left( \rho\hspace{1pt}\Lambda \right) \, : \,
z \ (\MOD \, \Lambda) \ \mapsto \
\rho \hspace{1pt} z \ (\MOD \, \rho\hspace{1pt}\Lambda)
\, \
& \hspace{-10pt} \\ & \hspace{-8pt}
( \,
\left( x : y : 1 \right)
\, \mapsto \,
\left( \rho^2 x : \rho^3 y : 1 \right) =
\left(
\wfun \left( \rho \hspace{1pt} z;\,
\rho \hspace{1pt} \omega_1,\, \rho \hspace{1pt} \omega_2 \right) :
\di_z \wfun \left( \rho \hspace{1pt} z;\,
\rho \hspace{1pt} \omega_1,\, \rho \hspace{1pt} \omega_2 \right) : 1
\right)
\, ) .
& \hspace{-10pt}
\nonumber
\eeqa
On the other hand, the choice of
basis $(\omega _{1},\omega _{2})$ in a given lattice $\Lambda$ is not
unique. Any linear transformation of the form
\beqa
&
\left( \omega_1,\, \omega_2 \right) \, \mapsto \,
\left( \omega_1',\, \omega_2' \right) :=
\left( a\,\omega _{1}+b\,\omega _{2},\,
c\,\omega_{1}+d\,\omega _{2} \right)
, & \nn &
a, b, c, d \, \in \, \Z
\, , \quad
ad-bc \, = \, \pm 1
&
\label{2.12}
\eeqa
gives rise to a new basis $(\omega_{1}',\, \omega_{2}')$
in $\Lambda$ which is as good as the original one.
Had we been given a basis $\left( \omega_1,\, \omega_2 \right)$
for which
\(\IM \, \txfrac{\omega_{1}}{\omega _{2}} < 0\),
we could impose (\ref{2.11}) for
$(\omega_{1}',\, \omega_{2}')$ $= \left( \omega_2,\, \omega_1 \right)$.
Orientation preserving transformations~(\ref{2.12})
form the \textit{modular group}
\begin{equation}
\Gamma (1):=\SL (2,\Z)=\left\{ \gamma =
\left(\hspace{-4pt} \begin{array}{cc} a \hspace{-4pt} & b \\
c \hspace{-4pt} & d \end{array} \hspace{-3pt}\right) :
\ a,\, b,\, c,\, d \, \in \, \Z
,\ \det \gamma \, = \, ad-bc \, = \, 1 \right\}
. \
\label{2.13}
\end{equation}
Thus, on one hand, we can define an elliptic curve, up to isomorphism,
factorizing $\C$ by the lattice $\Z\tau + \Z$ with \(\tau \in \hcom\) and
on the other, we can pass by a modular transformation
\(\gamma =
\left(\hspace{-4pt} \begin{array}{rr} a \hspace{-4pt} & b \\
c \hspace{-4pt} & d \end{array} \hspace{-3pt}\right)
\in \Gamma (1)\)
to an equivalent basis
\(\left( a\tau+b,c\tau+d \right)\).
Normalizing then the second period to $1$
we obtain the classical action of $\Gamma (1)$
on
$\hcom$~(\ref{2.11})
(mapping the upper half plane onto itself),
\beq\label{hcon_act}
\tau \, \mapsto \, \frac{a\tau+b}{c\tau+d}  \ .
\eeq
This action obviously has a $\Z_2$ kernel \(\{\pm 1\} \in \Gamma (1)\).

Note that the series~(\ref{zfun}) and (\ref{2.2}),
as well as~(\ref{p_rough}) for \(k \geqslant 3\),
are absolutely
convergent for \(z \notin \Lambda\).
This implies,
in particular, their independence of the choice of basis,
\beqa\label{o_oodinv}
& \zfun (z;\omega_1,\omega_2) & \ppodr = \,
\zfun (z;a\hspace{1pt}\omega_1+b\hspace{1pt}\omega_2,c\hspace{1pt}\omega_1+d\hspace{1pt}\omega_2)
\, , \ \nn
& \wfun (z;\omega_1,\omega_2) & \ppodr = \,
\wfun (z;a\hspace{1pt}\omega_1+b\hspace{1pt}\omega_2,c\hspace{1pt}\omega_1+d\hspace{1pt}\omega_2)
\, , \ \nn
& \pfun_k (z;\omega_1,\omega_2) & \ppodr = \,
\pfun_k (z;a\hspace{1pt}\omega_1+b\hspace{1pt}\omega_2,c\hspace{1pt}\omega_1+d\hspace{1pt}\omega_2)
\quad (k \geqslant 3) .
\eeqa
for \(\gamma =
\left(\hspace{-4pt} \begin{array}{rr} a \hspace{-4pt} & b \\
c \hspace{-4pt} & d \end{array} \hspace{-3pt}\right)
\in \Gamma (1)\).
Using the homogeneity
\beqa\label{homog}
& \zfun (\rho\hspace{1pt}z;\rho\hspace{1pt}\omega_1,\rho\hspace{1pt}\omega_2) \, = \,
\rho^{-1} \,
\zfun (z;\omega_1,\omega_2)
\, , \quad
\wfun (\rho\hspace{1pt}z;\rho\hspace{1pt}\omega_1,\rho\hspace{1pt}\omega_2) \, = \,
\rho^{-2} \,
\wfun (z;\omega_1,\omega_2)
\, , & \nn
& \pfun_k (\rho\hspace{1pt}z;\rho\hspace{1pt}\omega_1,\rho\hspace{1pt}\omega_2) \, = \,
\rho^{-k} \,
\pfun_k (z;\omega_1,\omega_2) &
\eeqa
(\(\rho \in \C \backslash \{0\}\)) and setting
\beqa\label{setting}
&
\zfun (z,\tau) \, := \, \zfun (z;\tau,1)
\, , \quad
\wfun (z,\tau) \, := \, \wfun (z;\tau,1)
\, , \quad
\pfun_k (z,\tau) \, := \, \pfun_k (z;\tau,1)
& \nn &
\pfun_k^{\kappa\lambda} (z,\tau) \, := \, \pfun_k^{\kappa\lambda} (z;\tau,1)
\quad
(\kappa,\lambda = 0,1)
, \quad
\pfun_k (z,\tau) \, \equiv \, \pfun_k^{00} (z,\tau)
&
\eeqa
(see~(\ref{2.5nnn}))
we find as a result, the modular transformation laws
\beqa\label{p_modinv}
& (c\tau +d)^{-1} \, \zfun \left(\raisebox{12pt}{\hspace{-2pt}}\right.
\Txfrac{z}{c\tau +d},\Txfrac{a\tau +b}{c\tau +d}
\left.\raisebox{12pt}{\hspace{-2pt}}\right)
& \ppodr = \, \zfun (z,\tau)
\, , \ \nn
& (c\tau +d)^{-2} \, \wfun \left(\raisebox{12pt}{\hspace{-2pt}}\right.
\Txfrac{z}{c\tau +d},\Txfrac{a\tau +b}{c\tau +d}
\left.\raisebox{12pt}{\hspace{-2pt}}\right)
& \ppodr = \, \wfun (z,\tau)
\, , \ \nn
& (c\tau +d)^{-k} \, \pfun_k \left(\raisebox{12pt}{\hspace{-2pt}}\right.
\Txfrac{z}{c\tau +d},\Txfrac{a\tau +b}{c\tau +d}
\left.\raisebox{12pt}{\hspace{-2pt}}\right)
& \ppodr = \, \pfun_k (z,\tau)
\quad (k \geqslant 3) .
\eeqa
The functions $p_1 \! \left( z,\tau \right)$ and
$p_2 \! \left( z,\tau \right)$
obey \textit{inhomogeneous} modular transformation laws
since $G_2 (\omega_1,\omega_2)$ transforms
inhomogeneously (see Sect.~3.2).
This is the price for preserving the periodicity property for
\(z \mapsto z+1\) according to~(\ref{eqnA.19n2}).
Nevertheless, all the functions $p_k^{\kappa\lambda}$
for \(k \geqslant 1\) and \(\kappa+\lambda > 0\) transform
homogeneously among themselves:
\beq\label{eqnXX.15}
(c\tau +d)^{-k} \,
p_k^{\hspace{1pt}[a\kappa+b\lambda]_2\hspace{1pt},\hspace{1pt}[c\kappa+d\lambda]_2}
\left(\raisebox{10pt}{\hspace{-2pt}}\right.
\frac{z}{c\tau \! + \! d},\,
\frac{a\tau \! + \! b}{c\tau \! + \! d}
\left.\raisebox{10pt}{\hspace{-2pt}}\right)
\, = \,
p_k^{\kappa\lambda}
\left( z,\, \tau \right)
\, \
\eeq
where \([\lambda]_2=0,1\) stands for the $\lambda \, \MOD \, 2$.

\begin{mexercise}
Prove the relation~(\ref{eqnXX.15}) for \(k \geqslant 3\) using
the absolute convergence of the series in Eq.~(\ref{2.5nnn}).
(For \(k=1,2\) one should use the uniqueness property of
the functions $p_k^{\kappa\lambda}$ given in
Appendix~\ref{ap:1}.)
\end{mexercise}

\begin{mexercise}\label{p_1}
(\textit{a})
Prove the representations
\beqa\label{p_1-rep1}
\hspace{-20pt}
p_1 \left( z,\tau \right) \, = && \podr
\mathop{\lim}\limits_{N \, \to \, \infty} \,
\Su_{k \, = \, -N}^{N}
\pi \, \cotg \, \pi (z \hspace{-1pt} + \hspace{-1pt} k\tau)
\, = \, \\ \label{p_1-rep2}
\, = && \podr
\pi \, \cotg \, \pi z + 4\pi \sum_{n \, = \, 1}^{\infty}
\frac{q^n}{1-q^n} \, \sin 2 \pi n z
\, \
\eeqa
where \(q := e^{2\pi i\tau}\) and
the series~(\ref{p_1-rep1}) absolutely converges for all
\(z \notin \Z\tau + \Z\) and \(\tau \in \hcom\)
while~(\ref{p_1-rep2}) absolutely converges for
\(|q|\) \(< |e^{-2\pi i z}|\) \(< 1\).
\textit{Hint}: take the difference between the two sides of~(\ref{p_1-rep1})
and prove that it is an entire, odd, elliptic function
using~(\ref{eqnA.19n1}) and~(\ref{eqnA.19n2});
to derive~(\ref{p_1-rep2}) from~(\ref{p_1-rep1}) use the
expansion
\beqa
&
\cotg \, \pi \left( z \hspace{-1pt} + \hspace{-1pt} k\tau \right)
\hspace{1pt} + \hspace{1pt}
\cotg \, \pi \left( z \hspace{-1pt} - \hspace{-1pt} k\tau \right)
\, = \, - \hspace{1pt} i\,\Txfrac{1+e^{2\pi i z}q^k}{1-e^{2\pi i z}q^k}
\hspace{1pt} + \hspace{1pt}
i\,\Txfrac{1+e^{-2\pi i z}q^k}{1-e^{-2\pi i z}q^k}
\, = \, & \nn & \, = \,
4\mathop{\dsum}\limits_{n \, = \, 1}^{\infty} q^{nk} \, \sin 2\pi nz .
& \nonumber
\eeqa

\noindent
(\textit{b}) Find similar representations for $p_2 (z,\tau)$,
$p_1^{11} (z,\tau)$ and $p_2^{11} (z,\tau)$.
\end{mexercise}

\subsection{Modular groups}

As an abstract discrete group, $\Gamma \left( 1 \right)$ has
two generators $S$ and $T$ satisfying the relations
\beqa\label{rel_1}
S^2 \, = && \podr \left( ST \right)^3
\, , \
\\ \label{rel_2}
S^4 \, = && \podr 1
\, ; \
\eeqa
their $2\times 2$ matrix realization is
\beq\label{ST}
S \, = \, \left(\hspace{-4pt} \begin{array}{rr} 0 \hspace{-4pt} & -1 \\
1 \hspace{-4pt} & 0 \end{array}\hspace{-3pt}\right)
\, , \quad
T \, = \,
\left(\hspace{-4pt} \begin{array}{rr} 1 \hspace{-4pt} & 1 \\
0 \hspace{-4pt} & 1 \end{array}\hspace{-3pt}\right)
\, . \
\eeq
This can be established by the following Exercise.

\begin{mexercise}\label{exr:3}
A subset \(D \subset \hcom\) is called a
\textit{fundamental domain} for $\Gamma(1)$ if each orbit
$\Gamma (1) \tau$ of a \(\tau\in \hcom\)
has at least one point in $D$,
and if two points of $D$ belong to the same orbit, they should
belong to the boundary of $D$.
Let
\beq\label{fun_domD}
D =
\left\{
\tau \in \hcom
:
-\frac{1}{2} \leqslant
\RE \left( \tau \right) \leqslant
\frac{1}{2}
,\
\left| \tau \right| \geqslant 1
\right\}
\, ; \
\eeq
prove that $D$ is a
\textit{fundamental domain} of $\Gamma (1)$.
Moreover, prove that
\begin{plist}
\item[1)]
$\tau\in \hcom$ then there exists a $\gamma \in \Gamma (1)$, such that
$\gamma \tau\in D$;
\item[2)]
if $\tau\neq \tau'$ are two points in $D$ such
that $\tau'=\gamma \tau$ then either
$\RE (\tau)=\mp \txfrac{1}{2}$
and $\tau'=\tau\pm 1$ or $\left| \tau \right| =1$ and
$\tau'=S\tau=-\txfrac{1}{\tau}$.
\item[3)]
Let
\(P_{1} := PSL(2,\Z) = \SL \left( 2,\, \Z \right)/\Z_{2}\)
be the (projective) modular group acting faithfully on $\hcom$
and let
\(I(\tau) = \left\{ \gamma \in P_{1} : \gamma \tau = \tau \right\}\)
be the stabilizer of $\tau$ in $P_{1}$.
Then if \(\tau \in D\), \(I(\tau)=1\) with the following three exceptions:
\(\tau=i\), then $I(\tau)$ is a
$2$--element subgroup of $P_{1}$ generated by $S$;
if \(\tau= \varrho :=e^{\frac{2\pi i}{3}}\)
then $I(\tau)$ is a $3$--element subgroup of $P_{1}$ generated by~$ST$;
if \(\tau=-\overline{\varrho}:=e^{\frac{\pi i}{3}}\) then $I(\tau)$ is a
3--element subgroup of $P_{1}$ generated by~$TS$.
\end{plist}
(See Sect.~1.2 of Chapter VII of \cite{Se}.)
Derive, as a corollary, that $S$ and $T$ generate~$P_{1}$.
\end{mexercise}

\begin{mexercise}\label{exr:3-1}
Verify that there are six images of the fundamental domain $D$~(\ref{fun_domD})
under the action of $\Gamma \left( 1 \right)$
incident with the vertex $e^{\frac{i\pi}{3}}$:
they are obtained from $D$ by applying the modular transformations
$1$, $T$, $TS$, $TST$, $ST^{-1}$ and $S$.
Note that all these domains are \textit{triangles in the Lobachevsky's plane}~\footnote{%
We thank Stanislaw Woronowicz for drawing our attention to this property.

Nikolai Ivanovich Lobachevsky (1793--1856) publishes his work
on the non-euclidean geometry in 1829/30.
Jules--Henri Poincar\'e (1854--1912) proposes his interpretation
of Lobachevsky's plane in 1882: it is the closed unit disk whose
boundary is called \textit{oricycle}
with straight lines corresponding to either diameters of the disk
or to circular arcs intersecting the oricycle under right angles.
The upper half plane
is mapped on the Poincar\'e disk by the complex
conformal transformation $\tau$ $\mapsto z$ $= \txfrac{1+i\tau}{\tau+i}$.}
with two $60^{\circ{}}$ ($= \txfrac{\pi}{3}$) angles and a zero angle vertex
at the \textit{oricycle}.
They split into two orbits under the $3$--element cyclic subgroup
of $P_1$ generated by $TS$.
Their union is the fundamental domain fo the index six subgroup $\Gamma(2)$
(defined in (\ref{2.15n}) below; cp. \cite{McKM} Sect.~4.3).
\end{mexercise}

\begin{mremark}\label{nrem2:2}
$\Gamma (1)$ can be viewed, alternatively, as a homomorphic image of the
braid group $B_{3}$ on three strands.
Indeed, the group $B_{3}$ can be characterized in terms of the elementary braidings
$b_{i},i=1,2$, which interchange the end points $i$ and $i+1$ and are subject to the
\textit{braid relation}
\begin{equation}
b_{1}b_{2}b_{1}=b_{2}b_{1}b_{2}
\, . \  \label{2.15}
\end{equation}
On the other hand,
the group $\widetilde{\Gamma}$  with generators $\widetilde{S}$
and $\widetilde{T}$ satisfying only the relation~(\ref{rel_1}) is isomorphic
to the group $B_3$ since the mutually inverse maps
\beq\label{}
\widetilde{S} \, \mapsto \, b_{1}b_{2}b_{1}
\, , \ \
\widetilde{T} \, \mapsto \, b_{1}^{-1}
\quad \text{and} \quad
b_1 \, \mapsto \, \widetilde{T}^{-1}
\, , \ \
b_2 \, \mapsto \, \widetilde{T}\widetilde{S}\widetilde{T}
\eeq
convert the relations~(\ref{rel_1}) and~(\ref{2.15})
into one another.
The element $\widetilde{S}^{2}$
is mapped to the generator of the (infinite) centre of $B_{3}$.
Its image $S^2$ in $\Gamma (1)$ satisfies the additional constraint~(\ref{rel_2}).
It follows that $B_{3}$ appears as a central extension of $\Gamma (1)$.
\end{mremark}

\medskip

We shall also need in what follows some \textit{finite} \textit{index subgroups}
$\Gamma \subset \Gamma (1)$ (i.~e. such that $\Gamma (1)/\Gamma $ has
a finite number of cosets).

\medskip

Let $\Lambda'$ be a sublattice of $\Lambda$ (\(:=\Z\omega_1+\Z\omega_2\))
of a finite index $N$.
This means that the factor group $\Lambda / \Lambda'$ is a finite abelian group
of order \(\left| \Lambda / \Lambda' \right| = N\).
The set of such sublattices,
\(\left\{ \Lambda' : \left| \Lambda / \Lambda' \right| = N \right\}\),
is \textit{finite} and
the group $\Gamma \left( 1 \right)$ acts on it via
\(\Lambda' \mapsto \gamma \left( \Lambda' \right)\) for
\(\gamma \in \Gamma \left( 1 \right)\)
(here we assume that
\(\gamma\hspace{1pt}\omega :=\)
$\left( a\hspace{1pt}m+b\hspace{1pt}n \right)\omega_1$
$+ \left( c\hspace{1pt}m+d\hspace{1pt}n \right)\omega_2$
$\equiv$ \(\left( \omega_1,\omega_2 \right) \gamma
\left(\hspace{-5pt} \begin{array}{c} m \\ n \end{array} \hspace{-5pt}\right)\)
for $\gamma =$
\(\left(\hspace{-4pt} \begin{array}{cc} a \hspace{-4pt} & b \\
c \hspace{-4pt} & d \end{array} \hspace{-3pt}\right)\) and $\omega =$
$m\hspace{1pt}\omega_1+n\hspace{1pt}\omega_2 $ $\in \Lambda$).

\begin{mexercise}\label{nex2.3}
Find all index $2$ sublattices $\Lambda'$ of the lattice $\Lambda$~(\ref{2.3}).
\textit{Answer}:
\(
\Lambda_{01} :=
\Z\hspace{1pt}\omega_1 + 2\hspace{1pt}\Z\hspace{1pt}\omega_2\),
\(
\Lambda_{10} :=
2\hspace{1pt}\Z\hspace{1pt}\omega_1 + \Z\hspace{1pt}\omega_2\)
and
\(\Lambda_{11} :=
\Z \left( \omega_1+\omega_2 \right) + 2\hspace{1pt}\Z\hspace{1pt}\omega_2\).
Prove that the stabilizer of
$\Lambda_{11}$,
denoted further by $\Gamma_{\theta}$
($:= \left\{\raisebox{9pt}{\hspace{-3pt}}\right.
\gamma \in \Gamma \left( 1 \right) :\)
\(\gamma \left( \Lambda_{11} \right)\)
\(\subseteq \Lambda_{11}
\left.\raisebox{9pt}{\hspace{-3pt}}\right\}\)), is
\begin{equation}
\Gamma _{\theta}=\left\{
\left(\hspace{-4pt} \begin{array}{cc} a \hspace{-4pt} & b \\
c \hspace{-4pt} & d \end{array} \hspace{-3pt}\right)
\in \Gamma (1) :
ac \ \text{and} \ bd \ \text{even} \, \right\}
.
\label{2.23}
\end{equation}
The group $\Gamma_{\theta}$ can be also characterized as
the index $3$ subgroup of $\Gamma \left( 1 \right)$
generated by $S$ and $T^{2}$
(see \cite{K90} Sect.~13.4).
\end{mexercise}

Other important finite index subgroups of $\Gamma \left( 1 \right)$
are the (normal) \textit{principal congruence subgroups}
\begin{equation}
\Gamma (N) \, = \,
\left\{
\left(\hspace{-4pt} \begin{array}{cc} a \hspace{-4pt} & b \\
c \hspace{-4pt} & d \end{array} \hspace{-3pt}\right)
\in \Gamma (1) \ : \
a \, \equiv \, 1 \ \MOD \, N \, \equiv \,
d,\ b \, \equiv \, 0 \ \MOD \, N \, \equiv \, c
\right\}
\, , \  \label{2.15n}
\end{equation}
(which justifies the notation $\Gamma \left( 1 \right)$ for $\SL (2,\Z)$)
and the subgroups
\begin{equation}
\Gamma _{0}(N) \, = \,
\left\{
\left(\hspace{-4pt} \begin{array}{cc} a \hspace{-4pt} & b \\
c \hspace{-4pt} & d \end{array} \hspace{-3pt}\right)
\in \Gamma (1) \ : \ c \, \equiv \, 0 \ \MOD \, N
\right\}
.  \label{2.20}
\end{equation}

\begin{mproposition}\label{pr:2.2}
{\rm (\cite{Sh94} Lemma 1.38).}
Let $\SL (2,\Z_N)$
be the (finite) group of $2\times 2$ matrices $\gamma$
whose elements belong to the finite ring
\(\Z_N \cong \Z/N\Z\)
of integers $\MOD \, N$ (and such that
$\det \gamma \equiv 1$ $\MOD \, N$).
If $f:\Gamma (1)\rightarrow \SL (2,\Z_N)$
is defined by $f(\gamma )=\gamma $ $\MOD \, N$, then the sequence
\begin{equation}
1\longrightarrow \Gamma (N)\longrightarrow \Gamma (1)
\mathop{\longrightarrow}\limits^{f}\SL (2,\Z_N)\longrightarrow 1  \label{2.21}
\end{equation}
is exact, i.e.,
the factor group \(\Gamma \left( 1 \right) / \Gamma \left( N \right)\)
is isomorphic to $\SL  \left( 2,\Z_N \right)$.
\end{mproposition}

We note that in the case \(N=2\) the factor group $\SL (2,\Z_2)$
is isomorphic to the permutation group $\mathcal{S}_3$ with the identification
\beq
s_{1}=f(T)=
\left(\hspace{-4pt} \begin{array}{rr} 1 \hspace{-4pt} & 1 \\
0 \hspace{-4pt} & 1 \end{array}\hspace{-3pt}\right)
\, , \quad
s_{2}=f(TST)=
\left(\hspace{-4pt} \begin{array}{rr} 1 \hspace{-4pt} & 0 \\
1 \hspace{-4pt} & 1 \end{array} \hspace{-3pt}\right)
\quad (s_{1}^{2}\equiv 1\, \MOD \, 2\equiv s_{2}^{2})
.  \label{2.22}
\end{equation}
In general, the number of elements of the factor group
$\SL \left( 2,\, \Z_N \right)$
(i.e., the index of $\Gamma (N)$ in $\Gamma (1)$, by
Proposition~2.3)
is
\beq\label{mu_N}
\mu \, = \,
N^3
\mathop{\prod}\limits_{p | N}
\left(\raisebox{10pt}{\hspace{-2pt}}\right. 1 -
\frac{1}{p^2} \left.\raisebox{10pt}{\hspace{-2pt}}\right)
\eeq
(the product being taken over the primes
$p$ which divide $N$,~\cite{Sh94} Sect.~1.6).

\begin{mremark}\label{nrem2}
The (invariant) commutator subgroup of the braid group $B_{3}$
is the monodromy group
$M_{3}$, which can, alternatively, be defined as the kernel of the group
homomorphism of $B_{3}$ onto the 6--element symmetric group $S_{3}$
realized by the map $b_{i}\mapsto s_{i}$, $i=1,2$, where
$s_{i}$ are the elementary transpositions satisfying (\ref{2.15}) and
$s_{i}^{2}=1$.
In other words, we have an exact sequence of groups and group homomorphisms,
\begin{equation}
1\rightarrow M_{3}\rightarrow B_{3}\rightarrow S_{3}\rightarrow 1
,\quad \text{i.~e.} \quad S_{3}=B_{3}/M_{3}
\, . \  \label{2.16}
\end{equation}
\end{mremark}

\begin{mexercise}\label{nex2.4}
Prove that the stabilizer of the sublattice
$\Lambda_{01} =$
\(\Z\hspace{1pt}\omega_1
+2\hspace{1pt}\Z\hspace{1pt}\omega_2\)
$\subseteq \Lambda$ is the subgroup $\Gamma_0 \left( 2 \right)$.
Thus $\Gamma_0 \left( 2 \right)$ and $\Gamma_{\theta}$ are
mutually conjugate subgroups of $\Gamma \left( 1 \right)$.
Prove also that the action of $\Gamma \left( 1 \right)$
on the three element set
\(\left\{ \Lambda_{10},\Lambda_{01},\Lambda_{11} \right\}\)
of index $2$ sublattices of $\Lambda$ is equivalent to the above homomorphism
\(\Gamma \left( 1 \right)
\, \mathop{\longrightarrow}\limits^{f} \, \SL  \left( 2,\Z_2 \right)\)
\(\cong \mathcal{S}_3\).
In fact, this action is given by the formula,
\(\Lambda_{\kappa\lambda} \to
\Lambda_{[a\kappa+b\lambda]_2\hspace{1pt},\hspace{1pt}[c\kappa+d\lambda]_2}\)
for \(\gamma =
\left(\hspace{-4pt} \begin{array}{rr} a \hspace{-4pt} & b \\
c \hspace{-4pt} & d \end{array} \hspace{-3pt}\right)
\in \Gamma (1)\)
(note that this is precisely the action of $\gamma$ on the upper indices of
\(p_{k}^{\kappa\lambda}\) in~(\ref{eqnXX.15})).
\end{mexercise}

\begin{mexercise}
Let $\Gamma$ be a finite index subgroup of $\Gamma (1)$.
Prove that there exists a nonzero power $T^h$ (i.e., $h \neq 0$)
belonging to $\Gamma$.
(\textit{Hint}: since there are finite number of right cosets
$\rcoset{\Gamma}{\Gamma(1)}$ there exist \(\gamma \in \Gamma\) and
$h_1,$ $h_2$ $\in \Z$, \(h_1\neq h_2\) such that \(T^{h_1} = \gamma \, T^{h_2}\).)
\end{mexercise}

\section{Modular forms and $\vartheta$--functions}
\setcounter{equation}{0}\setcounter{mtheorem}{1}\setcounter{mremark}{1}\setcounter{mdefinition}{1}\setcounter{mexample}{1}\setcounter{mexercise}{1}

\subsection{Modular forms}

Using the equivalence of proportional lattices we shall, from now on,
normalize the periods as ($\omega _{1},\omega _{2})=(\tau ,1)$ with $\tau $
belonging to the upper half plane $\hcom$ (\ref{2.11}).

Let $\Gamma$ be a subgroup of the modular group $\Gamma \left( 1 \right)$.
An analytic function $G_k \left( \tau \right)$ defined on
the upper half plane $\hcom$ \((\ni \tau)\) is called a
\textit{modular form} of weight $k$ and level $\Gamma$
if
\begin{plist}
\item[(\textit{\textit{i}})]
it is $\Gamma $--covariant:
\beq
(c\tau +d)^{-k} \,
G_{k} \!
\left( \frac{a\tau +b}{c\tau +d} \right)
\, = \,
G_{k}(\tau)
\quad \text{for} \quad
\left(\hspace{-4pt} \begin{array}{cc} a \hspace{-4pt} & b \\
c \hspace{-4pt} & d \end{array} \hspace{-3pt}\right) \in \Gamma
,  \label{3.2}
\eeq
i.e., the expression $G_{k}(\tau) \, (d\tau )^{\frac{k}{2}}$
is $\Gamma $--invariant:
\beq
G_{k}(\gamma \tau ) \, (d(\gamma\tau))^{\frac{k}{2}}
=
G_{k}(\tau) \, (d\tau)^{
\frac{k}{2}}
\quad \text{for} \quad
\gamma \tau = \frac{a\tau +b}{c\tau +d}
\label{3.1}
\eeq
(in view of the identity
$d(\gamma\tau)=\Txfrac{d\tau}{(c\tau+d)^{2}}$);
\item[(\textit{ii})]
$G_{k}(\tau)$ admits a \textit{Fourier}\footnote{%
Jean Baptiste Joseph Fourier (1768--1830).}
\textit{expansion} in non--negative
powers of
\beq
q=e^{2\pi i\tau} \quad (\,\left| q \right| <1 \,).  \label{3.3}
\eeq
\end{plist}
The coefficients $g_{2}(\tau)$ and $g_{3}(\tau)$ (\ref{2.4}) of the
Weierstrass equation provide examples of modular forms of level $\Gamma (1)$
and weights $4$ and~$6$, respectively.

\begin{mremark}\label{mrem3.1n}
The prefactor \(j \hspace{-1pt} \left( \gamma,\tau \right) =
\left( c\tau + d \right)^{-k}\) in~(\ref{3.2}),
called ``an automorphy factor'',
can be replaced by
a general \textit{cocycle}:
\(j \hspace{-1pt} \left( \gamma_1,\tau \right)
j \hspace{-1pt} \left( \gamma_2,\gamma_1\tau \right)
= j \hspace{-1pt} \left( \gamma_1\gamma_2,\tau \right)\)
(\(\gamma_1,\, \gamma_2 \in \Gamma\)).
If we stick to the prefactor \(\left( c\tau+d \right)^{-k}\), then
there are no non--zero modular forms of odd weights and level
$\Gamma$ provided
\(-1=
\left(\hspace{-4pt} \begin{array}{rr} -1 \hspace{-4pt} & 0 \\
0 \hspace{-4pt} & -1 \end{array}\hspace{-3pt}\right)\)
$\in \Gamma$.
Indeed, applying (\ref{3.2}) to this element
we find \(G_{k}(\tau )=(-1)^{k}G_{k}(\tau)\),
i.~e. \(G_{k}(\tau )=0\) for odd $k$.
For this reason we will mainly consider the case of even weights
(for an example of a modular form of weight one~--~see
Proposition~3.8).
\end{mremark}

\begin{mremark}\label{mrem3.2nn}
If the modular group \(\Gamma \subset \Gamma \! \left( 1 \right)\)
contains a subgroup of type $\Gamma \! \left( N \right)$~(\ref{2.15n})
we shall also use \textit{level} $N$ for the minimal such $N$
instead of ``level $\Gamma$''.
In particular, a modular form of level $\Gamma \! \left( 1 \right)$
is commonly called a \textit{level one form}.
\end{mremark}

Remarkably, the space of modular forms of a given weight and level is
finite dimensional.
This is based on the fact that every such modular form can be
viewed as a \textit{holomorphic section} of a line bundle
over a \textit{compact} \textit{Riemann}
\textit{surface}.
To explain this let us introduce the \textit{extended} upper half plane
\beq\label{hcom_st}
\hcom^* \, := \, \hcom \, \cup \, \mathbb{Q} \, \cup \, \left\{ \infty \right\}
\eeq
on which the modular group $\Gamma \left( 1 \right)$ acts
so that \(\mathbb{Q} \cup \left\{ \infty \right\}\) is a single orbit.
The set $\hcom^*$ can be endowed with a Hausdorff\footnote{%
Felix Hausdorff (1868--1942).}
topology,
extending that of $\hcom$, in such a way that the quotient space\footnote{%
Following the custom we will use the left coset notation
for the discrete group action
while $\hcom$ can be viewed as a right coset,
\(\hcom = \SL  \left( 2,\R \right) / \SO \left( 2 \right)\),
the maximal compact subgroup $\SO \left( 2 \right)$ of
$\SL  \left( 2,\R \right)$ being the stabilizer of the point $i$
in the upper half--plane.}
\(\rcoset{\Gamma \left( 1 \right)}{\hcom^*}\)
is isomorphic, as a topological space (i.e., it is homeomorphic),
to the Riemann sphere with a distinguished point,
the orbit \(\mathbb{Q} \cup \left\{ \infty \right\}\).
The points of the set \(\mathbb{Q} \cup \left\{ \infty \right\}\)
are called \textit{cusps} of the group\footnote{%
The cusps $\tau$ of $\hcom^*$
(with respect to some subgroup $\Gamma$ of $\Gamma (1)$)
are characterized by the property that they are
left invariant by an element of $\Gamma$
conjugate to $T^n$~(\ref{ST}) for some \(n \in \Z\).}
$\Gamma \! \left( 1 \right)$
as well as of any finite index subgroup $\Gamma$ in $\Gamma (1)$.
Then the quotient space
\(\rcoset{\Gamma}{\hcom^*}\)
is homeomorphic to a compact Riemann surface with distinguished points,
the cusps' orbits (with respect to $\Gamma$).
For more details on this constructions we refer
the reader to~\cite{Sh94}.

\begin{mproposition}\label{hol_sect}
{\rm (\cite{Mil97} Chapt.~4)}
Every modular form of weight $2k$ and level $\Gamma$,
for a finite index subgroup $\Gamma$ of $\Gamma \! \left( 1 \right)$,
can be extended to a meromorphic section of the line bundle of
$k$--differentials
$g \left( \tau \right) \left( d\tau \right)^{k}$
over the compact Riemann surface
\(\rcoset{\Gamma}{\hcom^*}\).
The degree of the pole of the resulting meromorphic section at every cusp is
not smaller than
\(-k\)
and the degree of the pole at an image \([\tau]_{\Gamma} \in
\rcoset{\Gamma}{\hcom}\)
of a point \(\tau \in \hcom\) is
not smaller than
\(- \!\left\llbracket\raisebox{10pt}{\hspace{-2pt}}\right.
k \left(\raisebox{10pt}{\hspace{-2pt}}\right.
1 - \txfrac{1}{e_{\tau,\Gamma}}
\left.\raisebox{10pt}{\hspace{-2pt}}\right)
\left.\raisebox{10pt}{\hspace{-2pt}}\right\rrbracket\)
where $e_{\tau,\Gamma}$ is the order of the stabilizer
of $\tau$ in $\Gamma/\{\pm 1\}$
and $\Dbrackets{a}$ stands for the integer part of the real num\-ber~$a$.
\end{mproposition}

Note that for the points \(\tau \in \hcom\)
having unit stabilizer in $\Gamma/\{\pm 1\}$ (i.e., \(e_{\tau,\Gamma} = 1\))
the corresponding holomorophic sections of
Proposition~3.1
have no poles at $[\tau]_{\Gamma}$.
This is because then the canonical projection
\(\hcom^* \to \rcoset{\Gamma}{\hcom^*}\)
is local (analytic) diffeomorphism around $[\tau]_{\Gamma}$.
On the other hand, if \(e_{\tau,\Gamma} > 1\) then
$[\tau]_{\Gamma}$ is a \textit{ramification} point for the projection
\(\hcom^* \to \rcoset{\Gamma}{\hcom^*}\), so that a holomorphic
(invariant) differential is projected, in general, to a meromorphic
differential.
For example, the weight $2$ holomorphic differential $\left( dz \right)^2$
is invariant under the projection $z \mapsto w = z^2$ and it is
projected to \(\txfrac{1}{4} \, w^{-1} \left( dw \right)^{2}\)
(\(= \txfrac{1}{4} \, z^{-2} \left( 2z \right)^{2} \left( dz \right)^{2}\)).

Exercise~2.12
implies that $e_{\tau,\Gamma}$ for any subgroup
$\Gamma$ of $\Gamma (1)$ is either $1$ or $2$, or $3$.
Let us set $\nu_{\ell}$ to be the number of points
\([\tau]_{\Gamma} \in \rcoset{\Gamma}{\hcom^*}\) with \(e_{\tau,\Gamma}=\ell\)
for \(\ell=2,3\) and let $\nu_{\infty}$ be the number of cusps' images in
$\rcoset{\Gamma}{\hcom^*}$.

\begin{mcorollary}\label{cr3.2n}
For \(k = 0\)
Proposition~3.1
implies that every modular
form of weight $0$ is represented by holomorphic function
over a compact Riemann surface and therefore, it is constant
by Liouville's theorem.
\end{mcorollary}

The Liouville theorem has a generalization to meromorphic
sections of line bundles over compact complex surfaces~--~this is
the Riemann--Roch\footnote{%
Gustav Roch (1839--1866).}
theorem (\cite{Mil97}, Chapt.~1)
stating that the vector space of such
sections with fixed singularities is \textit{finite dimensional}.

\begin{mtheorem}\label{fin_dim}
{\rm (See \cite{Mil97} Theorem~2.22 and Theorem~4.9, and \cite{Sh94} Proposition~1.40.)}
The vector space of modular forms of weight $2k$ and level $\Gamma$
(a finite index subgroup of $\Gamma (1)$)
has finite dimension
\beq\label{dim_kG}
d_{2k,\Gamma} \, = \,
\left\{
\begin{array}{ll}
0
& \text{for} \ \, k < 0 \\
1 & \text{for} \ \, k = 0 \\
\left( 2k - 1 \right) \left( g_{\Gamma} - 1 \right)
+ \nu_{\infty} k +
\left\llbracket\raisebox{10pt}{\hspace{-2pt}}\right.
\txfrac{k\,\nu_2}{2}
\left.\raisebox{10pt}{\hspace{-2pt}}\right\rrbracket
+
\left\llbracket\raisebox{10pt}{\hspace{-2pt}}\right.
\txfrac{2 \hspace{1pt} k\,\nu_3}{3}
\left.\raisebox{10pt}{\hspace{-2pt}}\right\rrbracket
& \text{for} \ \, k > 0 ,
\end{array}
\right.
\eeq
where
$g_{\Gamma}$ is the genus of the Riemann surface
$\rcoset{\Gamma}{\hcom^*}$
which can be calculated using the index $\mu$
of the subgroup $\Gamma$ in $\Gamma (1)$ by the formula
\beq\label{genus}
g_{\Gamma} \, = \,
1 + \frac{\mu}{12} - \frac{\nu_2}{4} - \frac{\nu_3}{3} - \frac{\nu_{\infty}}{2}
\, . \
\eeq
\end{mtheorem}

\begin{mremark}\label{rm3.1nn}
In the case of level $1$ modular forms:
\(g_{\Gamma \! \left( 1 \right)} = 0\),
\(\nu_{\infty} = 1\) and $e_{\tau,\Gamma \! \left( 1 \right)}$
takes nonunit values only at the images $[\tau]_{\Gamma \! \left( 1 \right)}$
of \(\tau = i\) and \(\tau = e^{\frac{2\pi i}{3}}\) which are
$2$ and $3$, respectively, i.e., $\nu_2$ $=\nu_3$ $=1$
(see Exercise~2.12).
Then Eq.~(\ref{dim_kG}) takes for \(k = 1,\dots,17,\dots\) the form
\beq\label{dim_kG1}
\hspace{2pt}
d_{2k,\Gamma \! \left( 1 \right)} =
1 \hspace{-1pt} - \hspace{-1pt} k \hspace{-1pt} + \!
\left\llbracket\raisebox{10pt}{\hspace{-2pt}}\right.
\hspace{-1pt}
\frac{k}{2}
\hspace{-1pt}
\left.\raisebox{10pt}{\hspace{-2pt}}\right\rrbracket
\! + \!
\left\llbracket\raisebox{10pt}{\hspace{-2pt}}\right.
\hspace{-1pt}
\frac{2k}{3}
\hspace{-1pt}
\left.\raisebox{10pt}{\hspace{-2pt}}\right\rrbracket
=
1,0,1,1,1,1,2,1,2,2,2,2,3,2,3,3,3,3,\dots
\eeq
(in the next subsection we will derive independently this formula
in a more direct fashion,
establishing on the way the recurrence relation
\(d_{2k+12,\Gamma (1)}\) \(=d_{2k,\Gamma (1)}\) \(+1\)).
For the principal congruence subgroups $\Gamma (N)$~(\ref{2.15n}),
we have, when \(N > 1\) (see~\cite{Sh94}, Sect.~1.6): \(\nu_2 = \nu_3 = 0\),
\(\nu_{\infty} = \Txfrac{\mu}{N}\) and $\mu$ is given by Eq.~(\ref{mu_N}).
In particular, $g_{\Gamma (N)}$ $=0$ for \(1 \leqslant N \leqslant 5\),
$g_{\Gamma (6)}$ $=1$, $g_{\Gamma (7)}$ $=3$, $g_{\Gamma (8)}$ $=5$,
$g_{\Gamma (9)}$ $=10$, $g_{\Gamma (10)}$ $=13$, $g_{\Gamma (11)}$ $=26$.
\end{mremark}

\subsection{Eisenstein series. The discriminant cusp form}

We proceed to describing the modular forms of level one.
Let $\mathcal{M}_{k}$ be the space of all such modular forms.
As a consequence of
Corollary~3.2 and Theorem~3.3,
$\mathcal{M}_{0}$ is $1$--dimensional (it consists of constant functions)
and $\mathcal{M}_{1}$ $= \mathcal{M}_{2k+1}$ $= \{0\}$.

Examples of non--trivial modular forms are given by the
\textit{Eisenstein series}
\beqa
G_{2k}(\tau) \, = && \podr
\frac{(2k-1)!}{2(2\pi i)^{2k}}
\mathop{\sum}\limits_{(m,n)}{\!\!}'\,(m\tau
+n)^{-2k} \, :=
\nn
\, = && \podr
\frac{(2k-1)!}{(2\pi i)^{2k}}
\left\{\dsum\limits_{n=1}^{\infty}\frac{1}{n^{2k}}
+
\dsum\limits_{m=1}^{\infty}\dsum\limits_{n\in \Z} \,
(m\tau +n)^{-2k}\right\}
\, . \  \label{3.5}
\eeqa
Note that for \(2k \geqslant 4\) we have:
\beq\label{n3.6}
G_{2k} \! \left( \tau \right) = G_{2k} \left( \tau,1 \right)
, \quad
G_{2k} \! \left( \omega_1, \omega_2 \right) :=
\frac{(2k-1)!}{2(2\pi i)^{2k}} \!
\Su_{\omega \in \Lambda \backslash \{0\}} \!
\omega^{-2k}
\quad (\Lambda = \Z \omega_1 + \Z \omega_2)
\
\eeq
where the series is absolutely convergent and therefore, it does not depend on the
basis $\left( \omega_1,\omega_2 \right)$:
\beq\label{n3.7}
G_{2k} \! \left( a\omega_1+b\omega_2, c\omega_1+d\omega_2 \right) \, = \,
G_{2k} \! \left( \omega_1,\omega_2 \right)
\quad \text{for} \quad
\left(\hspace{-4pt} \begin{array}{cc} a \hspace{-4pt} & b \\
c \hspace{-4pt} & d \end{array} \hspace{-3pt}\right) \in \Gamma (1)
\, . \
\eeq
It follows that, $G_{2k} \! \left( \tau \right)$ satisfies
for \(2k \geqslant 4\)
the conditions (\textit{i})
for modular forms since we have
\beq\label{n3.8}
G_{2k} \! \left( \rho\omega_1,\rho\omega_2 \right) \, = \,
\rho^{-2k} G_{2k} \! \left( \omega_1,\omega_2 \right)
\, . \
\eeq
For \(2k = 2\) the series~(\ref{3.5}) is only conditionally convergent
and it is not modular invariant (the sum depends on the choice of lattice basis,
see below).
To verify the second condition
one can use the \textit{Lipschitz formula}
(see, e.~g., Zagier in \cite{FNTP} Appendix)
\beq
\frac{(k-1)!}{(-2\pi i)^{k}}\,\dsum\limits_{n\in \Z}\,\frac{1}{(z+n)^{k}}
\, = \,
\dsum\limits_{l=1}^{\infty} \, l^{k-1}e^{2\pi ilz}
\label{3.6}
\eeq
and deduce the Fourier expansion of $G_{2k}$ (for \(k \geqslant 1\)):
\beq
G_{2k}(\tau)=
-\frac{B_{2k}}{4k}+\dsum\limits_{n=1}^{\infty}
\frac{n^{2k-1}}{1-q^{n}}q^{n}=
\frac{1}{2}\,\zeta (1-2k)+\dsum\limits_{n=1}^{\infty}\,\sigma_{2k-1}(n)q^{n}
\label{3.7}
\eeq
where \(\sigma _{l}(n)=\dsum\limits_{r\mid n}r^{l}\)
(sum over all positive divisors $r$ of $n$), $B_{l}$ are the
\textit{Bernoulli numbers}\footnote{%
Jacob Bernoulli (1654--1705) is the first in the great family of Basel
mathematicians (see \cite{Bell}, pp. 131--138 for a brief but lively
account). The Bernoulli numbers are contained in his treatise Ars
Conjectandi on the theory of probability, published posthumously in 1713.}
which are generated by the Planck\footnote{Max Planck (1858--1947)
proposed his law of the spectral distribution of the black--body radiation
in the fall of 1900 (Nobel Prize in Physics, 1918)~--~see M.J. Klein in~\cite{His77}.}
distribution function:
\beqa
\hspace{-16pt} &
\Txfrac{x}{e^{x}-1}=\dsum\limits_{l=0}^{\infty}B_{l}\,\Txfrac{x^{l}}{l!}
;\
B_{0}=1
, \
B_{1}=-\Txfrac{1}{2}
, \
B_{2}=\Txfrac{1}{6}
, \
B_{3}=\dots=B_{2k+1}=0 ,\,
& \hspace{-20pt} \nn \hspace{-16pt} &
B_{4}=B_{8}=-\Txfrac{1}{30}
, \hspace{3pt}
B_{6}=\Txfrac{1}{42}
, \hspace{3pt}
B_{10}=\Txfrac{5}{66}
, \hspace{3pt}
B_{12}=-\Txfrac{691}{2730}
, \hspace{3pt}
B_{14}=\Txfrac{7}{6} ,\hspace{3pt} \dots
\, , & \quad \label{3.8}
\eeqa
$\zeta (s)$ is the Riemann $\zeta$--function\footnote{%
The functional equation
\(\Gamma (\frac{s}{2}) \pi^{-\frac{s}{2}} \zeta (s) =
\Gamma (\frac{1-s}{2}) \pi^{\frac{s-1}{2}} \zeta (1-s)\),
which allows the analytic continuation of
\(\zeta (s)= \Su_{n=1}^{\infty} n^{-s}\)
as a meromorphic function (with a pole at \(s=1\)) to the entire complex plane
$s$, was proven by Riemann in 1859 (see also Cartier's lecture in~\cite{FNTP}).}.
Remarkably, for $n\geqslant 1$, all
Fourier coefficients of $G_{2k}$ are positive integers.

Thus,
for $k \geqslant 2$ the functions $G_{2k}$ are modular forms of weight $2k$
(satisfying (\ref{3.2})). For $k=1$, however, we have instead
\begin{equation}
(c\tau +d)^{-2}G_{2}(\frac{a\tau +b}{c\tau +d})
\, = \,
G_{2}(\tau )+\frac{i}{4\pi}
\frac{c}{c\tau +d}  \label{3.9}
\end{equation}
so that only $G_{2}^{\ast}(\tau )d\tau $ is modular invariant where
$G_{2}^{\ast}$ is the non--holomorphic function
\begin{equation}
G_{2}^{\ast} (\tau ) :=
- \frac{1}{8\pi^2} \, \mathop{\lim}\limits_{\varepsilon \searrow 0}
\left(\raisebox{16pt}{\hspace{-2pt}}\right.
\Su_{(m,n) \, \neq \, (0,0)}
\left( m\tau+n \right)^{-2} \left| m\tau+n \right|^{-\varepsilon}
\left.\raisebox{16pt}{\hspace{-2pt}}\right)
= G_{2}(\tau )+\frac{1}{8\pi \tau _{2}}
\, ,
\label{3.10}
\end{equation}
\(\tau _{2}=\IM \, \tau\).
(As the Eisenstein series~(\ref{n3.7}) is divergent for \(k=1\),
Eq.~(\ref{3.10}) can be taken as an alternative definition
of $G_{2} (\tau)$ which can be shown to agree with~(\ref{G_2}).)
In fact, there is no non--zero (level $1$, holomorphic) modular form of weight $2$
as a consequence of
Theorem~3.3.
There exist, on the other hand, level two forms of weight~2.
We shall use in applications to CFT the fact that
\begin{equation}
F_2 (\tau) \, := \, 2\hspace{1pt}G_{2}(\tau) - G_{2}(\frac{\tau +1}{2})
\label{3.11}
\end{equation}
is a modular form of weight $2$ and level $\Gamma _{\theta}$ (\ref{2.23}).

\begin{mexercise}\label{nex3.1}
Prove that the functions
$p_k^{\kappa\lambda} (z,\tau)$~(\ref{2.5nnn}) (\ref{setting})
(\(k=1,2,\dots\), \(\kappa,\lambda=0,1\))
have the Laurent\footnote{%
P.A.~Laurent (1813--1854) introduces his series in 1843.}
expansions
\beq\label{nwn3.19}
p_k^{\kappa\lambda} (z,\tau) \, = \,
\frac{1}{z^k} + (-1)^{k} \Su_{n \, = \, 1}^{\infty}
\BINOMIAL{2n-1}{k\hspace{-1pt}-\hspace{-1pt}1}
\frac{2(2\pi i)^{2n}}{(2n-1)!}
\, G_{2n}^{\kappa\lambda} (\tau) \,
z^{2n-k}
\, , \
\eeq
where $G_{2k}^{00} (\tau)$ coincides with the above introduced $G_{2k} (\tau)$
for \(k=1,2,\dots\), $G_{2}^{11} (\tau)$ coincides with $F_2 (\tau)$~(\ref{3.11})
and $G_{2k}^{\kappa\lambda} (\tau)$ has the following
absolutely convergent, Eisenstein series representation:
\beq\label{n3.15}
G_{2k}^{\kappa\lambda} \hspace{-1pt} \left( \tau \right) \, := \,
\frac{(2k\hspace{-1pt}-\hspace{-1pt}1)!}{2(2\pi i)^{2k}} \!
\Su_{(m,n) \, \in \, \Z\hspace{-1pt}\times\Z \hspace{1pt}\backslash\hspace{-1pt}
\{\hspace{-1pt}(0,0)\hspace{-1pt}\}}
\!\!
\left( -1 \right)^{\kappa m+\lambda n}
(m\tau+n)^{-2k}
\eeq
for \(k \geqslant 2\).
Using this prove that all $G_{2k}^{11} \! \left( \tau \right)$
(including $F_2 (\tau)$ $\equiv G_2^{11} (\tau)$)
are modular forms of weight $2k$ and level $\Gamma _{\theta}$
for every \(k = 1,2,\dots\).
(See Sect.~III.7 of~\cite{Weil} where the case \(\kappa=\lambda=0\) is considered.)
\end{mexercise}

If there are $d_{k}>1$ modular forms of weight $k$
and a fixed level, then one can form $d_{k}-1$ linearly independent
linear combinations $S_{k}$
of them, which have no constant term in their Fourier expansion. Such
forms, characterized by the condition $S_{k}\rightarrow 0$ for
$q\rightarrow 0$, are called \textit{cusp forms}\footnote{%
The notation $S_{k}$ for the cusp forms comes from the German word Spitzenform.
(The term ``parabolic form'' is used in the Russian literature.)}.
We denote by $\mathcal{S}_{k}$ the subspace of cusp forms.
The first nonzero cusp form of level one appears for weight 12 and,
as we shall see, its properties allow to determine the general structure
of level one modular forms.

\begin{mproposition}\label{pr:3.3}
The 24th power of the Dedekind\footnote{%
Richard Dedekind (1831--1916) also introduced
(in 1877) the absolute invariant $j$~(\ref{j-fun})~--~as well as
the modern concepts of a ring and an ideal.}
$\eta$--function
\beq
\Delta (\tau)
\, = \,
\left[ \eta (\tau )\right]^{24}
\, = \,
q \, \dprod\limits_{n=1}^{\infty} \, (1-q^{n})^{24}
\label{3.12}
\eeq
is a cusp form of weight 12.
\end{mproposition}

\begin{proof}
As $\Delta (\tau)$ clearly vanish for \(q=0\) we have just to show that it is
a modular form of degree $12$. To this end we compute the logarithmic derivative
\beq
\frac{\Delta' (\tau)}{\Delta (\tau)}
\, = \,
2\pi i \hspace{1pt}
\left(
1-24\dsum\limits_{n=1}^{\infty}\frac{nq^{n}}{1-q^{n}}
\right)
\, = \,
-48\pi i \hspace{1pt} G_{2}(\tau)
.
\label{3.14}
\eeq
It then follows from (\ref{3.9}) that
\beq
\frac{d}{d\tau}
\left( \log \Delta \left( \frac{a\tau +b}{c\tau +d} \right) \right)
\, = \,
\frac{d}{d\tau}
\left( \log \left[ (c\tau+d)^{12}
\Delta (\tau) \right] \right)
\label{3.15}
\eeq
and hence, noting that $\Delta (\tau)$ (\ref{3.12}) is $T$--invariant
(i.e., periodic of period 1 in $\tau$), we conclude that it is indeed
a modular form of weight $12$.
\end{proof}

\begin{mtheorem}\label{oth3.5}
The only non--zero dimensions \(d_{k} = \dim \mathcal{M}_{k}\) are given
(recursively) by
\beq
d_{0}=d_{4}=d_{6}=d_{8}=d_{10}=1
\, ; \quad
d_{12+2k}=d_{2k}+1
\, , \quad
k=0,1,2,\dots
. \
\label{3.17}
\eeq
In particular,
\(d_k = 0\) for \(k < 1\) and
$\dim \mathcal{S}_{k}=0$ for $k<12$.
\end{mtheorem}

\begin{proof}
1. If there were a modular form $f$ of weight $-m$ ($m>0$) then the
function $f^{12}\bigtriangleup ^{m}$would have had weight $0$ and a Fourier
expansion with no constant term, which would contradict the Liouville theorem
(cf. Theorem~3.3).

2. There is no cusp form of weight smaller than $12$.
Had there been one, say $S_{k}(\tau)$, with $k<12$, then
$\Txfrac{S_{k}(\tau)}{\Delta (\tau)}$
would be a modular form of weight $k-12<0$ in
contradiction with the above argument.
(Here we use the fact that
$\Txfrac{S_{k}}{\Delta}$ is holomorphic in $\hcom$
since the product formula (\ref{3.12})
shows that $\txfrac{1}{\Delta}$
has no poles in the upper half plane.)

The theorem follows by combining these results with
Proposition~3.3
and the argument that there is no level $1$ modular form of weight $2$.
\end{proof}

\begin{mremark}\label{nwrm3.4}
In fact, the linear span of all level $1$ modular forms of an arbitrary weight
is the free commutative algebra generated by $G_4 (\tau)$ and $G_6 (\tau)$,
i.e., it is the polynomial algebra $\C \left[ G_4,G_6 \right]$ (D.~Zagier \cite{FNTP}).
\end{mremark}

\begin{mcorollary}
The Dedekind $\eta$--function (\ref{3.12})
is proportional to the discriminant of the right hand side of (\ref{2.9}):
\beq
\Delta (\tau)
\, = \,
(2\pi)^{-12} \,
\left[ g_{2}^{3}(\tau)-27g_{3}^{2}(\tau)\right]
\, = \,
\left[ 20\hspace{1pt}G_{4}(\tau)\right]^{3}-
3\hspace{1pt}(7\hspace{1pt}G_{6}(\tau))^{2}
\, ( \, \in \, \mathcal{S}_{12})
. \
\label{3.13}
\eeq
\end{mcorollary}

\begin{proof}
The difference
\((20\hspace{1pt}G_{4})^{3}-3\hspace{1pt}(7\hspace{1pt}G_{6})^{2}\) also
belongs to $\mathcal{S}_{12}$ since, due to (\ref{3.8}),
\begin{equation*}
\left( -20\frac{B_{4}}{8}\right){\raisebox{12pt}{\hspace{-3pt}}}^{3}
-3\left( -7\frac{B_{6}}{12}\right){\raisebox{12pt}{\hspace{-3pt}}}^{2}
\, = \,
\frac{1}{(12)^{3}}-\frac{3}{(72)^{2}} \, = \, 0
\, .
\end{equation*}
Noting further that $\dim \mathcal{S}_{12}=1$
(Theorem 3.5)
and comparing the coefficient to $q$ in (\ref{3.13})
(\(1=60\left( 20\Txfrac{B_{4}}{8}\right){\raisebox{12pt}{\hspace{-3pt}}}^{2}
+21\left( \Txfrac{7B_{6}}{6}\right)\)),
we verify the relation
\(\Delta (\tau)=\left( 20\hspace{1pt}G_{4}\right)^{3}
-3\left( 7\hspace{1pt}G_{6}\right)^{2}\).
The first equation (\ref{3.13})\ then follows from the relations
\beq
20\hspace{1pt}G_{4}(\tau)
\, = \,
(2\pi)^{-4}g_{2}(\tau)
\, ; \quad
7\hspace{1pt}G_{6}(\tau)
\, = \,
-\frac{3}{(2\pi)^{6}} \, g_{3}(\tau)
\, . \
\label{3.16}
\eeq
\end{proof}

\begin{mremark}\label{nrem3.1}
$\mathcal{S}_{k}$ is a Hilbert space, equipped with the
\textit{Peterson scalar product}
\begin{equation}
(f,g)=\diint\limits_{B/\Gamma (1)}\tau _{2}^{k}\overline{f}(\tau )g(\tau)
d\mu
\quad \text{where} \quad
\tau = \tau _{1}+i\tau _{2}
\, , \quad
d\mu =\frac{d\tau _{1}d\tau _{2}}{\tau _{2}^{2}}
\, , \
\label{3.4}
\end{equation}
$d\mu $ being the $\SL (2,\R)$--invariant measure on $\hcom$.
\end{mremark}

\begin{mremark}\label{nrm3.2}
Comparing the constant term in the expansion~(\ref{3.7}) of $G_{2k}$
with the first few dimensions $d_{2k}$ we notice that $\mathcal{S}_{2k}$
$= \left\{ 0 \right\}$ for exactly those values of $k$ for which
$-\Txfrac{B_{2k}}{4k}$ is the reciprocal of an integer
(namely for $2k$ $=2,$ $4,$ $6,$ $8,$ $10,$ and $14$).
The curious reader will find a brief discussion of this (nonaccidental)
fact in Sect.~1B of Zagier's lectures in~\cite{FNTP}.
\end{mremark}

The existence of the discriminant form
$\Delta \hspace{-1pt} \left( \tau \right)$~(\ref{3.12})~--~whose zeros are precisely
the cusps of $\Gamma \hspace{-1pt} \left( 1 \right)$~--~allows to define the
modular invariant function
\beq\label{j-fun}
j \hspace{-1pt} \left( \tau \right) \, = \,
\frac{\left[ 240 \hspace{2pt} G_4 \hspace{-1pt} \left( \tau \right) \right]^4}{
\Delta \hspace{-1pt} \left( \tau \right)}
\, = \,
q^{-1} + 744 + \text{196\hspace{1.5pt}884} \hspace{3pt} q +
\text{21\hspace{1.5pt}493\hspace{1.5pt}760} \hspace{3pt} q^2 + \dots
\eeq
that is analytic in the upper half plane $\hcom$ but grows exponentially
for $\tau \to i \infty$.
The following proposition shows that $j$ is, in some sense,
the unique function with these properties.

\begin{mproposition}\label{nwpr3.7}
If $\Phi (\tau)$ is any modular invariant analytic function in $\hcom$
that grows at most exponentially for \(\IM \, \tau \to \infty\)
then $\Phi (\tau)$ is a polynomial in $j (\tau)$.
\end{mproposition}

\begin{proof}
The function \(f (\tau) = \Phi (\tau) \left[ \Delta (\tau) \right]^m\)
transforms as a modular form of weight $12 \hspace{1pt} m$ and if $m$
is large enough, it is bounded at infinity, hence $f (\tau)$
$\in \mathcal{M}_{12m}$.
It then follows from
Theorem~3.5 and Remark~3.4
that $f$ is a homogeneous polynomial
of degree $m$ in $G_4^3$ and $\Delta$.
Therefore, $\Phi$ $=\Txfrac{f}{\Delta^m}$ is a polynomial of degree
not exceeding $m$ in $j$.
\end{proof}

In fact, $j$ can be viewed as a (complex valued) function on the
set of $2$--dimensional Euclidean lattices invariant under
rotation and rescaling~--~see the thought provoking discussion in
Sect.~6 of~\cite{Ma04}.

\begin{mremark}
The function \(j (\tau) - 744 = q^{-1} + \text{196\hspace{1.5pt}884} \, q + \dots\)
is called \textit{Hauptmodul} of $\Gamma (1)$ (see~\cite{Ga04}, Sect.~2).
\end{mremark}

\begin{mremark}
As observed by McKey in 1978 (see~\cite{Ga04} for a review and references)
\beq\label{}
\left[ j (\tau) \right]^{\frac{1}{3}} \, = \,
q^{-\frac{1}{3}} \left( 1+ 248 \, q + \text{4\hspace{1pt}124} \, q^2 + \dots \right)
\eeq
is the character of the level $1$ affine Kac--Moody algebra
$(\hat{E}_8)_1$ (see also Sect.~5 below).
\end{mremark}

\subsection{$\vartheta$--functions}

Each meromorphic (and hence each elliptic) function can be presented as a ratio
of two entire functions.
According to property (2) of Sect.~2.1
these cannot be doubly periodic but, as we shall see, they
may satisfy a twisted periodicity condition.

We shall construct a family of entire analytic functions which allow for a
multiplicative cocycle defining the twisted periodicity condition.
A classical example of
this type is provided by the
\textit{Riemann} $\vartheta$--\textit{function}\footnote{%
$\vartheta$ functions appear before Riemann in Bernoulli's
{\it Ars Conjectandi} (1713), in the number theoretic studies of Euler (1773)
and Gauss (1801), in the study of the heat equation of Fourier (1826), and,
most importantly, in Jacobi's {\it Fundamenta Nova} (1829).}:
\beq
\vartheta (z,\tau) \, = \, \Su_{n \in \Z} \,
q^{\frac{1}{2}n^{2}}e^{2\pi inz}
\, = \, 1 + 2\Su_{n \, = \, 1}^{\infty} \, q^{\frac{1}{2}n^2} \, \cos 2\pi nz
\, , \quad
q^{\frac{1}{2}}=e^{i \pi\tau}
,\
\label{3.19}
\eeq
which belongs to the family of four Jacobi
$\vartheta$--functions\footnote{%
A more common notation for the Jacobi $\vartheta$--functions is
\(\vartheta_{11} = \vartheta_1\),
\(\vartheta_{10} = \vartheta_2\),
\(\vartheta_{00} = \vartheta_3\), and
\(\vartheta_{01} = \vartheta_4\).
Many authors also write \(q=e^{i \pi\tau}\) instead of
\(q=e^{2\pi i\tau}\); with our choice the exponent of $q$ will coincide
with the conformal dimension~--~see Sects.~4-7 below.
The function $\vartheta_{11}$ $=\vartheta_1$ plays an important role both in the study
of the elliptic Calogero--Sutherland model~\cite{L01} and in the study
of thermal correlation functions (Sect.~4.4 below).}
\beq
\vartheta_{\mu\nu}(z,\tau)
\, = \,
e^{i\pi \frac{\mu}{2}(\frac{\mu}{2}\tau-2z+\nu)} \,
\vartheta (z-\frac{\mu\tau}{2}-\frac{\nu}{2},\tau)
,\quad
\mu ,\nu =0,1
\label{3.20}
\eeq
(\(\vartheta _{00}=\vartheta\)).
They satisfy the ``twisted periodicity'' conditions
\beqa
\vartheta_{\mu\nu}(z+1,\tau) = && \podr (-1)^{\mu} \, \vartheta_{\mu\nu}(z,\tau)
,\,
\label{nw3.31}
\\
\vartheta_{\mu\nu}(z+\tau ,\tau ) = && \podr
(-1)^{\nu}q^{-\frac{1}{2}}e^{-2\pi iz} \,
\vartheta _{\mu \nu}(z,\tau)
. \quad  \label{3.21}
\eeqa
Note, in particular, that $\vartheta_{11}$ is the only odd in $z$
among the four $\vartheta$--functions and it can be written in the form
\beq\label{nw3.33}
\vartheta_{11} (z,\tau) \, = \,
2 \Su_{n \, = \, 0}^{\infty} \, \left( -1 \right)^n \,
q^{\frac{1}{2}( n + \frac{1}{2})^2} \, \sin \left( 2n+1 \right)\pi z
\quad
\eeq
while the others are even and can be written as follows (together with~(\ref{3.19}))
\beqa\label{nw3.34}
\vartheta_{01}
(z,\tau)
= && \podr
1 + 2\Su_{n \, = \, 1}^{\infty} \, \left( -1 \right)^n q^{\frac{1}{2}n^2} \,
\cos 2\pi nz
\, , \ \nn
\vartheta_{10}
(z,\tau)
= && \podr
2
\Su_{n \, = \, 1}^{\infty} \, q^{\frac{1}{2}( n - \frac{1}{2})^2} \,
\cos \left( 2n-1 \right)\pi z
\, . \
\eeqa
Thus $\vartheta_{11}$ has an obvious zero for \(z=0\) and hence
vanishes (due to the twisted periodicity) for all \(z = m\tau+n\).
In fact, this is the full set of zeros in $z$ of $\vartheta_{11}$
(which one can prove applying the Cauchy theorem to the logarithmic
derivative of $\vartheta_{11}$).
Using~(\ref{3.20}) we can then also find the zeros of all four Jacobi
$\vartheta$--functions.
This allows to deduce the following infinite product expression
for~\(\vartheta_{\mu\nu}\):
\beqa\label{nw3.35}
\vartheta_{\raisebox{0pt}{\scriptsize 00}\raisebox{-6pt}{\scriptsize\hspace{-6.5pt} 1}}
(z,\tau)
= && \podr
\mathop{\prod}\limits_{n \, = \, 1}^{\infty}
\left( 1 \hspace{-1pt} - \hspace{-1pt} q^n \right) \hspace{-2pt}
\left(\raisebox{10pt}{\hspace{-3pt}}\right.
1 \pm 2 \hspace{1pt} q^{n-\frac{1}{2}} \hspace{1pt} \cos 2\pi z + q^{2n-1}
\left.\raisebox{12pt}{\hspace{-3pt}}\right)
\, , \ \nn
\vartheta_{10}
(z,\tau)
= && \podr
2 \hspace{1pt} q^{\frac{1}{8}} \cos 2\pi z
\mathop{\prod}\limits_{n \, = \, 1}^{\infty}
\left( 1 \hspace{-1pt} - \hspace{-1pt} q^n \right) \hspace{-2pt}
\left(\raisebox{10pt}{\hspace{-3pt}}\right.
1 + 2 \hspace{1pt} q^{n} \hspace{1pt} \cos 2\pi z + q^{2n}
\left.\raisebox{12pt}{\hspace{-3pt}}\right)
\, , \ \nn
\vartheta_{11}
(z,\tau)
= && \podr
2 \hspace{1pt} q^{\frac{1}{8}} \sin 2\pi z
\mathop{\prod}\limits_{n \, = \, 1}^{\infty}
\left( 1 \hspace{-1pt} - \hspace{-1pt} q^n \right) \hspace{-2pt}
\left(\raisebox{10pt}{\hspace{-3pt}}\right.
1 - 2 \hspace{1pt} q^{n} \hspace{1pt} \cos 2\pi z + q^{2n}
\left.\raisebox{12pt}{\hspace{-3pt}}\right)
\, . \
\eeqa

One is naturally led to the above definition by considering
(as in~\cite{M83}) the action on $\vartheta$ of the
\textit{Heisenberg--Weyl}\footnote{%
Werner Heisenberg (1901--1976), Nobel Prize in physics 1932;
Hermann Weyl (1885--1955).}
\textit{group}
$U(1)\times \R^{2}$ that appears as a central extension of the $2$--dimensional
abelian group $\R^{2}$.
It is generated by the two $1$--parameter subgroups $U_a$ and $V_{b}$ acting on
(say, entire analytic) functions $f(z)$ as:
\beq
(U_{a}f)(z) = e^{\pi i(a^{2}\tau +2az)} \, f(z+a\tau)
\, , \quad
(V_{b}f)(z) = f(z+b)
\, . \quad
\label{3.22}
\eeq
A simple calculation gives \(U_{a_{1}+a_{2}}=U_{a_{1}}U_{a_{2}}\),
\beq
e^{2\pi iab}U_{a}V_{b}=V_{b}U_{a}
\, . \quad
\label{3.23}
\eeq
The function $\vartheta (z,\tau)$ (\ref{3.19}) is invariant under the discrete
subgroup $\Z^{2}=\{(a,b): a,b\in\Z\}$ which is commutative since
$e^{2\pi iab}=1$ for integer $ab$.
If \(a=\txfrac{\mu}{2}\), \(b=\txfrac{\nu}{2}\) then the action of
$V_{b}U_{a}$ gives rise to the four functions (\ref{3.20})
\beq
\vartheta_{\mu\nu}(z,\tau) \, = \,
\left( V_{-\frac{\nu}{2}}U_{-\frac{\mu}{2}}\vartheta \right) \! (z,\tau)
\, . \quad
\label{3.24}
\eeq
Similarly, for \(a,b\in \txfrac{1}{l}\Z\)
we obtain $l^{2}$ $\vartheta$--functions (that are encountered
in CFT applications).

The functions $\vartheta_{\mu\nu}(z,\tau)$ (\ref{3.20}) are solutions of the
Schr\"{o}dinger equation\footnote{%
Eq.~(\ref{3.25}) was known to J. Fourier (as ``the heat equation''),
long before Erwin Schr\"{o}dinger (1887--1961, Nobel Prize in Physics 1933)
was born.}
\beq
i\frac{\partial}{\partial \tau}\vartheta_{\mu\nu}(z,\tau)
\, = \, \frac{1}{4\pi}
\frac{\partial^{2}}{\partial z^{2}} \, \vartheta_{\mu \nu}(z,\tau)
\label{3.25}
\eeq
and so are, in fact, all functions of the type $V_{b}U_{a}\vartheta$.

The following fact is basic in the general theory of $\vartheta $--functions:

\begin{mproposition}\label{pr_Zag}
{\rm (See D. Zagier \cite{FNTP} Sect.~1C p. 245).}
Given an $r$--di\-men\-sio\-nal lattice $\Lambda _{r}$ in which the length
squared $Q(x)$ of any vector $x\in \Lambda _{r}$ is an integer, the
multiplicities of these lengths are the Fourier coefficients of a modular
form
\begin{equation}
\Theta _{Q}(\tau )=\dsum\limits_{x\in \Lambda _{r}}q^{Q(x)}  \label{3.26}
\end{equation}
of weight $\txfrac{r}{2}$.
More precisely, there exists a positive integer $N$ and a character $\chi$
of $\Gamma_0 \left( N \right)$ such that
\beq\label{eqnn1}
\Theta_{Q} \left( \gamma \tau \right) \, = \, \chi \left( d \right)
\left( c\tau + d \right)^{\frac{r}{2}}
\Theta_{Q} \left( \tau \right)
\quad
\text{\it for} \quad
\gamma \, = \,
\left(\hspace{-4pt} \begin{array}{cc} a \hspace{-4pt} & b \\
c \hspace{-4pt} & d  \end{array} \hspace{-3pt}\right)
\in \Gamma_0 \left( N \right)
\, . \
\eeq
For \(Q \left( x \right) = \txfrac{1}{2} x A x\)
where $A$ is an even symmetric
$r \times r$ matrix, the level $N$ is the smallest positive integer such that
$NA^{-1}$ is again even.
\end{mproposition}

Here is an \(r = 2\) example:
\beqa\label{n2}
&&
Q = x_1^2 + x_2^2
\, , \quad
\Theta_{Q} \, = \, 1 +4 q +4 q^2 +4q^4 +8 q^5 + 4q^8 + \dots
\, , \quad
\nn
&&
A \, = \, \left( \begin{array}{cc} 2 & 0 \\ 0 & 2  \end{array} \right)
\, , \quad
N=4
\, , \quad
\chi \left( d \right) \, = \, \left( -1 \right)^{\frac{d-1}{2}}
\, . \
\eeqa

\begin{mremark}
The modular form $G_4 (\tau)$ can be expressed in terms of
the Jacobi $\vartheta$--functions as:
\beq\label{nw3.43}
240 \, G_4 (\tau) \, = \,
\frac{1}{2} \hspace{1pt}
\left\{\raisebox{10pt}{\hspace{-2pt}}\right.
\vartheta_{00}^8 (0,\tau) \hspace{1pt} + \hspace{1pt}
\vartheta_{10}^8 (0,\tau) \hspace{1pt} + \hspace{1pt}
\vartheta_{01}^8 (0,\tau)
\left.\raisebox{10pt}{\hspace{-2pt}}\right\}
\, = \,
1 + 240 \, \frac{q}{1-q} + \dots \, . \
\eeq
\end{mremark}

\section{Quantum field theory and conformal invariance (a synopsis)}
\setcounter{equation}{0}\setcounter{mtheorem}{1}\setcounter{mremark}{1}\setcounter{mdefinition}{1}\setcounter{mexample}{1}\setcounter{mexercise}{1}

For the benefit of mathematician readers we shall give a
brief summary of the general properties of quantum fields (see \cite{SW},
\cite{BLOT}, \cite{QFS} for more details and proofs), and of the role of the
conformal group. (Our review will be necessarily one-sided: such
central concepts of real world quantum field theory (QFT) as
perturbative expansions, Feynman graphs, and Feynman path integral
won't be even mentioned.)

\subsection{Minkowski space axioms. Analyticity in tube domains}

Quantum fields generate~--~in the sense of \cite{SW}~--~an operator algebra in a
vacuum state space ${\cal V}$. A closely related approach, \cite{H}, \cite{Ar},
starts with an abstract $C^*$-algebra - the algebra of local
observables~--~and constructs different state spaces as Hilbert space
representations of this algebra, defining the different superselection sectors
of the theory.
(Important recent progress relating Haag's algebraic approach to
2D CFT~--~see \cite{KL}~--~is beyond the scope of the present notes.)

In the Wightman approach the fields are described as operator
valued distributions over \textit{Minkowski}\footnote{%
Hermann Minkowski (1864--1909) introduces the $4$--dimensional space-time
(in 1908 in G\"ottingen), thus completing the creation of the special
theory of relativity of Hedrik Antoon Lorentz (1853-1928, Nobel Prize
in physics, 1902), Henri Poincar\'e, and Albert Einstein (1979-1955, Nobel
Prize in physics, 1921).}
\textit{space--time} $M$.
It is a $D$-dimensional real
affine space equipped with a Poincar\'{e} invariant interval, which
assumes, in Cartesian coordinates, the form
\beq\label{x4.1}
x^2_{12} = \mbf{x}^2_{12} - (x^0_{12})^2, \quad x_{12}= x_1 - x_2, \quad
\mbf{x}^2 = \sum^{D-1}_{i=1} x^2_i
.
\eeq

The \emph{state space} ${\cal V}$ is a \emph{pre-Hilbert space}
carrying a (reducible) \emph{unitary positive energy representation}
$U(a,\widetilde{\Lambda})$ of the (quantum mechanical) \emph{Poincar\'{e} group}
$\Spin (D-1,1) \ltimes \mathbb{R}^D$.
This means that the joint spectrum of the
(hermitian, commuting) translation generators $P_0, \mbf{P}$ in
(the Hilbert space completion of) ${\cal V}$ belongs to the positive
light-cone $V_+$ (\textit{spectral condition}):
\beq\label{x4.2}
V_+ \, := \, \left\{ P \in \mathbb{R}^D \, : \, P_0 \ge |\mbf{P}| \equiv
\sqrt{\mbf{P}^2} \right\} \qquad (U(a,\ID) \, = \, e^{-iaP} ).
\eeq
(Boldface letters, $\mbf{x}$, $\mbf{P}$, denote throughout
$\left( D-1 \right)$--vectors.)
Furthermore, ${\cal V}$ is assumed to have a \emph{1-dimensional
translation invariant subspace} spanned by the \emph{vacuum vector}
$|0\rangle$ (which is, as a consequence, also Lorentz invariant):
\beq\label{x4.3}
|0\rangle \in {\cal V}, \quad
P^\mu |0\rangle = 0 \, ( \,  = ( U(a,\widetilde{\Lambda})-1) |0\rangle \, )
, \quad
\langle 0|0\rangle = 1.
\eeq

The field algebra is generated by a finite number of (finite component)
spin-tensor fields $\phi (x)$. Each $\phi$ is an \emph{operator valued
distribution on} ${\cal V}$: the smeared field $\phi(f)$ for any
Schwartz\footnote{
Laurent Schwartz (1915-2002).}
test function $f(x)$ is defined on ${\cal V}$ and leaves it
invariant. The fields $\phi$ obey the \emph{relativistic covariance condition}:
\beq\label{x4.4}
U (a,\widetilde{\Lambda}) \, \phi (x) \, U^{-1} (a,\widetilde{\Lambda}) \, = \,
S(\widetilde{\Lambda}^{-1}) \, \phi (\Lambda x + a).
\eeq
Here $S(\widetilde{\Lambda})$ is a finite dimensional representation of
the spinorial (quantum mechanical) Lorentz group $\Spin (D-1,1)$ of
$2^{d_0} \times 2^{d_0}$ matrices; for even $D$, the case of interest
here, the exponent $d_0$ coincides with the canonical dimension of a
free massless scalar field,
\beq\label{x4.5}
d_0 = \frac{D-2}{2}
\eeq
(in general, the spinorial representation has dimension
$2^{\Dbrackets{\frac{D-1}{2}}}$ where $\Dbrackets{\rho}$ stands for the integer part of the
positive real $\rho$); $\Lambda \in \SO (D-1,1)$ is the (proper) Lorentz transformation corresponding
to the matrices $\pm \widetilde{\Lambda}$ ($-\ID$ belonging to
the centre of the group $\Spin (D-1,1)$). We assume that
$S(-\ID)$ is a multiple (with a sign factor) of the identity operator:
\beq\label{x4.6}
S (-\ID) \phi (x) = \varepsilon_{\phi} \phi(x), \quad
\varepsilon_{\phi} = \pm 1.
\eeq

The sign $\varepsilon_\phi$ is related to the \emph{valuedness} of
$S(\widetilde{\Lambda})$: $\varepsilon_\phi = 1$ for single valued (tensor)
representations of $\SO (D-1,1); \varepsilon_\phi = -1$, for double valued
(spinor) representations.
The field $\phi$ and its hermitian conjugate $\phi^*$ (which is
assumed to belong to the field algebra whenever $\phi$ does) satisfy
the \emph{locality condition}
\beq\label{x4.7}
\phi(x_1) \phi^* (x_2) - \varepsilon_\phi \phi^*(x_2) \phi(x_1) = 0
\quad \text{for} \quad
x^2_{12} >0,
\eeq
which reflects the independence of the ``operations'' $\phi(x_1)$
and $\phi^*(x_2)$ at space-like separated points.
Finally, we assume that the \emph{vacuum} is a \emph{cyclic vector} of
the field algebra. In other words, (smeared) vector valued monomials
$\phi_1(x_1)\phi_2(x_2) \dots \phi_n(x_n) |0\rangle$ span ${\cal V}$.

A QFT is fully characterized by its \textit{correlation} (or Wightman)
\textit{functions}
\beq\label{wf}
w_n \left( x_{12},\dots,x_{n-1n} \right) \, = \,
\lvac \phi_1 (x_1) \dots \phi_n(x_n) \rvac
\eeq
which are, in fact, tempered distributions in $M^{\times \left( D-1 \right)}$,
only depending (due to translation invariance) on the independent
coordinate differences $x_{a\hspace{1pt}a+1}$, \(a=1,\dots,n-1\).

The spectral condition allows to view the above vector valued monomials
and Wightman distributions as boundary values of analytic functions.

{\samepage
\begin{mproposition}\label{npr4.1}
\begin{plist}
\item[{\rm (}a{\rm )}]
The vector valued distribution $\phi \left( x \right) \rvac$ is
the boundary value (for $y \to 0$, $y^0$ $> \left|\mbf{y}\right|$)
of a vector-valued function analytic in the forward tube domain
$\mathfrak{T}_{+}$ where
\beq\label{nw4.8}
\mathfrak{T}_{\pm} \, := \,
\left\{ \mzeta = x+iy \in M + i M : \pm y^0 > \left|\mbf{y}\right| \right\} .
\eeq
\end{plist}
\end{mproposition}}
\vspace{-12pt}
{\it
\begin{plist}
\item[{\rm (}b{\rm )}]
The Wightman distribution~(\ref{wf})
is a boundary value of an analytic function
$w_{n} \left( \mzeta_{12},\dots,\mzeta_{n-1n} \right)$
holomorphic in the product of backward tubes
$\mathfrak{T}_{-}^{\times \left( n-1 \right)}$
and polynomially bounded on its boundary
\beq\label{nw4.9}
\left| w_n \left( \mzeta_{12},\, \dots,\mzeta_{n-1n} \right) \right|
\, \leqslant \, A \left( 1 + \Su_{a \, = \, 1}^{n-1}
\left| \mzeta_{aa+1} \right|^{2} \right)
\left( \mathop{\min}\limits_{a} \left|y_{a\hspace{1pt}a+1}^{\, 2}\right| \right)^{-l}.
\eeq
\end{plist}}

The \textit{proof} \cite{SW}, \cite{BLOT} uses the spectral condition and
standard properties of Laplace transform of tempered distributions.

Each of the tube domains $\mathfrak{T}_{\varepsilon}$, $\varepsilon=+,-$
is clearly invariant under Poincar\'e transformations and uniform dilations
\(\mzeta \mapsto \varrho \mzeta\) (\(\varrho > 0\)).
A straightforward calculation shows that it is also invariant under
the \textit{Weyl inversion} $w$,
\beq\label{nw4.10}
\mzeta \, \mapsto \, w \mzeta \, = \, \frac{I_s \mzeta}{\mzeta^{\, 2}}
\, , \qquad
I_s \left( \mzeta^{0},\, \bzeta \right) \, := \,
\left( \mzeta^{0},\, -\bzeta \right) .
\eeq
It follows that $\mathfrak{T}_{\varepsilon}$ is actually
\textit{conformally invariant}~--~as $w$ and the (real) translations
generate the full
\(\left( \hspace{-5pt} \begin{array}{c}D\! + \! 2 \\
2\end{array} \hspace{-3pt} \right)\)
parameter \textit{conformal group} $\confgr$.
Moreover, each $\mathfrak{T}_{\varepsilon}$ is a homogeneous space
of $\confgr$~\cite{U}\footnote{%
This has been known earlier~--~e.g. to the late Vladimir Glaser
(1924--1984) who communicated it to I. Todorov back in 1962.}
(see also Sect.~4.2 below).
In fact, $\mathfrak{T}_{\varepsilon}$ is a \textit{coaddjoint orbit}
of $\confgr$ equipped with a \textit{conformally invariant symplectic form}
proportional to $dx^{\mu} \wedge d \Txfrac{y_{\mu}}{y^{\, 2}}$
(see \cite{T86} Sect.~3.3).

\begin{mremark}\label{nrm4.1}
Note that the $n$--point tubes of
Proposition~4.1~(\textit{b}),
\(\left\{\raisebox{9pt}{\hspace{-2pt}}\right.
\left(\raisebox{9pt}{\hspace{-2pt}}\right. x_1+iy_1,\)
\(\dots,\) \(x_n+iy_n \left.\raisebox{9pt}{\hspace{-2pt}}\right) :\)
\(x_{a\hspace{1pt}a+1}\) \(+ i y_{a\hspace{1pt}a+1}\)
\(\in \mathfrak{T}_{\varepsilon} \left.\raisebox{9pt}{\hspace{-2pt}}\right\}\),
are not conformally invariant for \(n > 1\).
\end{mremark}

From now on we shall consider conformally invariant QFT models.
Apart from the free Maxwell\footnote{%
James Clerk Maxwell (1831--1879) wrote his
\textit{Treatise on Electricity and Magnetism} in 1873.}
(photon) field and the massless neutrino,
real world fields are not conformally invariant.
The interest in unrealistic higher symmetry models comes from the fact
that the only (mathematically) existing so far QFT in four space time
dimensions~--~after three quarters of a century of vigorous efforts~--~are
the free field theories.
Conformal QFT has the additional advantage to provide (at least,
conjecturably) the short distance behaviour of more realistic (massive)
theories (for a discussion of this point~--~see~\cite{TMP})

\subsection{Conformal compactification of space--time.
The conformal Lie algebra}\label{ssec2.1}

The quantum mechanical conformal group $\confgr$ of $D$-dimensional space--time
can be defined as (a finite covering of) the group of real rational
coordinate transformations $g:$ $x \rightarrow x' (x)$ (with singularities)
of Minkowski space $M$ for which
\beq\label{x4.9}
d x'^{2} = \omega^{-2} (x,g) dx^2
\, ,\quad
dx^2 = \mbf{dx}^2 - (dx^0)^2
\, , \quad
\mbf{dx}^{\, 2} \, = \, \Su_{i \, = \, 1}^D \left( dx^i \right)^2,
\eeq
where $\omega(x,g)$ is found below to be a polynomial in $x^\mu$ of
degree not exceeding 2. An extension of the classical Liouville
theorem says that, for $D>2$, $\confgr$ is locally isomorphic to
the $\BINOMIAL{D+2}{2}$ parameter (connected) group \(\confgr = \Spin (D,2)\),
a double cover of the of pseoudo--rotation $\SO_0 \left( D,2 \right)$ of $\R^{D,2}$.
In fact,
the action of $\confgr$ on $M$ having singularities can be extended to
entirely regular action on a compactification of $M$ called
\textit{conformal compactification} or just \textit{compactified Minkowski space} $\M$.
A classical manifestly covariant description of compactified Minkowski space
is provided by the projective quadric in $\R^{D,2}$, introduced by Dirac\footnote{
Paul Adrian Maurice Dirac (1902--1984), Nobel Prize in physics 1933,
known for his equation and for the prediction of antiparticles,
speaks (in Varenna~--~\cite{His77}) of his great appreciation of projective
geometry since his student years at Bristol.}~\cite{Di 36}
(it generalizes to Lorentzian metric and to higher dimension of a construction
of Klein\footnote{Felix Klein (1849--1925),
a believer in a preestablished harmony between physics and mathematics,
has outlined this construction without formulae in his famous 1872 Erlanger program.}):
\beq\label{nn4.15}
\M = Q \left/\right.\! \R^*
, \quad
Q =
\left\{\raisebox{10pt}{\hspace{-2pt}}\right.
\xx \!\in \R^{D, 2}
\backslash \{0\}
:
\xx{\hspace{-1pt}}^{\, 2} := \mbf{\mxi}^{\, 2} +
\mxi_D^2 - \mxi_0^2 - \mxi_{-1}^2
\hspace{1pt}
(= \mxi^{a} \eta_{ab} \mxi^{b}) = 0
\left.\raisebox{10pt}{\hspace{-2pt}}\right\}
.
\eeq
The \textit{conformal Lie algebra} $\confalg$,
generated in the projective picture,
by the infinitesimal pseodo--rotations
\(X_{ab} = \mxi_b \Txfrac{\di}{\di \mxi^{a}} -
\mxi_b \Txfrac{\di}{\di \mxi^{a}}\) is characterized by the
commutation relations
\beq\label{cm_rel}
\left[ X_{ab}, X_{cd} \right] \, = \,
\eta_{ac} X_{bd} \, - \, \eta_{bc} X_{ad} \, + \, \eta_{bd} X_{ac} \, - \,
\eta_{ad} X_{bc}
\eeq
for \(a,b,c,d = -1,0,\dots,D\)
(\(\eta_{11} = \dots = \eta_{DD} = 1 = - \eta_{00} = - \eta_{-1-1}\),
\(\eta_{ab} = 0\) for \(a \neq b\)).
Then the embedding of $M$ in $\M$ is given by
\beqa\label{nw4.15}
&
x \mapsto
\left\{\raisebox{10pt}{\hspace{-3pt}}\right. \lambda \xx_{x}
\left.\raisebox{10pt}{\hspace{-3pt}}\right\} \in \M_{\C}
, \quad
\xx_x \, = \,
x^{\mu} \ee_{\mu} + \Txfrac{1+x^{\, 2}}{2} \, \ee_{-1} +
\Txfrac{1-x^{\, 2}}{2} \, \ee_{D}
\quad \text{or,} & \nn &
x^{\mu} \, = \, \Txfrac{\mxi^{\mu}}{\kappa}
\, , \quad
\kappa = \mxi^{D} + \mxi^{-1}
\, , &
\eeqa
where \(\left\{ \ee_{a} : -1 \leqslant a \leqslant D \right\}\)
is an orthonormal basis in $\R^{D,2}$,
so that the conformal structure on $M$, or the isotropy relation,
is encoded on $\M$ by the ($\SO (D,2)$--invariant relation of)
orthogonality of the rays because of the simple
formula:
\beq\label{nw4.16}
x_{12}^{\, 2} \, = \,
-2 \, \xx_{x_1} \spr \xx_{x_2} \, = \,
\left(\raisebox{9pt}{\hspace{-2pt}}\right.
\xx_{x_1}-\xx_{x_2}
\left.\raisebox{9pt}{\hspace{-2pt}}\right)^{\, 2}
\, . \
\eeq
Since the vectors $\xx_x$ of the map~(\ref{nw4.15}) can be
characterized by the condition
\beq\label{xx_x}
\kappa := \xx_{\infty} \spr \xx_x \, = \, 1
\, , \
\eeq
where \(\xx_{\infty} = \left( -1,0,\Mbf{0},1 \right)\),
we conclude that the complement set \(K_{\infty} := \M \backslash M\),
the set of points at ``infinity'', is the $(D-1)$--cone with tip
\(\left\{\raisebox{10pt}{\hspace{-3pt}}\right. \lambda \xx_{\infty}
\left.\raisebox{10pt}{\hspace{-3pt}}\right\}\):
\beq\label{k_inf}
K_{\infty} \, = \,
\left\{\raisebox{10pt}{\hspace{-3pt}}\right.
\left\{\raisebox{9pt}{\hspace{-3pt}}\right. \lambda \xx
\left.\raisebox{9pt}{\hspace{-3pt}}\right\} \in \M
: \,
\xx_{\infty} \spr \xx \
( = \kappa = \mxi^{D} + \mxi^{-1} = \mxi_{D} - \mxi_{-1})
= 0
\left.\raisebox{10pt}{\hspace{-3pt}}\right\}
.
\eeq
Note also that the Weyl inversion~(\ref{nw4.10})
is a proper conformal transformation given by a rotation of angle
$\pi$ in the \(\left( -1,0 \right)\) plane:
\(w \left( \mxi_{-1},\mxi_{0},\mbf{\mxi},\mxi_{D} \right)\)
\(= \left( -\mxi_{-1},-\mxi_{0},\mbf{\mxi},\mxi_{D} \right)\).

\begin{mremark}
One can, sure, identify the circle and the
\(\left(D\hspace{-1pt}-\hspace{-1pt}1\right)\)--sphere in the
definition~(\ref{nn4.15}) of the quadric $Q$, as well.
Indeed, the quotient space $Q/\R_+$ can be defined by the equations
\beq\label{nwn4.18}
Q/\R_+ \, = \,
\left\{\raisebox{10pt}{\hspace{-2pt}}\right.
\xx \!\in \R^{D, 2}
\backslash \{0\}
:
\mxi_0^2 + \mxi_{-1}^2 = 1 = \mbf{\mxi}^{\, 2} + \mxi_D^2
\left.\raisebox{10pt}{\hspace{-2pt}}\right\} \, \cong \,
\Sr^1 \times \Sr^{D-1} .
\eeq
Going from $Q/\R_+$ to $\M$ $= Q/\R^*$ amounts to dividing
$Q/\R_+$ by $\Z_2$ $=\Z/2\Z$, i.e., by identifying $\xx$ and $-\xx$
in the product of the circle and the sphere.
Thus we conclude that $\M$ is diffeomorphic to
\(\Sr^1 \times \Sr^{D-1} /\Z_2\).
It also follows that $\M$ is a \textit{nonorientable} manifold for
odd~\(D>1\).
\end{mremark}

The (linear) $\SO (D,2)$ action $g\xx$ $= (g^a_b\xi^b)$ on $Q$
induces the nonlinear transformations~(\ref{x4.9}) by
$g \xx_{x}$ $\sim \xx_{g (x)}$ where the proportionality coefficient
turns out to be equal to the (square root of the) conformal factor, $\omega (x,g)$:
\beq\label{omega}
g \xx_{x} \, = \, \omega (x,g) \xx_{g (x)} ,
\eeq
since the
relation~(\ref{xx_x}) together with
$\xx_{\infty} \spr \xx_{g(x)}$ $=1$ and Eq.~(\ref{nw4.16}) imply
\beq\label{gx_12}
\left( g (x_1) -g (x_2) \right)^2 \, = \,
\frac{x_{12}^{\, 2}}{\omega (x_1,g) \, \omega (x_2,g)} \,
\eeq
(proving, in particular, the conformal property of the $SO (D,2)$--action).
Note that the \textit{Weyl subgroup}, the Poincar\'e group with dilations, is
characterized by the condition that $\omega (g,x)$ does not depend on $x$ and
$\omega (g,x)$ $=1$ iff $g$ is a Poincar\'e transformation.

There is a natural basis of the conformal Lie algebra $\confalg$ generating the
simplest transformations in Minkowski space:
\begin{plist}
\item[$\bullet$]
\textit{Poincar\'e translations}
\(e^{i \, a \spr P} \left( x \right) \, ( \, \equiv
e^{i \, a^{\mu} P_{\mu}} \left( x \right) \, ) \, = x + a\)
(for \(x,a \in M\)),
\item[$\bullet$]
\textit{Lorentz transformations}
\(e^{t X_{\mu\nu}}\),
\(0 \leqslant \mu < \nu \leqslant D-1\)
(\(X_{\nu\mu} = -X_{\mu\nu}\)),
\item[$\bullet$]
\textit{dilations} \(x \mapsto \rho x\), \(\rho > 0\),
\item[$\bullet$]
\textit{special conformal transformations}
\(e^{i \, a \spr K} \left( x \right) = \Txfrac{x+x^{\, 2} \, a}{
1 + 2 \, a \spr x + a^{\, 2} \, x^{\, 2}}\).
\end{plist}
The generators
$iP_{\mu}$ ($\mapsto \txfrac{\di}{\di x^{\mu}}$), $iK_{\mu}$
and the dilations are expressed in terms of $X_{ab}$ as:
\beq\label{e2.3n}
i \hspace{1pt} P_{\mu} = -X_{-1\mu} - X_{\mu D}
, \ \
i \hspace{1pt} K_{\mu} = -X_{-1\mu} + X_{\mu D}
, \ \
\rho^{X_{-1D}} \left(  x\right) = \rho \, x
\ \ (\rho > 0)
\,
\eeq
(see Appendix~\ref{ap:2});
the Lorentz generators \(X_{\mu\nu}\)
correspond to \(0 \leqslant \mu,\nu \leqslant D-1\).

There is a remarkable complex variable parametrization of $\M$
given by:
\beq\label{nw4.18}
\M
\hspace{1pt} = \hspace{1pt}
\left\{ z_{\cmu}
\hspace{-1pt} = \hspace{-1pt}
e^{2\pi i \hspace{1pt} \czeta} \hspace{1pt} u_{\cmu} :
\czeta \in \R,\hspace{1pt}
u^{\, 2}
\hspace{-1pt} := \hspace{-1pt}
\mbf{u}^{\, 2} \hspace{-2pt} + u_D^2
\hspace{-1pt} = \hspace{-1pt} 1,\hspace{1pt}
u \in \R^D \right\}
\hspace{1pt} \cong \hspace{1pt}
\Sr^1 \hspace{-1pt} \times \Sr^{D-1}
\hspace{-2pt}\raisebox{-1pt}{$\left/\raisebox{9pt}{}\right.$}
\raisebox{-4pt}{$\Z_2$}
\hspace{1pt},
\eeq
where $z_{\cmu}$ can be
extended to the whole complex Euclidean space $\E_{\C}$ ($\cong \C^D$)
thus defining a chart in the \textit{complexification} $\M_{\C}$
of compactified Minkowski space;
they are connected to the (complex) Minkowski coordinates
$\mzeta$ $= x$ $+ i y$ $\in M_{\C}$ ($:= M$ $+ i M$)
by the rational conformal transformation
(\cite{T86} \cite{NT02} \cite{N03}):
\beq\label{nw4.19}
\mbf{z} \hspace{1pt} = \hspace{1pt}
\frac{\bzeta}{\omega \! \left( \mzeta \right)}
\hspace{1pt} , \hspace{8pt}
z_D \hspace{1pt} = \hspace{1pt}
\frac{1 \hspace{-1pt} - \hspace{-1pt} \mzeta^{\, 2}}{
2\, \omega \! \left( \mzeta \right)}
\hspace{1pt} , \hspace{8pt}
\omega \! \left( \mzeta  \right) \hspace{1pt} = \hspace{1pt}
\frac{1 \hspace{-1pt} + \hspace{-1pt} \mzeta^2}{2} - i \mzeta^0
.
\eeq

\begin{mproposition}\label{opr4.2}
The rational complex coordinate transformation
$g_c: M_{\C}$ $( \ni \mzeta )$ $\to E_{\C}$ $( \ni z)$,
defined by~(\ref{nw4.19}),
is a complex conformal map (with singularities)
between the complex Minkowski and Euclidean spaces,
such that
\beqa\label{t2.3}
&&
z_{12}^{\,2} = \frac{\mzeta_{12}^{\, 2}}{
\omega \left( \mzeta_1 \right)\omega \left( \mzeta_2 \right)}
\, , \quad
d z^{\, 2} \left(= d \mbf{z}^{\, 2} + d z_D^{\, 2} \right)
= \frac{d\mzeta^{\, 2}}{
\omega \left( \mzeta \right)^2}
\, . \quad
\eeqa
The transformation
$g_c$
is regular in the tube
domain \(\mathfrak{T}_+ =
\left\{\raisebox{9pt}{\hspace{-2pt}}\right. \mzeta = x + i \, y :\)
\(y^0 > \left| \mbf{y} \right| \left.\raisebox{9pt}{\hspace{-2pt}}\right\}\)
and on the real Minkowski space $M$.
The image $T_+$ of $\mathfrak{T}_+$ under $g_c$
\beq\label{t2.4}
T_+ \hspace{-1pt} := \hspace{-1pt}
\left\{\raisebox{12pt}{\hspace{-3pt}}\right. z \in \mathbb{C}^D
\hspace{-2pt} : \hspace{2pt}
\left| \hspace{1pt} z^{\, 2} \right| \hspace{-1pt} < \hspace{-1pt} 1,\
z \cdot \overline{z} =
\left| \hspace{1pt} z^1 \right|^2 \hspace{-2pt} + \dots +
\left| \hspace{1pt} z^D	 \right|^2
\hspace{-2pt} < \hspace{-1pt} \frac{1}{2}
\left( 1 + \left| \hspace{1pt} z^{\, 2} \right|^2 \right)
\left.\raisebox{12pt}{\hspace{-3pt}}\right\}
\eeq
is a precompact submanifold of $E_{\C}$.
The closure $\M$ of the precompact image of the real Minkowski
space $M$ in $E_{\C}$ has the form~(\ref{nw4.18}).
\end{mproposition}

The statement is verified by a direct
computation~(see~\cite{NT02}, \cite{N03}).

The transformation~(\ref{nw4.19}) generalizes the Cayley\footnote{%
Arthur Cayley (1821--1895) has introduced, in 1843, the notion of
$n$--dimensional space and is a pioneer of the theory of invariants.}
transformation
\beq\label{Cayley}
\mzeta \, \mapsto \, z \, = \, \frac{1+i\mzeta}{1-i\mzeta}
\qquad (\mzeta,z \in \C)
\eeq
arising in the description of the chiral (1-dimensional light-ray)
projection of the 2D CFT (see Sect.~5).
In the $D=4$ dimensional case~(\ref{nw4.19}) can be viewed as the
\textit{Cayley compactification map} $u(2)$ $\to U(2)$
in the space of (complex) quaternions $\C \otimes_{\R} \H$
(see~\cite{U} \cite{T86}):
\beqa\label{To86}
& \hspace{-10pt}
M \hspace{-1pt} \ni x \hspace{1pt} \mapsto \hspace{1pt}
i\widetilde{x} := ix^0 \ID \hspace{-1pt} + \hspace{-1pt}
\mbf{Q} \!\spr\! \mbf{x} \in \hspace{-1pt} u(2)
, \ \
\M \hspace{-1pt} \ni z \hspace{1pt} \mapsto \hspace{1pt}
\dirz := z^{\alpha} Q_{\alpha} \equiv z^4 \ID
\hspace{-1pt} + \hspace{-1pt} \mbf{Q} \!\spr\! \mbf{z} \in \hspace{-1pt} U(2) ,
\hspace{-10pt} & \hspace{-60pt} \nn & \hspace{-10pt}
i \widetilde{x} \, \mapsto \,
\dirz = \Txfrac{1+i \widetilde{x}}{1-i\widetilde{x}}
\, ,
\hspace{-10pt} &
\eeqa
where $Q_k$ (\(k=1,2,3\)) are the \textit{quaternion units}
(expressed in terms of the Pauli matrices, see Sect.~\ref{Sec.7}).
Another point of view on the transformation~(\ref{nw4.19})
is developed in \cite{N03} (see Appendix~A):
to each pair of mutually nonisotropic points, say $q_0$,
$q_{\infty}$ $\in \M_{\C}$, one assigns an affine chart of $\M_{\C}$
with a distinguished centre $q_0$ and a centre $q_{\infty}$
of the infinite light cone.
This is provided by the fact that the stability subgroup in $\confgr$
of a point, say $q_{\infty}$,
is isomorphic (conjugate) to the Weyl group (the group of affine conformal
transformations).
In the above $1$--dimensional case \(q_0 = i\), \(q_{\infty} = -i\),
while in a general dimension $D$, $q_0$ and $q_{\infty}$ are
mutually conjugate points in the forward and backward tubes,
respectively.
(Remarkably, only when $q_{\infty}$ $\in T_{\pm}$ the corresponding
affine chart entirely cover the real compact space $\M$;
this has no analog for signatures different from the Lorentz type
$(D-1,1)$ or $(1,D-1)$.)

Note that all pairs of mutually nonisotropic points of $\M$
(or $\M_{\C}$) form a single orbit under the action of the
(complex) conformal group (\cite{NT01}, Proposition~2.1).
In particular, the transformation $g_c$ of
Proposition~4.2
can be considered as an element of $\confgr_{\C}$ such that
$g_c (p_{0,\infty})$ $= q_{0,\infty}$.
Thus the stabilizer of the pair $p_0$, $p_{\infty}$, which is
the Lorentz group with dilations, is conjugate to the stabilizer
of $q_0$, $q_{\infty}$.
Since $q_0$ and $q_{\infty}$ are complex conjugate to one another,
it turns out that their stabilizer in $\confgr_{\C}$ is $*$--invariant.
Moreover, its real part coincides with the
\textit{maximal compact subgroup} $\compgr$ of $\confgr$,
\beq\label{nw4.24}
\compgr \, = \, \Spin (D) \times U(1) / \Z_2
\, , \
\eeq
generated by $X_{\alpha\beta}$ ($\alpha,\beta$ $=1,$ $\dots, D$) and
the \textit{conformal hamiltonian},
\beq\label{H}
H \, := \, i \, X_{-10} \, \equiv \, \frac{1}{2} \, (P_0+K_0) \, .
\eeq
As noted by Segal\footnote{%
Irving Ezra Segal (1918--1998).}
\cite{S71} $H$ is
positive whenever $P_0$ is (since \(K_0=wP_0w^-1\)
with $w$ defined in (\ref{nw4.10})).
The factor
$\Spin (D)$ of $\compgr$ acts on the coordinates $z$ by
(Euclidean) rotations while the $U(1)$ subgroup multiplies
$z$ by a phase factor.
Thus, $\compgr$ appears as the stability group of the
origin $z=0$ (i.e., $q_0$) in the \textit{real} conformal group $\confgr$.
Noting further the transitivity of the action of $\confgr$
on either $T_+$ or $T_-$ we conclude that the forward tube
is isomorphic to the coset space
\beq\label{nw4.25}
T_+ \, \cong \, \confgr / \compgr
\, . \
\eeq

We will need the complex Lie algebra generators \(T_{\cmu}\)
and \(\SCT_{\cmu}\) for \(\cmu = 1,\, \dots,\, D\) of
$z$--\textit{translations}
\(e^{w \spr T} (z)\) \(= z + w\)
and \textit{special conformal transformations}
\(e^{w \spr \SCT} (z)\)
\(= \Txfrac{z+z^2 \, w}{1 + 2 \, w \spr z + w^{\, 2} \, z^{\, 2}}\)
(\(w,\, z \in \C^D\))
which
are conjugate by $g_c$ to the analogous generators
$-iP_{\mu}$ and $-iK_{\mu}$.
This new basis of generators
($T_{\cmu}$, $\SCT_{\cmu}$, $H$ and $X_{\cmu\cnu}$)
is expressed in terms of
$X_{ab}$~as:
\beq\label{n2.10n}
T_{\cmu} = i X_{0\cmu} - X_{-1\cmu}
, \ \,
\SCT_{\cmu} = - i X_{0\cmu} - X_{-1\cmu}
\ \, \text{for} \ \cmu = 1,\dots,D
\eeq
(\(\left[ T_{\cmu}, C_{\cnu} \right] =
2 \left( \delta_{\cmu\cnu} H - X_{\cmu\cnu} \right)\),
\(\left[ H, C_{\cmu} \right] = - C_{\cmu}\),
\(\left[ H, T_{\cmu} \right] = T_{\cmu}\),
see Appendix~\ref{ap:2}).
The generators
$T_{\cmu}$, $\SCT_{\cmu}$ for \(\cmu = 1,\dots,D\),
together with the above introduced
$X_{\cmu\cnu}$ (\(\cmu,\cnu = 1,\dots,D\)) and $H$
span an \textit{Euclidean real form}
(\(\cong \spin \left( D+1,1 \right)\))
of the complex conformal algebra.
The generators $T_{\cmu}$, $X_{\cmu\cnu}$ and $H$ span the subalgebra
$\confalg_{\infty}$ of the complexified conformal Lie algebra $\cconfalg$,
the {\it stabilizer of the central point at infinity} in the
$z$--chart which is isomorphic
to the complex Lie algebra of Euclidean transformations with dilations.
The stabilizer of \(z=0\) in $\cconfalg$
is its conjugate $\confalg_0$,
\beq\label{c0}
\confalg_0 \, :=  \, \mathit{Span}_{\C} \,
\left\{ \SCT_{\cmu},\, X_{\cmu\cnu},\, H \right\}
\, .
\eeq

\begin{mremark}
The $z$--coordinates are expressed in terms of $\xi_a$
as \(z_{\alpha} = \Txfrac{\xi_{\alpha}}{i\xi_0-\xi_{-1}}\)
(in contrast with~(\ref{nw4.15}) the denominator \(i\xi_0 -\xi_{-1}\)
never vanishes for real $\xx$) .
We could have introduced a length scale $R$ replacing the numerator
$\xi_{\alpha}$ by $R\xi_{\alpha}$.
We shall make use of the parameter $R$ in
Sect.~7, 
where it is viewed as ``the radius of the Universe''
(in the sense of Irving Segal~\cite{S82}) and the thermodynamic
(\(R \to \infty\)) limit is studied.
\end{mremark}

\subsection{The concept of GCI QFT.
Vertex algebras, strong locality, rationality}

We proceed with a brief survey of the axiomatic QFT with GCI.
The assumptions of the GCI~QFT
are the Wightman axioms~\cite{SW}, briefly sketched in Sect.~4.1,
and the condition of GCI for the correlation functions~\cite{NT01}.
The latter means that the Wightman functions\footnote{
The superscript ``$M$'' will further mean that the corresponding
objects are considered over the Minkowski space (chart in $\M$).}
\(\lvac \phi_{\aa_1}^{\MINK} \left( x_1 \right)\)
\(\dots \phi_{\aa_n}^{\MINK} \left( x_n \right) \rvac\)
are invariant (in the sense of distributions) under the substitution
\beq\label{tr_law}
\phi_{\aa}^{\MINK} \left( x \right) \, \mapsto \,
\left[\raisebox{9pt}{\hspace{-2pt}}\right.
\pi^{\MINK}_{x} \left( g \right)^{-1}
\left.\raisebox{9pt}{\hspace{-2pt}}\right] \mtrx^{\bb}_{\aa} \
\phi_{\bb}^{\MINK} \left( g\left( x \right) \right)
\eeq
for every conformal transformation \(g \in \confgr\), outside
its singularities. The matrix valued function
$\pi_{x}^{M} \left( g \right)$ is called (Minkowski) cocycle
and is characterized by the properties
\beq\label{M-coc}
\pi_x^M \! \left( g_1 g_2 \right) =
\pi_{g_2 \left( x \right)}^M \! \left( g_1 \right)
\pi_x^M \! \left( g_2 \right)
\, , \quad
\pi_x^M \! \left(\raisebox{9pt}{\hspace{-2pt}}\right.
e^{i a \spr P}
\left.\raisebox{9pt}{\hspace{-2pt}}\right)
\mtrx_{\aa}^{\bb}
\, = \,
\delta_{\aa}^{\bb}
\, . \
\eeq
The transformation law~(\ref{tr_law}) extends
the Poincar\'e covariance~(\ref{x4.4})
for $S(\widetilde{\Lambda})$ $\equiv \pi^{M}_{0}(\widetilde{\Lambda})$
to the case of (nonlinear) conformal transformations.
An example of such a transformation law is given by the
\textit{electromagnetic} field that is transforming as a $2$--form
\beq\label{FM}
F^{\MINK}_{\mu\nu} \left( x \right) \, dx^{\mu} \! \wedge dx^{\nu}
\, = \,
F^{\MINK}_{\mu\nu} \left( g(x) \right) \, dg(x)^{\mu} \! \wedge dg(x)^{\nu}
\, . \
\eeq

As proven in \cite{NT01}, Theorem~3.1, GCI
is equivalent to the rationality of
the (analytically continued) Wightman functions.
Moreover (\cite{N03}~Theorem~9.1),
the product of fields acting on the vacuum,
\(\phi_{\aa_1}^{\MINK} \left( x_1 \right)\)
\(\dots \phi_{\aa_n}^{\MINK} \left( x_n \right) \rvac\),
are boundary values of analytic functions
\(\phi_{\aa_1} \left( \mzeta_1 \right)\)
\(\dots \phi_{\aa_n} \left( \mzeta_n \right) \rvac\)
defined for all sets of mutually nonizotropic
points \(\mzeta_1,\dots,\mzeta_n \in \mathfrak{T}_+\)
and the limit is obtained for
\(\IM \, \mzeta_{kk+1} \in \mathfrak{T}_-\).
This makes possible to consider the
$n$--point vacuum correlation functions
of the theory as \textit{meromorphic sections}
of the $n$th tensor power (over $\M_{\C}^{\times \hspace{1pt} n}$)
of a bundle.
This bundle is defined over the complex compactified Minkowski space
$\M_{\C}$ by the cocycle~(\ref{M-coc}) and is called the \textit{field bundle}.
It is then naturally endowed with an action of the conformal group
$\confgr_{\C}$ via (bundle) automorphisms.

\begin{mremark}
Trivializing the bundle over every affine chart on $\M_{\C}$,
for instance, in the $z$--coordinates~(\ref{nw4.19}),
by the action of the corresponding abelian group of affine translations,
\(\trn_w \left( z \right) \, ( \, \equiv e^{w \spr T} \left( z \right) ) \, = z+w\),
the action of $\confgr_{\C}$ will take the form
\beq\label{fib_act}
\left( z = \{z^{\alpha}\},\, \phi = \{\phi_{\aa}\} \right)
\, \mathop{\longmapsto}\limits_{g} \,
\left( g \left( z \right),\,
\pi_z \! \left( g \right) \phi =
\{\pi_z \! \left( g \right)\mtrx^{\bb}_{\aa} \phi_{\bb}\} \right)
\, \in \, \C^D \times F
\eeq
where $F$ is the standard fibre and $\pi_z \left( g \right)$ is
the $z$--picture cocycle.
This provides the general scheme for the passage from the GCI~QFT
over Minkowski space to the theory over
a complex affine chart which contains
the forward tube $\mathfrak{T}_+$~(\ref{nw4.9})~--~see \cite{N03} Sect.~9.
The fibre $F$ is the space of (classical) \textit{field values}
and the coordinates $\phi_{\aa}$ correspond to the
collection of local fields in the theory.
\end{mremark}

The (analytic) $z$-\textit{picture} of a GCI~QFT is obtained
by transformation of Minko\-w\-s\-ki space fields $\phi_{\aa}^{\MINK}$ to the
$z$--coordinates~(\ref{nw4.19}) as (operator valued) sections
of the field bundle:
\begin{equation}\label{WF}
\phi_{\aa} \left( z \right) \, = \,
\pi_{g_c^{-1} \! (z)}^{\MINK} \!
\left(\raisebox{9pt}{\hspace{-2pt}}\right. g_c
\left.\raisebox{9pt}{\hspace{-2pt}}\right)\mtrx^{\bb}_{\aa} \
\phi_{\bb}^{\MINK} \!
\left(\raisebox{9pt}{\hspace{-2pt}}\right. g_c^{-1} \! \left( z \right)
\left.\raisebox{9pt}{\hspace{-2pt}}\right)
\quad
(z = g_c \! \left( x \right))
\end{equation}
where $g_c$ is the transformation~(\ref{nw4.19})
viewed as an element of $\confgr_{\C}$.
Different normalization conventions in the $x$ and $z$
picture require an additional numerical factor in~(\ref{WF}).
For instance, the canonical commutation relations yield
an extra factor $2\pi$ in~(\ref{WF})
for a free massless scalar field $\varphi$, the standard
conventions for the $2$--point function being
\beq\label{18e}
\lvac \varphi^M (x_1) \varphi^M (x_2) \rvac =
\frac{1}{4\pi^2 \left( x_{12}^{\, 2} + i 0 x_{12}^0 \right)}
\quad \text{while} \quad
\lvac \varphi (z_1) \varphi (z_2) \rvac =
\frac{1}{z_{12}^{\, 2}}
\, . \
\eeq
The resulting theory is equivalent to the theory of vertex algebras
(\cite{Bo86} \cite{Ka96} \cite{Bo97} \cite{FBZ01})
extended to higher dimensions (see \cite{N03}).
We proceed to sum up the properties of $z$-picture fields
and of the more general \textit{vertex operator fields}
arising in their \textit{operator product expansion} (OPE).

\medskip

1) The {\it state space} $\VA$ of the theory is a ({\it pre-Hilbert})
{\it inner product space} carrying a (reducible) unitary
{\it vacuum representation} $U(g)$ of the conformal group $\confgr$, for which:

\smallskip

\noindent
1a) the corresponding representation of the complex Lie algebra $\cconfalg$
is such that the spectrum of the $U(1)$ generator $H$~(\ref{H})
belongs to \(\left\{\raisebox{10pt}{\hspace{-2pt}}\right.
0,\, \txfrac{1}{2},\, 1,\,
\txfrac{3}{2},\, \dots
\left.\raisebox{10pt}{\hspace{-2pt}}\right\}\)
and has a finite degeneracy:
\begin{equation}
\label{eq2.1new}
\VA = \bigoplus_{\rho \, = \, 0 , \frac{1}{2} , 1 , \ldots}
\VA_{\rho} \, , \ (H-\rho) \, \VA_{\rho} = 0
\, , \quad
\dim \, \VA_{\rho} < \infty \, ,
\end{equation}
each $\VA_{\rho}$ carrying a fully reducible representation
of $\Spin \left( D \right)$ (generated by \(X_{\cmu\cnu}\)).
Moreover, the central element $-\ID$ of the subgroup
$\Spin \left( D \right)$
is represented on $\VA_{\rho}$ by $\left( -1 \right)^{2\rho}$.

\smallskip

\noindent
1b) The {\it lowest energy space} $\VA_0$ is 1-dimensional:
it is spanned by the (norma\-lized) {\it vacuum vector} $\rvac$,
which is invariant under the full conformal group $\confgr$.

\medskip

As a consequence (see~\cite{N03}, Sect.~7) the Lie subalgebra
$\confalg_0$~(\ref{c0})
of $\cconfalg$ has locally finite action on $\VA$, i.e.,
every \(v \in \VA\) belongs to a finite dimensional
subrepresentation of $\confalg_0$.
Moreover, the action of $\confalg_0$ can be integrated to an action
of the complex Euclidean group with dilations $\pi_0 \left( g \right)$
and the function
\beq\label{def_pi}
\pi_z \left( g \right) \, := \,
\pi_0 \left( t_{g \left( z \right)}^{-1} \, g \, t_z \right)
\, \
\eeq
is rational in $z$ with values in $\mathit{End}_{\C} \, \VA$
(the space endomorphisms of $\VA$)
and satisfies the cocycle property
\beq\label{cocycle}
\pi_z \left( g_1 g_2 \right) \, = \,
\pi_{g_2 \left( z \right)} \left( g_1 \right)
\pi_z \left( g_2 \right)
\quad \text{iff} \quad
g_1 g_2 \left( z \right),\, g_2 \left( z \right) \, \in \, \C^D
\, . \
\eeq

\medskip

\noindent
2) The fields \(\phi \left( z \right)
\equiv \left\{ \phi_a \left( z \right) \right\}\)
(\(\psi \left( z \right)
\equiv \left\{ \psi_b \left( z \right) \right\}\), etc.)
are represented by infinite
power series of the type
\begin{equation}
\label{eq1.2}
\phi \left( z \right) \, = \, \sum_{n \, \in \, {\Z}} \, \sum_{m \, \geqslant \, 0}
\left( z^{\, 2} \right)^n \, \phi_{\{n , m \}} \left( z \right)
\, , \quad
z^{\, 2} \, := \, \sum_{\cmu \, = \, 1}^D z_{\cmu}^2
\, ,
\end{equation}
$\phi_{\left\{ n,m \right\}} \left( z \right)$ being
(in general, multicomponent) operator valued polynomial in
$z$ which is {\it homogeneous of degree} $m$ \textit{and harmonic},
\begin{equation}
\label{n1.3}
\phi_{\{n , m \}} \left( \lambda z \right) \, = \,
\lambda^m \, \phi_{\{n , m \}} \left( z \right)
\, , \quad
\Delta \, \phi_{\{n , m \}} \left( z \right) \, = \, 0
\, , \quad
\Delta \, = \, \sum_{\cmu=1}^D
\frac{\partial^2}{\partial z_{\cmu}^2} \,
\end{equation}
and
\beq\label{equ2.4}
\phi_{\left\{ -n,m \right\}} \left( z \right) \, v \, = \, 0
\eeq
for all \(m = 0,1,\dots\) if \(n > N_v \in \Z\).

\medskip

\noindent
3) {\it Strong locality}:
The fields $\phi$, $\psi$,~$\dots$ are assumed to have $\Z_2$--parities
$p_{\phi}$, $p_{\psi}$,~$\dots$ such that
\begin{equation}
\label{equ2.5}
\rho_{12}^N \,
\left\{
\phi_a \left( z_1 \right) \psi_b \left( z_2 \right)
- (-1)^{p_{\phi} p_{\psi}} \,
\psi_b \left( z_2 \right) \phi_a \left( z_1 \right)
\raisebox{9pt}{\hspace{-2pt}}
\right\} \, = \, 0 \,
\quad (\rho_{12} \, := \, z_{12}^{\, 2})
\end{equation}
for sufficiently large $N$.

\medskip

In Minkowski space, the strong locality condition is implied by
the Huygens'\footnote{%
The Dutch physicist, mathematician, and astronomer Christian Huygens (1629--1695)
is the originator of the wave theory of light.}
principle and the Wightman positivity.
Recall that the Huygens' principle is a stronger form of the
locality condition~(\ref{x4.7}) stating that the left hand side vanishes
for all nonisotropic separations (\(x_{12}^{\, 2} \neq 0\)).
This is a consequence of GCI~\cite{NT01} since, as we already mentioned,
the mutually nonizotropic pairs of points form a single orbit for the
(connected) conformal group $\confgr$.
Let us also note that
the assumption that the field algebra is $\Z_2$--graded, which underlies 3),
fixes the commutation relations among different fields thus excluding
the so called ``Klein transformations''
(whose role is discussed e.~g. in~\cite{SW}).

Strong locality implies an analogue of the
\textit{Reeh--Schlieder theorem}, the separating property of the
vacuum, namely

{\samepage
\begin{mproposition}
\begin{plist}
\item[{\rm (}a{\rm )}] {\rm (\cite{NT03}, Proposition 3.2~(a).)}
The series $\phi_a \left( z \right) \rvac$ does not contain negative
powers of $z^{\, 2}$.
\end{plist}
\end{mproposition}}
\vspace{-12pt}
{\it
\begin{plist}
\item[{\rm (}b{\rm )}] {\rm (\cite{N03}, Theorem~3.1.)}
Every local field component $\phi_a \left( z \right)$ is
uniquely determined
by the vector \(v_a = \phi_a \left( 0 \right) \rvac\).
\item[{\rm (}c{\rm )}] {\rm (\cite{N03}, Theorem~4.1 and Proposition~3.2.)}
For every vector \(v \in \VA\) there exists unique
local filed $Y \left( v,\, z \right)$ such that
\(Y \left( v, 0 \right) \rvac = v\). Moreover, we have
\begin{equation}
\label{eq2.3new}
Y(v,z) \rvac \, = \, e^{z \spr T} \, v
\, , \quad
z \spr T \, = \, z^1 \, T_1 + \dots + z^D \, T_D
\, . \
\end{equation}
\end{plist}
}

The part (\textit{c}) of the above proposition is the higher dimensional
analogue of the {\it state field correspondence}.

\medskip

\noindent
4) \textit{Covariance}:
\beqa
\label{equ2.7}
\left[ \hspace{1pt} T_{\cmu} ,
Y \left( v, z \right) \, \right]
= && \podr
\frac{\di}{\di z^{\cmu}} \,
\, Y \left( v, z \right)
, \quad
\\ \label{equ2.8}
\left[ \hspace{1pt} H ,
Y \left( v, z \right) \, \right]
= && \podr
z \spr
\frac{\di}{\di z} \,
\, Y \left( v,\ z \right) +
Y \left( H\hspace{1pt} v, z \right)
, \quad
\\ \label{equ2.9}
\hspace{-30pt}
\left[ \hspace{1pt} X_{\cmu\cnu} ,
Y \left( v, z \right) \, \right]
= && \podr
z^{\cmu} \,
\frac{\di}{\di z^{\cnu}} \,
Y \left( v, z \right) -
z^{\cnu} \,
\frac{\di}{\di z^{\cmu}} \,
Y \left( v,\, z \right) +
Y \left( X_{\cmu\cnu}\hspace{1pt} v, z \right)
, \quad \mgvspc{12pt}
\\ \label{equ2.10}
\left[ \hspace{1pt} \SCT_{\cmu} ,
Y \left( v, z \right) \, \right]
= && \podr
\left(\raisebox{10pt}{\hspace{-2pt}}\right.
z^{\, 2} \,
\frac{\di}{\di z^{\cmu}} \,
- 2 \, z^{\cmu} \, z \spr
\frac{\di}{\di z}
\left.\raisebox{10pt}{\hspace{-2pt}}\right)
Y \left( v, z \right) -
2 \, z^{\cmu} \, Y \left( H\hspace{1pt} v, z \right) +
\mgvspc{12pt}
\nn && \podr
+ \, 2 \, z^{\cnu} \, Y \left( X_{\cnu\cmu} \hspace{1pt} v, z \right) +
Y \left( \SCT_{\cmu}\hspace{1pt} v,\, z \right)
. \qquad
\eeqa

\medskip

If \(v \in \VA\) is a minimal energy state in an irreducible representation
of $\confalg$ then
\(\SCT_{\cmu} v =0 \) (\(\cmu = 1,\dots,D\))
(as $C_{\alpha}$ plays role of a lowering operator:
\(H C_{\alpha} v = C_{\alpha} (H -1) \, v\)); such vectors
are called \textbf{quasiprimary}.
Their linear span decomposes into irreducible representations
of the maximal compact subgroup $\compgr$, each of them characterized by
weights \(\left(\raisebox{9pt}{\hspace{-2pt}}\right.
d;\, j_1,\dots,j_{\frac{D}{2}} \left.\raisebox{9pt}{\hspace{-2pt}}\right)\).
We assume that our basic fields $\phi_a$, $\psi_b$,~$\dots$, correspond to
such quasiprimary vectors so that the transformation
laws~(\ref{equ2.7})--(\ref{equ2.10}) give rise to $\compgr$--induced
representations of the conformal group~$\confgr$.

If \(v \in \VA\) is an eigenvector of $H$ with eigenvalue $d_v$
then Eq.~(\ref{equ2.8}) implies that the field $Y \left( v, z \right)$
has dimension $d_v$:
\beq\label{ne2.11}
\left[ \hspace{1pt} H \, , \,
Y \left( v,\, z \right) \, \right]
\, = \,
z \spr
\frac{\di}{\di z} \,
\, Y \left( v,\, z \right) \, + \,
d_v \, Y \left( v,\, z \right)
\qquad
(H \hspace{1pt} v \, = \, d_v \hspace{1pt} v)
\, . \quad
\eeq
It also follows from the correlation between the dimension and the spin
in the property~1a) and from the spin and statistics theorem
that the $Z_2$--parity $p_v$ of $v$ is related to its dimension
by \(p_v \equiv 2d_v \ \MOD \, 2\); therefore
\beq\label{loc}
\rho_{12}^{\mu \left( v_1,v_2 \right)}
\left\{
Y \! \left( v_1, z_1 \right) Y \! \left( v_2, z_2 \right)
- (-1)^{4 \, d_{v_1} d_{v_2}} \,
Y \! \left( v_2, z_2 \right) Y \! \left( v_1, z_1 \right)
\raisebox{10pt}{\hspace{-2pt}}
\right\} = 0
\eeq
where $\mu \left( v_1,v_2 \right)$ depends on the spin and dimensions of
$v_1$ and~$v_2$
while the cocycle for $Y (v,z)$ satisfies $\pi_z (e^{2\pi X_{\cmu\cnu}})$
$= (-1)^{2d_v}$.
(For a description of the spinor representation of
$\Spin(D,2)$ for any $D$~--~see
Appendix~\ref{ap:2n}.)

\medskip

\noindent
5) \textit{Conjugation}:
\beq\label{equ2.11}
\La v_1 \Vl Y \left( \cnj{v},\, z \right) v_2 \Ra \, = \,
\La Y \left(\raisebox{9pt}{\hspace{-2pt}}\right.
\pi_{z^*} \hspace{-2pt} \left( I \right)^{-1} v,\, z^*
\left.\raisebox{9pt}{\hspace{-2pt}}\right) v_1 \Vl v_2 \Ra
\,
\eeq
for every \(v,v_1,v_2 \in \VA\), where
\beq\label{star}
z^{\, *} \, := \, \frac{\overline{z}}{\overline{z}^{\, 2}}
\eeq
is the $z$--picture conjugation (leaving invariant the real space~(\ref{nn4.15}))
and $I$ is the
element of $\confgr_{\C}$ representing the complex inversion
\beq\label{equ2.12}
I \left( z \right) \, := \, \frac{R_D \left( z \right)}{z^{\, 2}}
\, , \quad
R_{\mu} \left( z^1,\, \dots ,\, z^D \right) \, := \,
\left( z^1,\, \dots ,\, -z^{\mu},\, \dots ,\, z^D \right)
\, \
\eeq
($I^2$ is a central element of $\confgr$ while $I$
does not belong to the real conformal group).

\medskip

The above properties also imply the {\it Borcherds' OPE relation}. The equality
\begin{equation}
\label{eq2.8new}
Y (v_1 , z_1) \, Y (v_2 , z_2)
\, \text{``} \! = \! \text{''} \,
Y (Y (v_1 , z_{12}) \, v_2 , z_2) \, ,
\end{equation}
is satisfied after applying some transformations to the formal power series
on both sides which are not defined on the corresponding series' spaces
(see~\cite{N03}, Theorem~4.3).
Moreover, the vector valued function
$$
Y (v , z_1) \, Y (v_2 , z_2) \rvac = Y (Y (v_1 , z_{12}) \, v_2 , z_2) \rvac
$$
is analytic with respect to the Hilbert norm topology for
$\vert z_2^{\, 2} \vert < \vert z_1^{\, 2} \vert < 1$
and sufficiently small $\rho_{12}$.

\medskip

\begin{mproposition}
Under the above assumptions the vacuum correlation functions are
(Euclidean invariant, homogeneous) rational functions of $z_{\cmu}$.
\end{mproposition}

\begin{proof}
If we take vertex operators $Y (v_k, z)$ $=: \phi_k (z)$,
\(k=1,\dots,n\), having
fixed dimensions $d_k$ then the strong locality~(\ref{loc}) implies
that for large enough $N \in \N$ the product
\beq\label{eq2.3}
F_{1 \ldots n} (z_1 , \ldots , z_n) \, := \,
\left(\raisebox{14pt}{\hspace{-2pt}}\right.
\prod_{1 \, \leqslant \, i \, < \, j \, \leqslant \, n} \,
\rho_{ij} \left.\raisebox{14pt}{\hspace{-2pt}}\right)^{\hspace{-2pt} N}
\lvac \phi_1 (z_1) \ldots \phi_n (z_n)
\rvac
\eeq
\(\rho_{ij} = z_{ij}^2 \equiv (z_i - z_j)^2\), is $\Z_2$ symmetric
under any permutation of the factors within the vacuum expectation value.
Energy positivity, on the other hand, implies that
$\lvac \phi_1 (z_1) \ldots \phi_n (z_n) \rvac$, and hence
$F_{1 \ldots n} (z_1 , \ldots , z_n)$ do not contain negative powers of $z_n^2$.
It then follows from the symmetry and the homogeneity of $F_{1 \ldots n}$
that it is a polynomial in all $z_i^{\mu}$.
Thus the (Wightman) correlation functions are rational functions of the coordinate
differences. (See for more detail \cite{N03} \cite{NT03}; an equivalent Minkowski
space argument based on the support properties of the Fourier transform of
(the $x$-space counterpart of) (\ref{eq2.3}) is given in \cite{NT01}.)
\end{proof}

\noindent
6) \textit{The concept of stress--energy tensor.}
The importance of assuming the existence of a stress--energy tensor $T$
along with the Wightman axioms in a conformal field theory has been
recognized long ago~\cite{MS} (see also~\cite{M88}).
It is simpler to introduce $T$ in a GCI theory extended on compactified Minkowski
space.
It is a \textit{rank two conserved symmetric traceless tensor} which
will be written in the analytic picture in the form:
\beq\label{s-et1}
T (z;v) \, := \, T_{\cmu\cnu} (z) \, v^{\cmu} v^{\cnu}
\, , \quad
\left(\raisebox{10pt}{\hspace{-2pt}}\right.
\frac{\di}{\di v} \spr \frac{\di}{\di v}
\left.\raisebox{10pt}{\hspace{-2pt}}\right)
T (z;v) \, = \, 0
\, . \
\eeq
Its scale dimension is equal to the space-time dimension $D$.
The \textit{conservation law} reads:
\beq\label{s-et2}
\frac{\di}{\di z_{\cmu}} T_{\cmu\cnu} (z) \, = \, 0
\quad \Leftrightarrow \quad
\left(\raisebox{10pt}{\hspace{-2pt}}\right.
\frac{\di}{\di z} \spr \frac{\di}{\di v}
\left.\raisebox{10pt}{\hspace{-2pt}}\right)
T (z;v) \, = \, 0
\, . \
\eeq
Finally, the conformal Lie algebra generators
should appear among its modes (see~(\ref{H-T}));
in particular, the generators of the Lie algebra
\(u(1) \times \spin(D)\) of the maximal compact
subgroup $\compgr$ of $\confgr$ can be expressed by
(finite) linear combinations of the zero modes of~$T$.

\subsection{Real compact picture fields. Gibbs states and the KMS condition}

The conjugation law~(\ref{equ2.11}) simplifies in the
{\it real compact picture} in which
\beq
\phi (\zeta ,u) \, := \, e^{2\pi id\zeta} \, \phi (e^{2\pi i\zeta}u)
\label{4.5}
\eeq
where $d$ is the conformal dimension of $\phi$.
The commutation relation~(\ref{ne2.11}) of the $z$--picture fields with
the conformal energy operator implies then
\beq
e^{2\pi itH}\phi (\zeta ,u)=\phi (\zeta +t,u)e^{2\pi itH}
,
\label{4.7}
\eeq
i.e., $H$ appears as the translation generator in $\zeta$ in this realization.
(While the $z$--picture fields correspond to the complex euclidean invariant
line element $dz^2$~(\ref{t2.3}), the compact picture fields correspond to the real
$\compgr$-invariant line element $\Txfrac{dz^2}{z^2}=\Txfrac{dx^2}{|{\omega}|^2}$.)
Since all dimensions of GCI fields
should be integer or half odd integer depending on their spin the corresponding
compact picture fields are periodic or antiperiodic, respectively,
\beq
\phi (\zeta +1,u)=(-1)^{2d}\phi (\zeta ,u)
\, . \qquad  \label{4.6}
\eeq
The (anti)periodicity property (\ref{4.6}) implies that $\phi$ has a Fourier
series expansion
\beq
\phi (\zeta ,u) \, = \, \Su_{\nu \, \in \, d+\Z} \Su_{m \, = \, 0}^{\infty}
\phi _{\nu m} (u) \, e^{-2\pi i\nu\zeta}
\label{4.8}
\eeq
where $\phi_{\nu m} (u)$ are operator valued homogeneous harmonic polynomials
of degree~$m$ restricted to the unit sphere
(we leave it to the reader to find the connection between
the expansions~(\ref{n1.3}) and~(\ref{4.8})).
Combined with (\ref{4.7}), this gives
\beq
\left[ H, \phi_{\nu m}(u) \right] \, = \, -\nu \phi_{\nu m}(u)
\quad
(\Leftrightarrow \ \,
q^H \phi_{\nu m} (u) q^{-H} \, = \, q^{-\nu} \phi_{\nu m} (u) , \ \, |q| < 1)
.
\label{4.9}
\end{equation}
For a scalar field (of integer dimension)
the hermiticity condition~(\ref{equ2.11}) reads:
\beq\label{eqn4.?}
\phi_{\nu m} (u)^*=\phi_{-\nu m}(u)
\, . \
\eeq

We assume the existence of \textit{Gibbs}\footnote{
Josiah Williard Gibbs (1839--1903) has published his last work
``Elementary Principles in Statistical Mechanics developed with
special reference to the Rational Foundations of Thermodynamics''
in 1902.}
\textit{temperature states}~--~i.~e., the
existence of all traces of the type
\beq
tr_{\DOM{}}(Aq^{H})
\quad \text{for} \quad
q=e^{2\pi i\tau}
\, , \quad
\IM \, \tau >0
\quad (\text{i.~e.} \ \left| q \right| <1 )
\label{4.10}
\eeq
where $A$ is any polynomial in the (local, GCI) fields (including $A=1$) and
$\DOM{}$ is the space of finite energy states (dense in the Hilbert state space).
We then define the temperature mean of $A$ by the standard relation:
\beq
\La A \Ra _{q} \, = \,
\frac{1}{Z(\tau)} \, tr_{\DOM{}}(Aq^{H})
\, , \quad
Z(\tau) \, = \, tr_{\DOM{}}(q^{H})
\quad (2\pi \, \IM \, \tau \, = \, \frac{h\nu}{kT})
\label{4.11}
\eeq
thus identifying the imaginary part of $\tau$ with the Planck's energy
quantum divided by the absolute temperature.
The parameter $\tau$ of the upper half plane $\hcom$ thus appears in two
guises:\ as a moduli labeling complex structures on a torus (that are
inequivalent on $\rcoset{\Gamma (1)}{\hcom}$) and as (inverse) absolute temperature.

For free fields the partition function $Z(\tau)$ can, in fact, be computed
given the dimensions $d_b(n)$ and $d_f(n)$ of $1$-particle bosonic and fermionic
states of energy $n$ and $n-\txfrac{1}{2}$, respectively.
The result is:
\beq
Z(\tau )
\, = \,
\mathop{\prod}\limits_{n \, = \, 1}^{\infty} \,
\frac{
\left(\raisebox{10pt}{\hspace{-3pt}}\right.
1 + q^{n-\frac{1}{2}}
\left.\raisebox{10pt}{\hspace{-3pt}}\right)^{
d_f
\left( n \right)}}{
\left(\raisebox{10pt}{\hspace{-3pt}}\right.
1 - q^n
\left.\raisebox{10pt}{\hspace{-3pt}}\right)^{d_b
\left( n \right)}
}
.
\label{4.11n}
\eeq
The mean thermal energy $\La \! H \! \Ra_{\! q}$ is given
by the logarithmic derivative of $Z (\tau)$,
\beq\label{mean_H}
\La \! H \! \Ra_{\! q} \, = \,
\frac{1}{2\pi i Z (\tau)} \frac{d Z (\tau)}{d\tau} \, = \,
q \, \frac{d}{dq} \, \ln Z
\eeq
so that for the generalized free field models we have
\beq\label{mean_H1}
\La \! H \! \Ra_{\! q} \, = \,
\Su_{n \, = \, 1}^{\infty}
\frac{n \, d_b (n) \, q^n
}{1 - q^n}
\, + \,
\Su_{n \, = \, 1}^{\infty}
\frac{\left( n- \frac{1}{2} \right) d_f (n) \, q^{n-\frac{1}{2}}
}{1 + q^{n-\frac{1}{2}}}
\, . \
\eeq

It has been established that the thermal (Gibbs) correlation functions
are finite linear combinations of a fixed set of elliptic functions
in each of the conformal time differences $\zeta_{kk+1}$;
the coefficients that are (depending on $u_k$) $q$--series whose convergence
is conjectured (see \cite{NT03} Theorem~3.5, where we have been
motivated by an intuitive argument of Zhu~\cite{Zh96}).

\begin{mtheorem}
\label{nth4.5}
{\rm (see \cite{NT03} Theorem~3.5 and Corollary~3.6)}
If the finite temperature correlation functions of a
set of Bose fields $\left\{ \phi _{a} \right\}$,
\beq
\La
\phi _{1}(\zeta _{1},u_{1}) \dots \phi _{n}(\zeta_{n},u_{n})
\Ra_{\! q} \, := \,
\frac{1}{Z(\tau)} \,
tr_{\DOM{}}\left\{ \phi_{1}(\zeta _{1},u_{1}) \dots \phi _{n}(\zeta _{n},u_{n})
\, q^{H}\right\}
\, , \
\label{4.12}
\eeq
are meromorphic and symmetric (as meromorphic functions)
with respect to permutations of the factors
${\phi_a}$, then the KMS condition (\cite{HHW67}~\cite{BB})
\beq\label{addXX}
\La
\phi \left( \zeta_2,\, u_2 \right)
\dots \phi \left( \zeta_n,\, u_n \right)
\phi \left( \zeta_1+\tau,\, u_1 \right)
\Ra_{\! q}
\, = \,
\La \phi \left( \zeta_1,\, u_1 \right)
\dots \phi \left( \zeta_n,\, u_n \right) \Ra_{\! q}
\, \
\eeq
implies that the functions~(\ref{4.12}) are elliptic with respect to the
$(n-1)$ independent
differences \(\zeta _{i \, i+1}=\zeta _{i}-\zeta _{i+1}\) \((i=1,\dots,n-1)\)
of periods $1$ and $\tau$.
\end{mtheorem}

We shall verify that the Gibbs correlation functions of (generalized)
free fields are indeed symmetric elliptic functions of $\zeta_{i \, i+1}$ thus
confirming the above conjecture.

\medskip

It follows from the Huygens principle, established in
\cite{NT01}, that all singularities of correlation functions are poles for
isotropic intervals
\beq
z_{ab}^{2} \, = \, 2 \, e^{2\pi i(\zeta _{a}+\zeta _{b})}
(\cos 2\pi \zeta _{ab}-\cos 2\pi \alpha _{ab}) \, (=0)
\label{4.14}
\eeq
where $2\pi \alpha _{ab}$ is the angle between the unit Euclidean $D$--vector
$u_{a}$ and $u_{b}:$
\beq
u_{a} \spr u_{b}=\cos 2\pi \alpha _{ab} .
\label{4.15}
\eeq
Noting the relation
\beq
\cos 2\pi \zeta -\cos 2\pi \alpha \, = \,
-2\sin \pi (\zeta +\alpha) \sin \pi (\zeta -\alpha)
\label{4.16}
\eeq
we deduce that the singularities of temperature means are poles in
$\zeta_{ab}$ for
\beq
\zeta _{ab}\pm \alpha _{ab}= n \in \Z
\, . \
\label{4.17}
\eeq
The decomposition formula
\beq
\frac{1}{\sin \pi \zeta_{+}\sin \pi \zeta_{-}} \, = \,
\frac{1}{\sin 2\pi \alpha} (\cotg \, \pi \zeta _{+}-\cotg \, \pi \zeta _{-})
\quad \text{for} \quad \zeta _{\pm} \, = \, \zeta \pm \alpha
\label{4.18}
\eeq
allows to separate the two poles arising in the vacuum correlation functions.
The Euler expansion for
$\cotg \, \pi\zeta$ and its fermionic counterpart for
$\txfrac{1}{\sin \pi \zeta}$ gives
\beq
\cotg \, \pi \zeta = \frac{1}{\pi}
\mathop{\lim}\limits_{N \to \infty} \,
\mathop{\sum}\limits_{n = -N}^{N}
(\zeta +n)^{-1}
\, , \quad
\frac{1}{\sin \pi \zeta} = \frac{1}{\pi}
\mathop{\lim}\limits_{N \to \infty} \,
\mathop{\sum}\limits_{n = -N}^{N}
\frac{(-1)^{n}}{\zeta +n} . \label{4.19}
\eeq
In the finite temperature $2$--point correlators the corresponding
Eisenstein-We\-i\-er\-s\-t\-rass type series (on the lattice $\Z\tau+\Z$)
are expressed as linear combinations of
derivatives of both sides of (\ref{4.19})
proportional to the functions~(\ref{2.5nnn}),
\beq
\pfun_{k}^{\kappa\lambda}(\zeta ,\tau )
=
\mathop{\lim}\limits_{M \to \infty} \, \mathop{\lim}\limits_{N \to \infty} \,
\mathop{\sum}\limits_{m = -M}^{M} \, \mathop{\sum}\limits_{n = -N}^{N} \,
\frac{(-1)^{\kappa m +\lambda n}}{(\zeta +m\tau +n)^{k}}
\, , \
\label{4.20}
\eeq
\(\kappa,\lambda = 0,1\), \(k=1,2,\dots\) (\(\zeta=\zeta_+,\zeta_-\))
introduced in Sect.~2.

For a generalized free field $\phi (\zeta,u)$ ($=\{\phi_A (\zeta,u)\}$)
of (half)integer dimension $d$,
characterized by its $2$--point vacuum function,
the Gibbs correlation functions can be also expressed in terms of
the $2$--point Wightman function.
The result looks very simple (\cite{NT03}, Theorem~4.1):
\beqa\label{nw4.77}
&
\La \phi (\zeta_1,u_1) \, \sphi (\zeta_2,u_2) \Ra_{\! q} \, = \,
\mathop{\dsum}\limits_{k \, = \, -\infty}^{\infty} \,
\left( -1 \right)^{2d \, k}
W (\zeta_{12} \hspace{-1pt} + \hspace{-1pt} k\tau; u_1,u_2)
\, , & \nn &
W (\zeta_{12}; u_1,u_2) \, := \,
\lvac \phi (\zeta_1,u_1) \, \sphi (\zeta_2,u_2) \rvac
\, . &
\eeqa
To derive~(\ref{nw4.77}) one combines the KMS condition with the
cannonical (anti)com\-mu\-ta\-ti\-on relations.
Using the fact the canonical (anti)commutator is a $c$--number
(i.e., proportional to the unit operator),
equal to its vacuum expectation value (and that \(\La 1 \Ra_q = 1\)),
we find the following relation for the thermal mean value of products
of $\phi$--modes:
\beqa\label{CCR2}
& \hspace{-28pt}
\La \! \phi_{\nu_1m_1} (u_1) \, \sphi_{\nu_2m_2} (u_2) \! \Ra_{\! q} -
(-1)^{2d} \La \! \sphi_{\nu_2m_2} (u_2) \, \phi_{\nu_1m_1} (u_1) \! \Ra_{\! q}
= &
\nn & \hspace{-28pt} =
\delta_{-\nu_1,\nu_2}
\lvac (\phi_{\nu_1m_1} (u_1) \, \sphi_{-\nu_1m_2} (u_2)
- (-1)^{2d} \sphi_{-\nu_1m_2} (u_2) \, \phi_{\nu_1m_1} (u_1)) \rvac
;
&
\eeqa
on the other hand, the KMS condition together with~(\ref{4.9}) gives
\beqa\label{KMS-phi1}
&
\La \! \sphi_{\nu_2m_2} (u_2) \, \phi_{\nu_1m_1} (u_1) \! \Ra_{\! q} \, = \,
\La \! \phi_{\nu_1m_1} (u_1) \, q^{H} \, \sphi_{\nu_2m_2} (u_2) \, q^{-H} \! \Ra_{\! q}
\, = &
\nn & = \,
q^{-\nu_2} \La \! \phi_{\nu_1m_1} (u_1) \, \sphi_{\nu_2m_2} (u_2) \! \Ra_{\! q}
,
\eeqa
and therefore,
\beqa\label{KMS-phi2}
& \hspace{-18pt}
\La \! \phi_{\nu_1m_1} (u_1) \, \sphi_{\nu_2m_2} (u_2) \! \Ra_{\! q}
= \Txfrac{\delta_{-\nu_1,\nu_2}}{1-(-1)^{2d} q^{\nu_1}} \,
\lvac (\phi_{\nu_1m_1} (u_1) \, \sphi_{-\nu_1m_2} (u_2) \rvac \, +
&
\nn & \hspace{-18pt}
+ \,
\Txfrac{\delta_{-\nu_1,\nu_2} q^{\nu_2}} {1-(-1)^{2d} q^{\nu_2}} \,
\lvac (\sphi_{\nu_2m_2} (u_2) \, \phi_{-\nu_2m_1} (u_1) \rvac
&
\eeqa
where the first term is nonzero for \(\nu_1 = - \nu_2 \geqslant 0\)
(\(|q^{\nu_1}| < 1\))
while the second is nonzero for \(\nu_2 = - \nu_1 \geqslant 0\)
(\(|q^{\nu_2}| < 1\)).
Further, we expand the prefactors $\Txfrac{1}{1 - (-1)^{2d} q^{\nu_{1,2}}}$
in the right hand side of~(\ref{KMS-phi2})
in \(|q| < 1\), as in
Exercise~2.11,
and take the corresponding sum~(\ref{4.8}) over the modes
which gives~(\ref{nw4.77}) since, for instance, the term
\(\left( (-1)^{2d} q^{\nu_1} \right)^k\) multiplying the vacuum expectation
in the fist term~(\ref{KMS-phi2}) gives (after summing in $\nu_1$)
the expression $(-1)^{2dk}$ $W (\zeta_{12}+ k\tau;$ $u_1,$ $u_2)$
(\(k = 0,1,\dots\)).

To illustrate the conclusions of
Theorem~4.5
in the case of a two point bosonic thermal correlation function
let us choose as a basis of ``bosonic'' elliptic functions
those of the generalized free scalar fields of dimension
\(k = 1,2,\dots\):
\beq\label{basic_ef}
P_k (\zeta_{12}; u_1, u_2; \tau) \, := \,
\mathop{\sum}\limits_{n \, = \, -\infty}^{\infty} \,
\frac{\pi^{2k}}{\sin^k \! \pi \left( \zeta_+ \hspace{-1pt} + \hspace{-3pt} n\tau \right)
\, \sin^k \! \pi \left( \zeta_- \hspace{-3pt} + \hspace{-1pt} n\tau \right)}
\eeq
(\(\zeta_{\pm} = \zeta_{12} \pm \alpha\), \(\cos 2\pi \alpha = u_1 \spr u_2\)).
Then a corollary of
Theorem~4.5
states that for any two GCI bosonic fields $\phi$ and $\psi$ which obey
the Huygens' principle~(\ref{equ2.5}) with some \(N \in \N\) the Gibbs two point
function can be presented as
\beq\label{th_exp}
\La \! \phi (\zeta_1,u_1) \, \psi (\zeta_2,u_2) \! \Ra_{\! q} \, = \,
\Su_{k \, = \, 1}^{N} \,
F_k (u_1,u_2;\tau) \, P_k \left( \zeta_{12};u_1,u_2;\tau \right)
\eeq
whose coefficients $F_k (u_1,u_2;\tau)$ are in general $q$-series.
We will see in Sect.~6 that these coefficients carry an additional
physical information that may recover quantities like the mean thermal energy.

\begin{mremark}
The compact picture stress-energy tensor has a mode expansion of the form
\beq\label{T-cp}
T (\zeta,u;v) \, = \,
e^{2 D \pi i \zeta} \, T (e^{2\pi i \zeta} u; v) \, = \,
\Su_{n \, = \, -\infty}^{\infty} \Su_{m \, = \, 0}^{\infty}
T_{nm} (u;v) \, e^{-2\pi i n\zeta}
\eeq
where $T_{nm} (u;v)$ are operator valued,
separately homogeneous and harmonic polynomials in
$u$ and $v$ of degrees $m$ and $2$, respectively.
Among all such polynomials there is exactly one, up to proportionality,
which is $\SO(D)$--invariant:
\(\left( u \spr v \right)^2 - \frac{1}{D} \, u^{\, 2} v^{\, 2}\).
Then the conformal hamiltonian corresponds to the operator coefficient
to this polynomial in the zero--mode part,
\beq\label{H-T}
T_{02} (u;v) \, = \,
N_0 \, H \left(\raisebox{10pt}{\hspace{-2pt}}\right.
\left( u \spr v \right)^2 - \frac{1}{D} \, u^{\, 2} v^{\, 2}
\left.\raisebox{10pt}{\hspace{-2pt}}\right) \, + \,
\Su_{\sigma} \, T_{02;\sigma} \, h^{(02)}_{\sigma} (u;v)
\, , \
\eeq
where $N_0=\txfrac{D}{D-1}$, if the volume of the
unite $(D-1)$--sphere is normalized to one, and
$h^{(02)}_{\sigma} (u;v)$ is a basis in the $\SO (D)$--nonscalar space.
It follows from the $\compgr$--invariance of the thermal expectation values that
\beq\label{MH}
\La \! T \left( \zeta,u;v \right) \! \Ra_{\! q} \, = \,
N_0 \, \La \! H \! \Ra_{\! q} \left(\raisebox{10pt}{\hspace{-2pt}}\right.
\left( u \spr v \right)^2 - \frac{1}{D} \, u^{\, 2} v^{\, 2}
\left.\raisebox{10pt}{\hspace{-2pt}}\right)
\, . \
\eeq
\end{mremark}

\section{Chiral fields in two dimensions}
\setcounter{equation}{0}\setcounter{mtheorem}{1}\setcounter{mremark}{1}\setcounter{mdefinition}{1}\setcounter{mexample}{1}\setcounter{mexercise}{1}

The simplest, long known example of a quantum field theory with elliptic
correlation functions is provided by $2$--dimensional ($2D$) (Euclidean) CFT
on a torus or equivalently, by finite temperature $2D$ CFT on compactified
Minkowski space $\overline{M}$
(for a rigorous discussion~--~see~\cite{Zh96}).
We adopt the latter point of view since it
is the one that extends to higher dimensions.

The variables $\zeta _{\pm}$ (\ref{4.18}) can be viewed for $D=2$ as global
coordinates on (the universal covering of)\ $\overline{M}$. Conserved
currents and higher rank tensors (including the stress--energy tensor)
split into chiral components depending on one of these variables.
(This is simpler to derive in Minkowski space coordinates~--~see~\cite{FST}.)
The vertex algebra corresponding to the full 2D theory then usually splits
into tensor product of two copies of a vertex algebra over the real line
satisfying the postulates of Sect.~4.3 with \(D=1\).
A GCI chiral field $\phi (z)$ in this case is a formal Laurent series in a single
complex variable $z$ (one of the compactified light ray variables $z_+$ or $z_-$
whose physical values belong to the circle $\Sr^{1}$),
and the strong locality condition~(\ref{equ2.5}) for a pair of
hermitian conjugate fields
of dimension $d$ assumes the form
\beq
\label{nw5.1}
z_{12}^{2d} \,
\left\{
\sphi \left( z_1 \right) \phi \left( z_2 \right)
- (-1)^{2d} \,
\phi \left( z_2 \right) \sphi \left( z_1 \right)
\raisebox{9pt}{\hspace{-2pt}}
\right\} \, = \, 0 \, .
\eeq

\subsection{$U(1)$ current, stress energy tensor, and the free Weyl field}

A conformal $U(1)$ current $j_{\mu}(x)$ in 2D behaves as the gradient of a
(dimensionless) free massless scalar field. Hence both its divergence and its
curl vanish implying its splitting into chiral components:
\beq
\partial _{\mu}j^{\mu}=0=\partial _{0}j_{1}-\partial _{1}j_{0}
\quad \implies \quad (\partial _{0}\pm \partial _{1})(j^{1}\pm j^{0})=0
\, . \
\label{5.1}
\eeq
Similarly, the symmetry and tracelessness of the conserved stress--energy
tensor imply
\beq
(\partial _{0}\pm \partial _{1})(T_{0}^{0}\pm T_{0}^{1})=0
\, . \
\label{5.2}
\eeq
We leave it to the reader to verify, on the other hand, that for $z$ given
by (\ref{nw4.18}) with $D=2$, $\mzeta =x$ and the inverse transformation
given, in general, by
\beq
\mbf{x}=\frac{2\,\mbf{z}}{1+z^{2}+2z_{D}}
\, , \quad
-i\, x^{0}=\frac{1-z^{2}}{1+z^{2}+2z_{D}}
\label{5.3}
\eeq
setting in the $D=2$ case $z_{2}\pm iz_{1}=e^{2\pi i\zeta_{\pm}}$
we find
\beq
x^{0} \pm x^{1}=\tan \pi \zeta _{\pm}
\quad (\text{for} \ D=2,\quad
u = (\sin 2\pi \alpha ,\cos 2\pi \alpha ) ) .
\label{5.4}
\eeq
Thus the ``left movers'' compact picture current and stress--energy tensor
are functions of a single chiral variable $\zeta _{-}$:
\beq
j(\zeta _{-}):=\frac{1}{2}(j^{0}+j^{1})
, \quad
\mathcal{T}(\zeta _{-})
:=
\frac{1}{2}(\mathcal{T}_{0}^{0}+\mathcal{T}_{0}^{1})=
\frac{1}{4}(\mathcal{T}_{0}^{0}+\mathcal{T}_{0}^{1}-\mathcal{T}_{1}^{0}
-\mathcal{T}_{1}^{1}) .
\label{5.5}
\eeq
The same is valid for the Weyl components of a free $D=2$ Dirac field $\Psi (x)$
(and its conjugate $\sPsi$).
Introducing real off--diagonal $2D$ $\gamma$--matrices
\beqa
&
\gamma ^{0}=\left(\hspace{-4pt} \begin{array}{rr} 0 \hspace{-4pt} & -1 \\
1 \hspace{-4pt} & 0 \end{array}\hspace{-3pt}\right)
\, , \quad \gamma ^{1}=\left(\hspace{-4pt}\begin{array}{rr} 0 \hspace{-4pt} & 1 \\
1 \hspace{-4pt} & 0 \end{array}\hspace{-3pt}\right) ,
& \nn &
\Psi (x) = \left(\hspace{-4pt}\begin{array}{c} \Psi _{\uparrow} \\
\Psi _{\downarrow} \end{array}\hspace{-3pt}\right)
, \quad
\gamma^{\mu}\partial _{\mu} \Psi = 0 =
(\partial_{0}+\partial_{1})\Psi _{\uparrow}
, \quad
\sPsi (x) := \left( \Psi _{\uparrow},\sPsi _{\downarrow} \right)
&
\label{5.6}
\eeqa
we find
\beq
j(\zeta_{-}) \, = \, \frac{i}{2} \widetilde{\Psi} \, (\gamma ^{0}+\gamma ^{1}) \, \Psi
\, = \, \frac{1}{2}\sPsi \, (1+\gamma _{0}\gamma ^{1}) \, \Psi \, = \,
\sPsi_{\uparrow} (\zeta_{-}) \Psi _{\uparrow}(\zeta_{-})
\label{5.7}
\eeq
(\(\widetilde{\Psi} (x) := \sPsi (x) \gamma_0\)).

Omitting from now on the arrow sign on the left mover's Weyl field
$\Psi(\zeta )$ ($\equiv \Psi _{\uparrow}(\zeta)$) we obtain
the chiral field $\Psi (\zeta)$ as a
$1$--component complex field which, together with its conjujate $\sPsi$,
can be written in the compact picture form
\beq
\bsPsi \hspace{-1pt} \left( \zeta \right) \, = \, \dsum\limits_{n}\bsPsi _{n-\frac{1}{2}}e^{i\pi (1-2n)\zeta}
\, , \quad
(\Psi _{n-\frac{1}{2}})^{\ast} \, = \, \sPsi _{\frac{1}{2}-n}
\,
\label{4.25}
\eeq
(here $\zeta$ plays the role of a (compactified, chiral) light cone
variable, say $\zeta_{-}$ of (\ref{4.18})).
The $\Psi$--modes obey the \textit{canonical anticommutation relations}
and their index labels the energy they
carry,
\beq
\left[ \Psi _{m-\frac{1}{2}},\sPsi _{\frac{1}{2}-n}\right]_{+} \!\! = \delta _{mn}
, \quad
\left[ \bsPsi _{m-\frac{1}{2}},\bsPsi _{\frac{1}{2}-n}\right]_{+} \!\! = 0
, \quad
\left[ L_{0},\Psi _{\frac{1}{2}-n}\right] \! =
(n-\frac{1}{2})\Psi _{\frac{1}{2}-n}
,
\label{4.26}
\eeq
where $L_{0}$ stands for the Virasoro energy operator,
\beq
L_{0}
= \Su_{n \, = \, 1}^{\infty} (n-\frac{1}{2}) \, ( \sPsi_{\frac{1}{2}-n} \Psi _{n-\frac{1}{2}}
+ \Psi_{\frac{1}{2}-n} \sPsi _{n-\frac{1}{2}} )
\, ,
\label{4.28}
\eeq
the counterpart of $H$ in \(D=1\)
(in fact,
the full 2D conformal hamiltonian is a sum of chiral energy operators,
\(H = L_0 + \overline{L}_0\),
see, e.g., \cite{DMS} or \cite{FST}).
Energy positivity implies that the negative frequency modes annihilate
the vacuum:
\beq
\bsPsi_{n-\frac{1}{2}}\rvac =0
\quad \text{for} \quad n=1,2,\dots
\, .
\label{4.27}
\eeq
Then the vacuum $2$--point correlation function is
\beq\label{nw5.5}
\lvac \Psi \hspace{-1pt} \left( \zeta_1 \right) \sPsi \hspace{-1pt} \left( \zeta_2 \right) \rvac \, = \,
\frac{1}{2i \sin \pi \zeta_{12}} \, . \
\eeq

If we formally replace the normal product sum in~(\ref{4.28}) by the divergent sum
over all integer $n$,
\beq
\widetilde{L}_{0} = \mathop{\dsum}\limits_{n\in\Z}(n-\frac{1}{2})
\sPsi_{\frac{1}{2}-n}\Psi _{n-\frac{1}{2}}
= \mathop{\dsum}\limits_{n\in\Z}
(n-\frac{1}{2}) \Psi _{\frac{1}{2}-n} \sPsi_{n-\frac{1}{2}} =
L_{0}-\frac{1}{2}\mathop{\dsum}\limits_{n=1}^{\infty}(2n-1) ,
\label{5.16}
\eeq
the last infinite term being understood by $\zeta$--function
regularization:
$\mathop{\sum}\limits_{n = 1}^{\infty} \, ( 2n$ $-$ $1)$ ``$=$''
\(\mathop{\sum}\limits_{n = 1}^{\infty} n -
\mathop{\sum}\limits_{n = 1}^{\infty} 2n\) ``$=$'' \(- \zeta \left( -1 \right)\),
we will obtain\footnote{%
Note that the passage from $L_0$ to $\widetilde{L}_0$
can be interpreted as the result of the non M\"obius
transformation \(z\mapsto \zeta\) under which $L_0$ acquires a
Schwarz derivative term:
\(\widetilde{L}_{0}\) \(=L_0\) \(+\txfrac{1}{12}\left\{z,2\pi i\zeta \right\}\)
where
\(\left\{ z,w \right\}\) \(:=\txfrac{z^{\prime \prime \prime}(w)}{z^{\prime}(w)}\)
\(-\txfrac{3}{2}\left(\txfrac{z^{\prime \prime}(w)}{z^{\prime}(w)}\right) ^{2}\).
(Hermann Amandus Schwarz, 1843--1921 introduces his derivative in 1872.)}
\beq
\widetilde{L}_{0} \, = \, L_{0}+\frac{1}{2}\,\zeta (-1) \, = \, L_{0}-\frac{1}{24}
\quad (\text{as} \ \, \zeta (1-2k) = -\frac{B_{2k}}{2k})
\, .
\label{4.29}
\eeq
The calculations~(\ref{CCR2})--(\ref{KMS-phi2}) in this case give
\beqa
&
\La \Psi _{m-\frac{1}{2}} \, q^{n-\frac{1}{2}} \, \sPsi _{\frac{1}{2}-n}\Ra _{q}
\, = \, \La \sPsi _{\frac{1}{2}-n}\Psi _{m-\frac{1}{2}}\Ra _{q}
\, = \, \delta_{mn}-\La \Psi _{m-\frac{1}{2}}\sPsi _{\frac{1}{2}-n}\Ra _{q}
\, , &
\nn &
\La \Psi _{m-\frac{1}{2}}\sPsi _{\frac{1}{2}-n}\Ra _{q} \, = \,
\Txfrac{\delta_{mn}}{1+q^{m-\frac{1}{2}}}
\, . \
\label{4.30}
\eeqa
Inserting (\ref{4.30}) into the Gibbs 2-point function of the local
Fermi field (\ref{4.25})
we find
(see Exercises~2.11 and 3.1, and
\(F_2 (\tau) = 2\hspace{1pt}G_{2}(\tau) -
G_{2} \! \left( \Txfrac{\tau +1}{2} \right)\)
(\ref{3.11}))
\beqa\label{new4x-1}
\La \! \Psi \! \left( \zeta_1 \right) \! \sPsi \left( \zeta_2 \right) \! \Ra_{\! q}
\hspace{-1pt} =
&& \podr
\Txfrac{1}{2i\sin \pi \zeta_{12}} +
2i \! \mathop{\dsum}\limits_{n = 1}^{\infty}
\Txfrac{q^{n-\frac{1}{2}}}{1 \hspace{-1pt} + \hspace{-1pt} q^{n-\frac{1}{2}}}
\sin (2n \! - \! 1) \pi \zeta_{12} \hspace{-1pt} =
\nn = && \podr
\Txfrac{1}{2\pi i} \hspace{1pt} \pfun_{1}^{11}(\zeta _{12},\tau)
\eeqa
(see Exercise~2.11
and the derivation of~(\ref{nw4.77})).
The temperature mean value of the chiral energy (\ref{4.28}) is computed by Eq.~(\ref{mean_H1})
with \(d_b (n) = 0\) and \(d_f (n) = 1+1 =2\) (for the two charges):
\beq
\La \widetilde{L} _{0}\Ra _{q} \, = \,
-\Txfrac{1}{24} + 2 \Su_{n=1}^{\infty}
\left( n - \frac{1}{2} \right)
\Txfrac{q^{n-\frac{1}{2}}}{1+q^{n-\frac{1}{2}}}
\, = \, F_2 (\tau) \, ( \, = \, 2\, G_2^{11} (\tau))
\, .
\label{4.31}
\eeq

\begin{mexercise}\label{exr:4}
Verify the relation~(\ref{4.31})
for $F_2(\tau)$ defined by Eq.~(\ref{3.11}).
(\textit{Hint}: use the relation
\(q \left(\raisebox{12pt}{\hspace{-3pt}}\right.
\txfrac{\tau +1}{2} \left.\raisebox{12pt}{\hspace{-2pt}}\right)
= -q^{\frac{1}{2}}\)
for $q^{\frac{1}{2}}$ $\equiv$ \(q(\tau)^{\frac{1}{2}} = e^{i\pi \tau}\)
and cancel the terms with even $n$ in the expansion of
\(\txfrac{1}{2} \, G_{2} \left(\raisebox{12pt}{\hspace{-3pt}}\right.
\txfrac{\tau +1}{2} \left.\raisebox{12pt}{\hspace{-2pt}}\right)\)
with the expansion of $G_{2}(\tau )$.)
\end{mexercise}

According to Sect.~3.2
(see Exercise~3.1),
the \textit{free energy} $F_2$ is a modular form of
weight two and level $\Gamma_{\theta}$.

For \(H:=\widetilde{L}_{0}\) we can integrate and exponentiate
(\ref{mean_H})
(\ref{4.31}) with the
result
\beq
Z(\tau ) \, = \, q^{-\frac{1}{24}}\dprod\limits_{n=1}^{\infty}
(1+q^{n-\frac{1}{2}})^2
\, . \
\label{4.35}
\eeq
It follows from the $\Gamma_{\theta}$ invariance of the $1$--form
$F_2(\tau )d\tau$ that the partition function (\ref{4.35}) is
$\Gamma_{\theta}$ invariant.
(It is, sure, not a modular form since it is not analytic for \(q=0\).)
This invariance property of the partition function is peculiar for
space--time dimension \(D=2\) since the leading term of the energy mean value
has weight $D$ and only for \(D=2\) the fact, that $G_{D}(\tau)$ is a modular
form of weight $D$, implies that the $1$--form $G_{D}(\tau)d\tau$ is invariant.

\begin{mremark}\label{nrm5.1}
Minkowski space of dimension $D=8n+2$, $\ n=0,1,\dots$ is singled
out by the property of admitting \textit{Majorana-Weyl spinors}
(i.~e. $2^{\frac{D-2}{2}}$-component real semispinors).
Thus, the Majorana-Weyl chiral field is a hermitian field
$\Psi (\zeta)$ obeying (\ref{4.25}) with \(\Psi = \sPsi\),
and similar results can be obtained in this case, too.
In particular, the free energy is
$F_I(\tau)$ $\equiv \txfrac{1}{2} F_2(\tau)$
and the partition function
is $Z_I(\tau)$ $:=Z(\tau)^{\frac{1}{2}}$, the partition function for the
\textit{Neveu-Schwarz sector} of the
\textit{chiral Ising model} (hence the index $I$ on $F$ and $Z$).
Note that $F_I$ and $Z_I$ are transformed under the modular transformation $ST$ to
the free energy and partition function of the \textit{Ramond sector}
\beqa
&
F_{I}^{R}(\tau) \, := \, (ST \, F_{I})(\tau) \, = \,
\Txfrac{1}{\tau^{2}}F_{I} \left(\raisebox{12pt}{\hspace{-3pt}}\right.
1-\Txfrac{1}{\tau} \left.\raisebox{12pt}{\hspace{-2pt}}\right)
\, = \, G_{2} (\tau)-2G_{2}(2\tau)
,
& \nn &
Z_I^{R} \left( \tau \right)
\, = \, q^{\frac{1}{24}}\mathop{\prod}\limits_{n=1}^{\infty}(1+q^{n}) .
&
\label{4.36}
\eeqa
The Neveu-Schwarz $0$--point energy $E_{\text{\it NS}}$
is the $q$--independent term $-\txfrac{1}{48}$ in $F_I$
and it is uniquely determined by the modular covariance of $F_{I}$
which thus selects Weyl symmetrization
(accompanied by $\zeta$--function regularization)
rather than the normal ordering.
The $0$--point energy $E_{R}$ in the Ramond sector differs from $E_{\text{\it NS}}$
by the minimal conformal weight (eigenvalue of $L_{0}$) in that sector,
$\Delta_{R}$. It is calculated from (\ref{3.7}) and (\ref{4.36}) with the result
\beq
E_{R} \, = \, -
\left(\raisebox{12pt}{\hspace{-3pt}}\right.
-\frac{B_{2}}{4} \left.\raisebox{12pt}{\hspace{-2pt}}\right)
\, = \, \frac{1}{24}=E_{\text{\it NS}}+\Delta_{R}
\quad \Longrightarrow \quad
\Delta _{R} \, = \, \frac{1}{24}+\frac{1}{48} \, = \, \frac{1}{16}
\, . \
\label{4.39}
\eeq
the magnetization field of the $2$--dimensional Ising model has (left, right)
conformal weight
($\Txfrac{1}{16},\Txfrac{1}{16})$~--~i.~e., dimension $\Txfrac{1}{8}$ and
``spin'' $0$.
\end{mremark}

The knowledge of the energy eigenvalues does not suffice to label the states
of the complex Weyl field. We also need the \textit{charge operator}
which appears as the zero mode of a composite field, the current
\beq
J(\zeta )=\frac{1}{2} \, [ \sPsi (\zeta ),\Psi (\zeta) ]
\, = \, :\hspace{-3pt}\sPsi (\zeta) \, \Psi (\zeta)\hspace{-3pt} : \, = \,
\mathop{\dsum}\limits_{n} J_{n}e^{-2\pi in\zeta}
\label{5.25}
\eeq
where
\beq
J_{0}=\dsum\limits_{n=1}^{\infty}(\Psi _{\frac{1}{2}-n}^{+}
\Psi _{n-\frac{1}{2}}-\Psi _{\frac{1}{2}-n}\Psi _{n-\frac{1}{2}}^{+})
\, , \quad
J_{n}=\dsum\limits_{\rho \in \Z + \frac{1}{2}}
\Psi_{-\rho}^{+}\Psi_{n+\rho}
\quad \text{for} \quad n\neq 0
\, .
\label{5.26n}
\end{equation}
The current modes are characterized by their commutation relations
\beq\label{eqn5.27}
\left[ J_{n},  \sPsi \left( \zeta \right) \right]
=
e^{2\pi i n \zeta} \sPsi \left( \zeta \right)
, \ \
\left[ J_n, \Psi \left( \zeta \right) \right]
=
- e^{2\pi i n \zeta} \Psi \left( \zeta \right)
, \ \
\left[ J_n, J_m \right] = n \, \delta_{n,-m}
\, . \
\eeq
The conformal energy~(\ref{5.16}) can be reexpressed in terms
of the current modes~--~providing a special case of the so called
Sugwara formula:
\beq\label{eqn5.28}
\widetilde{L}_0 =
\frac{1}{2}
\sum_{n} J_{-n} J_n = L_0 +
\frac{1}{2} \zeta \left( -1 \right)
, \ \
L_0 =
\frac{1}{2} J_0^2 + \sum_{n = 1}^{\infty} J_{-n} J_n
, \ \
\frac{1}{2} \, \zeta \left( -1 \right) = - \frac{1}{24}
.
\eeq

While the energy is coupled in the partition function with the
(complexified) inverse temperature $\tau$ we shall express
the charge distribution by a parameter $\mu$ called the \textit{chemical potential}
introducing the generalized partition function of Neveu--Schwarz sector
\beq\label{eqn5.29}
Z_{NS} \left( \tau,\, \mu \right) =
\TR_{\DOM{}} \left( q^{\widetilde{L}_0} q_{\mu}^{J_0} \right)
\, , \quad
q_{\mu} = e^{2\pi i \mu}
\, . \
\eeq

\begin{mremark}
The complex Weyl field model can be viewed as
``the square of the chiral Ising model'' of
Remark~5.1.
If we split $\Psi$ into its real and imaginary parts
\(\sqrt{2} \Psi \left( \zeta \right) =
\Psi^1 \left( \zeta \right) - i \Psi^2 \left( \zeta \right)\), then
we shall have (setting \(\langle
\Psi^1 \left( \zeta_1 \right) \Psi^2 \left( \zeta_2 \right) \rangle_0 = 0\))
\beq\label{eqn5.30}
\langle
\Psi \left( \zeta_1 \right) \sPsi \left( \zeta_2 \right) \rangle_0
\, = \,
\langle
\Psi^1 \left( \zeta_1 \right) \Psi^1 \left( \zeta_2 \right) \rangle_0
\, = \,
\langle
\Psi^2 \left( \zeta_1 \right) \Psi^2 \left( \zeta_2 \right) \rangle_0
\, , \
\eeq
the energy of the charged field~(\ref{5.16}) being twice
the energy of the Ising model.
\end{mremark}

\begin{mexercise}
Set \(\langle A \rangle_{q,\mu} :=
\Txfrac{1}{Z \left( \tau,\mu \right)}
\TR_{\DOM{}} \left( A q^{\widetilde{L}_0}
q^{J_0} \right)\); use the generalized KMS condition
\(\langle A B_{\tau,\mu} \rangle_{q\mu} =
\langle BA \rangle_{q\mu}\) for
\(q^{\widetilde{L}_0} q_{\mu}^{J_0} \, B =
B_{\tau,\mu} q^{\widetilde{L}_0} q^{J_0}\) to prove
\beq\label{eqn5.31}
\langle
\Psi_{n-\frac{1}{2}} \Psi^{+}_{\frac{1}{2}-m}
\rangle_{q,\mu} \, = \,
\frac{\delta_{mn}}{1+q_{\mu}q^{n-\frac{1}{2}}}
\, , \quad
\langle
\Psi_{n-\frac{1}{2}}^+ \Psi_{\frac{1}{2}-m}
\rangle_{q,\mu} \, = \,
\frac{\delta_{mn}}{1+q_{\mu}^{-1}q^{m-\frac{1}{2}}}
\, , \
\eeq
\beq\label{???}
\hspace{0pt}
2{\pi}i\La {\Psi} \left( {\zeta_1} \right) \,
{\sPsi} \left( {\zeta_2} \right) \Ra_{q,\mu}
\! =
\pfun_1^{11}(\zeta_{12},\tau,\mu) \! :=
\frac{{\pi}}{\sin{\pi\zeta_{12}}}+
\mathop{\sum}\limits_{m}
\frac{e^{i{\pi}m(2{\mu}+{\kappa})}}{
\sin{\pi}({\zeta_{12}}+m{\tau})}
\eeq
(cf. Appendix~\ref{ap:1}).
\end{mexercise}

Inserting (\ref{eqn5.31}) into the mean value of $\widetilde{L}_0$~(\ref{5.16}),
we find the following generalization of (\ref{4.31}):
\beq\label{eqn5.32}
\langle \widetilde{L}_0 \rangle_{q,\mu} \, = \, - \frac{1}{24}
+ \sum_{n = 1}^{\infty}
\left( n - \frac{1}{2} \right) \!
\left(
\frac{q_{\mu}^{-1}q^{n-\frac{1}{2}}}{1+q_{\mu}^{-1}q^{n-\frac{1}{2}}} +
\frac{q_{\mu}q^{n-\frac{1}{2}}}{1+q_{\mu}q^{n-\frac{1}{2}}}
\right)
\, . \
\eeq
Integrating and exponentating the relation
\beq\label{eqn5.33}
\langle \widetilde{L}_0 \rangle_{q,\mu}
\, = \,
\frac{1}{2\pi i} \frac{\partial}{\partial \tau}
\log \, Z \left( \tau,\, \mu \right) \, = \,
\frac{1}{Z} q \frac{\partial}{\partial q} Z
\eeq
which generalizes~(\ref{mean_H}) we find
\beq\label{eqn5.34}
Z_{NS} \left( \tau,\, \mu \right) \, = \,
q^{-\frac{1}{24}} \prod_{n = 1}^{\infty}
\left( 1 + q^{-1}_{\mu} q^{n-\frac{1}{2}} \right)
\left( 1 + q_{\mu} q^{n-\frac{1}{2}} \right)
\, . \
\eeq

\begin{mexercise}
Use (\ref{eqn5.28}) and the KMS condition to derive the expression
\beq\label{eqn5.35}
Z_{NS} \left( \tau,\, \mu \right) \, = \, \frac{\vartheta \left( \mu,\, \tau \right)}{
\eta \left( \tau \right)} \, = \, q^{-\frac{1}{24}} \,
\mathop{\prod} \limits_{m \, = \, 1}^{\infty} \,
\left( 1-q^m \right)^{-1} \,
\sum_n q^{\frac{1}{2}\, n^2} \, q_{\mu}^n
\, \
\eeq
(\(\vartheta (= \vartheta_{00})\)) being the Riemann theta function~(\ref{3.19});
$\eta$, the Dedekind $\eta$--function of~(\ref{3.12})).
\end{mexercise}

Comparison between (\ref{eqn5.34}) and (\ref{eqn5.35}) yields a nontrivial
identity called the \textit{Jacobi triple product formula}.

We can repeat the study of modular properties of the chiral
Ising model
(Remark~5.1)
for the model at hand of a complex Weyl field
with the following results:
\beq\label{eqn5.36}
Z_{NS} \left( \tau+1,\, \mu \right) \, = \,
e^{-\frac{i\pi}{12}} \, Z_{NS} \left(  \tau,\, \mu+\frac{1}{2} \right)
\, , \quad
Z_{NS} \left( -\frac{1}{\tau},\, \frac{1}{2} \right) \, = \,
Z_R \left( \tau,\, 0 \right)
\, ; \
\eeq
here
\beq\label{eqn5.37n}
Z_R \left( \tau,\, \mu \right) \, = \, q^{\frac{1}{12}} \,
\left( q_{\mu}^{-\frac{1}{2}}+q_{\mu}^\frac{1}{2} \right)
\mathop{\prod}\limits_{n\, = \, 1}^{\infty} \,
\left( 1 +q_{\mu}^{-1} q^n \right)\left( 1+q_{\mu} q^n \right)
\eeq
where $Z_R$ is the partition function of the Ramond sector
characterized by charge and conformal weight of its lowest
energy state
\beq\label{eqn5.39}
e_R \, = \, \pm \, \frac{1}{2}
\, , \quad
\Delta_R \, = \, \frac{1}{2} \, e_R^2 \, = \, \frac{1}{8}
\, . \
\eeq

We conclude this section with a discussion of the possibility
to reconstruct the Gibbs mean energy from
the thermal $2$--point functions.
Let $\psi_d (z)$ and $\spsi_d (z)$ be
a pair of hermitian conjugate complex chiral fileds
of (half)in\-te\-ger dimension $d$ in the analytic picture.
Let us assume that the stress--energy tensor $T (z)$ contributes to the OPE
of $\spsi_d \psi_d$:
\beqa\label{to-1}
&&
\frac{1}{2}
\left( \spsi_d \left( z_1 \right) \psi_d \left( z_2 \right) +
\psi_d \left( z_1 \right) \spsi_d \left( z_2 \right) \right) \, = \,
\frac{N_0}{z_{12}^{2d}} + \frac{N_1}{z_{12}^{2d-2}} \ T (\sqrt{z_1z_2})
\, + \, \dots
\, \equiv \,
\nn && \quad \, \equiv \,
\frac{N_0}{z_{12}^{2d}} \left( 1 + \frac{2d}{c} \, z_{12}^2 \, T(\sqrt{z_1z_2})
+ O (z_{12}^4) \right)
,
\eeqa
where the Ward identities (property 7) of Sect.~4.3 imply that
\(N_1 = \txfrac{2d}{c} \, N_0\), \(c := 2 z_{12}^4 \lvac T(z_1) T(z_2) \rvac\)
being the Virasoro central charge (see~\cite{FST}, Sect.~3.5).
It is important for the validity of~(\ref{to-1}) that the product
$z_{12}^{2d}$
$\left(\raisebox{10pt}{\hspace{-1pt}}\right.\spsi_d (z_1) \,\psi_d (z_2)$ $+$
$\psi_d (z_1) \, \spsi_d (z_2)\left.\raisebox{10pt}{\hspace{-1pt}}\right)$
is a \textit{symmetric} dimensionless bilocal field of \((z_1,z_2)\)
for both even and odd $2d$.
Passing to the compact picture fields,
\(\psi_d (\zeta_k) = e^{2d\pi i \zeta_k} \psi_d (z_k)\),
\(T (\zeta_k) = e^{4\pi i \zeta_k} T (z_k)\),
\(z_k=e^{2\pi i \zeta_k}\)  (\(k=1,2\))
Eq.~(\ref{tm2}) takes the form
\beqa\label{tm2}
&& \hspace{-27pt}
\frac{1}{2}
\left( \spsi_d \left( \zeta_1 \right) \psi_d \left( \zeta_2 \right) +
\psi_d \left( \zeta_1 \right) \spsi_d \left( \zeta_2 \right) \right) \, = \,
\nn && \hspace{-27pt} \quad = \,
\frac{N_0}{\left( 2i\sin \pi \zeta_{12} \right)^d}
\left( 1
+
\frac{2d}{c} \left( 2i\sin \pi \zeta_{12} \right)^2
T \left(\raisebox{10pt}{\hspace{-2pt}}\right.
\frac{\zeta_1+\zeta_2}{2}
\left.\raisebox{10pt}{\hspace{-2pt}}\right)
+
O (\sin^4 \pi\zeta_{12}) \right)
.
\qquad
\eeqa
This implies the following Laurent expansions in $\sin \pi \zeta_{12}$
and $\zeta_{12}$
of the (symmetric under charge conjugation) thermal $2$--point function:
\beqa\label{tm3}
\hspace{0pt}
&& \hspace{-34pt} \podr
\frac{1}{2}
\La \!
\spsi_d \left( \zeta_1 \right) \psi_d \left( \zeta_2 \right) +
\psi_d \left( \zeta_1 \right) \spsi_d \left( \zeta_2 \right)
\! \Ra_{\! q} \! =
\nn \hspace{0pt}
&& \hspace{-34pt} \quad \podr =
\frac{N_0}{\left( 2i \sin \pi \zeta_{12} \right)^{2d}}
\left( 1 +
\frac{2d}{c} \La \! L_0 \! \Ra_{\! q} \left( 2i \sin \pi\zeta_{12} \right)^2
+
O (\sin^4 \pi \zeta_{12})
\right) =
\nn \hspace{0pt}
&& \hspace{-34pt} \quad \podr =
\frac{N_0}{\left( 2\pi i\zeta_{12} \right)^{2d}} \left(
1 + \frac{2d}{c} \, \La \! \widetilde{L}_0 \! \Ra_{\! q} \, \left( 2\pi i \zeta_{12} \right)^2 + O (\zeta_{12}^4)
\right)
\, ,
\quad
\widetilde{L}_0 :=
L_0 - \txfrac{c}{24}
.
\eeqa
(We have already encountered a Laurent expansion of this type in
Exercise~3.1
for the case of our basic elliptic functions.)
In the case of the complex Weyl field we have \(d = \txfrac{1}{2}\),
\(c=1\) and
\(N_0 = 1\) (\(N_1 = 1\))
reproduces the result for
$\La \! \widetilde{L}_0 \! \Ra_{\! q}$ (cp. with~(\ref{4.29}) and~(\ref{4.31})).
This gives an interpretation of the passage \(L_0 \mapsto \widetilde{L}_0\)
as an exchange, \(\sin \pi \zeta_{12} \mapsto \zeta_{12}\),
of the expansion variable
which thus makes all coefficients \textit{modular invariant}
if the left hand side is.
We can also write (\ref{tm3}) as an expansion in homogeneous elliptic
functions:
for \(2d \geqslant 3\) the stress--energy tensor $T (z)$
occurs in the singular part of the OPE of $\psi$ and then
(\ref{tm3}) implies
\beqa\label{tm6}
&& \hspace{-30pt}
\frac{1}{2}
\La \!
\spsi_d \! \left( \zeta_1 \right) \psi_d \! \left( \zeta_2 \right)
+
\psi_d \! \left( \zeta_1 \right) \spsi_d \! \left( \zeta_2 \right)
\! \Ra_{\! q}
\, = \,
\mathop{\dsum}\limits_{k \, = \, 0}^{\Dbrackets{d-\frac{1}{2}}}
C_k \, p_{2d-2k}^{\kappa\kappa} (\zeta_{12}, \tau)
\, = \,
\nn && \hspace{-30pt} \quad = \,
(2\pi i)^{-2d}
N_0 \, p_{2d}^{\kappa\kappa} (\zeta_{12},\tau) +
(2\pi i)^{-2d+2}
N_1 \La \! \widetilde{L}_0 \! \Ra_{\! q}
p_{2d-2}^{\kappa\kappa} (\zeta_{12},\tau) + \dots
\qquad
\eeqa
for some constants $C_k$
(\(C_0 = (2\pi i)^{-2d} N_0\),
\(C_1 = (2\pi i)^{-2d+2} N_1 \La \! \widetilde{L}_0 \! \Ra_{\! q}\), etc.)
and \(\kappa := 2d \, \MOD 2\) \(=0,1\)
(this follows from the fact that the singular part of each
$p_k^{\kappa\lambda} (\zeta_{12},\tau)$
contains only the term $\zeta_{12}^{-k}$).

\subsection{Lattice vertex algebras}

An important class of vertex algebras involved as building
blocks in most known examples of 2D CFT is based on the theory
of affine Kac--Moody algebras associated with connected
compact Lie groups (see~\cite{Ka96} and Sect.~1 of \cite{KT97}).
Each such group is \textit{reductive}: it can be written as
the direct product of a (rank~$r$) abelian group \(G = U(1)^r\)
and a semi-simple factor which can be (and often are) treated
separately.
We shall briefly review here the simpler case of a lattice vertex
algebra corresponding to $G$ (or rather, to the affine extension of its
Lie algebra $\widehat{\mathfrak{g}}$ $=u(1)^{\otimes r}$).
The case of nonabelian current algebra has been surveyed by both
mathematicians~--~including Kac's books~\cite{K90} \cite{Ka96} \cite{KR87}
and physicist \cite{GO86} \cite{FST} \cite{DMS}.)

We will construct first the associative algebra $\Alg$ generated by the
fields' modes together with its vacuum representation $\VA$
which will be the linear space of the lattice vertex algebra.
Let $\hh$ be an $r$-dimensional real vector space: denote by
$\hhc$ $:= \hh$ $+ i \hh$ its complexification.
We assume that $\hh$ is endowed with an Euclidean scalar product
$\la h \Vl h' \ra$ $\in \R$ for \(h,h' \in \hh\)
and with a rank $r$ $= \DIM \, \hh$ lattice \(\LAT \subset \hh\)
(an additive subgroup)
such that $\la \alpha \Vl \beta \ra$ $\in \Z$ for all
\(\alpha,\beta \in \LAT\).
Introduce the Heisenberg current Lie algebra $\hhhc$
with generators $h_n$ for \(n \in \Z\)  and \(h \in \hhc\) such that
\beq\label{neq5.45}
\left[ h_n, h'_m \right] \, = \, n \, \delta_{n,-m} \la h \Vl h' \ra
\quad ((\lambda h+ \mu h')_n = \lambda h_n + \mu h'_n , \ \,
\lambda,\mu \in \C)
\, . \
\eeq
Its universal enveloping algebra $\Alg_{\hh}$ contains three
subalgebras $\Ss (\hhhc^{(0,\pm)}\hspace{-0.5pt})$,
generated by $h_0$ and by $h_n$
for \(\mp n = 1,2,\dots\) (\(h \in \hhc\)), respectively.
Note that each of $\Ss (\hhhc^{(0,\pm)})$ is a commutative algebra isomorphic to
the algebra of symmetric polynomials over the underlying vector spaces
and $\Ss (\hhhc^{(0)})$ is central.
Moreover, every element $A$ in $\Alg_{\hh}$ can be represented
as a (finite) sum of \textit{normal order products}:
\(h^1_{n_1} \dots h^m_{n_m}\) with \(n_1 \leqslant \dots \leqslant n_m\),
so that we have the isomorphism of vector spaces
\beq\label{neq5.46}
\Alg_{\hh} \, \cong \,
\Ss (\hhhc^{(+)}) \otimes \Ss (\hhhc^{(0)}) \otimes \Ss (\hhhc^{(-)})
\, .
\eeq
For every \(\alpha\) belonging to the lattice $\LAT$ we will now define
the Fock space representation $\VA_{\alpha}$ generated by
a vector $\ralpha$ and the relations
\beq\label{neq5.47}
h_{n} \ralpha \, = \, 0 ,
\quad h_0 \ralpha \, = \, \la h \Vl \alpha \ra \ralpha
\quad (n = 1,2,\dots , \ \, h \in \hhc)
\, . \
\eeq
In other words, $\VA_{\alpha}$ is isomorphic
to the quotient of $\Alg_{\hh}$ by the left ideal
generated by
$(\{ h_0$ $-$ $\la h \Vl \alpha \ra :$  $h$ $\in$ $\hhc\}$
$\oplus$ $\hhhc^{(-)})$;
so that it is isomorphic, as a vector space, to the symmetric
subalgebra $\Ss (\hhhc^{(+)})$.
The vector space of the lattice vertex algebra is defined as
the sum of all Fock spaces $\VA_{\alpha}$:
\beq\label{neq5.48}
\VA \, = \, \mathop{\oplus}\limits_{\alpha \in \LAT} \, \VA_{\alpha}
\, . \
\eeq

Clearly, the algebra $\Alg_{\hh}$, which is a part of the
full algebra $\Alg$, acts on $\VA$ as the direct sum of its
actions on $\VA_{\alpha}$.
To complete the definition of the algebra $\Alg$ we further introduce
\textit{intertwining operators} $E^{\alpha}$ on $\VA$
for all \(\alpha,\beta \in \LAT\), defined by
\beq\label{neq5.49}
E^{\alpha} (\VA_{\beta}) \subseteqq \VA_{\alpha+\beta}
, \quad
\left[ h_n , E^{\alpha} \right] \, = \,
\delta_{n,0} \, \la h \Vl \alpha \ra \, E^{\alpha}
, \quad
E^{\alpha} \rbeta \, = \,
\epsilon (\alpha,\beta) \,
\rv{\alpha+\beta}
,
\eeq
where
\(h \in \hhc\), \(n \in \Z\) and
$\epsilon (\alpha,\beta)$ are $U(1)$--factors\footnote{%
$\epsilon (\alpha,\beta)$ are, in general,
nonzero complex numbers but in view of the hermitian
structure which we will further define they can
be restricted to $U(1)$--factors.}.
Eqs.~(\ref{neq5.49}) completely determine $E^{\alpha}$
since the $\Alg_{\hh}$--representations $\VA_{\alpha}$
are \textit{irreducible}.
We will require that the products $E^{\alpha} E^{\beta}$
are proportional to~$E^{\alpha+\beta}$, \(E^{0} = \ID_{\VA}\)
and \(\epsilon (\alpha,0) = 1\) (i.e., \(E^{\alpha} \rvac = \ralpha\)).
Computing then $E^{\alpha} E^{\beta} \rvac$ and
$E^{\alpha+\beta} \rvac$
by~(\ref{neq5.49})
one finds
\beq\label{neq5.51}
E^{\alpha} \, E^{\beta} \, = \, \epsilon(\alpha,\beta)
\, E^{\alpha+\beta}
.
\eeq
The algebra $\Alg$ is defined as the algebra generated
by all $E^{\alpha}$ and $\Alg_{\hh}$.
The associativity implies the
\textit{$2$--cocycle relation}:
\beq\label{neq5.52}
\epsilon (\alpha,\beta) \, \epsilon (\alpha+\beta,\gamma) \, = \,
\epsilon (\alpha,\beta+\gamma) \, \epsilon (\beta,\gamma)
\quad (\alpha,\beta,\gamma \, \in \LAT) .
\eeq
The \textit{gauge transformation}
\(E^{\alpha} \mapsto \eta (\alpha) E^{\alpha}\)
(\(\ralpha \mapsto \eta (\alpha) \ralpha\), \(\eta (0) = 1\)),
will give rise to a change of the $2$-cocycle $\epsilon$ by a
\textit{coboundary},
\beq\label{neq5.53}
\epsilon (\alpha,\beta) \, \mapsto \,
\frac{\eta (\alpha) \, \eta (\beta)}{\eta (\alpha+\beta)} \,
\epsilon (\alpha,\beta)
.
\eeq

There are further restrictions on $\epsilon (\alpha,\beta)$
coming from the physical requirements of locality and unitarity.
The first one means that we should define a vertex algebra
structure on the space $\VA$ such that the field modes
span the algebra $\Alg$.
By the Kac's existence theorem (\cite{Ka96} Chapt.~4)
it is enough to introduce
a system of mutually local fields whose modes generate $\Alg$.
Note that $\Alg$ is isomorphic, as a vector space, to the tensor
product $\Alg_{\hh}$ $\otimes$
\(\Span_{\C} \{ E^{\alpha} :\) $\alpha$ $\in$ \(\LAT \}\)
due to the
relations~(\ref{neq5.49})
and~(\ref{neq5.51}).
A system of mutually local fields whose modes generate $\Alg_{\hh}$
is given by the abelian currents
\beq\label{neq5.54}
h (z) \, \equiv \, Y (h,z) \, = \, \Su_{n \, \in \, \Z} \, h_n \, z^{-n-1}
\quad \text{for} \quad h \, \in \hhc
\, , \
\eeq
which obey the canonical commutation relations
\beq\label{neq5.55}
\left[ h (z), h' (w) \right] \, = \,
\la h \Vl h' \ra \, \di_{w} \delta (z-w)
\, , \quad
\delta (z-w) \, := \, \Su_{n \in \Z} \, z^n w^{-n-1}
\eeq
in accord to~(\ref{neq5.45}).
The fields whose modes will contain the operators $E^{\alpha}$ are
defined as follows:
\beq\label{neq5.56}
Y_{\alpha} (z) \, \equiv \, Y (E^{\alpha}, z) \, := \,
E^{\alpha} \, z^{\alpha_{0}} \,
e^{\,\text{\footnotesize \(-\Su_{n > 0} \frac{\alpha_{n}}{n} \, z^n\)}}
\, e^{\text{\footnotesize \(\Su_{n > 0} \frac{\alpha_{-n}}{n} \, z^n\)}}
\eeq
where $\alpha_n$ are the modes in the $\hhh$ corresponding to an element
\(\alpha \in \LAT\) $\subset$ $\hh$.
Note that if we introduce the ``integral'' field $\int \alpha (z)$ by
\beq\label{neq5.57}
\int \alpha (z) \, := \, - \Su_{n \, \neq \, 0} \, \frac{\alpha_{n}}{n} z^{-n}
\, = \, \Su_{n \, \neq \, 0} \, \frac{\alpha_{-n}}{n} z^{n}
\eeq
then (\ref{neq5.56}) is, by definition the Wick \textit{normal ordered exponent}
\beq\label{neq5.58m}
Y_{\alpha} (z) \, = \,
E^{\alpha} \, z^{\alpha_{0}} \,
: \hspace{-2pt}
e^{\,\text{\footnotesize \(\int \alpha (z)\)}} \hspace{-2pt} :
.
\eeq
One derives the commutation relations
\beq\label{neq5.58}
\left[ h (z), Y_{\alpha} (w) \right] \, = \,
\la h \Vl \alpha \ra \, \delta (z-w)
\eeq
and the operator product expansion formula
\beq\label{neq5.59}
Y_{\alpha} (z) \, Y_{\beta} (w) \, = \,
\epsilon (\alpha,\beta) \left( z-w \right)^{\la \alpha \Vl \beta \ra} \,
E^{\alpha+\beta}
z^{\alpha_0} w^{\beta_0}
: \hspace{-2pt}
e^{\,\text{\footnotesize \(\int \alpha (z) + \int \beta (w)\)}} \hspace{-2pt} :
\, . \qquad
\eeq
(Eq.~(\ref{neq5.59}) follows from the \textit{Weyl property}
of the normal ordered exponents
\beq\label{neq5.60}
: \hspace{-2pt}
e^{A (z)} \hspace{-3pt} : \,
: \hspace{-2pt}
e^{B (w)} \hspace{-3pt} :
\ = \,
e^{\La A (z) B (w) \Ra_0} \,
: \hspace{-2pt}
e^{A (z) + B (w)} \hspace{-3pt} :
\eeq
for every two fields $A (z)$ and $B (w)$ whose modes belong to
a Heisenberg Lie algebra, $\La A (z) B (w) \Ra_0$ standing
for their Wick contraction.)
Now the locality assumes that there exists a $\Z_2$--factor
$\left( -1 \right)^{p_{\alpha}}$, \(p_{\alpha} = 0,1\),
such that
\beq\label{neq5.61}
\left( z-w \right)^{N_{\alpha\beta}}
\left(
Y_{\alpha} (z) \, Y_{\beta} (w) - \left( -1 \right)^{p_{\alpha} p_{\beta}}
Y_{\beta} (w) \, Y_{\alpha} (z) \right)
\, = \, 0
\eeq
for \(N_{\alpha\beta} \mgrt 0\)
(in fact, for \(N_{\alpha\beta} =
\Txfrac{\left| \alpha \right|^2+\left| \beta \right|^2}{2}\))
and this combined with~(\ref{neq5.60}) implies that
\beq\label{neq5.62}
p_{\alpha} = \left| \alpha \right|^2 \MOD \, 2
\quad
(\left| \alpha \right| := \la \alpha \Vl \alpha \ra)
, \quad
s (\alpha,\beta) :=
\frac{\epsilon (\alpha,\beta)}{\epsilon (\beta,\alpha)} \, = \,
\left( -1 \right)^{\la \alpha \Vl \beta \ra
+ \left| \alpha \right|^2 \left| \beta \right|^2}
\, . \
\eeq
Note that the gauge transformations~(\ref{neq5.53}) leave invariant the
statistical factor $s (\alpha,\beta)$.
The \textit{stress--energy tensor} of the lattice vertex algebra
is defined by the sum of normal products
\beq\label{neq5.63n}
T (z) = \frac{1}{2} \Su_{j = 1}^r \!
:\hspace{-2pt} \lambda_j (z) \alpha^j (z) \hspace{-2pt}:
\quad (:\hspace{-2pt} \lambda (z) \alpha (z) \hspace{-2pt}:
\, = \! \Su_{n > 0} \! \left( \lambda_{-n} z^{n-1} \alpha (z) +
\alpha (z) \, \lambda_{n-1} z^{-n} \right))
\eeq
where \(\left\{ \alpha^j \right\}_{j \, = \, 1}^r\) is basis of
the lattice $\LAT$ and \(\left\{ \lambda_j \right\}_{j \, = \, 1}^r\)
is the dual basis
\beq\label{neq5.64n}
\la \lambda_j \Vl \alpha^k \ra \, = \, \delta_{jk}
\eeq
spanning the \textit{dual lattice} $\LAT^*$,
which consists of all \(\lambda \in \R^r\) such that \(\la \lambda \Vl \alpha \ra \in \Z\)
for all \(\alpha \in \LAT\).
Eq.~(\ref{neq5.63n}) represents the so called ``\textit{Sugawara formula}''
(for a review and references see~\cite{FST}).
The modes $L_n$ in the Laurent expansion of $T$,
which are expressed in terms of the current modes,
\beq\label{neq5.65n}
T (z) \, = \, \Su_{n \, \in \, \Z} \, L_n \, z^{-n-2}
, \quad
L_n \, = \, \frac{1}{2} \, \Su_{j \, = \, 1}^r \Su_{m} \,
:\hspace{-2pt} \lambda_{jmn-m} \alpha^j_m \hspace{-2pt}:
\eeq
generate the \textit{Virasoro algebra},
characterized by the commutation relations
\beq\label{neq5.66n}
\left[ L_{n},L_{m}\right] \, = \,
(n-m)L_{n+m}+\frac{c}{12}(n^{3}-n)\delta _{n,-m}
\, , \quad
\left[ c,L_{m}\right] \, = \, 0
\eeq
We infer that
\beq\label{neq5.66nn}
\left[ L_0 , Y_{\alpha} (z) \right] \, = \,
\left( z \frac{d}{dz} + \frac{1}{2} \, \left| \alpha \right|^2 \right)
\, , \
\eeq
i.e., the conformal dimension of $Y_{\alpha}$ is
$d_{\alpha}$ $= \frac{1}{2} \, \left| \alpha \right|^2$
(in accord with the value of $p_{\alpha}$ in~(\ref{neq5.62}) and
the spin--dimension connection~(\ref{loc})).

We proceed to define a hermitian structure on a lattice vertex algebra.
The fields $h (z)$ having an interpretation
of currents whose zero modes $h_0$ correspond to real \textit{charges}
(spanning $Q$)
are, hence, assumed hermitian:
\beq\label{neq5.63}
\left( h_n \right)^* \, = \, h_{-n} \quad (n \in \Z) .
\eeq
In particular, the hermiticity of $h_0$ together with the commutation
relations (\ref{neq5.49})
will require that
\(\left( E^{\alpha} \right)^* = \sigma (\alpha) E^{-\alpha}\)
but we can make all
\(\sigma (\alpha)\) equal to $1$ by a suitable gauge transformation
\(E^{\alpha} \mapsto \eta (\alpha) E^{\alpha}\);
so that we have
\beq\label{neq5.64}
\left( E^{\alpha} \right)^* \, = \, E^{-\alpha}
\, . \
\eeq
This completely determines the hermitian structure on $\VA$.
It imposes the following additional properties on the 2-cocycle:
\beq\label{to5.67}
\epsilon(-\beta,-\alpha) \, = \, \overline{\epsilon(\alpha,\beta)}
\, , \quad
\epsilon(\alpha,-\alpha) \, = \, 1
\eeq
(the latter formula uses the fact that
\(E^{\alpha} E^{-\alpha} = E^{\alpha} (E^{\alpha})^*
( = \epsilon(\alpha,-\alpha) \, \ID_V)\) is a positive operator).
To summarize, the 2-cocycle e satisfies the following conditions: (1) \(\epsilon \left( \alpha,\beta \right) \in U(1)\);
(2) \(\epsilon (\alpha,0) \, = \, \epsilon (0,\alpha) = 1\);
(3) Eq.~(\ref{neq5.52}); (4) Eq.~(\ref{neq5.62});
and
(5)~Eq.~(\ref{to5.67}).
A nontrivial example of such cocycle can be given using an ordering
in the lattice $\LAT$.
An equivalent but different choice (with the same symmetry
factor~(\ref{neq5.62})) is made in \cite{BK04}:
\(\epsilon (\alpha,\beta)\) is assumed real,
\(\epsilon : \LAT \times \LAT\) $\to$ $\left\{ \pm 1 \right\}$,
and bimultiplicative (but it does not satisfy
the above condition~(5), and the conjugation law reads
\(\left( E^{\alpha} \right)^* = \epsilon (\alpha,-\alpha)^{-1} E^{-\alpha}\)).

The irreducible positive energy (local field) representations of
the lattice vertex algebra are labeled by elements of the dual
lattice $\LAT^*$ ($:=$ $\{ \lambda$ $\in$ $\hh :$
$\la \lambda \Vl \alpha \ra$ $\in$ $\Z$ for all $\alpha$ $\in$ $\LAT \}$)
modulo $\LAT$;
in other words, they are in one--to--one correspondence with the elements
of the \textit{finite abelian group} $\LAT^* / \LAT$.
The \textit{character} of a representation of weight
\(\lambda \in \LAT^*\) is given by
\beq\label{neq5.68}
\chi_{\lambda} (\tau,\mu) =
\left[ \eta (\tau) \right]^{-r} \Theta_{\lambda}^Q (\tau,\mu) =
\frac{1}{\left[ \eta (\tau) \right]^r} \,
\Su_{\gamma \, \in \, \lambda + \LAT} \,
q^{\frac{1}{2} \la \gamma \Vl \gamma \ra} \,
e^{2\pi i \la \gamma \Vl \mu \ra}
, \quad
\mu \in \hh
\eeq
(where $\eta$ is the Dedekind $\eta$-function~(\ref{3.12}), \(q = e^{2\pi i \tau}\)).
If the lattice $\LAT$ is \textit{even}
(i.e. if the norm square, $\left| \alpha \right|^2$,
of any \(\alpha \in \LAT\) is an even integer)
then $\left\{ \chi (\tau,\mu) \right\}$ span a
finite dimensional representation of $\Gamma (1)$:
\beqa\label{neq5.69}
&
\chi_{\lambda} (\tau+1,\mu)
\, = \,
e^{2\pi i \left(
\text{\footnotesize $\frac{\left|\lambda \right|^2}{2} \! - \! \frac{r}{24}$}
\right)} \,
\chi_{\lambda} (\tau,\mu)
& \nn &
e^{-i \pi \frac{\left| \mu \right|^2}{\tau}} \,
\chi_{\lambda} \left( -\txfrac{1}{\tau}, \txfrac{\mu}{\tau} \right)
\, = \,
\mathop{\dsum}\limits_{\lambda' \, \in \, \LAT^* / \LAT} \,
\left| \LAT^* / \LAT \right|^{-\frac{1}{2}} \,
e^{-2\pi i \la \lambda \Vl \lambda' \ra} \chi_{\lambda'} (\tau,\mu)
\, , & \qquad
\eeqa
where $\left| \LAT^* / \LAT \right|$ is the number of elements of
(the finite group) $\LAT^* / \LAT$.
(For odd lattices $\left\{ \chi_{\lambda} \right\}$ span
a representation of the index three subgroup $\Gamma_{\theta}$~(\ref{2.23})
of~$\Gamma (1)$.)

The case of the (even) \textit{self--dual} lattice $\LAT$ $=$ $E_8$ $(=Q^*)$
is particularly interesting (see~\cite{MH73}, \cite{KP85});
we have a single modular invariant character, $\chi_0$,
in this case
\beq\label{neq5.70}
\chi_0^{E_8} (\tau,0) \, = \, \frac{1}{\left[ \eta (\tau) \right]^8} \,
\Theta_0^{E_8} (\tau,0) \, = \,
\left[ j (\tau) \right]^\frac{1}{3}
\, \
\eeq
where $j (\tau)$ is the absolute invariant~(\ref{j-fun}).
We have, in particular,
\beq\label{neq5.71}
\Theta_0^{E_8} (\tau,0) = \!
\Su_{\gamma \, \in \, E_8} q^{\frac{1}{2} \la \gamma \Vl \gamma \ra}
= \frac{1}{2}
\left[ \vartheta_{00} (0,\tau)^8 \hspace{-2pt} + \vartheta_{10} (0,\tau)^8
\hspace{-2pt} + \vartheta_{01} (0,\tau)^8  \right] = 240 \hspace{1pt} G_4 (\tau)
.
\eeq

\begin{mremark}
A lattice $\LAT$ is self dual iff it is {\it unimodular}, i.e.
iff the volume of its fundamental cell (defined as the square
root of the absolute value of the determinant of the matrix
of inner products of basis vectors (the Gram determinant)
of any given basis of $\LAT$) is one.
Even unimodular lattices only exist in inner products
spaces of signature divisible by $8$ (see~\cite{MH73} Theorem~5.1).
Moreover, even unimodular lattices with indefinite inner product
(i.e. with a non-degenerate symmetric bilinear form such that there
exist vectors of positive and negative square lengths) are determined up to
isomorphism by their rank and signature (\cite{MH73} Theorem~5.3).
In particular, there are unique (isomorphism classes of) even self-dual
lattices of type $(25,1)$ and $(9,1)$ corresponding to bosonic
and super-string theories, respectively.
This is not true for lattices equipped with positive definite (integral)
bilinear form.
For instance, there are two non-isomorphic positive definite
even unimodular lattices of rank $16$: $\Gamma_{16}$
(having a basis of vectors of length squares $2$ and $4$~--~see \cite{MH73}
Lemma~6.5) and $E_8 \oplus E_8$;
there are $24$ such lattices of rank $24$.
By contrast, $E_8$ is the unique (up to isomorphism) even unimodular lattice of signature~$(8,0)$.
A canonical basis in $E_8$ is given by the {\it roots} \(\alpha_1,\dots,\alpha_8\)
whose scalar products are given by the {\it Cartan matrix}:
\(\la \alpha_i \Vl \alpha_j \ra = c_{ij} = c_{ji}\),
\(c_{ii}=2\), \(c_{58}=c_{ii+1}=-1 \, (=c_{i+1i})\) for \(i=1,\dots,6\),
\(c_{ij}=0\) otherwise.
The reader will find more information about the $E_8$ lattice, its automorphism group,
and the associated Lie algebra e.g. in~\cite{Bou68} and in~\cite{K90}, Chapters~4 and~6.
\end{mremark}

\subsection{The $N=2$ superconformal model}

The \(N=2\) (extended) superconformal model \cite{BFK86}
considered as a vertex algebra (in the sense of \cite{Ka96})
is generated by a pair of conjugate to each other local Fermi
fields of dimension $\txfrac{3}{2}$,
\beq\label{eqn5.40n}
G^{\pm} \left( \zeta \right) \, = \,
\sum_{\rho \in \Z + \frac{1}{2}} \,
G^{\pm}_{\rho} \, e^{-2\pi i \rho \zeta}
\, , \quad
\left[ G_{\rho}^{\epsilon},\,  G_{\sigma}^{\epsilon} \right]_+ \, = \, 0
\quad \text{for} \quad \epsilon \, = \, \pm
\, . \
\eeq
Regarded as an (infinite dimensional) Lie superalgebra,
the \(N=2\) extended super--Virasoro algebra,
$SV \left( 2 \right)$, is spanned by $G^{\pm}$, a
$U \left( 1 \right)$ current $J$, the stress--energy tensor $T$
(of modes $L_n$) and a central element $c$.
The non--trivial (anti) commutation relations among their
modes read:
\beqa\label{eqn5.41n}
&& \hspace{-15pt}
\left[ G^{\pm}_{n-\frac{1}{2}}, G^{\mp}_{\frac{1}{2}-m} \right]_+
=
2L_{n-m} \pm \left( n+m-1 \right) J_{n-m} +
\frac{c}{3} n \left( n-1 \right) \delta_{nm}
\, , \ \,
\nn && \hspace{-15pt}
\left[ J_n, J_m \right]
= \frac{c}{3} n \delta_{n, -m}
\, , \ \,
\left[ J_n, G^{\pm}_{\rho} \right] = \pm G^{\pm}_{n+\rho}
\, , \ \,
\left[ J_n, L_m \right] = n J_{n+m}
\, , \ \,
\nn && \hspace{-15pt}
\left[ L_n, G_{\rho}^{\pm} \right]
=
\left( \frac{n}{2} - \rho\right) G^{\pm}_{n+\rho}
\, , \ \,
\left[ L_{n}, L_m \right] =
\left( n-m \right) L_{n+m} +
\frac{c}{12} \left( n^3-n \right) \delta_{n,-m}
\, . \ \nn
\eeqa
Applying the KMS condition and using the first equation~(\ref{eqn5.41n})
and \(\La \! J_0 \! \Ra_{\! q} \, = \, 0\),
we find the following non--zero Gibbs average of products of
$G^{\pm}$--modes
\beq\label{eqn5.42n}
\La \! G_{n-\frac{1}{2}}^{+} G_{\frac{1}{2}-n}^- \! \Ra_{\! q} \, = \,
\frac{q^{n-\frac{1}{2}}}{1+q^{n-\frac{1}{2}}} \,
\left\{ \frac{c}{3} \left( n-\frac{1}{2} \right)^{\! 2}
\! + \La \! 2\widetilde{L}_0 \! \Ra_{\! q} \right\}
, \quad
\widetilde{L}_0 = L_0 - \frac{c}{24}
.
\eeq
This gives the following $q$--expansion of the
$2$--point thermal correlation function
\beqa\label{eqn5.46n}
\hspace{-18pt} &&
\La \! G^{+} \left( \zeta_1 \right) G^- \left( \zeta_2 \right) \! \Ra_{\! q}
\, = \,
\nn
\hspace{-18pt} && \hspace{10pt} =
\frac{2c}{3}
\left\{\raisebox{16pt}{\hspace{-3pt}}\right.
\frac{3+\cos2\pi\zeta_{12}}{4 \left( 2i\sin \pi\zeta_{12} \right)^3}
+
2i \sum_{n \, = \, 1}^{\infty}
\frac{\left( n \! - \! \frac{1}{2} \right)^2 q^{n-\frac{1}{2}}}{
1+q^{n-\frac{1}{2}}} \, \sin \left(  2n \! - \!1 \right) \pi \zeta_{12}
\left.\raisebox{16pt}{\hspace{-2pt}}\right\}
\! +
\nn \hspace{-18pt} && \hspace{10pt} \hspace{10pt} + \,
\La \! \widetilde{L}_0 \! \Ra_{\! q}
\left\{
\frac{1}{i\sin \pi \zeta_{12}} +
4i \sum_{n \, = \, 1}^{\infty} \frac{
q^{n-\frac{1}{2}}}{1+q^{n-\frac{1}{2}}} \,
\sin \left( 2n \! - \! 1 \right) \pi \zeta_{12}
\right\}
\, . \
\eeqa
It can be expressed (as in the example of the Weyl field) in terms of
$p^{11}_k(\zeta_{12},\tau)$ (\ref{2.5nnn}):
\beq\label{eqn5.45n}
\La \! G^{+} \left( \zeta_1 \right) G^- \left( \zeta_2 \right) \! \Ra_{\! q}
\, = \,
\frac{ic}{12\pi^3} \
\pfun^{11}_3 \left( \zeta_{12},\, \tau \right) \, - \,
\frac{i}{\pi} \, \La \! \widetilde{L}_0 \! \Ra_{\! q} \,
\, \pfun_1^{11} \left( \zeta_{12},\, \tau \right)
\, . \
\eeq
In this case the Laurent expansion of type~(\ref{tm3}) takes the form:
\beq\label{neq5.80n}
\La \! G^+ (\zeta_1) G^- (\zeta_2) \! \Ra_{\! q} \, = \,
\frac{ic}{12 \pi^3} \, \zeta_{12}^{-3} - \frac{i}{\pi} \,
\La \! \widetilde{L}_0 \! \Ra_{\! q} \, \zeta_{12}^{-1} + \dots
.
\eeq

As we are no longer dealing with a free field theory the energy mean
$\La \! \widetilde{L}_0 \! \Ra_{\! q}$
is not determined from the thermal $2$--point function of $G^{\pm}$.
It can be computed, however, using our knowledge of the representation
theory of $SV \left( 2 \right)$ (see~\cite{BFK86}).

The Neveu--Schwarz sector of positive energy
unitary irreducible representations (UIR)
of $SV \left( 2 \right)$ are described as follows.
For each of the discrete set of values of the central charge,
\beq\label{eqn5.48n}
c \, = \, c_k \, = \, 3 - \frac{6}{k+2}
\, , \quad
k \, = \, 1,\, 2,\, \dots,\
\eeq
there are \(\BINOMIAL{k+2}{k}\) UIR
\(\left( k;l,m \right)\) with representation spaces
\beq\label{eqn5.49n}
\mathcal{H}_{lm} ( = \mathcal{H}_{lm}^{\left( k \right)})
, \quad
l = 0,1,\dots,k
, \quad
\frac{1}{2} \left( l \! - \! m \right) = 0,1,\dots,l
\quad (m = -l,-l+2,\dots,l)
\, .
\eeq
They are characterized by a \textit{charge} $e_m$ and a \textit{lowest weight}
\(\Delta_{lm}\) given by
\beqa\label{eqn5.50n}
&&
e_m \, = \, \frac{m}{k+2}
\, , \quad
\Delta_{lm} \, = \, \frac{l \left( l+2 \right)-m^2}{4 \left( k+2 \right)}
\, , \quad \text{so that} \nn &&
\left[ e^{2\pi i \left( J_0 -e_m \right)} - 1 \right]
\mathcal{H}_{lm} \, = \, 0 \, = \,
\left[ e^{2\pi i \left( L_0 -\Delta_{lm} \right)} - 1 \right]
\mathcal{H}_{lm}
\, . \
\eeqa
Let $\chi_{lm} \left( \tau,\, k \right)$  be the (restricted) \textit{character}
of the UIR \(\left( k;l,m \right)\):
\beq\label{eqn5.51n}
\chi_{lm} \left( \tau,\, \mu;\, k \right) \, = \,
\TR_{\mathcal{H}_{lm}} \left( q^{\widetilde{L}_0} q_{\mu}^{J_0} \right)
\quad (q_{\mu} = e^{2\pi i \mu})
, \quad
\chi_{lm} \left( \tau,\, k \right) \, = \,
\chi_{lm} \left( \tau,\, 0;\, k \right)
\, . \
\eeq

\begin{mproposition}
The character~(\ref{eqn5.51n}) span a $\BINOMIAL{k+2}{k}$
dimensional representation of the modular group
\(\Gamma_{\theta} (\subset \Gamma \left( 1 \right))\)~(\ref{2.23}).
They are, in particular, eigenvectors of $T^2$,
\beq\label{eqn5.52n}
T^2 \, \chi_{lm} \left( \tau,\, k \right) \, := \,
\chi_{lm} \left( \tau+2,\, k \right) \, = \,
e^{2\pi i \, \left( \Delta_{lm} - c_k \right)} \chi_{lm} \left( \tau\, , k \right)
\, . \
\eeq
Each ``finite ray'' \(\left\{ \eta\chi_{00} \left( \tau,\, k \right):\,
\eta \in \C,\, \eta^{4\left( k+2 \right)} = 1 \right\}\)
is left invariant by the finite index subgroup \(\Gamma_{\theta}^{\left( k \right)}\)
of \(\Gamma_{\theta}\) such that
\beq\label{eqn5.53n}
(\Gamma_{\theta} \, \supset \, ) \, \Gamma_{\theta}^{\left( k \right)} \, := \,
\Gamma_0 \left( 2k+4 \right) \cap \Gamma_{\theta} \, \supset \,
\Gamma \left( 2k+4 \right)
\,  \
\eeq
where \(\Gamma\left( N \right)\) and \(\Gamma_0 \left( N \right)\) are defined
by (\ref{2.15n}) and (\ref{2.20}), respectively.
The group \(\Gamma_{\theta}^{\left( k \right)}\) is generated by $T^2$,
\(ST^{2k+4}S\) and the central element $S^2$ of $\Gamma \left( 1 \right)$.
\end{mproposition}

{\samepage
\begin{Skproof}
We shall first prove that $T^{2k+4}$ acts as a multiple of the unit operator
in the (finite dimensional) space spanned by $\chi_{lm}$:
\beq\label{eqn5.54n}
T^{2\left( k+2 \right)}
\chi_{lm} \left( \tau,\, k \right) \, = \,
e^{i\pi \left\{ l \left(  l+2 \right) - m^2 - \frac{k}{2} \right\}} \,
\chi_{lm} \left( \tau,\, k \right)
\, . \
\eeq
This follows from the explicit form of $c_k$~(\ref{eqn5.48n})
and \(\Delta_{lm}\)~(\ref{eqn5.50n}), and from the observation that
\(l \left( l+2 \right) - m^2\) is even in the range~(\ref{eqn5.49n}).
It implies that each \(\chi_{lm}\) is an eigenvector of
$\Gamma_{\theta}^{\left( k \right)}$.
One can prove using~\cite{G88} that the characters~(\ref{eqn5.51n})
transform among themselves under the modular inversion
according to the law \(\chi_{lm} \left( - \frac{1}{\tau},\, k \right)
\, = \,
\mathop{\sum}\limits_{l'm'} S_{lml'ml} \chi_{l'm'} \left( \tau,\, k \right)\)
with
\beq\label{eqn5.55n}
S_{lml'm'} \, = \,
\frac{2}{k+2} \, \sin \left( \frac{\pi \left( l+1 \right)
\left( l'+1 \right)}{k+2} \right) e^{i\pi \frac{mm'}{k+2}}
\, . \
\eeq
\end{Skproof}}

In order to get a glimpse of the rich variety of physical models
captured by the above series of representations of $SV \left( 2 \right)$
we shall briefly discuss the first two of them,
corresponding to \(k = 1,\, 2\).

The \(k=1\) model is an example of an one dimensional lattice
current algebra (with \(Q=\Z \sqrt{3}\)) considered in Sect.~5.2.
It can be viewed as a local extension
of the $U \left( 1 \right)$ current algebra~--~see~\cite{BMT88}.
There are just three Neveu--Schwarz representations in this case
(corresponding to \(l=m=0\) and to \(l=1\), \(m=\pm 1\))
which can be labeled by a single quantum number $m$ (giving the
charge).
Their conformal dimensions are proportional to the squares of the
corresponding charges:
\beq\label{eqn5.56n}
e_m \, = \, \frac{m}{3}
\, , \quad
m \, = \, 0,\,  \pm 1
\, , \quad
\Delta_l \, = \, \frac{3}{2} e^2 \, ( \, = \, \frac{m^2}{6} )
\, . \
\eeq
The latter formula also applies to the basic
local fermionic fields $G^{\pm}$ (with \(e = \pm\), \(\Delta = \txfrac{3}{2}\)).
(The factor $3$ in the numerator of $\Delta_l$ is the reciprocal of the central
term, \(\txfrac{c}{3} = \txfrac{1}{3}\), in the current commutation relations~(\ref{eqn5.41n}).)
The characters~(\ref{eqn5.51n}) can be computed explicitly in this case in terms
of $\theta$--like functions
that is a special case of~(\ref{neq5.68}):
\beq\label{eqn5.57n}
\chi_{lm} \left( \tau,\, \mu ;\, 3 \right) \, = \,
K_m \left( \tau,\, \mu;\, 3 \right)
\, , \quad
\eta \left( \tau \right) K_m \left( \tau,\, \mu;\, l \right) \, = \,
\sum_{n \in \Z} \,
q^{\frac{l}{2} \left( n+\frac{m}{l} \right)^2} \, q_{\mu}^{n+\frac{m}{l}}
\, . \
\eeq
The modular $S$--matrix~(\ref{eqn5.55n}) (for \(k=1\)) is
then recovered from the known transformation law~(\ref{neq5.69}) for \(K_m\):
\beq\label{eqn5.58n}
K_m \left( -\frac{1}{\tau},\,  \frac{\mu}{\tau};\, l \right) \, = \,
\frac{e^{\frac{2\pi i \mu^2}{l \tau}}}{\sqrt{l}} \,
\sum_{m'=1}^l \, e^{-2\pi i \frac{mm'}{l}} \,
K_{m'} \left( \tau,\, \mu;\, l \right)
\, . \
\eeq
The maximal bosonic subalgebra $\mathcal{B}_2$ of the $N=2$
vertex operator algebra is generated by a pair of
opposite charged fields of charges $\pm 2$
(and dimension \(\txfrac{3}{2} \, 2^2 = 6\)).
The representations of $\mathit{SV}_1 \left( 2 \right)$
(where we denote by $\mathit{SV}_k \left( 2 \right)$ the Lie superalgebra
with (anti)commutation relations~(\ref{eqn5.41n}), in which the central charge $c$
is replaced by its numerical value $c_k$~(\ref{eqn5.48n})) splits
into two pieces with
respect to $\mathcal{B}_2$  and so do the characters $K_m$~(\ref{eqn5.57n}):
\beq\label{eqn5.59n}
K_m \left( \tau,\, \mu;\,3 \right) \, = \,
K_{2m} \left( \tau,\, 2\mu;\,12 \right) +
K_{2m+6} \left( \tau,\, 2\mu;\, 12 \right)
\, \
\eeq
corresponding to minimal charge and conformal weight
\beq\label{eqn5.60n}
e \left( 2m \right) =
\frac{m}{3}
, \
e \left( 2m+6 \right) =
\delta_{m}^{0} - \frac{2m}{3}
, \
\Delta \left( 2m \right) = \frac{m^2}{6}
, \
\Delta \left( 2m+6 \right) = \frac{1}{2} + \frac{m^2}{6}
\, .
\eeq

For $k=2$ the fields $G^{\pm} \left( \zeta \right)$ can be factorized into two
commuting factors
\beq\label{eqn5.61n}
G^{\pm} \left( \zeta \right) \, = \, J^{\pm} \left( \zeta \right)
\psi \left( \zeta \right)
\, \
\eeq
where $J^{\pm} \left( \zeta \right)$ are $\mathit{su} \left( 2 \right)$ currents of
charge $\pm 1$ and dimension $1$ and $\psi \left( \zeta \right)$ is the Majorana--Weyl
fermion of Remark~5.1.
The Neveu--Schwarz representations of $\mathit{SV}_2 \left( 2 \right)$
involve products of $\Z_2$ twisted representation of the
$\widehat{\mathit{su}}_1 \left( 2 \right)$ current algebra.
Their characters can be written in the form:
\beqa\label{eqn5.62n}
\chi_{00} \left( \tau,\, \mu;\,2 \right) \, = && \hspace{-15pt}
\frac{q^{-\frac{1}{48}}}{2}
\left\{\raisebox{15pt}{\hspace{-3pt}}\right. K_0 \left( \tau,\, \mu;\, 2 \right)
\mathop{\prod}\limits_{n\, = \, 1}^{\infty} \left( 1+q^{n-\frac{1}{2}} \right)
+
\nn && \hspace{17pt}
+ K_0 \left( \tau,\, \mu+\frac{1}{2};\, 2 \right)
\mathop{\prod}\limits_{n\, = \, 1}^{\infty} \left( 1-q^{n-\frac{1}{2}} \right)
\left.\raisebox{15pt}{\hspace{-3pt}}\right\} \, = \,
\nn
\, = && \hspace{-15pt} \frac{q^{-\frac{1}{48}}}{\eta \left( \tau \right)}
\left\{ 1 + \left( q_{\mu} +q_{\mu}^{-1} \right) q^{\frac{3}{2}} + q^2 + \dots \right\}
\\ \label{eqn5.63n}
\chi_{1m} \left( \tau,\, \mu;\,2 \right) \, = && \hspace{-15pt}
K_{\frac{m}{2}} \left( \tau,\, \mu;\,2 \right)
q^{-\frac{1}{24}}
\mathop{\prod}\limits_{n\, = \, 1}^{\infty} \left( 1+q^n \right)
\, , \quad m \, = \, \pm 1
\\ \label{eqn5.64n}
&& (e_m \, = \, \frac{m}{4}\, , \quad \Delta_m \, = \,
\left( \frac{m}{4} \right)^2 + \frac{1}{16} \, = \, \frac{1}{8})
\, , \
\nn
\chi_{20} \left( \tau,\, \mu;\,2 \right) \, = && \hspace{-15pt}
\frac{q^{-\frac{1}{48}}}{2}
\left\{\raisebox{15pt}{\hspace{-3pt}}\right. K_0 \left( \tau,\, \mu;\, 2 \right)
\mathop{\prod}\limits_{n\, = \, 1}^{\infty} \left( 1+q^{n-\frac{1}{2}} \right)
-
\nn && \hspace{17pt}
- K_0 \left( \tau,\, \mu+\frac{1}{2};\, 2 \right)
\mathop{\prod}\limits_{n\, = \, 1}^{\infty} \left( 1-q^{n-\frac{1}{2}} \right)
\left.\raisebox{15pt}{\hspace{-3pt}}\right\} \, = \,
\nn
\, = && \hspace{-15pt} \frac{q^{-\frac{1}{48}}}{\eta \left( \tau \right)}
\left\{  q^{\frac{1}{2}} + \left( q_{\mu} +q_{\mu}^{-1} \right) + \dots \right\}
\\ \label{eqn5.65n}
\chi_{2,2m} \left( \tau,\, \mu;\,2 \right) \, = && \hspace{-15pt}
K_{m} \left( \tau,\, \mu;\,2 \right)
q^{-\frac{1}{48}}
\mathop{\prod}\limits_{n\, = \, 1}^{\infty} \left( 1+q^{n-\frac{1}{2}} \right)
\, , \quad m \, = \, \pm 1
\\ \label{eqn5.64nn}
&& (e_{2m} \, = \, \frac{m}{2}\, , \quad \Delta_{2m} \, = \,
\frac{m^2}{4} \, = \, \frac{1}{4})
\, . \
\eeqa

We leave it to the reader to read off the operator content of the
corresponding representations and derive (using~(\ref{eqn5.58n})
and~(\ref{eqn5.37n}))
Eq.~(\ref{eqn5.55n}) for \(k=2\).

\section{Free massless scalar field for even $D$. Weyl and Maxwell fields for $D=4$}
\setcounter{equation}{0}\setcounter{mtheorem}{1}\setcounter{mremark}{1}\setcounter{mdefinition}{1}\setcounter{mexample}{1}\setcounter{mexercise}{1}
\label{sec:6}

\subsection{Free scalar field in $D=2d_0+2$ dimensional space--time}\label{Ssec.4.1}

Generalized free fields \cite{SW} in a QFT with a unique vacuum
can be characterized by having
correlation functions expressed as sums of products of 2-point ones.
It is important for our purposes that this property remains true,
as a corollary, for finite temperature expectation values. We shall
accordingly only deal with 2-point functions and the related energy mean
values in this section.

A canonical z-picture scalar field $\varphi (z)$ (of conformal dimension
$d_0=\txfrac{D-2}{2}$) satisfies the Laplace equation $\Delta_z \varphi (z)=0$
(\(\Delta_z = \di_z^{\, 2}\)),
which assumes, in the real compact picture, the form
\beqa\label{6.1}
&&
\left( \Delta_u -
\left( \frac{\partial}{2\pi\partial \zeta} \right)^2 - d_0^2 \right)
\varphi \left( \zeta,\, u;\, d_0 \right) \, = \, 0
\, , \quad
\nn &&
\Delta_u \, = \,
\partial_u^{\, 2} \, - \, \left( u \cdot \partial_u \right)
\left(
\left( u \cdot \partial_u \right) + 2d_0 \right)
\, , \quad
\partial_u \, = \, \frac{\partial}{\partial u}
\, . \
\eeqa
(Note that the compact picture parametrization \(z = e^{2\pi i \zeta} u\)
can be interpreted as spherical coordinates with a ``logarithmic radius'' $\zeta$.)

\begin{mremark}
The operator $\Delta_u$ in (6.1) is an interior differentiation on
the $(D-1)$-sphere $u^2=1$ since \(
\Delta_u \left\{ \left( u^2 -1 \right) f \left( u \right) \right\}
= \left( u^2-1 \right) \left( \Delta_u \, f -
4 \left( u\cdot \partial_u \right) f \right)\)
\(= 0
\) for \(u^2=1\).
\end{mremark}

The Fourier modes of $\varphi (\zeta, u)$ are eigenfunctions of $\Delta_u$:
\beq\label{6.2}
\varphi \left( \zeta,\, u \right) \, = \,
\sum_{n \in \Z} \, \varphi_n \left( u \right)
e^{-2\pi \, i \, n \, \zeta}
\, , \quad
\left( \Delta_u + n^2 - d_0^2 \right)
\varphi_n \left( u \right) \, = \, 0
\, . \
\eeq
Note that $\varphi_n (z)$ is a homogeneous function of $z$ of degree \(-d_0-n\);
for \(n\geqslant d_0\) $\varphi_{-n} (z)$ is a homogeneous harmonic polynomial of degree
\(n-d_0\). Moreover, we have \(\varphi_n=0\) for \(|n|<d_0\). The compact picture
$2$-point vacuum expectation value,
\beq\label{6.3}
\lvac \varphi \left( \zeta_1,\, u_1 \right)
\varphi \left( \zeta_2,\, u_2 \right) \rvac \, = \,
e^{-2\pi i \, d_0 \, \zeta_{12}} \,
\left( 1-2\cos 2\pi\alpha \,
e^{-2\pi i \, \zeta_{12}} + e^{-4\pi i \, \zeta_{12}} \right)^{-d_0}
\eeq
(\(\cos 2 \pi \alpha := u_1 \cdot u_2\))
is proportional to the generating function for the Gegenbauer polynomials in
$x=\cos 2\pi \alpha$,
\beq\label{6.4}
\left( 1-2xt +t^2 \right)^{-\lambda} \, = \,
\sum_{n \, = \, 0}^{\infty}
\, t^n \, C_n^{\lambda} \left( x \right)
\quad
(\, C_n^{\lambda} \left( x \right) =
\sum_{k \, = \, 0}^{\left[ \frac{n}{2} \right]}
\left( -1 \right)^k \frac{\left( \lambda \right)_{n-k}}{k!}
\frac{\left( 2x \right)^{n-2k}}{\left( n-2k \right)!} \, )
\, \
\eeq
that satisfy the differential equation
\beq\label{6.5}
\left[ \left( 1-x^2 \right) \frac{d^2}{d x^2} -
\left( 2\lambda+1 \right) x \frac{d}{dx} +
n \left( n+2\lambda \right) \right] C_n^{\lambda} \left( x \right)
\, = \, 0
\, \
\eeq
and the orthogonality and normalization conditions
\beqa\label{6.6}
&
\Txfrac{2^{2\lambda-1}}{\pi}
\INTG{-1}{1}
C_n^{\lambda} \left( x \right)
C_m^{\lambda} \left( x \right)
\left( 1-x^2 \right)^{\lambda-\frac{1}{2}} \, dx \, = \,
\Txfrac{\Gamma \left( 2\lambda+n \right) \delta_{mn}}{
\Gamma^2 \left( \lambda \right) n! \left( n+\lambda \right)}
\, , &
\nn &
C_n^{\lambda} \left( 1 \right) \, = \,
\BINOMIAL{2\lambda+n-1}{n}
\, . &
\eeqa
It follows that $\varphi_n (u)$ obey the commutation relations
\beq\label{6.7}
\left[ \varphi_n \left( u_1 \right),\, \varphi_m \left( u_2 \right) \right]
\, = \,
\frac{n}{\left| n \right|}
\delta_{n,-m} \, C_{\left| n \right|-d_0}^{d_0}
\left( \cos 2\pi \alpha \right)
\quad \text{for} \quad
\left| n \right| \geqslant d_0
\, . \
\eeq
For $D=4, d_0=1$ (the case studied in \cite{T86})
these {\it canonical commutation relations}
assume an elementary explicit form:
\beq\label{6.8}
\left[ \varphi_n \! \left( u_1 \right),
\varphi_m \! \left( u_2 \right) \right] =
\frac{\sin 2\pi n \alpha}{\sin 2\pi \alpha} \,
\delta_{n,-m}
\quad ( D=4,\ d_0 = 1,\
\cos 2 \pi \alpha = u_1 \cdot u_2)
.
\eeq

The calculations~(\ref{CCR2})--(\ref{KMS-phi2}) in this case give
\beqa
&
q^n \La \varphi_n \left( u_1 \right)
\varphi_{-n} \left( u_2 \right) \Ra_{\! q}
=
\La \varphi_{-n} \left( u_2 \right)
\varphi_{n} \left( u_1 \right) \Ra_{\! q}
=
& \nn & =
\La \varphi_n \! \left( u_1 \right)
\varphi_{-n} \! \left( u_2 \right) \Ra_{\! q}
\! - C^{d_0}_{n-d_0} \! \left( u_1 \!\cdot\! u_2 \right)
&
\nonumber
\eeqa
and hence
\beq\label{6.14}
\La \! \varphi_n \! \left( u_1 \right)
\varphi_{-n} \! \left( u_2 \right) \! \Ra_{\! q}
=
\frac{C^{d_0}_{n-d_0} \! \left( u_1 \!\cdot\! u_2 \right)}{1-q^n}
=
q^{-n} \La \! \varphi_{-n} \! \left( u_2 \right)
\varphi_{n} \! \left( u_1 \right) \! \Ra_{\! q}
\eeq
for \(n \geqslant d_0\). Inserting this in the Fourier expansion
of the 2-point function,
\beq\label{6.13}
\La \varphi \left( \zeta_1,\, u_1 \right)
\varphi \left( \zeta_2,\, u_2 \right) \Ra_{\! q}
\, = \,
\Su_{|n| \, \geqslant \, d_0} \,
\La \varphi_n \left( u_1 \right)
\varphi_{-n} \left( u_2 \right) \Ra_{\! q}
\, e^{-2\pi i \, n \, \zeta_{12}}
\, . \
\eeq
we find
\beqa\label{6.15}
& \hspace{-10pt}
\La \varphi \left( \zeta_1,\, u_1 \right)
\varphi \left( \zeta_2,\, u_2 \right) \Ra_{\! q}
\, = \,
\left( -4\sin \pi\zeta_+ \sin \pi\zeta_- \right)^{-d_0}
+
& \nn & \hspace{-10pt}
+ 2 \mathop{\dsum}\limits_{n \, = \, d_0}^{\infty}
\Txfrac{q^n}{1-q^n}
\cos 2 \pi n \zeta_{12} \
C^{d_0}_{n-d_0} \! \left( \cos 2\pi\alpha \right)
=
\Txfrac{1}{(4\pi)^{2d_0}}
P_{d_0} (\zeta_{12}; u_1,u_2; \tau)
& \qquad
\eeqa
where according to Eq.~(\ref{nw4.77}) $P_{d_0}$
is the basic elliptic function~(\ref{basic_ef}).
We thus have found in particular, the $q$--expansion of the functions
$P_k$ since the above arguments are valid for any field dimension~$d$.

In the case \(D=4\) (\(d_0 = 1\)), using the canonical normalization~(\ref{18e})
we deduce
\beqa\label{neq6.12n}
&&
\lvac \varphi (\zeta_1,u_1) \varphi (\zeta_2,u_2) \rvac \, = \,
-\frac{1}{8 \pi \sin \pi \zeta_+ \sin \pi \zeta_-} \, = \,
\nn && \quad = \,
\frac{1}{4\pi \sin 2\pi \alpha}
\left( \cotg \pi \zeta_+ - \cotg \pi \zeta_- \right) \, = \,
\nn && \quad = \,
\frac{1}{4\pi \sin 2\pi \alpha}
\Su_{n \, \in \, \Z}
\left( \frac{1}{\zeta_+ + n} - \frac{1}{\zeta_- + n}  \right)
\eeqa
and the passage to the thermal $2$--point function consists
of replacing the sum with the doubly periodic Eisensten-Weierstrass series
\beq\label{6.11}
\La \varphi \left( \zeta_1,\, u_1 \right)
\varphi \left( \zeta_2,\, u_2 \right) \Ra_{\! q}
\, = \,
\frac{1}{4\pi \sin 2\pi\alpha}
\left( \pfun_1 \left( \zeta_+,\, \tau \right) -
\pfun_1 \left( \zeta_-,\, \tau \right) \right)
\quad \!
\text{for} \quad \! D=4
\eeq
(which corresponds to Eq.~(\ref{p_1-rep2})).
Similarly, for \(D = 6\) (\(d_0 = 2\)) we have
\beqa\label{6.10}
&&
\lvac \varphi \left( \zeta_1,\, u_1 \right)
\varphi \left( \zeta_2,\, u_2 \right) \rvac
\, = \,
\left( 2\sin \pi\zeta_+ \sin \pi\zeta_- \right)^{-2}
\, = \,
\nn && \quad
= \,
\frac{1}{16 \sin^2 2\pi\alpha}
\left\{\raisebox{10pt}{\hspace{-3pt}}\right.
\left( \sin \pi\zeta_- \right)^{-2}
+
\left( \sin \pi\zeta_+ \right)^{-2}
+ \nn && \quad
+ \,
2 \cotg \, 2\pi\alpha
\left(\raisebox{9pt}{\hspace{-2pt}}\right.
\cotg \, \pi\zeta_- - \cotg \, \pi\zeta_+
\left.\raisebox{9pt}{\hspace{-2pt}}\right)
\left.\raisebox{10pt}{\hspace{-3pt}}\right\}
\, ,\,
\nn &&
\La \varphi \left( \zeta_1,\, u_1 \right)
\varphi \left( \zeta_2,\, u_2 \right) \Ra_{\! q}
\, = \,
\left( 4 \pi \sin 2\pi\alpha \right)^{-2}
\left\{\raisebox{10pt}{\hspace{-3pt}}\right.
\pfun_2 \left( \zeta_-,\, \tau \right)
+
\pfun_2 \left( \zeta_+,\, \tau \right)
+ \nn && \quad
+ \,
2 \pi \cotg \, 2\pi\alpha
\left(\raisebox{9pt}{\hspace{-2pt}}\right.
\pfun_1 \left( \zeta_-,\, \tau \right) -
\pfun_1 \left( \zeta_+,\, \tau \right)
\left.\raisebox{9pt}{\hspace{-2pt}}\right)
\left.\raisebox{10pt}{\hspace{-3pt}}\right\}
\, . \qquad
\eeqa

To compute the mean thermal energy by Eq.~(\ref{mean_H1}) we have to find
the dimensions $d_b (n)$ of the space
$\Span_{\C} \left\{ \varphi_{-n} (u) \rvac \right\}$
of the energy $n$ one particle states (\(d_f (n) = 0\) in this case).
According to the properties of the mode expansion of $\varphi (\zeta,u)$
every such space is isomorphic to the space of homogeneous harmonic polynomials
in $u$ of degree $n-d_0$ for \(n = d_0, d_0+1,\dots\).
Recalling that the generating function for the dimensions of the spaces is
$\Txfrac{1-t^2}{(1-t)^D}$
($=$ $\mathop{\dsum}\limits_{n \, = \, d_0}^{\infty} d_b (n)$ $t^{n-d_0}$,
\(D=2d_0+2\)) we find
\(d_b (n) = n^2\) for \(D = 4\) and for even \(D > 4\):
\beq\label{neq6.15n}
d_b (n) \, = \,
\frac{2n^2}{\left( 2d_0 \right)!} \, \mathop{\prod}\limits_{k \, = \, 1}^{d_0 -1}
\left( n^2 - k^2 \right)
\, = \,
\mathop{\sum}\limits_{k \, = \, 0}^{d_0} \,
c_k^{\left( D \right)} n^{2k}
\, . \
\eeq
Thus, using the $q$--expansions~(\ref{3.7}) we find
\beq\label{neq6.16n}
\La \! H \! \Ra_{q}
( \equiv
\frac{\mathit{tr}_{\DOM{}} \left( H \, q^H \right)}{
\mathit{tr}_{\DOM{}} \left( q^H \right)} \, ) =
\mathop{\sum}\limits_{n \, = \, 1}^{\infty}
\frac{n \, d \left( n \right) \, q^n}{1-q^n}
=
\mathop{\sum}\limits_{k \, = \, 1}^{d_0+1}
c_{k-1}^{\left( D \right)} \,
\frac{B_{2k}}{4k}
\, +
\mathop{\sum}\limits_{k \, = \, 1}^{d_0+1}
c_{k-1}^{\left( D \right)}
G_{2k} \left( \tau \right)
\eeq
for even \(D > 4\) ($B_{2k}$ being the Bernoulli numbers).
In particular, for the physical 4-dimensional case we have:
\beq\label{neq6.17n}
\La \! H + E_0 \! \Ra_{\! q} \, = \, G_4 (\tau)
\quad \text{for} \quad E_0 \, = \, \frac{1}{240}
\, . \
\eeq

\begin{mremark}
All thermal correlation functions have well defined
restrictions to coinciding $u_a$ on the $D-1$ sphere.
This is seen directly
from our formulae and is a consequence of a general observation by
Borchers~\cite{Bo64}.
The result is an expression for the corresponding
correlation functions in a chiral CFT. For instance the chiral (2D)
restriction of the 2-point function~(\ref{6.11}),
\beq\label{old6.18}
\mathcal{W} \! \left( \zeta_{12}, \tau \right) =
- \left( 2\pi \right)^{-2}
\pfun_2 \! \left( \zeta_{12}, \tau \right) =
- \left( 2\sin \pi\zeta_{12} \right)^{-2} \! + 2
\sum_{n \, = \, 1}^{\infty}
\frac{nq^n}{1-q^n}
\cos 2 \pi n \, \zeta_{12}
\eeq
coincides with the thermal 2-point function of a chiral $U(1)$ current.
The importance of this remark stems from the fact that it is easier to verify,
say, Wightman positivity in the 1-dimensional (chiral) case, thus
obtaining a necessary condition for the existence of a consistent higher
dimensional theory.
\end{mremark}

\subsection{Weyl fields}\label{Sec.7}

We begin by introducing the $2\times 2$ matrix representation of the quaternionic algebra
(see also Appendix~\ref{ap:2n})
which will prove useful for studying both spinor and antisymmetric tensor z-picture fields:
\beqa\label{6.33}
&
Q_k \, = \, -i \, \sigma_k \, = \, - Q_k^+ \quad (k \, = \, 1,\, 2,\, 3)
\, , \quad
Q_4 \, = \, \ID &
\nn &
\sigma_{1} \, = \, \left( \begin{array}{cc} 0 & 1 \\ 1 & 0 \end{array}  \right)
\, , \quad
\sigma_{2} \, = \, \left( \begin{array}{cc} 0 & -i \\ i & 0 \end{array}  \right)
\, , \quad
\sigma_{3} \, = \, \left( \begin{array}{cc} 1 & 0 \\ 0 & -1 \end{array}  \right)
, & \qquad
\eeqa
characterized by the anticommutation relations
\beq\label{6.34}
Q_{\cmu}^+ \, Q_{\cnu} + Q_{\cnu}^+ \, Q_{\cmu} \, = \,
2 \delta_{\cmu\cnu} \, = \,
Q_{\cmu} \, Q_{\cnu}^+ + Q_{\cnu} Q_{\cmu}^+
\quad \text{for} \quad \cmu,\, \cnu = 1,\, \dots,\, 4
\, . \
\eeq
Here and below we denote (as in Sect.~4.4) the hermitean matrix conjugation
by a superscript ``+''. The matrices
\beq\label{6.35}
i \, \sigma_{\cmu\cnu} \, = \,
\frac{1}{2} \left( Q_{\cmu}^+ \, Q_{\cnu} - Q_{\cnu}^+ \, Q_{\cmu} \right)
\, , \quad
i \, \widetilde{\sigma}_{\cmu\cnu} \, = \,
\frac{1}{2} \left( Q_{\cmu} \, Q_{\cnu}^+ - Q_{\cnu} \, Q_{\cmu}^+ \right)
\, \
\eeq
are the selfdual and antiselfdual antihermitean $\spin (4)$ Lie
algebra generators. We shall also use the notation
\beq\label{6.36}
\dirz = \mathop{\sum}\limits_{\cmu = 1}^{4} \, z^{\cmu} \, Q_{\cmu}
, \ \
\dirz^+ = \, \mathop{\sum}\limits_{\cmu = 1}^{4} \, z^{\cmu} \, Q_{\cmu}^+
, \ \
\dirdi_z = \mathop{\sum}\limits_{\cmu = 1}^{4} \, Q_{\cmu} \, \di_{z^{\cmu}}
, \ \
\dirdi_z^+ = \mathop{\sum}\limits_{\cmu = 1}^{4} \, Q_{\cmu}^+ \, \di_{z^{\cmu}}
, \
\eeq
Note that in the definition of $\dirz^+$ we do not conjugate the coordinates~$z^{\cmu}$.
Then Eqs.~(\ref{6.34}) are equivalent to
\beq\label{6.37}
\dirz_1^+ \, \dirz_2 + \dirz_2^+ \, \dirz_1 \, = \,
\dirz_1 \, \dirz_2^+ + \dirz_2 \, \dirz_1^+ \, = \,
2 \, z_1 \spr z_2
\quad
(\dirz^+ \, \dirz \, = \, \dirz \, \dirz^+ \, = \, z^{\, 2})
\, . \
\eeq

The Weyl generalized free fields of dimension
\(d = \txfrac{1}{2},\,
\txfrac{3}{2},\, \dots\) are two mutually conjugate
complex $2$--component fields,
\beq\label{6.38}
\chi \left( z \right)^+ \, = \,
\left( \chi^*_1 \left( z \right),\, \chi^*_2 \left( z \right) \right)
\quad \text{and} \quad \chi \left( z \right) \, = \,
\left( \begin{array}{c} \chi_1 \left( z \right) \\ \chi_2 \left( z \right) \end{array} \right)
\, , \
\eeq
transforming under the
elementary induced representations of \(\spin \left( 4 \right)\)
corresponding to the selfdual and antiselfdual representations $\sigma$
and $\widetilde{\sigma}$~(\ref{6.35}),
respectively.
In particular, the action of the Weyl reflection $j_W$ is,
\beqa\label{6.39}
\chi \left( z \right) \ \longmapsto &&
\frac{\dirz}{\left( z^{\, 2} \right)^{
d+\frac{\raisebox{0pt}{\small $1$}}{\raisebox{0pt}{\small $2$}}}} \ \chi \left( z \right)
\, ( \, \equiv \, \pi(z,R) \, \chi \left( z \right) \, )
\, , \quad \nn
\chi^+ \left( z \right) \ \longmapsto &&
\chi^+ \left( z \right) \,
\frac{\dirz}{\left( z^{\, 2} \right)^{
d+\frac{\raisebox{0pt}{\small $1$}}{\raisebox{0pt}{\small $2$}}}}
\, ( \, \equiv \, \pi^+(z,R) \, \chi^+ \left( z \right) \, )
\, . \
\eeqa
The conformal invariant $2$--point functions, characterizing the fields, have
the following matrix representation
\beq\label{6.40}
\lvac \chi \left( z_1 \right) \chi^+ \left( z_2 \right) \rvac \, = \,
\frac{\dirz^+_{12}}{\left( z_{12}^{\, 2} \right)^{
d+\frac{\raisebox{0pt}{\small $1$}}{\raisebox{0pt}{\small $2$}}}}
\, , \quad
\eeq
\beqa
\lvac \chi_{\alpha} \left( z_1 \right) \chi_{\beta} \left( z_2 \right) \rvac \, = \,
\lvac \chi_{\alpha}^+ \left( z_1 \right) \chi^+_{\beta} \left( z_2 \right) \rvac \, = \,
0
\, . \
\nonumber
\eeqa
In particular, the invariance under the reflection $I(z)$~(\ref{equ2.12}) is ensured by the
equality
\beq\label{6.42}
\frac{\dirz_1}{z^{\, 2}_1} \, \dirz_{12}^+ \, \frac{\dirz_2}{z^{\, 2}_2} \, = \,
\frac{\dirz_1^+}{z^{\, 2}_1} - \frac{\dirz_2^+}{z^{\, 2}_2}
\, . \
\eeq
The conjugation law reads
\beq\label{6.43}
\chi^+ \left( \overline{z} \right)^+ \, = \,
\frac{\dirz}{\left( z^{\, 2} \right)^{
d+\frac{\raisebox{0pt}{\small $1$}}{\raisebox{0pt}{\small $2$}}}}
\ \,
\chi \left( \frac{z}{z^{\, 2}} \right)
\, . \
\eeq

The canonical, $d=\txfrac{3}{2}$,
$z$-picture Weyl field $\psi$ and its subcanonical counterpart $\chi$ satisfy a
first and a third order partial differential equation, respectively:
\beq\label{6.44}
\dirdi_z \, \psi \left( z \right) \, = \, 0
\, , \quad
\Delta_z \, \dirdi_z \, \chi \left( z \right) \, = \, 0
\, . \
\eeq
Their vacuum correlation functions are diagonal in
``the moving frame'' representation
defined as follows.
For given non-collinear unit real vectors \(u_1,\, u_2 \in \Sr^{D-1} ( \subset \R^D)\)
such that \(u_1 \spr u_2 = \cos 2\pi \alpha\)
let $v$ and $\overline{v}$ be the unique complex vectors (in $\C^D$)
for which
\beq\label{6.45}
u_1 \, = \, e^{\pi \, i \, \alpha} \, v + e^{-\pi \, i \, \alpha} \, \overline{v}
\, , \quad
u_2 \, = \, e^{-\pi \, i \, \alpha} \, v + e^{\pi \, i \, \alpha} \, \overline{v}
\, . \
\eeq
Their compact picture 2-point functions have the form:
\beq\label{6.46}
\lvac \chi \left( \zeta_1,\, u_1 \right) \chi^+ \left( \zeta_2,\, u_2 \right) \rvac
\, = \,
\frac{1}{2\hspace{1pt}i} \left( \frac{\dirv^+}{\sin \pi \zeta_-} +
\frac{\overline{\dirv}{\hspace{1pt}}^+}{\sin \pi \zeta_+} \right)
\, , \
\eeq
\beqa\label{6.46n}
&& \hspace{-15pt}
\lvac \psi \left( \zeta_1,\, u_1 \right) \psi^+ \left( \zeta_2,\, u_2 \right) \rvac
=
\frac{1}{2\hspace{1pt}i
\sin \pi\zeta_- \, \sin \pi \zeta_+} \left( \frac{\dirv^+}{\sin \pi \zeta_-} +
\frac{\overline{\dirv}{\hspace{1pt}}^+}{\sin \pi \zeta_+} \right)
\, = \,
\nn
&& \hspace{-15pt} =
\frac{i}{8\ \sin 2\pi\alpha}
\left(\raisebox{12pt}{\hspace{-2pt}}\right.
{\dirv^+}
\left(\raisebox{12pt}{\hspace{-2pt}}\right.
\frac{\cos \pi\zeta_-}{\sin^2 \pi\zeta_-}
-\frac{\cotg \, 2\pi\alpha}{\sin \pi\zeta_-}
+\frac{1}{\sin 2\pi\alpha \, \sin \pi\zeta_+}
\left.\raisebox{12pt}{\hspace{-2pt}}\right) -
\nn && \hspace{-15pt} \hspace{68pt}
-\overline{\dirv}^+
\left(\raisebox{12pt}{\hspace{-2pt}}\right.
\frac{\cos \pi\zeta_+}{\sin^2 \pi\zeta_+}
+
\frac{\cotg \, 2 \pi\alpha}{\sin \pi\zeta_+}
-\frac{1}{\sin 2\pi\alpha \, \sin \pi\zeta_-}
\left.\raisebox{12pt}{\hspace{-2pt}}\right)
\left.\raisebox{12pt}{\hspace{-2pt}}\right)
\, , \
\eeqa
where \(\zeta_{\pm} = \zeta_{12} \pm \alpha\) (as in previous sections).
Let $N$ be the ``charge operator'' defined by
\(\left[ N, \chi^+ \left( z \right) \right]\) $=$ \(\chi^+ \left( z \right)\),
\(\left[ N,\, \chi \left( z \right) \right] = - \chi \left( z \right)\)
(and similarly for $\psi$).
We introduce the {\it grand canonical mean value}
\beq\label{6.47}
\La A \Ra_{q,\,\mu}
\, := \,
\frac{\TR_{\mathcal{H}} \left( A \, q^H \,
e^{2 \pi i\hspace{1pt}\mu\hspace{1pt}N}
\right)}{
\TR_{\mathcal{H}}
\left( q^H \, e^{2\hspace{1pt}\pi\hspace{1pt}i\hspace{1pt}\mu\hspace{1pt}N} \right)}
\, . \
\eeq
Then the grand canonical 2-point correlation functions assume the form:
\beq\label{6.48}
\La \hspace{-2pt}
\chi \hspace{-1pt} \left( \zeta_1,\, u_1 \right)
\chi^+ \hspace{-2pt} \left( \zeta_2,\, u_2 \right) \hspace{-2pt}
\Ra_{q,\mu}
=
\frac{1}{2\pi\hspace{1pt}i}
\left(
\pfun_1^{11} \hspace{-2pt} \left( \zeta_-,\, \tau,\, \mu \right) \, \dirv^{\hspace{1pt}+}
\hspace{-2pt} +
\pfun_1^{11} \hspace{-2pt} \left( \zeta_+,\, \tau,\, \mu \right) \, \overline{\dirv}{\hspace{1pt}}^+
\right)
\, \
\eeq
\beqa\label{6.48n}
&& \hspace{-15pt}
\La \psi \left( \zeta_1,\, u_1 \right) \psi^+ \left( \zeta_2,\, u_2 \right) \Ra_{q,\mu}
=
\frac{i}{8\pi \sin 2\pi\alpha}
\left(\raisebox{12pt}{\hspace{-2pt}}\right.
{\dirv^+}
\left(\raisebox{12pt}{\hspace{-2pt}}\right.
\pfun_2^{11} \left( \zeta_-,\tau,\mu \right) -
\nn && \hspace{-15pt}
-\cotg \, 2\pi\alpha \ \pfun_1^{11} \left( \zeta_-,\tau,\mu \right)
+\frac{\pfun_1^{11} \left( \zeta_+,\tau,\mu \right)}{\sin 2\pi\alpha}
\left.\raisebox{12pt}{\hspace{-2pt}}\right) -
\overline{\dirv}^+
\left(\raisebox{12pt}{\hspace{-2pt}}\right.
\pfun_2^{11} \left( \zeta_+,\tau,\mu \right) +
\nn && \hspace{-15pt}
+\cotg \, 2\pi\alpha \ \pfun_1^{11} \left( \zeta_+,\tau,\mu \right)
-\frac{\pfun_1^{11} \left( \zeta_-,\tau,\mu \right)}{\sin 2\pi\alpha}
\left.\raisebox{12pt}{\hspace{-2pt}}\right)
\left.\raisebox{12pt}{\hspace{-2pt}}\right)
\, , \
\eeqa
where
\beq\label{6.49}
\pfun_k^{\kappa\lambda} \left( \zeta,\tau,\mu \right)
\, = \,
\mathop{\lim}\limits_{M \to \infty} \, \mathop{\lim}\limits_{N \to \infty} \,
\mathop{\sum}\limits_{m = -M}^{M} \, \mathop{\sum}\limits_{n = -N}^{N} \,
\frac{e^{\pi i \hspace{1pt} m \hspace{1pt}
	\left( 2 \hspace{1pt}\mu+\kappa \right)} \,
e^{\pi i \hspace{1pt} n \hspace{1pt} \lambda}}{
\left( \zeta+m\hspace{1pt}\tau+n \right)^k}
\, , \quad
k = 1,2,\dots
\, \
\eeq
(for the properties of these extended $p$-functions,
see Appendix~\ref{ap:1}).

We observe that the \({\alpha} \to 0\)
(\(u_1=u_2\)) limit of~(\ref{6.48}) reproduces the
Gibbs $2$-point function~(\ref{???}) of the chiral Weyl field for \(D=2\) (as
\(\dirv^++\overline{\dirv}^+=\diru^+\) is equal to the unit matrix in the frame
\(u=(\mbf{0},1)\).
It is noteworthy that it is modular invariant for
\(\mu=0,\txfrac{1}{2}\); for instance,
\(\pfun_1^{11}({\zeta,\tau},0)=\pfun_1^{11}(\zeta,\tau)\)
obeys~(\ref{p_modinv}) with \(k=1\) and
\({\gamma} \in {\Gamma_{\theta}}\) (see~(\ref{eqnXX.15})
and Exercise~2.14).
By contrast, the \(d=\txfrac{3}{2}\) restricted 2-point function
\beq\label{7.19new}
\La {\psi}({\zeta_1},u){\psi}^+({\zeta_2},u) \Ra_{\! q}
=\frac{i}{(2{\pi})^3} \,
\left( \pfun_3^{11}({\zeta_{12},\tau})+\frac{{\pi}^2}{2} \,
\pfun_1^{11}({\zeta_{12},\tau}) \right)
\eeq
is a linear superposition of modular functions of weight 3 and 1, and
hence, is not modular invariant.

\begin{mremark}
Comparing with the \(N=2\) superconformal model of Sect.~5.3 we
observe that one would have had the same problem had one assumed that the pair
of \(d=\txfrac{3}{2}\) fields
\(G^{\pm}\) satisfy $c$-number anticommutation relations (instead
of~(\ref{eqn5.41n})).
In \(D=4\), however, one is bound to consider the
\(d=\txfrac{3}{2}\) field
${\psi}$ as a free field unless one is prepared to give up Wightman
positivity. Indeed, the unique conformally invariant 2-point
function~(\ref{6.40}) for \(d=\txfrac{3}{2}\) implies (in a positive metric
Hilbert space framework) the free field equation~(\ref{6.44})
for ${\psi}$. It appears intriguing to try to
construct an indefinite metric model for a \(d=\txfrac{3}{2}\)
Weyl field with a modular invariant 2-point function.
\end{mremark}

\medskip

We conclude this subsection with a consideration of
the Gibbs energy distributions for Weyl fields.

The temperature 2-point function~(\ref{7.19new}) can be used to derive the
meanvalue of the conformal Hamiltonian in the equilibrium (Gibbs) state at
hand. Denote by $\Rw (\zeta_{12},{\tau})$ the function
$\W_{\frac{3}{2}}$~(\ref{7.19new}),
regularized by subtracting its (third and first order) poles at
\(\zeta_{12}=0\). Then the canonical expression for the conformal Hamiltonian
of a free Weyl field tells us that its Gibbs energy distribution is given
by
\beq\label{7.19neww}
\La \! H \! \Ra_{\! q}
=
\frac{1}{2{\pi}i} \,
\mathop{\lim}\limits_{\zeta \to 0} \ \frac{\di}{\di\zeta} \, \Rw({\zeta,\tau})
\, . \
\eeq
We shall verify this formula by a direct computation that will also apply
to deriving the energy distribution of the subcanonical field ${\chi}$.

The positive charge 1--particle state--space is a direct sum of energy
eigenspa\-ces corresponding to eigenvalues
\(E_0+n+\txfrac{3}{2}\), \(n=0,1,\dots\),
where $E_0$ is the vacuum energy. Each such eigenspace is spanned by
vectors of the form \({\psi}_{-n-\frac{3}{2}}^+(u)\rvac\) and carry the irreducible
representation
\(\left(
\txfrac{n \! + \! 1}{2},\,
\txfrac{n}{2}
\right)\)
of \(\text{\it Spin}(4)\) of dimension \((n+2)(n+1)\).
The dimension of the full
$1$--particle space, including charge $-1$ states, is twice as big.
It follows that
\beq\label{eqn6.26}
\La \! H \! \Ra_{q}
\, = \,
E_0 +
\mathop{\sum}\limits_{n \, = \, 0}^{\infty} \,
\frac{2
\left(\raisebox{9pt}{\hspace{-2pt}}\right.
n \hspace{-1pt} + \hspace{-1pt}
\frac{\raisebox{0pt}{\small $3$}}{\raisebox{0pt}{\small $2$}}
\left.\raisebox{9pt}{\hspace{-2pt}}\right)
\left( n \hspace{-1pt} + \hspace{-1pt} 1 \right)
\left( n \hspace{-1pt} + \hspace{-1pt} 2 \right)
q^{n+\frac{\raisebox{0pt}{\small $3$}}{\raisebox{0pt}{\small $2$}}}
}{
1 + q^{n+\frac{\raisebox{0pt}{\small $3$}}{\raisebox{0pt}{\small $2$}}}
}
\eeq
which is verified to coincide with~(\ref{7.19neww})
(for a suitable choice of $E_0$).
To express this Gibbs average in terms of modular forms we use the
identity \(n(n+1)\left( n+\txfrac{1}{2} \right)={\txfrac{1}{8}}
\left( (2n+1)^3-(2n+1) \right)\) with the result
\beq\label{newww}
\La \! H \! \Ra_{\! q}
\, = \,
\frac{1}{4}
\left(\raisebox{10pt}{\hspace{-2pt}}\right.
G_4 \! \left(\raisebox{9pt}{\hspace{-2pt}}\right.
\frac{\tau \! + \! 1}{2}
\left.\raisebox{9pt}{\hspace{-2pt}}\right) - 8 \, G_4 \! \left( \tau \right)
\left.\raisebox{10pt}{\hspace{-2pt}}\right)
\, - \,
\frac{1}{4}
\left(\raisebox{10pt}{\hspace{-2pt}}\right.
G_2 \! \left(\raisebox{9pt}{\hspace{-2pt}}\right.
\frac{\tau \! + \! 1}{2}
\left.\raisebox{9pt}{\hspace{-2pt}}\right) - 2 \, G_2 \! \left( \tau \right)
\left.\raisebox{10pt}{\hspace{-2pt}}\right)
\, , \quad
\eeq
where we have set
\beq\label{7.22new}
E_0 \, ( \, = \, \lvac H \rvac \, ) \, = \,
\frac{7}{4} \, \frac{B_4}{8}-\frac{1}{4} \, \frac{B_2}{4} \, = \,
-\frac{17}{960}
.
\eeq

We now proceed to derive the Gibbs distribution of the conformal
Hamiltonian $H_{\frac{1}{2}}$ of the subcanonical \(d=\txfrac{1}{2}\)
Weyl field.
Using the implication of the third order equation~(\ref{6.44})
on the modes of ${\chi}$ we
deduce that the energy $n+\txfrac{1}{2}$ eigenspace
of positive charge
is isomorphic in this case to the
(pseudoorthogonal) direct sum of three irreducible $\text{\it Spin}(4)$
representations,
\beq\label{eqn6.28}
\left(
\frac{n \! + \! 1}{2},\,
\frac{n}{2}
\right)
\oplus
\left(
\frac{n \! - \! 1}{2},\,
\frac{n}{2}
\right)
\oplus
\left(
\frac{n \! - \! 1}{2},\,
\frac{n \! - \! 2}{2}
\right)
\, \
\eeq
of total dimension
\beq\label{7.24new}
d_{\frac{1}{2}}(n)=(n+2)(n+1)+n(n+1)+n(n-1)=3n(n+1)+2=
\frac{3(2n+1)^2 \! + \! 5}{4}.
\eeq
It follows that the Gibbs energy average is given in this case by
\beqa\label{7.25new}
\La H_{\frac{1}{2}} \Ra_{\! q}
= && \hspace{-15pt}
\lvac H_{\frac{1}{2}} \rvac
+\mathop{\sum}\limits_{n>0}
\frac{2
\left( n+\frac{1}{2} \right) d_{\frac{1}{2}}(n)q^{n+\frac{1}{2}}}{1+q^{n+\frac{1}{2}}}
\, = \,
\nn \, = && \hspace{-15pt}
\frac{3}{4}
\left(\raisebox{10pt}{\hspace{-2pt}}\right.
G_4 \! \left(\raisebox{9pt}{\hspace{-2pt}}\right.
\frac{\tau \! + \! 1}{2}
\left.\raisebox{9pt}{\hspace{-2pt}}\right) - 8 \, G_4 \! \left( \tau \right)
\left.\raisebox{10pt}{\hspace{-2pt}}\right)
\, + \,
\frac{5}{4}
\left(\raisebox{10pt}{\hspace{-2pt}}\right.
G_2 \! \left(\raisebox{9pt}{\hspace{-2pt}}\right.
\frac{\tau \! + \! 1}{2}
\left.\raisebox{9pt}{\hspace{-2pt}}\right) - 2 \, G_2 \! \left( \tau \right)
\left.\raisebox{10pt}{\hspace{-2pt}}\right)
\, \qquad
\eeqa
for
\beq\label{eqn6.31}
\lvac H \rvac \, = \,
- \frac{3}{4} \, \frac{B_4}{8} \left( 1 \! - \! 2^3 \right)
- \frac{5}{4} \, \frac{B_2}{4} \left( 1 \! - \! 2 \right)
\, = \, \frac{29}{960}
\, . \
\eeq

\subsection{The free Maxwell field}\label{Ssec.5.2}

It is convenient to write the Maxwell field as a 2-form,
\beq\label{eqn6.34}
F \left( z \right) \, = \, \frac{1}{2} \, F_{\cmu\cnu} \! \left( z \right)
\, dz^{\cmu} \!\wedge dz^{\cnu}
\, , \
\eeq
which makes clear not just its transformation properties but also its
conjugation law:
\beq\label{eqn6.36}
\hspace{0pt}
\left( F_{\cmu\cnu} \! \left( z \right) \, dz^{\cmu} \!\wedge dz^{\cnu} \right)^*
=
F_{\cmu\cnu} \! \left( z^* \right) \,
d\overline{z}^{\cmu} \!\wedge d\overline{z}^{\cnu}
\quad (z^* =
\text{\(
\frac{\textstyle \overline{z}}{\textstyle\overline{z}^{\, 2}}
\)},
\ \
\left( dz^{\cmu} \right)^* = d \left( z^* \right)^{\cmu})
.
\eeq
The free field is characterized by its 2-point function
\beq\label{eqn6.37}
\lvac F_{\cmu_1\cnu_1} \! \left( z_1 \right)
F_{\cmu_2\cnu_2} \! \left( z_2 \right) \rvac \, := \,
\frac{
r_{\cmu_1\cmu_2} \! \left( z_{12} \right) r_{\cnu_1\cnu_2} \! \left( z_{12} \right) -
r_{\cmu_1\cnu_2} \! \left( z_{12} \right) r_{\cnu_1\cmu_2} \! \left( z_{12} \right)}{
\left( z_{12}^{\, 2} \right)^2} \, ,
\eeq
\(r_{\cmu\cnu} \left( z \right) := \delta_{\cmu\cnu} -
2 \, \frac{\textstyle z_{\cmu} z_{\cnu}}{\textstyle z^{\, 2}}\).\gvspc{-7pt}
It is verified to satisfy the \textit{Maxwell equations}
\beq\label{eqn6.38}
d F \left( z \right) \, = \, 0
\, , \quad
d * \! \left( F \right) \left( z \right) \, = \, 0
\, , \
\eeq
$*$ being \textit{Hodge} conjugation
\(* \! \left( F \right)_{\cmu\cnu} \! \left( z \right) \, := \,
\varepsilon_{\cmu\cnu\rho\sigma} \, F^{\rho\sigma} \! \left( z \right)\).

To compute the (compact picture) finite temperature correlation functions
\linebreak
\(\La F_{\cmu_1\cnu_1} \left( \zeta_1, u_1 \right)\)
\(F_{\cmu_2\cnu_2} \left( \zeta_2, u_2 \right) \Ra_{\! q}\)
we use again the diagonal frame
in which, $2v$ $=$ $(0,$ $0,$ $-i,$ $1)$,
\(u_{1,2} = \left( 0,\right.\)
$0,$ $\pm\sin \pi\alpha,$ \(\left.\cos \pi\alpha \right)\); then there
exist
linear combinations of the field components
\beq\label{eqn6.40new}
\sqrt{2} \, F_1^{\pm} \! = \! F_{23} \hspace{-1pt} \pm \hspace{-1pt} F_{14}
, \ \,
\sqrt{2} \, F_2^{\pm} \! = \! F_{31} \hspace{-1pt} \pm \hspace{-1pt} F_{24}
, \ \,
\sqrt{2} \, F_3^{\pm} \! = \! F_{12} \hspace{-1pt} \pm \hspace{-1pt} F_{34}
, \ \,
\sqrt{2} \, F_{\pm}^{\varepsilon} \! = \!
F_1^{\varepsilon} \hspace{-1pt} \pm \hspace{-1pt} iF_2^{\varepsilon}
\eeq
(\(\varepsilon = \pm\))
such that
\beqa\label{eqn6.39}
& \hspace{-10pt}
\lvac F_+^+ \! \left( \zeta_1, u_1 \right)
F_-^- \! \left( \zeta_2, u_2 \right) \rvac
=: \W_0 \left( \zeta_{12}, \alpha \right) =
& \nn & \hspace{-10pt} =
\frac{\raisebox{1pt}{\(
1
\)}}{\raisebox{-4pt}{\(
4 \sin^3 \! 2 \pi \alpha
\)}}
\left(
\mathrm{cotg} \hspace{1pt} \pi \zeta_-
\hspace{-1pt} - \hspace{-1pt}
\mathrm{cotg} \hspace{1pt} \pi \zeta_+
\right)
-
\frac{\raisebox{1pt}{\(
1
\)}}{\raisebox{-4pt}{\(
4 \sin 2 \pi \alpha
\)}}
\left(
\frac{\raisebox{1pt}{\(
\cos \hspace{1pt} \pi\zeta_+
\)}}{\raisebox{-4pt}{\(
\sin^3 \hspace{-1pt} \pi\zeta_+
\)}}
\hspace{-1pt} - \hspace{-1pt}
\frac{\raisebox{1pt}{\(
\mathrm{cotg} \hspace{1pt} 2\pi\alpha
\)}}{\raisebox{-4pt}{\(
\sin^2 \hspace{-1pt} \pi\zeta_+
\)}}
\right)
;
& \nn & \hspace{-10pt}
\lvac F_-^+ \! \left( \zeta_1, u_1 \right)
F_+^- \! \left( \zeta_2, u_2 \right) \rvac
= \W_0 \left( \zeta_{12}, -\alpha \right);
\hspace{6pt} \mgvspc{12pt}
\lvac F_3^+ \! \left( \zeta_1, u_1 \right)
F_3^- \! \left( \zeta_2, u_2 \right) \rvac =
& \hspace{-45pt} \nn & \hspace{-10pt} \hspace{-5pt} =
\frac{\raisebox{1pt}{\(
1
\)}}{\raisebox{-4pt}{\(
4 \sin^2 \! 2 \pi \alpha
\)}}
\left(
\frac{\raisebox{1pt}{\(
1
\)}}{\raisebox{-4pt}{\(
\sin^2 \hspace{-1pt} \pi\zeta_+
\)}}
\hspace{-1pt} + \hspace{-1pt}
\frac{\raisebox{1pt}{\(
1
\)}}{\raisebox{-4pt}{\(
\sin^2 \hspace{-1pt} \pi\zeta_-
\)}}
\hspace{-1pt} + \hspace{-1pt}
2 \mathrm{cotg} \hspace{1pt} 2 \pi \alpha
\left(
\mathrm{cotg} \hspace{1pt} \pi \zeta_+
\hspace{-1pt} - \hspace{-1pt}
\mathrm{cotg} \hspace{1pt} \pi \zeta_-
\right)
\!\!\right)
& \!
\eeqa
(\(\zeta_{\pm} = \zeta_{12} \pm \alpha\)).
The corresponding finite temperature correlation functions are:
\beqa\label{eqn6.40}
& \hspace{-10pt}
\La F_+^+ \! \left( \zeta_1, u_1 \right)
F_-^- \! \left( \zeta_2, u_2 \right) \Ra_{\! q}
\! =: \! \W_q \left( \zeta_{12}, \alpha \right) \! = \!
\frac{\raisebox{1pt}{\(
1
\)}}{\raisebox{-4pt}{\(
4 \sin^3 \! 2 \pi \alpha
\)}}
\!
\left( \pfun_1 \left( \zeta_-, \tau \right)
\hspace{-1pt} - \hspace{-1pt}
\pfun_1 \left( \zeta_+, \tau \right) \right)
\! - \!\!
\hspace{-7pt}
& \nn & \hspace{-10pt} -
\frac{\raisebox{1pt}{\(
1
\)}}{\raisebox{-4pt}{\(
4 \sin 2 \pi \alpha
\)}}
\left(
\frac{\raisebox{1pt}{\(
1
\)}}{\raisebox{-4pt}{\(
2\pi
\)}}
\pfun_3 \left( \zeta_+, \tau \right)
\hspace{-1pt} - \hspace{-1pt}
\mathrm{cotg} \, 2\pi \alpha \, \pfun_2 \left( \zeta_+, \tau \right)
\right)
;
& \nn & \hspace{-10pt}
\La F_-^+ \! \left( \zeta_1, u_1 \right)
F_+^- \! \left( \zeta_2, u_2 \right) \Ra_{\! q}
= \W_q \left( \zeta_{12}, -\alpha \right);
\hspace{9pt} \mgvspc{12pt}
\La F_3^+ \! \left( \zeta_1, u_1 \right)
F_3^- \! \left( \zeta_2, u_2 \right) \Ra_{\! q} =
& \nn & \hspace{-10pt} \!\! = \!
\frac{\raisebox{1pt}{\(
1
\)}}{\raisebox{-4pt}{\(
4 \sin^2 \! 2 \pi \alpha
\)}}
\left(
\pfun_2 \left( \zeta_+, \tau \right)
\hspace{-2pt} + \hspace{-1pt}
\pfun_2 \left( \zeta_-, \tau \right)
\hspace{-2pt} + \hspace{-1pt}
2 \mathrm{cotg} \hspace{1pt} 2 \pi \alpha
\left(
\pfun_1 \left( \zeta_+, \tau \right)
\hspace{-2pt} - \hspace{-1pt}
\pfun_1 \left( \zeta_-, \tau \right)
\right)
\right)
\! .
&
\nn & &
\eeqa

In order to find the temperature energy mean value for the Maxwell field
we have to compute the dimension \(d_F(n)\)
of the $1$--particle state space of conformal energy $n$,
spanned by \(F_{\cmu\cnu;\, -n} \left( z \right)\rvac\) where the mode
\(F_{\cmu\cnu;\, -n} \left( z \right)\) is a homogeneous (harmonic) polynomial of
degree \(n-2\), satisfying the Maxwell equations.
To this end we display the $\SO (4)$ representation content of the modes
satisfying the Maxwell equations.
Decomposing the antisymmetric tensor $F_{\cmu\cnu}$ into selfdual and antiselfdual
parts, \((1,0) \oplus (0,1)\), we see that the full space of homogeneous
skewsymmetric--tensor valued polynomials in $z$ of degree \(n-2\) generically
splits into a direct sum of three conjugate pairs of \(SU(2) \times SU(2)\)
representations; for instance,
\(
\left( 1,0 \right) \otimes
\left(\frac{\textstyle n \hspace{-1pt} - \hspace{-1pt} 2}{\textstyle 2},\,
\frac{\textstyle n \hspace{-1pt} - \hspace{-1pt} 2}{\textstyle 2}\right)
=
\left(\frac{\textstyle n}{\textstyle 2},\,
\frac{\textstyle n \hspace{-1pt} - \hspace{-1pt} 2}{\textstyle 2}\right)
\oplus
\left(\frac{\textstyle n \hspace{-1pt} - \hspace{-1pt} 2}{\textstyle 2},\,
\frac{\textstyle n \hspace{-1pt} - \hspace{-1pt} 2}{\textstyle 2}\right)
\oplus
\left(\frac{\textstyle n \hspace{-1pt} - \hspace{-1pt} 4}{\textstyle 2},\,
\frac{\textstyle n \hspace{-1pt} - \hspace{-1pt} 2}{\textstyle 2}\right)
\)
(for \(n>3\)).
Maxwell equations imply that only two of the resulting six
representations, those with maximal weights, appear in
the energy $n$ $1$--particle space:
\(\left(\frac{\textstyle n}{\textstyle 2},\,
\frac{\textstyle n \hspace{-1pt} - \hspace{-1pt} 2}{\textstyle 2}\right)
\oplus
\left(\frac{\textstyle n \hspace{-1pt} - \hspace{-1pt} 2}{\textstyle 2},\,
\frac{\textstyle n}{\textstyle 2}\right)\). Thus,
\beq\label{eqHFM1}
d_F \left( n \right) = 2 \left( n^2-1 \right)
\eeq
and then find
\beq\label{eqHFM2}
\La \hspace{-1pt} H_F \hspace{-1pt} \Ra_{\! q} \, = \, 2 G_4(\tau)-2G_2(\tau)
\, , \quad
\lvac \hspace{-1pt} H_F \hspace{-1pt} \rvac \, = \,
-2\,\frac{B_4}{8}+2\,\frac{B_2}{4} = \frac{11}{120}
\, . \
\eeq

If $A_\cmu$ is the gauge potential,
such that \(F_{\cmu\cnu}=\di_\cmu A_\cnu-\di_\cnu A_\cmu\), then
we say that that \(A_\cmu\mapsto A_\cmu+l_\cmu\)
is a {\it conformal gauge transformation} if $l_\cmu(z)$
is purely {\it longitudinal}  generalized free field,
\beq\label{7.37}
\di_\cmu \, l_\cnu=\di_{\cnu}l_\cmu,
\quad \text{so that} \quad
\lvac l_\cmu(z_1)l_\cnu(z_2)\rvac \, = \, C \,
\frac{r_{\cmu\cnu}}{z_{12}^2}
\eeq
which, hence, satisfies the third order equation
\beq\label{7.38}
\Delta \, \di \spr l \, = \, 0
\, . \
\eeq
One can write $l_\cmu(z)=\di_{\cmu}s(z)$
where $s$ has a logarithmic 2--point function
but there is no conformal scalar field whose
gradient is $l_\cmu(z)$.
The energy $n$ pure gauge 1--particle state space is spanned
by vectors of the form
$l_{-n}^{\cmu_1\cmu_2\dots\cmu_n} z_{\cmu_2}$ $\dots$ $z_{\cmu_n}$ $\rvac$
where $l_{-n}^{\cmu_1\dots\cmu_n}$ is a symmetric rank n tensor
(due to (\ref{7.38})).
(We are using interchangeably~--~for writing convenience~--~upper
and lower Euclidean indices.)
Taking (\ref{7.38}) into account one computes the dimension of
this space to be
\beq\label{7.39}
d_{l}(n) \, = \,
\BINOMIAL{n+3}{3}
-
\BINOMIAL{n-1}{3}
\, = \,
(n+1)^2+(n-1)^2
\, = \, 2(n^2+1)
\, . \
\eeq
It follows that the total energy mean value is a modular form of weight
four:
\beq\label{7.40}
\La H_F+H_l \Ra_{\! q} \, = \, 4 \, G_4(\tau)
\, . \
\eeq

\section{The thermodynamic limit}\label{Sect.7n}
\setcounter{equation}{0}\setcounter{mtheorem}{1}\setcounter{mremark}{1}\setcounter{mdefinition}{1}\setcounter{mexample}{1}\setcounter{mexercise}{1}

\subsection{Compactified Minkowski space as a ``finite box'' approximation}\label{SSect.7.1}

We shall now substitute $z$ in Eq.~(\ref{nw4.19})
by $\Txfrac{z}{R}$ thus treating $\Sr^{D-1}$ and $\Sr^{1}$ in
the definition of $\M$ as a sphere and a circle of radius \(R \, (> 0)\).
Performing further the Minkowski space dilation
\(\left( 2R \right)^{X_{-1D}} :\)
\(x^{\mu} \mapsto \Txfrac{x^{\mu}}{2R}\), \(\mu = 0,\dots,D-1\)
(see Eq.~(\ref{e2.3n}))
on the (real) variable \((\mzeta =) \, x\) in (\ref{nw4.19}) we find
\(z \! \left( x;R \right) \hspace{-2pt} = \hspace{-1pt}
R \, z \! \left( \Txfrac{x}{2R} \right)\) or
\beq\label{e7.1}
\mbf{z} \! \left( x;R \right) \hspace{-2pt} = \hspace{-1pt}
\frac{\mbf{x}}{
2 \omega \! \left( \hspace{-1pt}\Txfrac{x}{2R}\hspace{-1pt} \right) \mgvspc{13pt}}
, \ \,
z_D \! \left( x;R \right) - R
\hspace{-2pt} = \hspace{-2pt}
\frac{i x^0 \! - \! \Txfrac{x^{\, 2}}{2R \mgvspc{-2pt}}}{
2 \omega \! \left( \hspace{-1pt}\Txfrac{x}{2R}\hspace{-1pt} \right) \mgvspc{13pt}}
, \ \,
2 \omega \! \left( \hspace{-1pt}\frac{x}{2R}\hspace{-1pt} \right)
\hspace{-2pt} = \hspace{-1pt}
1 + \frac{x^{\, 2}}{4R^2} - i \hspace{1pt} \frac{x^{0}}{R}
. \ \
\eeq
The stability subgroup of \(z \left( x;R \right) = 0 \, ( \in T_+)\)
in $\confgr$ is conjugate to the maximal compact subgroup
\(\compgr \subset \confgr\):
\beq\label{e7.2}
\compgr \! \left( 2R \right) \, = \,
\left( 2R \right)^{X_{-1D}} \compgr \, \left( 2R \right)^{-X_{-1D}}
\, , \quad
\compgr \, \equiv \, \compgr \!
\left(\raisebox{9pt}{\hspace{-2pt}}\right.
1
\left.\raisebox{9pt}{\hspace{-2pt}}\right)
\, \cong \,
\nfrc{U \left( 1 \right) \times \Spin \left( D \right)}{\Z_2}
\, . \
\eeq
In particular, the hermitian $U \left( 1 \right)$--generator $H \left( 2R \right)$,
which acts in the $z$--coor\-di\-na\-tes~(\ref{e7.1})
as the Euler vector field $z \spr \Txfrac{\di}{\di z}$, is
conjugate to \(H \equiv H \! \left( 1 \right)\),
\beq\label{e7.3}
H \left( 2R \right) \, = \,
\left( 2R \right)^{X_{-1D}} H \,
\left( 2R \right)^{-X_{-1D}}
\, , \quad
H \, \equiv \,
H \!
\left(\raisebox{9pt}{\hspace{-2pt}}\right.
1
\left.\raisebox{9pt}{\hspace{-2pt}}\right)
\, . \
\eeq

For large $R$ and finite $x$ the variables \(\left( \mbf{z},z_D-R \right)\)
approach the (Wick rotated) Minkowski space coordinates
\(\left( \mbf{x},ix^0 \right)\).
In particular, for \(x^0 = 0 \, ( \, =\czeta)\),
the real $\left( D-1 \right)$--sphere \(z^{\, 2} = R^2\)
can be viewed as a $\mathit{SO} \! \left( D \right)$--invariant
``box'' approaching for \(R \to \infty\) the flat
space $\R^{D-1}$.
Thus the conformal compactification of Minkowski space also
plays the role of a convenient tool for studying the thermodynamic
limit of thermal expectation values.
This interpretation is justified in view of the following:

\begin{mproposition}
\prlabel{pr:7.1}
The asymptotic behaviour of \(z \left( x; R \right) -R e_D\)
(\(e_D = \left( \Mbf{0},1 \right)\))
and of the associated Hamiltonian
for large $R$ is:
\beqa\label{e7.4}
&
\mbf{z} \left( x;R \right) \, = \, \mbf{x} \, + \,
O \left(\raisebox{10pt}{\hspace{-2pt}}\right. \Txfrac{\|x\|^2}{R}
\left.\raisebox{10pt}{\hspace{-2pt}}\right)
\, , \quad
z_D \left( x;R \right) - R \, = \, i \hspace{1pt} x^0 \, + \,
O \left(\raisebox{10pt}{\hspace{-2pt}}\right. \Txfrac{\|x\|^2}{R}
\left.\raisebox{10pt}{\hspace{-2pt}}\right)
, &
\\ \label{e7.5} &
H_R \, := \, \Txfrac{H \left( 2R \right)}{R} \, = \,
P_0 \, + \, \Txfrac{1}{4R^2} \, K_0
\ ( \, = \, P_0 +
O \left(\raisebox{10pt}{\hspace{-2pt}}\right. \Txfrac{1}{R}
\left.\raisebox{10pt}{\hspace{-2pt}}\right)
\ \in i \hspace{1pt} \cnfalg \, )
\, , &
\eeqa
where
\(\left\| x \right\| := \sqrt{\left( x^0 \right)^2 + \left| \mbf{x} \right|^2}\)
for \(x = \left( x^0, \mbf{x} \right) \in M\) and
$i P_0$ is the real
conformal algebra generator of the Minkowski time ($x^0$) translation
(see Sect.~\ref{ssec2.1}).
The operator $H_R$ is the physical conformal Hamiltonian
(of dimension inverse length).
\end{mproposition}

\begin{proof}
Eq.~(\ref{e7.4}) is obtained by a straightforward computation.
To derive Eq.~(\ref{e7.5}) one should use (\ref{H})
and the equations
$\lambda^{X_{-1D}}$ $P_0$ $\lambda^{-X_{-1D}}$ $=$
$\lambda \hspace{1pt} P_0$,
$\lambda^{X_{-1D}}$ $K_0$ $\lambda^{-X_{-1D}}$ $=$
$\lambda^{-1} K_0$;
hence,
\(
H \left( 2R \right) =\) \(\left( 2R \right)^{X_{-1D}}\) $H$
\(\left( 2R \right)^{-X_{-1D}} =\)
\(R P_0 +\) \(\Txfrac{1}{4R} \, K_0\).
\end{proof}

\begin{mremark}
\rmlabel{rm:7.1}
The observation that the universal cover of $\M$,
the Einstein universe
\(\widetilde{M} =\) $\R \times \Sr^{D-1}$
(for \(D=4\)), which admits a
\textit{globally causal structure}, is locally undistiguishable
from $M$ for large $R$ has been emphasized over 30 year ago
by Irving Segal (for a concise expos\'e and further
references~--~see~\cite{S82}).
For a fixed choice, $X_{-1D}$, of the dilation generator
in~(\ref{e7.2}) he identifies the Minkowski energy $P_0$
with the scale covariant component of $H_R$.
With this choice $M$ is osculating $\M$ (and hence $\widetilde{M}$)
at the north pole \(\left( \mbf{z},z_D \right)\) $=$
\(\left( \Mbf{0},R \right)\) (respectively, \(\czeta = 0\),
\(u = e_D\)), identified with the origin \(x=0\) in $M$.
(The vector fields associated with $H_R$ and $P_0$ coincide
at this point.)
\end{mremark}

Using the Lie algebra limit \(\mathop{\lim}\limits_{R \to \infty} \, H_R = P_0\)
implied by~(\ref{e7.5}),
one can approximate the Minkowski energy operator $P_0$
for large $R$ by the physical conformal Hamiltonian $H_R$.
As we shall see below, the fact that in all considered free field models
in dimension \(D=4\) the conformal mean energy
is a linear combination of modular forms $G_{2k} \left( \tau \right)$
with highest weight \(2k=4\),
has a remarkable corollary: the \textit{density} $\MED$
of the physical mean energy has a limit reproducing the
\textit{Stefan--Boltzmann} law
\beq\label{e7.7n}
\MED \hspace{-1pt} \left( \beta \right) \, := \,
\mathop{\lim}\limits_{R \to \infty} \,
\frac{\La \hspace{-2pt} H_R \hspace{-2pt} \Ra_{q_{\beta}}}{\VOL{R}}
\, = \, \frac{C}{\beta^4}
\quad \text{for} \quad q_{\beta} \, := \, e^{-\beta}
\, \
\eeq
where $C$ is some constant,
\(\beta = \Txfrac{1}{k T}\) is the inverse absolute temperature $T$
(multiplied by the Boltzmann constant $k$) and
\(\VOL{R} := 2\pi^2 R^3\) is the volume of the $3$--sphere
of radius~$R$ at a fixed time (say \(x^0 = 0 = \czeta\)).
We will calculate this limit for two cases: the model of a free
scalar filed in \(D=4\) (see Sect.~\ref{Ssec.4.1})
which we will further denote by $\varphi$ and
the Maxwell free field model introduced in Sect.~\ref{Ssec.5.2}.

\begin{mproposition}
\prlabel{pr:7.2}
For the free scalar field $\varphi$ in dimension \(D=4\)
we have the following behaviour
of the mean energy density
for \(\Txfrac{R}{\beta} \mgrt 1\)
\beq\label{e7.8}
\MED_R^{\left( \hspace{-1pt} \varphi \hspace{-1pt} \right)}
\! \left( \beta \right) \, := \,
\frac{1}{\VOL{R}} \, \frac{\mathit{tr}_{\DOM{}} \,
H_R \hspace{2pt} e^{-\beta H_R}}{
\mathit{tr}_{\DOM{}} \, e^{-\beta H_R}}
\, = \, \left( \frac{\pi^2}{30} -
\frac{1}{480 \hspace{1PT} \pi^2} \, \frac{\beta^4}{R^4} +
O \! \left(\raisebox{10pt}{\hspace{-2pt}}\right.
e^{-4\pi^2 \frac{R}{\beta}}
\left.\raisebox{10pt}{\hspace{-2pt}}\right) \right)
\frac{1}{\beta^4}
\, .
\eeq
The corresponding result for of the Maxwell free field
$F_{\mu\nu}$ is
\beq\label{e7.9}
\MED_R^{\left( \hspace{-1pt} F \hspace{-1pt} \right)}
\! \left( \beta \right) \, = \,
\left( \frac{\pi^2}{15} -
\frac{1}{6} \, \frac{\beta^2}{R^2} +
\frac{1}{
4 \pi^3
} \, \frac{\beta^3}{R^3} -
\frac{11}{240 \hspace{1pt}\pi^2} \, \frac{\beta^4}{R^4} +
O \! \left(\raisebox{10pt}{\hspace{-2pt}}\right.
e^{-4\pi^2 \frac{R}{\beta}}
\left.\raisebox{10pt}{\hspace{-2pt}}\right) \right)
\frac{1}{\beta^4}
\, .
\eeq
\end{mproposition}

\begin{proof}
The hermitian operators $H$ and $H \left( 2R \right)$ are
unitarily equivalent due to Eq.~(\ref{e7.3}).
This leads to the fact that
\(\mathit{tr}_{\DOM{}} \,q^{H \left( 2R \right)}\) and
\(\mathit{tr}_{\DOM{}} \, H \left( 2R \right) q^{H \left( 2R \right)}\)
do not depend on $R$.
Then Eqs.~(\ref{neq6.17n}) and (\ref{eqHFM2}) imply that in the two
models under consideration we have
\beq\label{e7.11n}
\MED_R^{\left( \hspace{-1pt} \varphi \hspace{-1pt} \right)}
\! \left( \beta \right)
=
\frac{
G_4 \!
\left(\raisebox{10pt}{\hspace{-3pt}}\right.
\Txfrac{i \beta}{2 \pi R}
\left.\raisebox{10pt}{\hspace{-3pt}}\right)
\hspace{-1pt} -
\Txfrac{1}{240} \mgvspc{-10pt}
}{R \VOL{R}}
\, , \hspace{6pt}
\MED_R^{\left( \hspace{-1pt} F \hspace{-1pt} \right)}
\! \left( \beta \right)
=
\frac{2 G_4 \!
\left(\raisebox{10pt}{\hspace{-3pt}}\right.
\Txfrac{i \beta}{2 \pi R}
\left.\raisebox{10pt}{\hspace{-3pt}}\right)
\hspace{-1pt} -
2 G_2 \!
\left(\raisebox{10pt}{\hspace{-3pt}}\right.
\Txfrac{i \beta}{2 \pi R}
\left.\raisebox{10pt}{\hspace{-3pt}}\right)
\hspace{-1pt} -
\Txfrac{11}{120} \mgvspc{-10pt}
}{R \VOL{R}}
\, . \hspace{2pt}
\eeq
Using further the relations
\beq\label{e7.10}
G_2 \! \left( \tau \right) \, = \, \frac{1}{\tau^2}
\, G_2 \! \left(\raisebox{10pt}{\hspace{-2pt}}\right.
\frac{-1}{\tau}
\left.\raisebox{10pt}{\hspace{-2pt}}\right)
-
\frac{i}{4 \pi \tau}
\, , \quad
G_4 \! \left( \tau \right) \, = \, \frac{1}{\tau^4}
\, G_4 \! \left(\raisebox{10pt}{\hspace{-2pt}}\right.
\frac{-1}{\tau}
\left.\raisebox{10pt}{\hspace{-2pt}}\right)
\, \
\eeq
(which are special cases of~(\ref{3.9}))
we find
\beqa\label{e7.11}
& \hspace{-10pt}
\MED_R^{\left( \hspace{-1pt} \varphi \hspace{-1pt} \right)}
\! \left( \beta \right)
\, = \, \Txfrac{1}{\beta^4}
\left(\raisebox{10pt}{\hspace{-2pt}}\right.
8 \pi^2 G_4 \!
\left(\raisebox{10pt}{\hspace{-2pt}}\right.
\Txfrac{2 \pi i \hspace{1pt} R}{\beta}
\left.\raisebox{10pt}{\hspace{-2pt}}\right)
-
\Txfrac{\beta^4}{480 \pi^2 R^4}
\left.\raisebox{10pt}{\hspace{-2pt}}\right)
\, , & \\ \label{e7.12} & \hspace{-10pt}
\MED_R^{\left( \hspace{-1pt} F \hspace{-1pt} \right)}
\! \left( \beta \right)
\! = \! \Txfrac{1}{\beta^4}
\left(\raisebox{10pt}{\hspace{-2pt}}\right. \!
16 \pi^2
G_4 \!
\left(\raisebox{10pt}{\hspace{-2pt}}\right.
\Txfrac{2 \pi i \hspace{1pt} R}{\beta}
\left.\raisebox{10pt}{\hspace{-2pt}}\right)
\! + \!
\Txfrac{4 \beta^2}{R^2}
G_2 \!
\left(\raisebox{10pt}{\hspace{-2pt}}\right.
\Txfrac{2 \pi i \hspace{1pt} R}{\beta}
\left.\raisebox{10pt}{\hspace{-2pt}}\right)
\! + \!
\Txfrac{\beta^3}{
4 \pi^3
R^3}
\! - \!
\Txfrac{11 \beta^4}{240 \hspace{1pt} \pi^2 R^4}
\left.\raisebox{10pt}{\hspace{-2pt}}\right)
\! .
& \qquad
\eeqa
Finally, to obtain Eq.~(\ref{e7.8})
one should apply the expansion~(\ref{3.7})
implying that
\beqa
G_2 \! \left(\raisebox{10pt}{\hspace{-2pt}}\right.
\frac{2 \pi i \hspace{1pt} R}{\beta}
\left.\raisebox{10pt}{\hspace{-2pt}}\right)
= - \frac{1}{24} +
O \! \left(\raisebox{10pt}{\hspace{-2pt}}\right.
e^{-4\pi^2 \frac{R}{\beta}}
\left.\raisebox{10pt}{\hspace{-2pt}}\right)
, \quad
G_4 \! \left(\raisebox{10pt}{\hspace{-2pt}}\right.
\frac{2 \pi i \hspace{1pt} R}{\beta}
\left.\raisebox{10pt}{\hspace{-2pt}}\right)
= \frac{1}{240} +
O \! \left(\raisebox{10pt}{\hspace{-2pt}}\right.
e^{-4\pi^2 \frac{R}{\beta}}
\left.\raisebox{10pt}{\hspace{-2pt}}\right)
. \
\nonumber
\eeqa
\end{proof}

\begin{mremark}
\rmlabel{rm:7.2}
In order to make comparison with the familiar expression
for the black body radiation it is instructive to restore
the dimensional constants $h$ and $c$ setting
\(H_R = \Txfrac{hc}{R} \, H \hspace{-1pt} \left( 2R \right)\)
(instead of (\ref{e7.4})).
The counterpart of~(\ref{e7.11n}) and~(\ref{3.7}) then reads
\beq\label{bl_b}
\La H_R \Ra_q =
\frac{h c}{R}
\left(\raisebox{10pt}{\hspace{-3pt}}\right.
G_4 \!
\left(\raisebox{10pt}{\hspace{-3pt}}\right.
\frac{i \hspace{1pt} h \hspace{0.5pt} c \hspace{0.5pt} \beta}{R}
\left.\raisebox{10pt}{\hspace{-3pt}}\right)
\hspace{-2pt} - \hspace{-1pt} E_0
\left.\raisebox{10pt}{\hspace{-3pt}}\right)
=
\frac{hc}{R} \,
\mathop{\sum}\limits_{n \, = \, 1}^{\infty} \hspace{3pt}
\frac{n^3
e^{- n \text{\footnotesize \(\Txfrac{h \hspace{0.5pt} c \hspace{0.5pt} \beta}{R}\)}}
\mgvspc{-8pt}}{\mgvspc{19pt}
1 - e^{- n \text{\footnotesize \(\Txfrac{h \hspace{0.5pt} c \hspace{0.5pt} \beta}{R}\)}}}
\, . \
\eeq
Each term in the infinite sum in the right hand side is a constant
multiple of Plank's black body radiation formula for frequency
\beq\label{freq}
\nu = n \frac{c}{R} .
\eeq
Thus, for finite $R$, there is a minimal frequency, $\Txfrac{c}{R}$.
Using the expansion in~(\ref{bl_b}) one can also find an alternative
integral derivation of the limit mean energy density
\(\MED_R^{\left( \hspace{-1pt} \varphi \hspace{-1pt} \right)}
\! \left( \beta \right)\)~(\ref{e7.11n}):
\beq\label{e7.16}
\MED_R^{\left( \hspace{-1pt} \varphi \hspace{-1pt} \right)}
\! \left( \beta \right)
\, = \,
\frac{1}{2 \pi^2 \hspace{-0.3pt} h^3 \hspace{-0.3pt} c^3 \hspace{-0.3pt} \beta^4}
\mathop{\sum}\limits_{n \, = \, 1}^{\infty} \hspace{-1pt}
\frac{\left( n
\Txfrac{h \hspace{0.5pt} c \hspace{0.5pt} \beta}{R} \right)^{\hspace{-2pt}3}
e^{- n \text{\footnotesize \(\Txfrac{h \hspace{0.5pt} c \hspace{0.5pt} \beta}{R}\)}}
\mgvspc{-8pt}}{\mgvspc{19pt}
1 - e^{- n \text{\footnotesize \(\Txfrac{h \hspace{0.5pt} c \hspace{0.5pt} \beta}{R}\)}}}
\, \frac{h \hspace{0.5pt} c \hspace{0.5pt} \beta}{R}
\ \mathop{\longrightarrow}\limits_{R \to \infty} \
\frac{\pi^2}{30 \hspace{-0.3pt} h^3 \hspace{-0.3pt} c^3 \hspace{-0.3pt} \beta^4}
\, \
\eeq
since the sum in the right hand side goes to the integral
\(\mathop{\int}\limits_{0}^{\infty}
\Txfrac{t^3 e^{-t}}{1 - e^{-t} \mgvspc{9pt}} \, d t
= \Txfrac{\pi^4}{15}\).
\end{mremark}

\begin{mremark}
\rmlabel{rm:7.3}
We observe that the constant $C$ in~(\ref{e7.7n}) in both
considered models
is equal to $\Txfrac{c_1}{30 \hspace{1pt} \pi^2}$, where
$c_1$ is the coefficient to the $G_4$--modular form in
$\La \hspace{-1pt} H \hspace{-1pt} \Ra_q$ (see Eq.~(\ref{neq6.16n})).
If we use in the definition~(\ref{e7.5}) of $H_R$
the Hamiltonian \(H \left( 2R \right) + E_0'\) instead of
$H \left( 2R \right)$,
\(\widetilde{H}_R := \Txfrac{H \left( 2R \right) + E_0'}{R}\),
then this will only reflect on the (non--leading)
terms $c_4 \Txfrac{\beta^4}{R^4}$
in (\ref{e7.8}) (\ref{e7.9}) replacing them
by $\Txfrac{E_0' - E_0}{2 \pi^2} \Txfrac{\beta^4}{R^4}$,
where $E_0$ is the ``vacuum energy'' for the corresponding models
(i.e., $E_0$ is $\Txfrac{1}{240}$ and $\Txfrac{11}{120}$
for the fields $\varphi$ and $F_{\mu\nu}$, respectively).
\end{mremark}

\subsection{Infinite volume limit of the thermal correlation functions}\label{SSect.7.2}

We shall study the \(R \to \infty\) limit on the example of
a free scalar field $\varphi$ in four dimensions.

Denote by
$\varphi^{\MINK} \! \left( x \right)$
(the canonically normalized) \(D=4\) free massless scalar field
with $2$--point function
\beq\label{ea5.17}
\lvac \varphi^{\MINK} \! \left( x_1 \right)
\varphi^{\MINK} \! \left( x_2 \right) \rvac \, = \,
\left( 2\pi \right)^{-2} \left( x_{12}^{\, 2} + i \, 0 \, x_{12}^0 \right)^{-1}
\quad
\eeq
(\(x_{12} = x_1 - x_2\),
\(x_{12}^{\, 2} = \mbf{x}_{12}^{\, 2} - \left( x_{12}^0 \right)^2\), see also~(\ref{18e})).
We define, in accord with
Proposition~\ref{pr:7.1},
a finite volume approximation of its thermal correlation function by
\beq\label{e7.18n}
\La \hspace{-1pt}
\varphi^{\MINK} \! \left( x_1 \right) \varphi^{\MINK} \! \left( x_2 \right)
\hspace{-1pt} \Ra_{\hspace{-1pt} \beta, R}
\, := \,
\frac{\mathit{tr}_{\DOM{}} \
\varphi^{\MINK} \! \left( x_1 \right) \varphi^{\MINK} \! \left( x_2 \right) \,
e^{-\beta H_{R}}}{
\mathit{tr}_{\DOM{}} \
e^{-\beta H_{R}}}
\, \
\eeq
and will be interested in the thermodynamic limit,
\beq\label{e7.19n}
\La \hspace{-1pt}
\varphi^{\MINK} \! \left( x_1 \right) \varphi^{\MINK} \! \left( x_2 \right)
\hspace{-1pt} \Ra_{\hspace{-1pt} \beta, \infty} :=
\mathop{\lim}\limits_{R \to \infty}
\La \hspace{-1pt}
\varphi^{\MINK} \! \left( x_1 \right) \varphi^{\MINK} \! \left( x_2 \right)
\hspace{-1pt} \Ra_{\hspace{-1pt} \beta, R}
\, . \
\eeq

\begin{mproposition}
\prlabel{pr:7.3}
The limit (\ref{e7.19n}) (viewed as a meromorphic function) is given by
\beq\label{e7.20n}
\La \hspace{-1pt}
\varphi^{\MINK} \! \left( x_1 \right) \varphi^{\MINK} \! \left( x_2 \right)
\hspace{-1pt} \Ra_{\hspace{-1pt} \beta, \infty}
=
\Txfrac{
\sinh
\hspace{1pt} 2 \pi
\Txfrac{\left| \mbf{x}_{12} \right| \mgvspc{-4pt}}{\mgvspc{9pt} \beta}
}{
8 \pi \beta \left| \mbf{x}_{12} \right|}
\left(\raisebox{16pt}{\hspace{-3pt}}\right.
\cosh
\hspace{1pt} 2 \pi
\Txfrac{\left| \mbf{x}_{12} \right|}{\beta}
-
\cosh
\hspace{1pt} 2 \pi
\Txfrac{x_{12}^0}{\beta}
\left.\raisebox{16pt}{\hspace{-3pt}}\right)^{\hspace{-1pt}-1}
\!\! , \
\eeq
(\(\left| \mbf{x}_{12} \right| := \sqrt{\mbf{x}_{12}^{\, 2}} \equiv
\sqrt{\left( x^1_{12} \right)^2 + \left( x^2_{12} \right)^2 +
\left( x^3_{12} \right)^2}\)).
\end{mproposition}

We shall \textit{prove} this statement by relating
$\varphi^{\MINK} \! \left( x \right)$ to the compact picture field
$\varphi \left( \czeta, u \right)$
\((\equiv \phi^{\left( 1 \right)} \left( \czeta,u \right))\)
whose thermal $2$--point function was computed in Sect.~\ref{sec:6}.

First, we use Eq.~(\ref{WF}) to express
$\varphi^{\MINK} \! \left( x \right)$ in terms of the $z$--picture
field (corresponding to the $R$--depending chart~(\ref{e7.1}))
\beq\label{e7.17}
2 \pi \,\varphi^{\MINK} \! \left( x \right) \, = \,
\frac{1}{2\omega \!
\left(\raisebox{10pt}{\hspace{-3pt}}\right.
\Txfrac{x}{2R}
\left.\raisebox{10pt}{\hspace{-3pt}}\right)} \
\varphi_R \! \left( z \left( x; R \right) \right)
\, \
\eeq
(since \(dz^{\, 2} = \omega \!
\left(\raisebox{10pt}{\hspace{-3pt}}\right.
\Txfrac{x}{2R}
\left.\raisebox{10pt}{\hspace{-3pt}}\right)^{-\hspace{-1pt} 2}
\Txfrac{dx^{\, 2}}{4}\), cp.~(\ref{t2.3})).
The factor $2\pi$ in front of $\varphi^{\MINK}$ accounts for
the different normalization conventions for the $x$--
and $z$--picture fields (we set
\(\lvac \varphi \! \left( z_1 \right)\)
\(\varphi \! \left( z_2 \right) \rvac =\) $\left( z_{12}^{\, 2} \right)^{-1}$
instead of~(\ref{ea5.17})).

As a second step we express
$\varphi_R \! \left( z \right)$~--~and
thus $\varphi^{\MINK} \! \left( x \right)$~--~in
terms of the compact picture field
$\varphi_R \! \left( \czeta,u \right)$:
\beqa\label{e7.21n}
&
\varphi_R \! \left( \czeta, u \right) \, := \,
R \hspace{1pt} e^{2 \pi i \czeta} \varphi \!
\left(\raisebox{9pt}{\hspace{-3pt}}\right.
R \hspace{1pt} e^{2 \pi i \czeta} u
\left.\raisebox{9pt}{\hspace{-3pt}}\right)
, & \nn &
2 \pi \hspace{1pt}
\varphi^{\MINK} \left( x \right) =
\Txfrac{1}{2R \left|\raisebox{10pt}{\hspace{-2pt}}\right.
\omega \!
\left(\raisebox{10pt}{\hspace{-3pt}}\right.
\Txfrac{x}{2R}
\left.\raisebox{10pt}{\hspace{-3pt}}\right)
\left.\raisebox{10pt}{\hspace{-3pt}}\right|
\mgvspc{14pt}}
\hspace{1pt} \varphi_R \!
\left(\raisebox{10pt}{\hspace{-3pt}}\right.
\czeta \! \left(\raisebox{10pt}{\hspace{-3pt}}\right.
\Txfrac{x}{2R}
\left.\raisebox{10pt}{\hspace{-3pt}}\right)
,
u \! \left(\raisebox{10pt}{\hspace{-3pt}}\right.
\Txfrac{x}{2R}
\left.\raisebox{10pt}{\hspace{-3pt}}\right)
\left.\raisebox{10pt}{\hspace{-3pt}}\right)
. & \qquad
\eeqa
Here $\czeta$ and $u$ are determined as functions of
$\Txfrac{x}{2R}$ from \(e^{2 \pi i \czeta} u
= \Txfrac{z \left( x; R \right)}{R} =
z \! \left(\raisebox{10pt}{\hspace{-3pt}}\right.
\Txfrac{x}{2R}
\left.\raisebox{10pt}{\hspace{-3pt}}\right)\)
($z \! \left( x \right)$ is given by~(\ref{nw4.19}) for \(\mzeta=x \in M\));
in deriving the second equation in (\ref{e7.21n})
we have used the relation
\(
e^{4 \pi i \czeta} =
\Txfrac{z \left( x; R \right)^{2}}{R^2} =
\overline{\omega \!
\left(\raisebox{10pt}{\hspace{-3pt}}\right.
\Txfrac{x}{2R}
\left.\raisebox{10pt}{\hspace{-3pt}}\right)}
\,
\omega \!
\left(\raisebox{10pt}{\hspace{-3pt}}\right.
\Txfrac{x}{2R}
\left.\raisebox{10pt}{\hspace{-3pt}}\right)^{-1}
\).

Next we observe that
$\varphi_R \hspace{-1pt} \left( \czeta, u \right)$
are mutually conjugate (for different $R$) just as
$H \left( 2R \right)$ in Eq.~(\ref{e7.3}).
(To see this one can use an intermediate
``dimensionless'' coordinates
\(\widetilde{z} \hspace{-1pt} \left( x; R \right)
= \Txfrac{z}{R} =
z \! \left(\raisebox{10pt}{\hspace{-3pt}}\right.
\Txfrac{x}{2R}
\left.\raisebox{10pt}{\hspace{-3pt}}\right)\),
which differs from (\ref{nw4.19}) just by the dilation $\left( 2R \right)^{X_{-1D}}$.)
It follows that its vacuum and thermal $2$--point function
with respect to the Hamiltonian $H \! \left( 2R \right)$
do not depend on $R$ and coincide with~(\ref{6.11}).
Thus
\beq\label{e7.22}
4 \pi^2
\La \hspace{-2pt}
\varphi^{\MINK} \!\hspace{-1pt} \left( x_1 \right)
\hspace{-1pt}
\varphi^{\MINK} \!\hspace{-1pt} \left( x_2 \right)
\hspace{-2pt}
\Ra_{\hspace{-2pt} \beta,R}
\hspace{1pt} = \hspace{1pt}
\Txfrac{
p_1 \! \hspace{-1pt} \left( \czeta_{12} \hspace{-2pt} + \hspace{-2pt} \alpha
, \hspace{-1pt}
\tau_R \hspace{-1pt} \right)
\hspace{-1pt} - \hspace{-1pt}
p_1 \! \hspace{-1pt} \left( \czeta_{12} \hspace{-2pt} - \hspace{-2pt} \alpha
, \hspace{-1pt}
\tau_R \hspace{-1pt} \right)
}{16 \hspace{-0.5pt} \pi \hspace{-0.5pt} R^2 \hspace{-1pt}
\left| \omega_1 \omega_2 \right| \sin 2\pi\alpha}
\hspace{-1pt}
\eeq
for
\(\omega_k =
\omega \! \left(\raisebox{10pt}{\hspace{-3pt}}\right.
\Txfrac{x_k}{2R}
\left.\raisebox{10pt}{\hspace{-3pt}}\right)\),
\(\czeta_{12} =
\czeta \! \left(\raisebox{10pt}{\hspace{-3pt}}\right.
\Txfrac{x_1}{2R}
\left.\raisebox{10pt}{\hspace{-3pt}}\right)
-
\czeta \! \left(\raisebox{10pt}{\hspace{-3pt}}\right.
\Txfrac{x_2}{2R}
\left.\raisebox{10pt}{\hspace{-3pt}}\right)\),
\(\cos \hspace{1pt} 2\pi \alpha =
u \! \left(\raisebox{10pt}{\hspace{-3pt}}\right.
\Txfrac{x_1}{2R}
\left.\raisebox{10pt}{\hspace{-3pt}}\right)
\spr
u \! \left(\raisebox{10pt}{\hspace{-3pt}}\right.
\Txfrac{x_2}{2R}
\left.\raisebox{10pt}{\hspace{-3pt}}\right)\),
\(\tau_R = \Txfrac{i \beta}{2 \pi R}\).
In order to perform the \(R \to \infty\) limit we derive
the large $R$ behaviour of
$\left| \omega_k \right|$, $\czeta_{12}$ and $\alpha$:
\beqa\label{e7.23}
& \hspace{-7pt}
2 \pi \czeta_{12} \hspace{-1pt} = \hspace{-1pt}
\Txfrac{x_{12}^0 \mgvspc{-4pt}}{\mgvspc{9pt} R}
\hspace{-1pt}
\left(\raisebox{10pt}{\hspace{-3pt}}\right.
1 \hspace{-1pt} + \hspace{-1pt}
O \! \left(\raisebox{10pt}{\hspace{-3pt}}\right.
\Txfrac{\left\| x_1 \right\|^2 \hspace{-2pt} +
\hspace{-1pt} \left\| x_2 \right\|^2 \mgvspc{-4pt}}{\mgvspc{9pt} R^2}
\left.\raisebox{10pt}{\hspace{-3pt}}\right)
\left.\raisebox{10pt}{\hspace{-4pt}}\right)
\hspace{-1pt} , \quad \hspace{-1pt}
2 \pi \alpha \hspace{-1pt} = \hspace{-1pt}
\Txfrac{\left| \mbf{x}_{12} \right| \mgvspc{-4pt}}{\mgvspc{9pt} R}
\left(\raisebox{10pt}{\hspace{-3pt}}\right.
1 \hspace{-1pt} + \hspace{-1pt}
O \! \left(\raisebox{10pt}{\hspace{-3pt}}\right.
\Txfrac{\left\| x_1 \right\|^2 \hspace{-2pt} +
\hspace{-1pt} \left\| x_2 \right\|^2 \mgvspc{-4pt}}{\mgvspc{9pt} R^2}
\left.\raisebox{10pt}{\hspace{-3pt}}\right)
\left.\raisebox{10pt}{\hspace{-4pt}}\right)
\hspace{-1pt} ,
& \hspace{-10pt} \nn & \hspace{-7pt}
4 \left| \omega_k \right|^2 = 1 +
O \! \left(\raisebox{10pt}{\hspace{-2pt}}\right.
\Txfrac{\left\| x_k \right\|^2 \mgvspc{-4pt}}{\mgvspc{9pt} R^2}
\left.\raisebox{10pt}{\hspace{-2pt}}\right)
, & \hspace{-10pt}
\eeqa
(\(\left\| x \right\| := \sqrt{\left( x^0 \right)^2 + \left| \mbf{x} \right|^2}\))
following from
\beqa\label{nnnnn}
& \hspace{-9pt}
\cos 2 \pi \czeta_k \! = \!
\Txfrac{1 \hspace{-1pt} + \hspace{-1pt}
\left(\raisebox{12pt}{\hspace{-3pt}}\right.
\Txfrac{x_k \mgvspc{-3pt}}{\mgvspc{8pt} 2R}
\left.\raisebox{12pt}{\hspace{-3pt}}\right)^2
\mgvspc{-9pt}}{\mgvspc{9pt} 2
\left| \omega_k \right|}
, \ \
\sin 2 \pi \czeta_k \! = \!
\Txfrac{x^0_k \mgvspc{-4pt}}{\mgvspc{9pt} 2 R \left| \omega_k \right|}
, \ \
\mbf{u} \! = \!
\Txfrac{\mbf{x}_k \mgvspc{-4pt}}{\mgvspc{9pt} 2 R \left| \omega_k \right|}
, \ \
u_4 \! = \!
\Txfrac{1 \hspace{-1pt} - \hspace{-1pt}
\left(\raisebox{12pt}{\hspace{-3pt}}\right.
\Txfrac{x_k \mgvspc{-3pt}}{\mgvspc{8pt} 2R}
\left.\raisebox{12pt}{\hspace{-3pt}}\right)^2
\mgvspc{-9pt}}{\mgvspc{9pt} 2 R \left| \omega_k \right|}
, & \nn & \hspace{-9pt}
4 \sin^2 \pi \alpha =
\left( u_1 - u_2 \right)^2 =
\Txfrac{\left| \mbf{x}_{12} \right|^2 \mgvspc{-4pt}}{\mgvspc{9pt} R^2}
\left(\raisebox{10pt}{\hspace{-2pt}}\right.
1 +
O \! \left(\raisebox{10pt}{\hspace{-2pt}}\right.
\Txfrac{\left\| x_1 \right\|^2 \hspace{-2pt} +
\hspace{-1pt} \left\| x_2 \right\|^2 \mgvspc{-4pt}}{\mgvspc{9pt} R^2}
\left.\raisebox{10pt}{\hspace{-2pt}}\right)
\left.\raisebox{10pt}{\hspace{-2pt}}\right)
. & \nonumber
\eeqa
To evaluate the small $\tau_R$ (large $R$) limit of
the difference of $p_1$--functions in~(\ref{e7.22})
we use (\ref{eqnA.19n}), (\ref{p_modinv}) and (\ref{3.9})
to deduce
\beq\label{e7.24}
p_1 \left( \zeta, \tau \right) = \frac{1}{\tau}
\left(\raisebox{10pt}{\hspace{-2pt}}\right.
p_1
\left(\raisebox{10pt}{\hspace{-2pt}}\right.
\frac{\zeta}{\tau}, \frac{-1}{\tau}
\left.\raisebox{10pt}{\hspace{-2pt}}\right)
- 2 \pi i \zeta
\left.\raisebox{10pt}{\hspace{-2pt}}\right)
\, . \
\eeq
Eq.~(\ref{e7.24}) implies, on the other hand, that
\beq\label{e7.25}
p_1 \!
\left(\raisebox{10pt}{\hspace{-2pt}}\right.
\frac{\czeta_{12} \hspace{-1pt} \pm \hspace{-1pt} \alpha}{\tau_R},
\frac{-1}{\tau_R}
\left.\raisebox{10pt}{\hspace{-2pt}}\right)
\mathop{\approx}\limits_{R \to \infty}
p_1 \!
\left(\raisebox{10pt}{\hspace{-2pt}}\right.
\frac{x_{12}^0 \hspace{-1pt} \pm \hspace{-1pt}
\left| x_{12} \right|}{i \beta},
\frac{i 2 \pi R}{\beta}
\left.\raisebox{10pt}{\hspace{-2pt}}\right)
\mathop{\longrightarrow}\limits_{R \to \infty}
\pi i \,
\coth \!
\left(\raisebox{10pt}{\hspace{-3pt}}\right.
\pi \,
\frac{x_{12}^0 \hspace{-1pt} \pm \hspace{-1pt}
\left| x_{12} \right|}{\beta}
\left.\raisebox{10pt}{\hspace{-2pt}}\right)
.
\eeq
Inserting~(\ref{e7.23})--(\ref{e7.25}) into (\ref{e7.22})
we complete the proof of~(\ref{e7.20n}) and hence of Proposition~\ref{pr:7.3}.

\begin{mremark}
\rmlabel{rm:7.4}
The physical thermal correlation functions should be, in fact,
defined as distributions which amounts to giving
integration rules around the poles.
To do this one should view~(\ref{e7.20n}) as a boundary
value of an analytic function in $x_{12}$ for
\(x_{12}^0 \to x_{12}^0 - i \varepsilon\), \(\varepsilon > 0\),
\(\varepsilon \to 0\) (cf.~(\ref{ea5.17})).
It is not difficult to demonstrate that the limit \(\varepsilon \to +0\)
and \(R \to \infty\) in (\ref{e7.19n}) commute.
Using (\ref{nnnnn}) we can also compute the $\Txfrac{1}{R\beta}$
correction to~(\ref{e7.20n}):
\beq\label{e7.27}
\La \hspace{-1pt}
\varphi^{\MINK} \! \left( x_1 \right) \varphi^{\MINK} \! \left( x_2 \right)
\hspace{-1pt} \Ra_{\hspace{-1pt} \beta, R}
\ \, \mathop{\approx}\limits_{R \mgrt \beta} \ \,
\La \hspace{-1pt}
\varphi^{\MINK} \! \left( x_1 \right) \varphi^{\MINK} \! \left( x_2 \right)
\hspace{-1pt} \Ra_{\hspace{-1pt} \beta, \infty}
- \,
\frac{1}{4 \pi^2 \beta R}
.
\eeq
\end{mremark}

To obtain the Fourier expansion of the result
we combine Eqs.~(\ref{e7.22}) (\ref{e7.23}) with the
$q$--series (\ref{p_1-rep2}) and set
(as in Remark~\ref{rm:7.2})
\beq\label{e7.28}
\frac{n}{R} \, \to \, p
\, , \quad
\frac{1}{R} \, \to \, dp
\, , \quad
\mathop{\sum}\limits_{n \, = \, 1}^{\infty} \,
\frac{1}{R} \, f \!
\left(\raisebox{10pt}{\hspace{-2pt}}\right.
\frac{n}{R} ;\, x, \beta
\left.\raisebox{10pt}{\hspace{-2pt}}\right)
\ \mathop{\longrightarrow}\limits_{R \to \infty} \
\mathop{\int}\limits_{\hspace{-7pt} 0}^{\hspace{7pt} \infty} \!
f \! \left( p; x, \beta \right) \, dp
. \
\eeq
The result is
\beqa\label{e7.29}
&
\left( 2\pi \right)^2
\La \hspace{-1pt}
\varphi^{\MINK} \! \left( x_1 \right) \varphi^{\MINK} \! \left( x_2 \right)
\hspace{-1pt} \Ra_{\hspace{-1pt} \beta, \infty}
\, = \,
\Txfrac{1 \mgvspc{-3pt}}{\mgvspc{9pt}
x_{12}^{\, 2} + i \hspace{1pt} 0 x_{12}^0} \, + \,
& \nn & \, + \,
\Txfrac{2 \mgvspc{-3pt}}{\mgvspc{10pt} \left| \mbf{x}_{12} \right|} \,
\mathop{\text{\LARGE $\int$}}\limits_{\hspace{-7pt} 0}^{\hspace{7pt} \infty} \!
\Txfrac{e^{-\beta p} \mgvspc{-3pt}}{\mgvspc{10pt} 1 - e^{-\beta p}} \,
\cos \left( p x_{12}^0 \right)
\sin \left( p \left| \mbf{x}_{12} \right| \right)
\, dp
\, . \
&
\eeqa

\medskip

To conclude: the conformal compactification $\M$ of Minkowski space $M$
can play a dual role.

On one hand, it can serve as a \textit{symmetric} finite box approximation
to $M$ in the study of finite temperature equilibrium states.
In fact, any finite inverse temperature $\beta$ actually fixes a Lorentz
frame (cf.~\cite{Bu03}) so that the symmetry of a Gibbs state is described
by the $7$--parameter ``Aristotelian group'' of ($3$--dimensional) Euclidean
motions and time translations.
In the passage from $M$ to $\M$ the Euclidean group is deformed to the
(stable) compact group of $4$--dimensional rotations while the group
of time translations is compactified to $U \left( 1 \right)$.
Working throughout with the maximal ($7$--parameter) symmetry allows to
write down simple explicit formulae for both finite $R$ and the
``thermodynamic limit''.

On the other hand, taking $\M$ (and its universal cover,
\(\widetilde{M} = \R \times \Sr^{3}\)) not as an auxiliary
finite volume approximation
but as a model of a static space--time, we can view $R$
as a (large but) finite 	length and use the above discussion as
a basis for studding finite $R$ corrections to the
Minkowski space formulae.
It is a challenge from this second point of view to study the conformal
symmetry breaking by considering massive fields in $\widetilde{M}$.

\section*{Guide to references}\addcontentsline{toc}{section}{Guide to references}

Among the books on elliptic
functions and modular forms we have mostly referred in these notes to the
conference proceedings~\cite{FNTP} (see, in particular, the beginning of Sect.~2),
to the readers-friendly text~\cite{McKM} (which gives a flavour of the work
of the founding fathers on the subject) and, to a lesser extent, to Weil's
book~\cite{Weil} which provides both a valuable contribution to the history and
an elegant exposition of the theory of Eisenstein's series.

The reader who enjoys learning about history of mathematics for its own
sake is probably aware without our recommendation of the entertaining
essays of Bell~\cite{Bell} but may find also interesting the emotionally told
story of 19th century mathematics by one of its participants~\cite{KW41}.
We have also referred, in passing, to books on the history of topics that are
(to a varying degree) periferal to our subject (\cite{HHW67}\cite{Weil84}).

The books by Serge Lang \cite{L87} and \cite{L76} provide a systematic background on
the mathematical part of these lectures on a more advanced level.
The electronically available lecture notes by Milne~\cite{Mil97} are recommended
for a ready to use treatment of the Riemann-Roch theorem
(applied in Sect.~3.1 to the classification of modular forms).
The rather engaging exposition in \cite{PS97} is mostly directed
towards the solution of algebraic equations but also contains an elementary
introduction to the arithmetic theory of elliptic curves (Chapter 5)
including an idea about Wiles' proof of Fermat last theorem. More
systematic on number theoretic applications are the earlier
texts \cite{Se}, \cite{Sh94}.
Mumford's book \cite{M83} on
theta functions is a classic.

A rigorous treatment of the applications of elliptic functions and
modular forms to 2-dimensional conformal field theory has been given in
\cite{Zh96}. The mathematical theory of chiral vertex algebras was anticipated
in work by Frenkel and Kac \cite{FK80} and developed by Borcherds \cite{Bo86}
\cite{Bo97}. Nowadays, it is the subject of several books: \cite{FLM88} \cite{Ka96}
\cite{FBZ01}.

\bigskip

\section*{Acknowledgments}\addcontentsline{toc}{section}{Acknowledgments}
\noindent
\textbf{Acknowledgments.}
{\small
The present notes are gradually growing
into an extended version of lectures presented by one of the authors at the
University of Sofia in 2002--2004, at the Dubna International Advanced
Summer School in Modern Mathematical Physics (July 11-22, 2003), at the
International School for Advanced Studies (SISSA-ISAS), Trieste,
December 2003,
at the III~Summer School in Modern Mathematical Physics,
Zlatibor, Serbia (August 20--31, 2004), and at
the Institut f\"ur Theoretische Physik, Universit\"at G\"ottingen
in the fall of 2004.

The authors thank Seif Randjbar-Daemi and the Abdus Salam International
Centre for Theoretical Physics for invitation and support during an early stage
of their work on these notes.
We thank Alexander Filippov, Boris Dubrovin, and Branko Dragovich
for inviting us
to present these lactures at the Bogolubov Laboratory of Theoretical Physics
(JINR, Dubna), at the International School for Advanced Studies (Trieste), and at
the Zlatibor Summer School (Serbia), respectively.
We would like to thank Herbert Gangl for a critical reading of
a draft of these notes which stimulated us to undertake their
revision.
Discussions with Petko Nikolov in Sofia are also gratefully acknowledged.
Discussions with Detlev Buchholz in G\"ottingen led to including
the addition of the present
Sect.~7. 
I.T. thanks the Alexander von Humboldt Foundation for support
and the Institut f\"ur Theoretische Physik der Universit\"at G\"ottingen for
its hospitality.
We are grateful to Bogdan G. Dimitrov for preparing a first version of the notes.
This work is supported in part
by the Research Training Network within the Framework Programme 5
of the European Commission under contract HPRN-CT-2002-00325
and by the Bulgarian National Council for Scientific
Research under contract~PH-1406.}

\appendix

\apsection{Elliptic functions in terms of Eisenstein series}{ap:1}

The family of functions
\(\pfun_k^{\kappa\lambda} \left( \zeta,\, \tau \right)\),
\(k=1,2,\dots\), \(\kappa,\lambda = 0,1\),
encountered throughout these lectures, is
uniquely determined by the following set of properties.
\begin{plist}
\item[(\textit{i})\hspace{8pt}]
\(\pfun_k^{\kappa\lambda} \left( \zeta, \tau \right)\) are meromorphic functions
in \(\left( \zeta, \tau \right) \in \C \times \hcom\)
which have exactly one pole at \(\zeta = 0\)
of order $k$ and residue $1$
in the domain
\(\left\{\raisebox{9pt}{\hspace{-3pt}}\right.
\alpha\tau+\beta :\) \(\alpha, \beta \in\)
\(\left[ 0,\, 1 \right)
\left.\raisebox{9pt}{\hspace{-3pt}}\right\} \subset \C\)
for all \(\tau \in \hcom\) and \(k = 1,2,\dots\);\gvspc{10pt}
\item[(\textit{ii})\hspace{5pt}]
\(\pfun_{k+1}^{\kappa\lambda} \left( \zeta, \tau \right) =
- \frac{\textstyle 1}{\textstyle k\mgvspc{8pt}} \,
\frac{\textstyle \di}{\textstyle \di \zeta\mgvspc{8pt}} \,
\pfun_k^{\kappa\lambda} \left( \zeta, \tau \right)
\) \ for \ \(k = 1,2,\dots\);
\item[(\textit{iii})\hspace{2pt}]
\(\pfun_k^{\kappa\lambda} \left( \zeta+1, \tau \right) = \left( -1 \right)^{\lambda}
\pfun_k^{\kappa\lambda} \left( \zeta, \tau \right)\) \ for \ \(k = 1,2,\dots\);
\item[(\textit{iv})\hspace{3pt}]
\(\pfun_k^{\kappa\lambda} \left( \zeta+\tau, \tau \right) =
\left( -1 \right)^{\kappa} \pfun_k^{\kappa\lambda} \left( \zeta, \tau \right)\)
\ for \ \(k+\kappa+\lambda > 1\);\gvspc{10pt}
\item[(\textit{v})\hspace{6pt}]
\(\pfun_k^{\kappa\lambda} \left( -\zeta, \tau \right)
= \left( -1 \right)^k \, \pfun_k^{\kappa\lambda} \left( \zeta, \tau \right)\)
\ for \ \(k = 1,2,\dots\).\gvspc{10pt}
\end{plist}
One cannot require that the function $\pfun_1({\zeta},{\tau})(=\pfun_1^{00})$ is
doubly periodic (as it has a simple pole in the fundamental parallelogram)
and we have chosen it to have a single period, $1$, in ${\zeta}$.

For $k>2$ the above functions are readily determined: they can be written as
absolutely convergent Eisenstein(-Weierstrass) series,
\beq\label{A.1}
\pfun_k^{\kappa\lambda} \left( \zeta,\tau \right)
\, = \,
\mathop{\sum}\limits_{m,\, n \, \in \, \Z}
\,
\frac{\left( -1 \right)^{\kappa m+ \lambda n}}{
\left( \zeta+m\hspace{1pt}\tau+n \right)^k}
\quad (k > 2)
\, , \
\eeq
for ${\zeta}$ outside the lattice $\Z{\tau}+\Z$.
Condition (\textit{ii}) then determines each of the four functions
$\pfun_1^{\kappa\lambda}$ up to an additive linear in $\zeta$ term.
Conditons (\textit{iv}) and (\textit{v}) guarantee their uniqueness
(provided that they exist). The existence is established by the explicit
construction (\ref{4.20}) which can be rewritten in the form
\beq\label{A.3}
\pfun_1^{\kappa\lambda} \hspace{-2pt} \left( \zeta,\tau,\mu \right)
= \hspace{-1pt}
\mathop{\lim}\limits_{M \to \infty}
\mathop{\sum}\limits_{n \hspace{1pt} = \hspace{1pt} -M}^{M}
\hspace{-1pt}
\frac{\pi
\cos^{\hspace{1pt}1-\lambda} \hspace{-2pt}
\left[\raisebox{9pt}{\hspace{-2pt}}\right. \pi \hspace{-1pt}
\left(\raisebox{9pt}{\hspace{-3pt}}\right.
\zeta \hspace{-2pt} + \hspace{-1pt} n \hspace{1pt} \tau
\left.\raisebox{9pt}{\hspace{-2pt}}\right)
\left.\raisebox{9pt}{\hspace{-3pt}}\right]
}{
\sin \left[\raisebox{9pt}{\hspace{-2pt}}\right. \pi \hspace{-1pt}
\left(\raisebox{9pt}{\hspace{-3pt}}\right.
\zeta \hspace{-2pt} + \hspace{-1pt} n \hspace{1pt} \tau
\left.\raisebox{9pt}{\hspace{-2pt}}\right)
\left.\raisebox{9pt}{\hspace{-3pt}}\right]
} \
e^{\hspace{1pt} \pi i \hspace{1pt} n \hspace{0.5pt}
\left( 2 \mu + \kappa \right)}
\! .
\eeq
The resulting expressions for $\pfun_1$ and $\pfun_2$ are related to the corresponding
Weierstrass functions by (\ref{eqnA.19n}) (\ref{eqnA.21a}).

The above $p$-functions admit an extension, needed when dealing with a
chemical potential, in which the character $(-1)^{\kappa}$ in condition
(\textit{iv}) is replaced by a more general one:
\beq\label{A.4}
\pfun_k^{\kappa\lambda} \left( \zeta+\tau,\tau,\mu \right) \, = \,
\left( -1 \right)^k \, e^{-2\pi i \hspace{1pt} \mu}
\, \pfun_k^{\kappa\lambda} \left( \zeta,\tau,\mu \right)
\, . \
\eeq
The resulting $\pfun_1$-functions have a manifestly meromorphic representation
in terms of ratios of Jacobi ${\vartheta}$ functions:
\beq\label{A.5}
\pfun_1^{\kappa\lambda}\hspace{-1pt} \left( \zeta,\,\tau,\,\mu \right)
=
\frac{\left( \partial_{\zeta} \, \vartheta_{11} \right) \! \left( 0, \tau \right)}{
\vartheta_{1-\lambda \hspace{1pt} 1-\kappa} \hspace{-2pt} \left( \mu, \tau \right)}
\,
\frac{\vartheta_{1-\lambda \hspace{1pt} 1-\kappa} \hspace{-2pt}
\left( \zeta\! +\! \mu, \tau \right)}{
\vartheta_{11} \hspace{-2pt} \left( \zeta, \tau \right)}
\hspace{1pt} - \hspace{1pt}
\left( 1\! -\! \lambda \right) \pi \hspace{2pt}
\mathrm{cotg} \, \pi \hspace{-2pt}
\left(\raisebox{10pt}{\hspace{-3pt}}\right.
\mu \! + \frac{\kappa}{2}
\left.\raisebox{10pt}{\hspace{-2pt}}\right)
\eeq
(see Proposition A.1. of \cite{NT03}).

\apsection{The action of the conformal Lie algebra on different realizations of (compactified) Min\-kow\-ski space}{ap:2}

In this section we will sketch of the proof of the relations~(\ref{e2.3n}),
(\ref{H}) and (\ref{n2.10n}) between the three (complex) bases
of the conformal Lie algebra $\confalg$ used in Sect.~4:
the basis $\{X_{ab}\}$ in the projective description,
the familiar generators in the Minkowski ($x$-space) chart, and the
$T_{\cmu}$, $C_{\cmu}$ and $H$ generators of the $z$-picture.
We begin with an important observation.

\begin{mproposition}
The correspondence between the conformal Lie algebra
generators \(X \in \confalg\) and first order linear differential operators
$\OP{X}$, given by
\beq\label{neB.1}
\left[ X, \phi (u) \right] \, = \, \OP{X} \phi (u)
\, , \quad u = x,\, z
\eeq
($\phi (u)$ $=$ $\{\phi_{\aa} (u)\}$)
is a Lie algebra {\bf anti}homomorphism, i.e. for $X$, $Y$ $\in$ $\confalg$:
\beq\label{neB.2}
\left[ -\OP{X},-\OP{Y} \right] \, = \, -\OP{\left[ X,Y \right]}
\, . \
\eeq
As a corollary, the correspondence \(X \mapsto -\OP{X}\)
is a Lie algebra homomorphism.
\end{mproposition}

{\samepage
\begin{proof}
Using the Jacobi identity for the double commutator we find
\beqa\label{neB.3}
\left[ \left[ X,Y \right],\phi (u) \right] \hspace{-1pt} = \hspace{-1pt} && \!\podr
\left[ X, \left[ Y, \phi (u) \right] \right] \hspace{-1pt} - \hspace{-1pt}
\left[ Y, \left[ X, \phi (u) \right] \right] \hspace{-1pt} = \hspace{-1pt}
\left[ X, \OP{Y} \! \left( \phi (u) \right) \right]
\hspace{-1pt} - \hspace{-1pt}
\left[ Y, \OP{X} \! \left( \phi (u) \right) \right] \hspace{-1pt} = \hspace{-1pt}
\nn \hspace{-1pt} = \hspace{-1pt} && \! \podr
\OP{Y} \! \left( \left[ X, \phi (u) \right] \right)
\hspace{-1pt} - \hspace{-1pt}
\OP{X} \! \left( \left[ Y, \phi (u) \right] \right) \hspace{-1pt} = \hspace{-1pt}
- \left[ \OP{X},\OP{Y} \right] \phi (u)
.
\eeqa
\end{proof}}

Note that the derivatives' parts of the above operators $\OP{X,Y}$ are some
vector fields $\DOP{X,Y}$ and the correspondence $X$ $\mapsto$ $\DOP{X}$
is again antihomomorphism.

In order to derive Eq.~(\ref{e2.3n})
let us define the (physical) generators of translations $iP_{\mu}$,
special conformal transformations $iK_\mu$ and dilations $i\dlt$ by
\beqa\label{neB.4}
& \hspace{-11pt}
\left[ iP_{\mu}, \phi (x) \right] \! = \!
\tdd{x^{\mu}} \, \phi (x)
, \quad \!
\left[ iK_{\mu}, \phi (x) \right] \! = \!
\left(\raisebox{12pt}{\hspace{-4pt}}\right.
x^{\, 2} \tdd{x^{\mu}} - x_{\mu} x^{\nu} \tdd{x^{\nu}}
\left.\raisebox{12pt}{\hspace{-4pt}}\right)
\phi (x) + M (x) \phi (x)
, \hspace{-10pt} &
\nn &
\left[ i\dlt, \phi_{\aa} \right] \, = \,
x^{\nu} \tdd{x^{\nu}} \phi (x) + d_{\phi} \phi (x)
&
\eeqa
where $M(x)$ is some $x$-dependent (matrix) function
(whose explicit form is not essential for the present calculations)
and $d_{\phi}$ is the field dimension.
We note that $P_0$ is the physical (hermitian) energy operator
($e^{itP_0}$ generating the unitary time evolution) and is, hence,
positive definite in the state space $\VA$.
Using the Klein--Dirac compactification formulae~(\ref{nw4.15})
mapping the Minkowski space $M$ into the quadric $Q$~(\ref{nn4.15})
we can find a representation on $Q$ of the vector fields corresponding to
$iP_{\mu}$, $iK_{\mu}$ and $i\dlt$:
$-\tdd{x^{\mu}}$, $-x^{\, 2}\tdd{x^{\mu}}$ $+$ $2x_{\mu}x^{\nu}\tdd{x^{\nu}}$
and $-x^{\nu}\,\tdd{x^{\nu}}$, respectively (accordingly to Proposition~B.1).
In order to express the generators $iP_{\mu}$, $iK_{\mu}$ and $i\dlt$
in terms of $X_{ab}$
we should, in addition,
factorize with respect to the Euler field $\xi^{a} \tdd{\xi^{a}}$
(noting that the scalar functions on $\M$ (\ref{nn4.15}) are lifted
to homogeneous functions of degree zero on $Q$ on which the Euler
field vanishes).

Similarly, in order to prove that~(\ref{H}) and (\ref{n2.10n})
agree with the relations (\ref{equ2.7})--(\ref{equ2.10})
we use the imbedding analogous to~(\ref{nw4.15})
\beqa\label{nw4.15xyz}
&
z \mapsto
\left\{\raisebox{10pt}{\hspace{-3pt}}\right. \lambda \xx_{z}
\left.\raisebox{10pt}{\hspace{-3pt}}\right\} \in \M_{\C}
, \quad
\xx_z \, = \,
z^{\cmu} \ee_{\cmu} + \Txfrac{1+z^{\, 2}}{2} \, \ee_{-1} +
i\, \Txfrac{1-z^{\, 2}}{2} \, \ee_{0}
\quad \text{or,} & \nn &
z^{\cmu} \, = \, \Txfrac{\mxi^{\cmu}}{\mxi^{-1} - i \mxi^{0}}
\, . &
\eeqa
(Note that~(\ref{nw4.15}) and~(\ref{nw4.15xyz}) reproduce~(\ref{nw4.19}).)

\apsection{Clifford algebra realization of $\spin (D,2)$ and the centre of $\Spin (D,2)$}{ap:2n}

Let $\beta_{\mu}$, \(\mu = 0,1,\dots,2r-1\) be $2^r \times 2^r$ complex matrices
generating the Clifford algebra $\Cliff (2r-1,1)$; more precisely, we
assume the relations
\beq\label{nB.1}
\left[ \beta_{\mu},\beta_{\nu} \right]_+ \, = \,
2 \eta_{\mu\nu}
\quad
(\eta_{\mu\nu} \, = \, \diag (-1,1,\dots,1))
, \quad
\beta_{\mu}^* \, = \, \beta^{\mu} \, = \, \eta^{\mu\nu} \, \beta_{\nu}
\
\eeq
(where we skip~--~here and in what follows~--~the corresponding unit matrix).
These Clifford units give rise to two collections of $2r+2$ matrices
$\{\beta_{a}\}$ and $\{\vbeta_{a}\}$, \(a=-1,0,\dots,2r\),
by setting \(\vbeta_{\mu} = \beta_{\mu}\), for \(\mu =0,1,\dots,2r-1\),
\beq\label{nB.2}
\beta_{2r} \, = \,
i^{r-1} \, \beta_0 \, \beta_1 \dots \beta_{2r-1} \, = \, \vbeta_{2r}
, \quad
\beta_{-1} \, = \, \ID \, = \, - \vbeta_{-1}
. \
\eeq
The resulting pair of matrix valued vectors are characterized by the
bilinear relations
\beq\label{nB.3}
\beta_{a} \, \vbeta_{b} + \beta_{b} \, \vbeta_{a} \, = \,
2 \, \eta_{ab} \, = \,
\vbeta_{a} \, \beta_{b} + \vbeta_{b} \, \beta_{a}
. \
\eeq
The two $2^r$--dimensional spinorial representation $S_{\pm}$
of $\spin (2r,2)$ are then generated by
\beq\label{nB.4}
S_+ (X_{ab}) \, \equiv \, \gamma_{ab} \, := \,
\frac{1}{4} \, (\beta_{b} \, \vbeta_a - \beta_a \, \vbeta_b)
, \quad
S_- (X_{ab}) \, \equiv \, \vgamma_{ab} \, := \,
\frac{1}{4} \, (\vbeta_{b} \, \beta_a - \vbeta_a \, \beta_b)
. \
\eeq
The matrices $\beta_{a}$ and $\vbeta_{a}$ can be also used to
construct the single spinorial representation of the
conformal group $\Spin (2r+1,2)$ corresponding to an odd
dimensional space---time.
To this end we introduce the $\Cliff (2r+1,2)$ algebra generators
\beqa\label{nB.5}
&
\Gamma_{a} \, = \, \left(\hspace{-4pt} \begin{array}{cc} 0 \hspace{-4pt} & \beta_{a} \\
\vbeta_{a} \hspace{-4pt} & 0 \end{array} \hspace{-3pt} \right)
, \quad
a = -1,0,1,\dots,2r
, \quad
\Gamma_{2r+1} \, = \, \left(\hspace{-4pt} \begin{array}{cr} \ID \hspace{-4pt} & 0 \\
0 \hspace{-4pt} & -\ID \end{array} \hspace{-3pt} \right)
, & \nn &
\left[ \Gamma_{A},\Gamma_{B} \right]_+ \, = \, 2 \eta_{AB}
, \quad
A,B = -1,0,\dots,2r+1
\quad (\eta_{2r+1, 2r+1} = 1);
&
\eeqa
then the $2^{r+1}\times 2^{r+1}$--dimensional spinorial representation of the
$\spin (2r+1,2)$ generators assumes the form
\beq\label{nB.6}
S (X_{AB}) \, = \, \Gamma_{AB} \, = \,
\frac{1}{4} \, \left[ \Gamma_B,\Gamma_A \right]
. \
\eeq

For \(a \neq b\) the generators
\(\gamma_{ab} = \txfrac{1}{2} \, \beta_b \, \vbeta_a\)~(\ref{nB.4})
satisfy
\beq\label{nB.7}
(2\gamma_{ab})^2 \, = \, (\beta_b \, \vbeta_a)^2 \, = \,
- \eta_{aa} \, \eta_{bb}
\quad
(\text{in general}, \ (2\gamma_{ab})^2 = \eta_{ab}^2 - \eta_{aa} \, \eta_{bb}).
\eeq
It follows that the \textit{valence element} $v$ of the centre $Z$
of $\Spin (2r,2)$ is the common central element of $U (1)$ and $\Spin (2r)$:
\beq\label{nB.8}
v \, := \, e^{2\pi \gamma_{-10}} \, = \, \cos \pi + 2\gamma_{-10} \sin \pi \, = \,
- \ID \, = \, e^{2\pi \gamma_{\cmu\cnu}}
.
\eeq
Note that $v$ is mapped on the group unit by the corresponding (two-to-one)
map \(\Spin (2r,2) \to \SO (2r,2)\).
On the other hand, for even $r$ the centre of $\Spin (2r,2)$ is $\Z_4$
and it is generated by the product
\beq\label{nB.9}
c_1 \, = \, e^{\pi \gamma_{-10}} \, e^{\pi \gamma_{12}} \dots
e^{\pi \gamma_{2r-1,2r}} \, = \,
\beta_0 \, \beta_2 \, \beta_1 \dots \beta_{2r} \, \beta_{2r-1} \, = \, i^{r-3}
. \
\eeq
(In the last equation we have used the definition~(\ref{nB.2}) of $\beta_{2r}$.)
Clearly, in parallel with Bott periodicity property we have
\beq\label{nB.10}
c_1^2 \, = \, v \quad \text{for $r$ even}, \quad
c_1^2 \, = \, \ID \quad \text{for \(r = 1\) $\MOD$ $4$}, \quad
c_1 \, = \, \ID \quad \text{for \(r = 3\) $\MOD$ $4$}.
\eeq

\begin{mremark}
The term ``valence'' for the element $v$ of $Z (G)$ originates from
the fact that it coincides with the unit operator for single valued
representations of the pseudoorthogonal group $\SO_0 (2r,2)$ and
corresponds to multiplication by $-\ID$ on its double valued representations
(i.e. for exact representations of $\Spin (2r,2)$).
\end{mremark}

We note that the centre of $\Spin (2r+1,2)$ is $\Z_2$ with non--trivial
element $v$~(\ref{nB.8}).
The centre of $\Spin (2r,2)$ for odd $r$ is $\Z_2 \times \Z_2$ (see Appendix~A to~\cite{KT97}).

We end up with some additional remarks about the case \(\confgr = \SU (2,2)\)
of chief interest.

Here is a realization of the matrices $\beta_a$ and $\gamma_{ab}$ for \(D=4\)
in terms of the quaternion units of Sect.~7:
\beq\label{nB.11}
\beta_0 = \left(\hspace{-4pt} \begin{array}{rr} -i \ID \hspace{-4pt} & 0 \\
0 \hspace{-4pt} & i \ID \end{array} \hspace{-3pt} \right) = -\beta^0
, \quad
\beta_j = \left(\hspace{-4pt} \begin{array}{cc} 0 \hspace{-4pt} & Q_j \\
Q_j^+ \hspace{-4pt} & 0 \end{array} \hspace{-3pt} \right)
, \quad
\beta_4 = i \beta_0 \beta_1 \beta_2 \beta_3 =
\left(\hspace{-4pt} \begin{array}{rr} 0 \hspace{-4pt} & -\ID \\
-\ID \hspace{-4pt} & 0 \end{array} \hspace{-3pt} \right)
\eeq
(\(Q_j Q_k = \epsilon_{jkl} Q_l - \delta_{jk}\), \(Q_j^+ = -Q_j\),
\(j,k,l =1,2,3\));
\beq\label{nB.12}
\gamma_{31} \, = \, \frac{1}{2}
\left(\hspace{-4pt} \begin{array}{cc} Q_2 \hspace{-4pt} & 0 \\
0 \hspace{-4pt} & Q_2 \end{array} \hspace{-3pt} \right)
, \quad
\gamma_{j4} \, = \, \frac{1}{2}
\left(\hspace{-4pt} \begin{array}{cc} Q_j \hspace{-4pt} & 0 \\
0 \hspace{-4pt} & -Q_j \end{array} \hspace{-3pt} \right)
, \quad
\gamma_{-1\cmu} \, = \, \frac{1}{2} \beta_{\cmu}
\eeq
(\(j=1,2,3,\) \(\cmu=1,2,3,4\)).

The positive energy unitary irreducible representations $U (g)$
of $\SU (2,2)$~\cite{Ma77} are labeled by triples $(d;j_1,j_2)$
of non-negative half integers ($j_1$ and $j_2$ being the spins giving
the IR of the semisimple part \(\Spin (4) = \SU (2) \times \SU (2)\)
of the maximal compact subgroup $\compgr$).
The lowest energy subspace $\mathcal{V}_d (j_1,j_2)$ has dimension
$(2j_1+1)(2j_2+1)$.
The triples are restricted by the conditions:
\beq\label{nB.13}
d+j_1+j_2 \in \N
; \quad
d \geqslant 2 + j_1 + j_2 \ \ \text{for} \ \ j_1 j_2 \neq 0
; \quad
d \geqslant 1 + j_1 + j_2 \ \ \text{if} \ \ j_1j_2 = 0.
\eeq
The fields which give rise to representations with
\(d+j_1+j_2\) (\(j_1j_2 = 0\)) satisfy free field equations.
The symmetric tensor fields, corresponding to
\(j_1=j_2=\txfrac{\ell}{2}\), \(d = \ell + 2\), \(\ell = 1,2,\dots\),
are conserved.
It follows from the first relation~(\ref{nB.13}) that the representation
$U_{d;j_1,j_2} (v)$ of the valence element~(\ref{nB.8}) of $\SU (2,2)$
is given by
\beq\label{nB.14}
U_{d;j_1,j_2} (v) \, = \, (-1)^{2j_1+2j_2} \, = \, (-1)^{2d} . \
\eeq

\newpage

\addcontentsline{toc}{section}{References}


\end{document}